\documentclass[onehalfspace,a4paper,11pt]{thesis}
\usepackage{amsmath,amsthm,amssymb,euscript,eufrak,mathrsfs,epsf,epsfig,axodraw}
\usepackage{array,verbatim}
\usepackage[sort&compress]{natbib}
\makeatletter
\makeatletter
\@addtoreset{equation}{section}
\makeatother
\renewcommand{\theequation}{\thesection.\arabic{equation}}

\newcounter{multieqs}

\newcommand{\be}{\begin{equation}}
\newcommand{\ee}{\end{equation}}
\newcommand{\eq}[1]{(\ref{#1})}

\newcommand{\bm}[1]{\mbox{\boldmath $#1$}}

\newcommand{\kslash}{k \!\!\! / }
\newcommand{\qslash}{q \!\!\! / }
\newcommand{\lslash}{l \!\! / }
\newcommand{\Pslash}{P \!\!\!\! / }
\newcommand{\Qslash}{Q \!\!\!\! / }
\newcommand{\islash}{i \!\!\! / }
\newcommand{\jslash}{j \!\!\! / }
\newcommand{\aslash}{a \!\!\! / }
\newcommand{\bslash}{{b \hspace{-6pt} \slash} }
\newcommand{\partialslash}{\partial \!\!\! / }

\newcommand{\onslash}{1 \!\!\! / }
\newcommand{\twslash}{2 \!\!\!/ }
\newcommand{\thslash}{3 \!\!\!/ }
\newcommand{\foslash}{4 \!\!\! / }
\newcommand{\fislash}{5 \!\!\! / }

\newcommand{\mslash}{m \!\!\! / }

\def\bd{\begin{document}}
\def\ed{\end{document}}
\def\nn{\nonumber}
\def\bea{\begin{eqnarray}}
\def\eea{\end{eqnarray}}
\let\bm=\bibitem
\let\la=\label

\def\ab{(ijab)}
\def\ba{(ijba)}
\def\ijab{{\tr}_{+}(\islash\, \jslash\, \aslash \, \bslash)}
\def\ijba{{\tr}_{+}(\islash\, \jslash\, \bslash \, \aslash)}
\def\ijaP{{\tr}_{+}(\islash\, \jslash\, \aslash \, \Pslash)}
\def\ijPLa{{\tr}_{+}(\islash\, \jslash\, \Pslash_L \, \aslash)}
\def\ijaPL{{\tr}_{+}(\islash\, \jslash\, \aslash \, \Pslash_L)}
\def\ijPLza{{\tr}_{+}(\islash\, \jslash\, \Pslash_{L;z} \, \aslash)}
\def\ijaPLz{{\tr}_{+}(\islash\, \jslash\, \aslash \, \Pslash_{L;z})}
\def\ijPa{{\tr}_{+}(\islash\, \jslash\, \Pslash \, \aslash)}
\def\iaPb{{\tr}_{+}(\islash\, \aslash\, \Pslash \, \bslash)}
\def\ibPa{{\tr}_{+}(\islash\, \bslash\, \Pslash \, \aslash)}
\def\ijPmu{{\tr}_{+}(\islash\, \jslash\, \Pslash \, \mu)}
\def\ibmuP{{\tr}_{+}(\islash\, \bslash\, \mu \, \Pslash)}
\def\ibmua{{\tr}_{+}(\islash\, \bslash\, \mu \, \aslash)}
\def\iamub{{\tr}_{+}(\islash\, \aslash\, \mu \, \bslash)}
\def\jaPb{{\tr}_{+}(\jslash\, \aslash\, \Pslash \, \bslash)}
\def\ijmuP{{\tr}_{+}(\islash\, \jslash\, \mu \, \Pslash)}
\def\ijmum{{\tr}_{+}(\islash\, \jslash\, \mu \, \mslash)}
\def\ijmmu{{\tr}_{+}(\islash\, \jslash\, \mslash \, \mu)}
\def\ijmP{{\tr}_{+}(\islash\, \jslash\, \mslash \, \Pslash)}
\def\iabP{{\tr}_{+}(\islash\, \aslash\, \bslash \, \Pslash)}
\def\ijbP{{\tr}_{+}(\islash\, \jslash\, \bslash \, \Pslash)}
\def\jbPa{{\tr}_{+}(\jslash\, \bslash\, \Pslash \, \aslash)}
\def\ijPb{{\tr}_{+}(\islash\, \jslash\, \Pslash \, \bslash)}
\def\jbmua{{\tr}_{+}(\jslash\, \bslash\, \mu \, \aslash)}

\def\loablt{ {\tr}_{+}(\lslash_1\, \aslash \, \bslash\, \lslash_2)}

\def\ijlolt{{\tr}_{+}(\islash\, \jslash\, \lslash_1 \, \lslash_2)}
\def\ijltlo{{\tr}_{+}(\islash\, \jslash\, \lslash_2 \, \lslash_1)}
\def\ibloa{{\tr}_{+}(\islash\, \bslash\, \lslash_1 \, \aslash)}
\def\jaltb{{\tr}_{+}(\jslash\, \aslash\, \lslash_2 \, \bslash)}
\def\ialtb{{\tr}_{+}(\islash\, \aslash\, \lslash_2 \, \bslash)}
\def\bltloa{{\tr}_{+}(\bslash\, \lslash_2\, \lslash_1 \, \aslash)}
\def\jbloa{{\tr}_{+}(\jslash\, \bslash\, \lslash_1 \, \aslash)}
\def\ibPb{{\tr}_{+}(\islash\, \bslash\, \Pslash \, \bslash)}
\def\ijltb{{\tr}_{+}(\islash\, \jslash\, \lslash_2 \, \bslash)}

\def\ijloa{{\tr}_{+}(\islash\, \jslash\,  \lslash_1 \, \aslash)}
\def\ijblt{{\tr}_{+}(\islash\, \jslash\,  \bslash \, \lslash_2)}

\def\jakb{{\tr}_{+}(\jslash\, \aslash\, \kslash \, \bslash)}
\def\iakb{{\tr}_{+}(\islash\, \aslash\, \kslash \, \bslash)}

\def\tofo{{\tr}_{+}(\onslash\, \thslash\, \twslash \, \foslash)}
\def\foto{{\tr}_{+}(\onslash\, \thslash\, \foslash \, \twslash)}
\def\tofi{{\tr}_{+}(\onslash\, \thslash\, \twslash \, \fislash)}
\def\fito{{\tr}_{+}(\onslash\, \thslash\, \fislash \, \twslash)}

\def\lrangle#1#2{\langle #1\,#2\rangle}

\newcommand{\EQ}[1]{\begin{equation} #1 \end{equation}}
\newcommand{\AL}[1]{\begin{subequations}\begin{align} #1 \end{align}\end{subequations}}
\newcommand{\SP}[1]{\begin{equation}\begin{split} #1 \end{split}\end{equation}}
\newcommand{\ALAT}[2]{\begin{subequations}\begin{alignat}{#1} #2 \end{alignat}
                        \end{subequations}}
\def\beqa{\begin{eqnarray}}
\def\eeqa{\end{eqnarray}}
\def\beq{\begin{equation}}
\def\eeq{\end{equation}}
\def\N{{\cal N}}
\def\sst{\scriptscriptstyle}
\def\thetabar{\bar\theta}
\def\Tr{{\rm Tr}}
\def\one{\mbox{1 \kern-.59em {\rm l}}}
 \def\Nh{\hat{N}}

\def\a{\alpha}      \def\da{{\dot\alpha}}
\def\b{\beta}       \def\db{{\dot\beta}}
\def\c{\gamma}  \def\G{\Gamma}  \def\cdt{\dot\gamma}
\def\d{\delta}  \def\D{\Delta}  \def\ddt{\dot\delta}
\def\e{\epsilon}        \def\vare{\varepsilon}
\def\f{\phi}    \def\F{\Phi}    \def\vvf{\f}
\def\h{\eta}
\def\k{\kappa}
\def\l{\lambda} \def\L{\Lambda}
\def\m{\mu} \def\n{\nu}
\def\o{\omega}
\def\p{\pi} \def\P{\Pi}
\def\r{\rho}
\def\s{\sigma}  
\def\t{\tau}
\def\th{\theta} \def\Th{\Theta} \def\vth{\vartheta}
\def\X{\Xeta}
\def\z{\zeta}

\def\cA{{\cal A}} \def\cB{{\cal B}} \def\cC{{\cal C}}
\def\cD{{\cal D}} \def\cE{{\cal E}} \def\cF{{\cal F}}
\def\cG{{\cal G}} \def\cH{{\cal H}} \def\cI{{\cal I}}
\def\cJ{{\cal J}} \def\cK{{\cal K}} \def\cL{{\cal L}}
\def\cM{{\cal M}} \def\cN{{\cal N}} \def\cO{{\cal O}}
\def\cP{{\cal P}} \def\cQ{{\cal Q}} \def\cR{{\cal R}}
\def\cS{{\cal S}} \def\cT{{\cal T}} \def\cU{{\cal U}}
\def\cV{{\cal V}} \def\cW{{\cal W}} \def\cX{{\cal X}}
\def\cY{{\cal Y}} \def\cZ{{\cal Z}}

\def\ua{\underline{\alpha}}
\def\ub{\underline{\phantom{\alpha}}\!\!\!\beta}
\def\uc{\underline{\phantom{\alpha}}\!\!\!\gamma}
\def\um{\underline{\mu}}
\def\ud{\underline\delta}
\def\ue{\underline\epsilon}
\def\una{\underline a}\def\unA{\underline A}
\def\unb{\underline b}\def\unB{\underline B}
\def\unc{\underline c}\def\unC{\underline C}
\def\und{\underline d}\def\unD{\underline D}
\def\une{\underline e}\def\unE{\underline E}
\def\unf{\underline{\phantom{e}}\!\!\!\! f}\def\unF{\underline F}
\def\unm{\underline m}\def\unM{\underline M}
\def\unn{\underline n}\def\unN{\underline N}
\def\unp{\underline{\phantom{a}}\!\!\! p}\def\unP{\underline P}
\def\unq{\underline{\phantom{a}}\!\!\! q}
\def\unQ{\underline{\phantom{A}}\!\!\!\! Q}
\def\unH{\underline{H}}

\def\As {{A \hspace{-6.4pt} \slash}\;}
\def\bs {{b \hspace{-6.4pt} \slash}\;}
\def\Ds {{D \hspace{-6.4pt} \slash}\;}
\def\ds {{\del \hspace{-6.4pt} \slash}\;}
\def\ss {{\s \hspace{-6.4pt} \slash}\;}
\def\ks {{ k \hspace{-6.4pt} \slash}\;}
\def\ps {{p \hspace{-6.4pt} \slash}\;}
\def\pas {{{p_1} \hspace{-6.4pt} \slash}\;}
\def\pbs {{{p_2} \hspace{-6.4pt} \slash}\;}
\def\Ps {{P \hspace{-6.4pt} \slash}\;}
\def\Qs {{Q \hspace{-6.4pt} \slash}\;}

\def\Fh{\hat{F}}
\def\Vh{\hat{V}}
\def\Xh{\hat{X}}
\def\ah{\hat{a}}
\def\xh{\hat{x}}
\def\yh{\hat{y}}
\def\ph{\hat{p}}
\def\xih{\hat{\xi}}
\def\psit{\tilde{\psi}}
\def\Psit{\tilde{\Psi}}
\def\tht{\tilde{\th}}
\def\lt{\tilde{\lambda}}
\def\llt{\tilde{l}}
\def\At{\tilde{A}}
\def\Qt{\tilde{Q}}
\def\Rt{\tilde{R}}
\def\Nt{\tilde{N}}

\def\at{\tilde{a}}
\def\st{\tilde{s}}
\def\ft{\tilde{f}}
\def\pt{\tilde{p}}
\def\qt{\tilde{q}}
\def\vt{\tilde{v}}
\def\nt{\tilde{n}}

\def\delb{\bar{\partial}}
\def\bz{\bar{z}}
\def\bD{\bar{D}}
\def\bB{\bar{B}}

\def\bk{{\bf k}}
\def\bl{{\bf l}}
\def\bp{{\bf p}}
\def\bq{{\bf q}}
\def\br{{\bf r}}
\def\bx{{\bf x}}
\def\by{{\bf y}}
\def\bR{{\bf R}}
\def\bV{{\bf V}}

\def\d{\delta}\def\D{\Delta}\def\ddt{\dot\delta}
\def\pa{\partial} \def\del{\partial}
\def\xx{\times}
\def\uno{\mbox{1 \kern-.59em {\rm l}}}
\def\trp{^{\top}}
\def\inv{^{-1}}
\def\dag{{^{\dagger}}}
\def\pr{^{\prime}}
\def\lan{\langle}
\def\ran{\rangle}
\def\rar{\rightarrow}
\def\lar{\leftarrow}
\def\lrar{\leftrightarrow}
\newcommand{\0}{\,\!}      
\def\one{1\!\!1\,\,}
\def\im{\imath}
\def\jm{\jmath}
\newcommand{\tr}{\mbox{tr}}
\newcommand{\slsh}[1]{/ \!\!\!\! #1}
\def\vac{|0\rangle}
\def\lvac{\langle 0|}
\def\hlf{\frac{1}{2}}
\def\ove#1{\frac{1}{#1}}
\def\Box{\square}
\def\ZZ{\mathbb{Z}}
\def\CC#1{({\bf #1})}
\def\bcomment#1{}
\def\bfhat#1{{\bf \hat{#1}}}
\def\VEV#1{\left\langle #1\right\rangle}
\newcommand{\ex}[1]{{\rm e}^{#1}} \def\ii{{\rm i}}
\def\rr{{\rm r}} \def\rs{{\rm s}}\def\rv{{\rm v}}
\def\ri{{\rm i}}\def\rj{{\rm j}}
\newcommand{\lrbrk}[1]{\left(#1\right)}
\newcommand{\sfrac}[2]{{\textstyle\frac{#1}{#2}}}
\newcommand{\trace}{\textrm{tr}}

\def\Li2{{\rm Li}_2}
\newcommand{\al}{\alpha}
\newcommand{\bt}{\beta}
\newcommand{\ad}{\dot{\alpha}}

\newcommand{\doublet}[2]{\left(\begin{array}{c}#1\\#2\end{array}\right)}
\newcommand{\twobytwo}[4]{\left(\begin{array}{cc} #1&#2\\#3&#4\end{array}\right)}
\newcommand{\row}[2]{\left(\begin{array}{cc}#1&#2\end{array}\right)}
\newcommand{\bars}{\bar{\sigma}}

\newcommand{\mut}{\tilde{\mu}}
\newcommand{\onedot}{\dot{1}}
\newcommand{\two}{\dot{2}}
\newcommand{\sd}{\dot{\sigma}}
\newcommand{\sig}{\sigma}
\newcommand{\td}{\dot{\tau}}
\newcommand{\ep}{\epsilon}
\newcommand{\adt}{\dot{a}}
\newcommand{\bdt}{\dot{b}}

\newcommand{\SU}{\mathrm{SU}}
\newcommand{\SO}{\mathrm{SO}}
\newcommand{\Sp}{\mathrm{Sp}}
\newcommand{\U}{\mathrm{U}}
\newcommand{\SL}{\mathrm{SL}}
\newcommand{\PSU}{\mathrm{PSU}}

\font\mybb=msbm10 at 12pt
\def\bb#1{\hbox{\mybb#1}}

\font\myBB=msbm10 at 18pt
\def\BB#1{\hbox{\myBB#1}}

\begin{document}

\rightline{QMUL-PH-07-22} \rightline{CERN-PH-TH/2007-157}
\vspace{1truecm}

\centerline{\LARGE \bf On Perturbative Field Theory and\\}
\centerline{{\LARGE \bf Twistor String Theory}\footnote{PhD thesis presented in June 2007.}} \vspace{1truecm}
\thispagestyle{empty} \centerline{
    {\large \bf James Bedford${}^{a,b}$}}

\vspace{.4cm}
\centerline{{\it ${}^a$ Centre for Research in String Theory, Department of Physics}}
\centerline{{ \it Queen Mary, University of London}} \centerline{{\it Mile End Road, London E1 4NS, UK}}

\vspace{.4cm}
\centerline{{\it ${}^b$ Department of Physics, CERN - Theory
    Division}}
\centerline{{\it 1211 Geneva 23, Switzerland}}

\vspace{1.5truecm}

\thispagestyle{empty}

\centerline{\bf ABSTRACT}

\vspace{.5truecm}

\noindent It is well-known that perturbative calculations in field theory
can lead to far simpler answers than the Feynman diagram approach might suggest.
In some cases scattering amplitudes can be constructed for processes with any
desired number of external legs yielding compact expressions which are
inaccessible from the point of view of conventional perturbation theory.
In this thesis we discuss some attempts to address the nature of this
underlying simplicity and then use the results to calculate some previously
unknown amplitudes of interest. Witten's twistor string theory is
introduced and the CSW rules at tree-level and one-loop are described.
We use these techniques to calculate the one-loop gluonic MHV amplitudes in
$\cN\!=\!1$ super-Yang-Mills as a verification of their validity and then
proceed to evaluate the general MHV amplitudes in pure Yang-Mills with a scalar
running in the loop. This latter amplitude is a new result in QCD. In addition to this, we review some
recent on-shell recursion relations for tree-level amplitudes in gauge theory and apply them to gravity.
As a result we present a new compact form for the $n$-graviton MHV amplitudes in
general relativity. The techniques and results discussed are relevant to the
understanding of the structure of field theory and gravity and the
non-supersymmetric Yang-Mills amplitudes in-particular are pertinent
to background processes at the LHC. The gravitational recursion relations
provide new techniques for perturbative gravity and have some bearing on the
ultraviolet properties of Einstein gravity.

\vspace{.5cm}

\dedication
\begin{center}
\vspace{4.5 in}
\emph{~~To my parents and in loving memory of my grandfather Leonard Rogers.}
\end{center}

\topmatter{Acknowledgements}

First and foremost I would like to thank my supervisor Andreas Brandhuber
for all his guidance and support throughout this work and for doing so much to
mature my approach to physics. Similarly, I am especially grateful to Gabriele Travaglini and
Bill Spence who have also acted as advisors and to Lance Dixon for being an
inspirational mentor.

I am greatly indebted to David Mulryne, Constantinos Papageorgakis, John Ward and
Konstantinos Zoubos for their unquestioning friendship and support and without
whom I would not be the person I am today. I would also like to
express my gratitude to my other collaborators Narit Pidokrajt and Diego Rodr\'{i}guez-G\'{o}mez
for putting up with me. It is likewise a pleasure to thank all
these people for greatly helping the development of my physical understanding.

For interesting discussions and help with the subject-matter I am extremely grateful to
Luis \'{A}lvarez-Gaum\'{e}, Charalampos Anastasiou, Pascal Anastasopoulos, Lilya Anguelova,
Marcos Mari\~{n}o Beiras, David Berman, David Birtles, Emil Bjerrum-Bohr,
Vincent Bouchard, Jacob Bourjaily, Ruth Britto, Tom Brown, John Charap, Joe
Conlon, Cedric Delaunay, Sara Dobbins, Ellie
Dobson, Vittorio Del Duca, David Dunbar, David Dunstan, Dario Duo, John Ellis, Kazem Bitaghsir Fadafan,
Malcolm Fairbairn, Bo Feng,
Joshua Friess, Michael Green, Umut G\"{u}rsoy, Stefan Hohenegger,
Harald Ita, Bob Jones, Chris Hull, Valya Khoze, Johanna
Knapp, David Kosower, Theo Kreouzis,
Duc Ninh Le, Wolfgang Lerche, Simon McNamara, Alex Morisse, Michele Della Morte, Adele Nasti,
Georgios Papageorgiou, Alice Perrett, Roger Penrose,
Brian Powell, Are Raklev, Sanjaye Ramgoolam,
Christian Romelsberger, Rodolfo Rousso, Ricardo Schiappa, Shima Shams, Jonathan
Shock, Laura Tadrowski, Alireza Tavanfar,
Steve Thomas, David Tong,
Nicolaos Toumbas, Massimiliano Vincon, Daniel Waldram, Marlene Weiss, Arne Wiebalck and Giulia Zanderighi.

The hospitality of both Queen Mary, University of London \& CERN and their support via
a Queen Mary studentship and a Marie Curie early stage training grant are gratefully
acknowledged, as are the many staff including Terry Arter, Kathy Boydon, Denise Paige, Nanie Perrin,
Jeanne Rostant, Kate Shine, Diana De Toth and Suzy Vascotto who make these places
truly special.

For help with the manuscript I am especially grateful to Andi Brandhuber,
Johanna Knapp, Costis Papageorgakis, Jon Shock, Gabriele Travaglini, Marlene Weiss and Costas Zoubos.

The support of many friends and family members, the number of whom is too great to list is also gratefully acknowledged. Especially,
these include Sam Chamberlain, Jeremy and Sine Pickles, Barbara and Leonard Rogers and Christiane and Varos Shahbazian. Last, but by no means
least I owe a great debt to my parents Andrew and Linda, my brother Simon and to Claire.

\tableofcontents
\listoffigures

\mainmatter
\topmatter{Introduction}

In the realm of high energy physics, the \emph{standard model} of particle physics is
our crowning achievement to-date.\footnote{See \emph{e.g.} \cite{smodel} for an introductory text on the
standard model and \emph{e.g.} \cite{pc,wein,Weinberg:1996kr,Weinberg:2000cr} for treatises on
quantum field theory in general.}
It describes the fundamental forces of nature - excluding gravity -
 as a quantum (gauge) field theory with gauge symmetry group \mbox{$\SU(3)\!\times\! \SU(2)\!\times\! \U(1)$}. In this description,
the strong force - described by a gauge theory known as quantum chromodynamics with gauge group
$\SU(3)$ - is adjoined to
electro-weak theory which is itself a unification of quantum electrodynamics and the weak
interaction. The standard model is well-verified experimentally and will soon be put to even greater tests
by the large hadron collider at CERN which will start running later in 2007.

However, there are a number of features of the standard model (SM) which are not
fully understood. Most prominent of these is perhaps that it predicts the existence of a scalar
particle called the Higgs boson of mass $M_{\mathrm{H}}>114.4\,\mathrm{GeV}$ \cite{pdg} which is responsible for the generation of mass in electro-weak
symmetry breaking and which has not yet been observed, though few doubt that it will not be found. Indeed one of the central
goals of the large hadron collider (LHC) is to find such a
particle. There is also evidence that neutrinos should have (tiny) masses and mixings and the SM should be
extended to accomodate this.

On the other hand there are also theoretical issues that lead physicists to believe that the standard model is not the
final story. For a start, (quantum) gravity is not incorporated into the theory. In addition, the SM suffers from
a problem known as the \emph{hierarchy problem}. This problem asks why there is such a large hierarchy of scales
for the interaction strengths of the different forces present. It seems natural to theorists that just as the
electromagnetic and weak forces are unified into the electro-weak (EW) force at scales \mbox{$M_{\mathrm{EW}}\sim 100\,\mathrm{GeV}$},
so should EW
theory be unified with quantum chromodynamics (QCD) at some (higher) scale. As such it is generally believed that SM particles are coming
from a grand unified theory (GUT) that spontaneously broke to \mbox{$\SU(3)\times \SU(2)\times \U(1)$} at energies
$M_{\mathrm{GUT}}\sim 10^{16}\,\mathrm{GeV}$. Popular gauge groups that might unify those of the SM include $\SU(5)$ and $\SO(10)$.

Several ideas which try to deal with the hierarchy problem exist. One of these is a theory called \emph{technicolour}
\cite{Susskind:1978ms,Farhi:1979zx} which considers
all scalar fields in the SM to be bound states of fermions joined by a new set of interactions. Another idea is
that a new symmetry may exist such as \emph{supersymmetry} - see \emph{e.g.}
\cite{Wess:1992cp,Sohnius,Figueroa-O'Farrill:2001tr,Alvarez-Gaume:1996mv} for an
introduction. Supersymmetry (SUSY) relates bosons and fermions and predicts that many more particles exist than are
currently observed as each boson/fermion is associated with a partner fermion/boson. It can, however, unify the gauge couplings
of the various component theories of the standard model and thus solve the hierarchy problem. As such
the SM would be replaced by some supersymmetric version, the minimal realisation of which is
usually termed the minimally supersymmetric standard model (MSSM).\footnote{See \cite{Rosiek:1989rs,Rosiek:1995kg} for
an overview containing the action and Feynman rules.} The LHC is also geared towards searching for
physics beyond the standard model such as technicolour and supersymmetry.

The case for unification with gravity is very much more speculative at present. This is not least because its
tiny interaction strength compared with the other forces of nature makes
experimental tests of gravity on small length-scales difficult to perform with existing technology. As such
there is no accepted quantum theory of gravity at present let alone a unification of quantum gravity with
the SM. Currently studied theories that address the issue of the quantisation of gravity include causal set theory
\cite{Bombelli:1987aa,Henson:2006kf}, loop quantum gravity \cite{Ashtekar:1986yd,Ashtekar:2004eh} and string theory
\cite{gsw1,gsw2,Polchinski:1998rq,Polchinski:1998rr}. Of these, string theory has also emerged as a possible
framework for providing a complete unified theory of all the forces of nature or a theory of everything
(TOE) as it is sometimes called.

For string theory, the starting point is best understood as a generalisation of the world-line approach to
particle physics as opposed to the spacetime approach of quantum field theory. In this approach one
considers particles from the point of view of their world-volume or world-line (as their
trajectories are lines in spacetime) and describes this trajectory using an action of the
form
\be
S_{\mathrm{particle}}=\frac{1}{2}\int\!d\tau\left(e^{-1}\eta_{\mu\nu}\partial_{\tau}{X}^{\mu}
\partial_{\tau}{X}^{\nu}-em^2\right)\ ,\nonumber
\ee
where $\tau$ is a parameter along the world-line which can naturally be taken to be the
proper time. $e(\tau)$ is a function\footnote{Actually $e(\tau)$ is an einbein.} introduced to make the action valid
for zero particle mass ($m=0$) as well as $m\neq 0$ and $X^{\mu}(\tau)$ represents the position vector of the
particle in the `target' space in which it lives. For the sake of generality we may consider the
target space to be $d$-dimensional though of course four dimensions is what we're aiming for.
While $S_{\mathrm{particle}}$ describes a free particle, interactions may be included by adding
terms such as $\int dX^{\mu}A_{\mu}(X)$ for a coupling to the electromagnetic field.

To go from point-particles to strings we simply replace $S_{\mathrm{particle}}$ by an action appropriate
for describing the world-sheet of a string embedded in spacetime. An action which
naturally incorporates both massive and massless strings is the Brink-Di Vecchia-Howe or Polyakov
action
\be
S_{\mathrm{string}}=-\frac{T}{2}\int\!d^2\xi\sqrt{\det{|\gamma|}}\,\gamma^{\alpha\beta}
(\eta_{\mu\nu}\partial_{\alpha}X^{\mu}\partial_{\beta}X^{\nu})\ .\nonumber
\ee
Here the parameters of the world-sheet are $\xi=\tau,\sigma$, the
tension of the string is $T$ and $\gamma_{\alpha\beta}$ can be thought of as a
metric on the world-sheet.\footnote{Note that the string tension is usually written as
$T=1/(2\pi\alpha')$ where $\alpha'=l_s^2$ with $l_s$ the string length.} Consistently quantizing $S_{\mathrm{string}}$ leads
(eventually) to the many interesting consequences that string theory predicts,
not least of these being that gravity is quantized and the demand
that the dimension of the target space be 26-dimensional for the bosonic string
(the action of which is the one given by $S_{\mathrm{string}}$ above) or
$10$-dimensional for any of its supersymmetric extensions.

There are 5 of these consistent supersymmetric string theories that are known as
type I, type IIA, type IIB, heterotic $\SO(32)$ and heterotic $\mathrm{E}_8\times \mathrm{E}_8$ respectively, each of
which has its use in describing the physics of this 10-dimensional
universe in different scenarios. They are, however, intrinsically perturbative
constructions and as such it has been proposed that each of these theories is
just a different limit of a unique \mbox{11-dimensional} theory which describes the full
non-perturbative range of physics and is known as M-theory \cite{Witten:1995ex}.

The intrinsically higher-dimensional nature of these theories is clearly in
contrast with current experimental results, although such results do not extend
down to the Planck scale $M_P\sim 10^{19}\,\mathrm{GeV}$ where it is believed that the effects of
quantum gravity will be most prevalent. Nonetheless it is
hoped by many that a compactification down to four dimensions or a
realisation of string theory on a 4-dimensional submanifold such as a brane \cite{Polchinski:1995mt}
may provide a unified description of the standard model plus gravity in $3+1$ dimensions.

Aside from quantizing gravity or being a possible TOE, string theory has many other
facets. Not least among these is the capacity to provide alternative or `dual' descriptions of
many well-known 4-dimensional quantum field theories. In particular these quantum field theories include
highly symmetric gauge theories such as maximally supersymmetric ($\cN\!=\!4$) Yang-Mills, but also
extend to certain aspects of QCD for example.

It has long been thought that gauge theories may be described by string theories and the idea goes back
at least till 't Hooft's diagrammatic proposal \cite{thooft}. However, it wasn't until much more recently
that this proposal was realised in a concrete way by Maldacena \cite{Maldacena:1997re} who discovered
a duality between type IIB string theory with target space $\textrm{AdS}_5\times \mathbb{S}^5$ - the product of
5-dimensional anti-de-Sitter space and a 5-sphere - and a certain conformal field theory (CFT), namely $\cN\!=\!4$ super-Yang-Mills
theory in Minkowski space with gauge group $\SU(N)$.\footnote{Note that in order to treat the strings
perturbatively we must actually take $N\rightarrow\infty$.} The duality is a `weak-strong' one in the sense
that weakly coupled strings are describing the strong coupling regime of a gauge theory and as such this provides
a fascinating perturbative window into non-perturbative 4-dimensional physics.

In addition to this, the duality provides a concrete realisation of the so-called holographic principle
\cite{'tHooft:1993gx,bousso}
which asserts that physics in $d$-dimensional spacetimes that include gravity may be
describable by degrees of freedom in $d\!-\!1$ dimensions. One of the key ideas in this is that the
Bekenstein-Hawking entropy of a black hole (a system whose dominant force is gravity) is given by
$S_{BH}=A/4$ in `natural' units where $A$ is the area of the event horizon. This is in contrast with the fact that entropy is
an extensive variable and thus usually scales with the volume of the system concerned. In the case of the
Maldacena conjecture (also known as the AdS/CFT correspondence), the 5-sphere essentially scales to
a point and we are left with gravity (\emph{i.e.} closed strings) in 5 dimensions being described by
Yang-Mills (\emph{i.e.} open strings) in 4 dimensions.

In any case, it is not only the non-perturbative aspects of four-dimensional gauge theory that we would like to understand
better. Although weak-coupling perturbation theory is in-principle well understood for such theories, the complexity
is so great as to make many calculations intractable. The asymptotic freedom of QCD \cite{Gross:1973id,Politzer:1973fx}
means additionally that perturbative results become more important as the energy of interaction is increased, and
many of these will be necessary input for the discovery of new physics at colliders such as the LHC. As such it
would be very interesting from both a theoretical and a phenomenological perspective if a duality existed that might
describe a 4-dimensional gauge theory at weak coupling.

In fact a key step was taken in this direction by Witten at the end of 2003 \cite{witten}. He discovered a
remarkable new duality between \emph{weakly-coupled} $\cN\!=\!4$ super-Yang-Mills theory in Minkowski space
and a \emph{weakly-coupled} topological string theory (known as the B-model) whose target space is the
Calabi-Yau super-manifold $\mathbb{CP}^{3|4}$. This manifold has 6 real bosonic dimensions which are related to
the usual 4-dimensional spacetime of the quantum field theory by the twistor construction of Penrose \cite{Penrose}.

In \cite{witten}, it was observed that tree-level gluon-scattering amplitudes in $\cN\!=\!4$ super-Yang-Mills%
\footnote{The same applies to QCD at tree-level due to an effective supersymmetry - see \S\ref{susydecomp}.}
localise on holomorphically embedded algebraic curves in twistor space and proposed that they could be
calculated from a string theory by integrating over the moduli space of D1-brane instantons in the B-model on (super)-twistor
space. The localisation properties of these amplitudes helped to explain the unexpectedly simple
structure that often arises in their calculation from Feynman diagrams despite the large degree of complexity
at intermediate stages in the computation. In the simplest case the maximally helicity violating (MHV) amplitudes,
which describe the scattering of \mbox{2 gluons} of negative helicity with $n-2$ gluons of positive helicity, are
localised on simple straight lines in twistor space. Similarly, amplitudes which are known to vanish such as those involving $n$ gluons of
positive helicity or 1 gluon of negative helicity and $n-1$ gluons of positive helicity are explained
in this scheme.

The beautifully simple localisation properties of the MHV amplitudes led Cachazo, Svr{\v{c}}ek and Witten \cite{csw}
to propose a new diagrammatic way of calculating tree amplitudes in gauge theory using MHV amplitudes as
effective vertices. These are taken off-shell and glued together with simple scalar propagators to give
amplitudes with successively greater numbers of negative-helicity particles. These rules turn out to be
just the Feynman rules for light-cone Yang-Mills theory with a particular non-local change of
variables and have more recently been put on a firmer theoretical footing \cite{Risager:2005vk,Mansfield:2005yd}.

The situation at loop-level is not as clear. In \cite{BerkWitt2} it was shown that states of conformal supergravity
are present which do not decouple at one-loop and the procedure for calculating loop amplitudes in Yang-Mills
from a twistor string theory is not clear. Despite this, it is a remarkable result of Brandhuber, Spence and
Travaglini \cite{bst} that the so-called CSW rules can also be applied at loop-level. In \cite{bst} it was shown that
the one-loop MHV amplitudes originally found by Bern, Dixon, Dunbar and Kosower (BDDK) in \cite{Bern:zx} could be
calculated using MHV amplitudes as effective vertices in the same spirit as \cite{csw}. This strongly
hints at the existence of a full quantum duality between maximally supersymmetric Yang-Mills and a twistor
string theory, though the situation is unresolved at present.\footnote{An interesting possibility has recently
arisen in \cite{Abou-Zeid:2006wu} where a number of new dualities were constructed between field theories involving gravity
and twistor string theory, One of which is a duality between
$\cN\!=\!4$ Yang-Mills coupled to Einstein supergravity and a twistor string theory. An interesting feature of this appears to be
the existence of a decoupling limit giving pure Yang-Mills which might open the prospect of a
twistor string formulation of Yang-Mills at loop-level.}

A natural question now arises: Can the MHV rules be applied at loop level in \emph{any} gauge theory? The
answer to this is not \emph{a priori} clear as the duality in \cite{witten} applies to $\cN\!=\!4$ Yang-Mills
which is known to be very special due to its high degree of symmetry. Without the existence of a formal
proof of the MHV rules at loop level, one way to proceed is certainly to try a similar method to that in \cite{bst} in other
theories. To that end, the present author and the authors of \cite{bst} used the MHV rules to
calculate the one-loop MHV amplitudes in $\cN\!=\!1$ super-Yang-Mills \cite{bbst} (see also Chapter 2 of this thesis). This was independently
confirmed by Quigley and Rosali in \cite{quig} and both results found complete agreement with the
amplitudes first presented by BDDK \cite{Bern:1994cg}.

Following this, the authors of \cite{bbst} tackled the MHV amplitudes in pure Yang-Mills with a scalar
running in the loop \cite{bbst2}. There the amplitudes for arbitrary positions of the negative-helicity
gluons were derived for the first time and complete agreement was found with the existing
special cases \cite{Bern:1994cg,bdk9302280}. It was discovered, however, that the MHV-vertex formalism
calculates only the so-called `cut-constructible' part - that is, the part containing branch cuts - of the amplitudes
and thus misses possible rational terms. These rational terms are also present in the cases of the supersymmetric amplitudes,
but it turns out that they are intrinsically linked to the cut-constructible parts \cite{Bern:zx,Bern:1994cg} and thus
it is enough to know the cuts to fully determine the amplitudes. More recently, and building on the results
in \cite{bbst2}, the rational terms for the MHV amplitudes in pure Yang-Mills have been found \cite{Berger:2006vq} and
due to a supersymmetric decomposition of one-loop amplitudes described in \S\ref{susydecomp}, this means that
the complete $n$-gluon MHV amplitudes in QCD are now known. The calculation of the
cut-constructible part of the amplitudes in pure Yang-Mills will be the subject of Chapter 3.

In a different direction, various results emerging from twistor string theory \cite{rsv04,Britto} inspired
Britto, Cachazo and Feng to propose certain on-shell recursion relations for tree-level amplitudes in
gauge theory \cite{bcf} which were later proved more rigorously in a paper with Witten \cite{bcfw}. These
represent tree amplitudes as sums over amplitudes containing smaller numbers of external particles
connected by scalar propagators. Starting from amplitudes with 3 particles one can thus build up
all $n$-point tree-level amplitudes recursively.

Subsequently the present author, together with Brandhuber, Spence and Travaglini showed that
similar on-shell recursion relations for tree-level amplitudes in gravity could be
constructed \cite{bbst3}, where a new form for the $n$-graviton MHV amplitudes was also
proposed. Such recursion relations for gravity were independently found by Cachazo and
Svr{\v{c}}ek in \cite{Cachazo:2005ca} which has some overlap with \cite{bbst3}. One
striking feature of these recursion relations is that they require a certain behaviour
of the amplitudes ($\cM$) as a function of momenta in the ultraviolet (UV) such that when thought of as a
function of a complex parameter $z$, ${\lim}_{z\rightarrow\infty}\,\cM(z)=0$. For
Yang-Mills amplitudes this was proved to be the case in \cite{bcfw}, but it is \emph{a priori}
less clear how gravity might behave. In \cite{bbst3,Cachazo:2005ca} the particular amplitudes in
question were shown to have this behaviour and more recently it was established for all
tree-level gravity amplitudes in \cite{Benincasa:2007qj}. This unambiguously
establishes the validity of the recursion relation in gravity, the construction of
which is the subject of Chapter 4, and also lends support to the recent conjectures
that gravity as a field theory may not be as divergent as previously thought
\cite{Bern:2005bb,Bjerrum-Bohr:2005xx,Bjerrum-Bohr:2006sg,Bjerrum-Bohr:2006yw,Bern:2006kd,Bern:2007hh,
Green:2006gt,Green:2006yu}.

This thesis will be concerned with a few \cite{bbst,bbst2,bbst3} of the many developments arising from
twistor string theory \cite{witten}. These include the use of MHV vertices
to calculate many tree-level (and some one-loop) processes
\cite{Zhu,Georgiou:2004wu,WuZhutwo,WuZhuthree,dk,ggk,Abe:2004ep,higgs,a,Birthwright:2005ak,Birthwright:2005vi}
as well as the use of the so-called \emph{holomorphic anomaly} \cite{csw3} (which arose to solve a discrepancy
between the twistor space picture of one-loop amplitudes presented in \cite{csw2} with the
derivation in \cite{bst}) to evaluate one-loop amplitudes \cite{freddy,freddy2,BBDD}. MHV vertices have
also been found at tree-level in gravity \cite{Bjerrum-Bohr:2005jr} (after understanding how to deal with the
non-holomorphicity which stalled initial progress \cite{Giombi:2004ix}) and the CSW rules in gauge theory
at loop level have been more rigorously proved in \cite{Brandhuber:2005kd} together with
recent advances at elucidating the loop structure in pure Yang-Mills
\cite{Ettle:2006bw,Brandhuber:2006bf,Brandhuber:2007vm,Ettle:2007qc}.

Recent improvements \cite{Britto,Brandhuber:2005jw,Buchbinder:2005wp} to the unitarity method pioneered in
\cite{Bern:zx,Bern:1994cg,Bern:1995db,Bern:1996ja,Bern:1996je,Bern:2000dn,Bern:2002tk} use
\emph{complex} momenta (in similarity with the on-shell recursion relations presented
in \cite{bcf,bcfw,bbst3,Cachazo:2005ca}) which allows, for example, a simple and purely
algebraic determination of integral coefficients \cite{Britto,Bidder:2005ri}. In \cite{Britto:2005ha} Britto, Buchbinder,
Cachazo and Feng developed efficient techniques for evaluating generic one-loop unitarity
cuts which have since been applied in \cite{Britto:2006sj} and further developed in
\cite{Anastasiou:2006jv,Britto:2006fc,Anastasiou:2006gt}.

Stemming from the on-shell recursion relations written down at tree-level by Britto, Cachazo
and Feng \cite{bcf} (which have been successfully exploited in \cite{Britto:2005dg,Luo:2005rx,Luo:2005my}
and understood in terms of twistor-diagram theory in \cite{Hodges:2005bf,Hodges:2005aj,Hodges:2006tw})
is the application of on-shell recursion to one-loop
amplitudes which allows for a practical and systematic construction of their rational
parts. These have been pioneered in
\cite{Bern:2005hs,Bern:2005ji,Bern:2005cq,Bern:2005hh,Forde:2005hh,Berger:2006ci,Berger:2006sh,Badger:2007si}, leading to the
full expression for the rational terms of the one-loop MHV amplitudes in QCD in
\cite{Berger:2006vq}. Some success has also been had with such on-shell recursion in
one-loop gravity \cite{Brandhuber:2007up}.

Progress on the string theory side has been somewhat more limited after some promising
initial work. Alternative twistor string theories to that introduced by Witten \cite{witten}
to describe perturbative $\cN\!=\!4$ Yang-Mills have been put forward, though these have
generally seemed to be more formal and less practical than the original proposal. Most notably
there is that of Berkovits (and Motl) \cite{Berkovits,BerkMotl} which was also addressed at loop level in
\cite{BerkWitt2}, and which has been recently used to calculate loop amplitudes in
Yang-Mills coupled to conformal supergravity \cite{Dolan:2007vv}. Other proposals include
those of \cite{Siegel:2004dj,Neitzke:2004pf,Aganagic:2004yh,Bars:2004dg}.

Similarly, dual twistor string theories have been constructed for other field theories
including marginal deformations of $\cN\!=\!4$ (and non-supersymmetric theories) \cite{Kulaxizi:2004pa,Gao:2006mw},
orbifolds of Witten's original proposal to include theories with less supersymmetry
and product gauge groups \cite{Park:2004bw,Giombi:2004xv} as well as twistor string descriptions of
supergravity theories. This latter section of work includes twistor descriptions of $\cN\!=\!1,2$ conformal supergravity
\cite{Ahn0409,Ahn:2004ua}, as well as a more recent construction for Einstein supergravity \cite{Abou-Zeid:2005dg,Abou-Zeid:2006wu}
following initial observations of the special properties of graviton amplitudes \cite{witten,Giombi:2004ix,Nair:2005iv,Bjerrum-Bohr:2005jr}.
Additionally, twistor string dual constructions have been presented for truncations of self-dual $\cN\!=\!4$ super-Yang-Mills \cite{Popov:2004nk}, lower
dimensional theories \cite{Chiou:2005jn,Popov:2005uv,Saemann05,LechtenfeldSaemann05,Popov:2007hb} and $\cN\!=\!4$ SYM with a chiral mass
term \cite{Chiouetal0512}.

\pagebreak

Directly following from \cite{witten}, it was shown how to construct amplitudes that are more complex
than the MHV amplitudes from an integral over a suitable moduli space of curves in twistor string theory. Some
simple 5-point next-to-MHV (NMHV) amplitudes were addressed in \cite{rsv} as well as all n-gluon
$\overline{\mathrm{MHV}}$ amplitudes\footnote{\emph{i.e.} the amplitudes which are MHV amplitudes when the helicities of all
particles are reversed. They thus describe the scattering of 2 gluons of positive helicity with $n\!-\!2$ gluons
of negative helicity.} in \cite{rv} and all 6-gluon amplitudes in \cite{rsv2}.

Another avenue that has proved illuminating is the study of gauge and gravity theories in
twistor space. This includes \cite{Boels:2006ir} where the partition function of
$\cN\!=\!4$ Yang-Mills was examined in twistor space, \cite{Boels:2007qn} where the CSW
rules were treated from a purely gauge theoretic perspective in twistor space and
\cite{Boels:2007gv} where loops have been studied and other related work including
\cite{Mason:2005zm,Mason:2005kn}. Furthermore, self-dual supergravity theories have been
investigated from a twistor space perspective in \cite{Wolf:2007tx,Mason:2007ct}, relations between twistors, hidden
symmetries and integrability elucidated in \cite{Wolf:2004hp,Popov:2006qu}, and the connection with
string field theory developed in \cite{LechtenfeldPopov04}. Finally, twistor string theory has inspired
a great deal of work in understanding supermanifolds and their connections with string theory and
gauge theory such as that of \cite{Saemann04,GrassiPolicastro04,Tokunaga05,Ricci:2005cp} and references therein.

\vspace{0.5cm}

\newpage

\topmatter{Summary}

This thesis is organised as follows:

In Chapter \ref{chapter1} we discuss perturbative gauge theory and the unexpectedly
simple results that it can produce despite the huge number of Feynman
diagrams that have to be summed. We introduce various techniques for explaining
this simplicity including colour ordering, the spinor helicity formalism, supersymmetric decompositions,
supersymmetric ward identities and the use of twistor space. We go on
to review the twistor string theory introduced in \cite{witten} and show
how it can be used to calculate tree-level scattering amplitudes of gluons. Finally
we describe some key ideas in perturbative gauge theory that were inspired
by the twistor string theory. In particular we present an overview of the
CSW rules and their application at tree- and loop-level in $\cN\!=\!4$
super-Yang-Mills.

Chapter \ref{chapter2} is devoted to elucidating the calculation of MHV loop amplitudes in
$\cN\!=\!1$ Yang-Mills using a perturbative expansion in terms of
MHV amplitudes as vertices as was introduced for $\cN\!=\!4$ Yang-Mills
in \cite{bst}. We follow \cite{bbst} where the calculation was originally
performed and use the decomposition
of the integration measure advocated in \cite{bst,Brandhuber:2005kd} to
reconstruct the $n$-gluon MHV amplitudes in $\cN\!=\!1$ Yang-Mills
first given in \cite{Bern:1994cg}. This provides strong evidence that
the MHV diagram method is valid in general supersymmetric field theories at
loop level. Some technical details are relegated to Appendix \ref{neq1appendix}.

In much the same spirit, Chapter \ref{chapter3} describes the calculation of the
MHV amplitudes in pure Yang-Mills with a scalar running in the loop.
We take the same approach as in Chapter \ref{chapter2} and closely follow \cite{bbst2}.
This produces the first results for the (cut-constructible part of the)
$n$-gluon MHV amplitudes with arbitrary positions for the negative-helicity
particles in pure Yang-Mills. The results obtained are in complete agreement with
the previously known special cases in \cite{Bern:1994cg,bdk9302280} and as with
Chapter \ref{chapter2}, many technical details to do with the evaluation of integrals are omitted
and provided in Appendix \ref{nonsusyappendix}.

In Chapter \ref{chapter4} we describe some tree-level on-shell recursion relations in gravity as
constructed in \cite{bbst3} and highlight some of
their similarities with the on-shell recursion relations proposed for
gauge theory in \cite{bcf,bcfw}. The format followed is that of
\cite{bbst3} and as such we describe a new compact form for the $n$-graviton
MHV amplitudes arising from the recursion relation. We also comment
on the existence of recursion relations in other field theories such as
$\phi^4$ theory and mention the connection between the CSW rules at
tree-level and these on-shell recursion relations.

We conclude and discuss future directions in Chapter \ref{chapter5}. Additionally,
there are appendices describing the spinor helicity formalism and Feynman
rules for massless gauge theory in such a formalism, $d$-dimensional Lorentz-invariant
phase space, unitarity and the
Kawai-Lewellen-Tye (KLT) relations in gravity which relate tree amplitudes in
gravity to (products of) tree amplitudes in Yang-Mills.

\chapter{Perturbative gauge theory}
\label{chapter1}

In the traditional approach to quantum field theory, one writes
down a classical Lagrangian and can quantise the theory by
defining the Feynman path integral. Perturbative physics can then
be studied by drawing Feynman diagrams and using the Feynman rules
generated by the path integral to calculate scattering amplitudes.
For a non-Abelian gauge theory the classical theory is
well-described by the Yang-Mills Lagrangian \cite{YM}:
\begin{eqnarray}
\label{gtaction}
\mathcal{L}&=&\bar{\psi}(i {\partialslash}-m)\psi-\frac{1}{4}(\partial_\mu A_\nu^a-
\partial_\nu A_\mu^a)^2+gA_{\mu}^a\bar{\psi}\gamma^{\mu}T^a\psi\nonumber \\
& & -gf^{abc} (\partial_\mu A_\nu^a)A^{\mu b}A^{\nu
c}-\frac{1}{4}g^2(f^{eab}A^a_\mu A_\nu^b) (f^{ecd}A^{\mu c}A^{\nu
d})\ , \eeqa where $\psi$ is a fermion field, $A$ the gauge
boson field and $g$ is the coupling. Greek indices are associated
with spacetime, while Roman indices describe the structure in
gauge group space. This can then be used to construct the Feynman rules in the
usual way.

Although this construction is somewhat technical it is easy so see what these
interactions will be from a heuristic standpoint. The first two terms in
(\ref{gtaction}) will give the fermion and gauge boson propagators respectively.
The third term involves two $\psi$s and an $A$ and thus represents a vertex where
two fermions interact with a gauge boson. The fourth term involves 3 $A$s and
represents a 3-boson vertex while the fifth term gives a 4-boson vertex.

If we work everything out properly then we find that, in Feynman
gauge for example where we have set $\xi=1$ in a more general gauge boson
propagator of the form
\begin{eqnarray}
\label{propagator}
\frac{-i}{p^2+i\varepsilon}\left(g_{\mu\nu}-
(1-\xi)\frac{p_{\mu}p_{\nu}}{p^2}\right)\delta_{ab}\ ,
\end{eqnarray}
 the Feynman rules for an $\SU(N_c)$ gauge theory are:
\newpage
\hfill\\
\begin{picture}(300,70)(0,47)
\thicklines
\Text(80,50)[bp]{Gauge Boson Propagator:}
\Photon(180,50)(250,50){5}{3}
\LongArrow(199,40)(228,40)
\Text(215,32)[pp]{$p$}
\Text(170,50)[aa]{$a$}
\Text(260,47)[bb]{$b$}
\end{picture}
$\displaystyle \!\!\!\!\!\!=\,\, \frac{-ig_{\mu\nu}}{p^2+i\varepsilon}\delta_{ab}$
\\
\begin{picture}(300,50)(0,47)
\thicklines
\Text(80,50)[fp]{Fermion Propagator:}
\ArrowLine(180,50)(250,50)
\Text(215,40)[p]{$p$}
\Text(170,50)[i]{$i$}
\Text(260,47)[j]{$\bar{\jmath}$}
\end{picture}
$\displaystyle \!\!\!\!\!\!=\,\, \frac{i(p\!\!\!\slash+m)}{p^2-m^2+i\varepsilon}{\delta_i}^{\bar{\jmath}}$
\\
\begin{picture}(300,100)(0,47)
\thicklines
\Text(80,50)[tbv]{Fermion Vertex:}
\Photon(215,100)(215,50){5}{3}
\ArrowLine(215,50)(172,25)
\ArrowLine(258,25)(215,50)
\Vertex(215,50){2}
\Text(215,107)[amu]{$a,\mu$}
\end{picture}
$\displaystyle \!\!\!\!\!\!=\,\, ig\gamma^{\mu}T^a$
\\
\begin{picture}(300,100)(0,47)
\thicklines
\Text(80,50)[tbv]{3-Boson Vertex:}
\Photon(215,100)(215,50){5}{3}
\Photon(172,25)(215,50){5}{3}
\Photon(258,25)(215,50){5}{3}
\Vertex(215,50){2}
\Text(215,107)[amu]{$a,\mu$}
\Text(170,16)[crho]{$c,\rho$}
\Text(275,20)[bnu]{$b,\nu$}
\LongArrow(225,65)(225,85)
\Text(233,75)[p1]{$p_1$}
\LongArrow(198,51)(180,39)
\Text(186,50)[p3]{$p_3$}
\LongArrow(221,36)(240,25)
\Text(227,25)[p2]{$p_2$}
\end{picture}
\begin{tabular}{l}
$\displaystyle \!\!\!\!\!\!=\,\, -gf^{abc}[g^{\mu\nu}(p_1\!-\!p_2)^\rho$\\
$\displaystyle\quad\quad\quad\quad\,\!\!\!\!\!\!+g^{\nu\rho}(p_2\!-\!p_3)^{\mu}$\\
$\displaystyle\quad\quad\quad\quad\,\!\!\!\!\!\!+g^{\rho\mu}(p_3\!-\!p_1)^{\nu}]$
\end{tabular}
\\
\begin{picture}(300,100)(0,47)
\thicklines
\Text(80,50)[fbv]{4-Boson Vertex:}
\Photon(250,85)(215,50){5}{3}
\Photon(180,85)(215,50){5}{3}
\Photon(180,15)(215,50){5}{3}
\Photon(250,15)(215,50){5}{3}
\Vertex(215,50){2}
\Text(255,87)[bnu]{$b,\nu$}
\Text(176,92)[amu]{$a,\mu$}
\Text(170,8)[dsig]{$d,\sigma$}
\Text(270,13)[crho]{$c,\rho$}
\end{picture}
\begin{tabular}{l}
$\displaystyle \!\!\!\!\!\!=\,\, 2ig^2[f^{abe}f^{ecd}g^{\mu[\rho}g^{\sigma]\nu}$\\
$\displaystyle\quad\quad\quad\quad\!\!\!\!\!\!\!\!\!\!\!\!\! +f^{dae}f^{ebc}g^{\mu[\nu}g^{\sigma]\rho}$\\
$\displaystyle\quad\quad\quad\quad\!\!\!\!\!\!\!\!\!\!\!\!\! +f^{cae}f^{ebd}g^{\mu[\nu}g^{\rho]\sigma}]$
\end{tabular}
\vspace{10pt}\hfill \\
\begin{figure}[ht]
\label{feynrules}
\caption{\emph{Feynman rules for $\SU(N_c)$ Yang-Mills theory in Feynman gauge.}}
\end{figure}
\newpage

In the above rules we have taken all particles to be outgoing and we use the
convention that $C^{[\mu\nu]}=(C^{\mu\nu}-C^{\nu\mu})/2$ for some 2-index object $C$. We have also
ignored the contributions due to ghost fields and will stick to these choices
in what follows unless otherwise specified.
Amplitudes for physical processes are obtained by drawing all the ways that
the process can occur using the above rules and associating each of these
with a specific mathematical expression. They are then evaluated and added up
to produce the desired result. Classical results are obtained from diagrams
without any closed loops while quantum corrections involve an increasing number of
loops. For more details see e.g. \cite{pc}.

Even though gauge theories present many technical challenges, the way to proceed
(at least perturbatively) is in-principle well
understood. In practice, however, the calculational complexity
grows rapidly with the number of external particles (legs) and the
number of loops. For example, even at tree-level where there are
no loops to consider, the number of Feynman diagrams describing
$n$-particle scattering of external gluons in QCD grows faster
than factorially with $n$ \cite{kleisskuijf89,manganorev}:\footnote{Note
that the following numbers are relevant for the case where one is considering
a single colour structure only. The total number of diagrams after summing
over all possible colour structures is even greater still. For more on this
see \S \ref{colourordersection}}

\begin{center}
\begin{displaymath}
\begin{array}{|c|c|c|c|c|c|c|c|}
\hline n & 4 & 5 & 6 & 7 & 8 & 9 & 10 \\
\hline \textrm{\# diagrams} & 4 & 25 & 220 & 2,485 & 34,300 &
559,405 & 10,525,900 \\
\hline
\end{array}
\end{displaymath}
\begin{figure}[ht]
\caption{\emph{The number of Feynman diagrams required for
tree-level n-gluon scattering.}}
\end{figure}
\end{center}

\noindent Despite this, the final result is often simple and elegant.
A prime example is the so-called Maximally Helicity Violating (MHV)
amplitude describing the scattering of 2 gluons ($i$ and $j$) of negative helicity with $n-2$ gluons of positive helicity. At tree-level
the amplitude is given by:
\be
\label{MHV}
\mathcal{A}^{\textrm{tree}}_n=\frac{\langle i\,j\rangle^4}
{\langle 1\,2\rangle\langle 2\,3\rangle\ldots\langle n\!-\!1\,n\rangle\langle n\,1\rangle}\ ,
\ee
for any $n$. We will leave the explanation of the meaning of this expression
to later in the chapter, but the reader is nonetheless
able to appreciate its simplicity compared with the ever increasing
number of Feynman diagrams needed to produce it.

The question then arises of: Why is there this simplicity
underlying the apparently more complex perturbative expansion and how
does it arise. The
rest of this chapter is devoted to setting up a framework in which
these questions may be addressed.

\section{Colour ordering}

\label{colourordersection}

One prominent complication experienced by gauge theories is the
extra structure inherent in their gauge invariance. This means
that fields of the theory do not just carry spacetime indices but
also indices relating to their transformation under the gauge
group. In the standard model it has been found that $\SU(N_c)$
groups are the most appropriate ones for describing the gauge
symmetry and so unless otherwise specified we will consider gauge
groups of this type.

As is well-known, gluons carry an adjoint colour index
$a=1,2,\ldots ,N_c^2-1$, while quarks and antiquarks carry
fundamental $(\mathbf{N_c})$ or anti-fundamental $(\mathbf{\bar{N}_c})$ indices
$i,\bar{\jmath}=1,2,\ldots ,N_c$. The $\SU(N_c)$ generators
in the fundamental representation are traceless Hermitian $N_c\times N_c$
matrices, ${(T^a)_i}^{\bar{\jmath}}$ which we normalise to
$\trace(T^aT^b)=\delta^{ab}$.\footnote{This is different from the more
familiar $\trace(T^aT^b)=\delta^{ab}/2$, but is purely a convention used to avoid the
proliferation of factors of 2. Note that the Feynman rules
written down at the beginning of the chapter use $\trace(T^aT^b)=\delta^{ab}/2$. To rewrite the
diagrams in a way that is consistent with these `more natural' colour ordering
conventions one simply has to replace $T^a\rightarrow T^a/\sqrt{2}$ and $f^{abc}\rightarrow f^{abc}/\sqrt{2}$.
See also Appendix \ref{helicityrules}.}
The Lie-algebra
is defined by $[T^a,T^b]=if^{abc}T^c$, where the structure constants
$f^{abc}$ satisfy the Jacobi Identity:
\be
\label{jacobi}
f^{ade}f^{bcd}+f^{bde}f^{cad}+f^{cde}f^{abd}=0 \ .
\ee

Let us begin by considering a generic tree-level
scattering amplitude. It is apparent from the Feynman rules given in Figure \ref{feynrules}
that each quark-gluon vertex contributes a group theory factor of
${(T^a)_i}^{\bar{\jmath}}$ and each tri-boson vertex a factor of
$f^{abc}$, while four-boson vertices contribute more complicated contractions
involving pairs of structure constants such as $f^{abe}f^{cde}$. The
quark and gluon propagators will then contract many of the indices together
using their group theory factors of $\delta_{ab}$ and ${\delta_i}^{\bar{\jmath}}$.
We can now start to illuminate the general colour structure of the amplitudes
if we first use the definition of the Lie-algebra to re-write the
structure constants as
\be
\label{restructure}
f^{abc}=-i\,\trace(T^a[T^b,T^c])\ .
\ee
Doing this means that all colour factors in the Feynman rules can be
replaced by linear combinations of strings of $T^{a}$s, \emph{e.g.}
$\sum \trace(\ldots T^aT^b\ldots)\,\trace(\ldots T^bT^c\ldots)\ldots\trace(\ldots T^d\ldots)$
if we only have external gluons, or
$\ldots{(T^a\ldots T^b)_i}^{\bar{\jmath}}\,\trace(T^b\ldots T^c){(T^cT^d...)_k}^{\bar{l}}\ldots$
- where the strings are terminated by (anti)-fundamental indices - if external quarks are present.

In order to reduce the number of traces we make use of the identity
\be
\label{trid}
\sum_{a=1}^{N_c^2-1}{(T^a)_{i}}^{\bar{\jmath}}{(T^a)_{k}}^{\bar{l}}
={\delta_i}^{\bar{l}}{\delta_k}^{\bar{\jmath}}-\frac{1}{N_c}
{\delta_i}^{\bar{\jmath}}{\delta_k}^{\bar{l}}\ ,
\ee
which is just an algebraic statement of the fact that the generators
$T^a$ form a complete set of traceless Hermitian matrices. This in turn gives rise
to simplifications such as
\begin{eqnarray}
\label{trsimple1}
\!\!\!\!\!\!\!\!\!\!\sum_a \trace(T^{a_1}\ldots T^{a_{k}}T^a)\,\trace(T^{a}T^{a_{k+1}}\ldots T^{a_n})
\!\!\!&=&\!\!\!\trace(T^{a_1}\ldots T^{a_k}T^{a_{k+1}}\ldots T^{a_n})\nonumber\\
\!\!\!&-&\!\!\!\!\!\frac{1}{N_c}\trace(T^{a_1}\ldots T^{a_k})\,\trace(T^{a_{k+1}}\ldots T^{a_n})
\end{eqnarray}
and
\begin{eqnarray}
\label{trsimple3}
\!\!\!\!\!\!\!\!\!\!\sum_a \trace(T^{a_1}\ldots T^{a_k}T^a){(T^aT^{a_{k+1}}\ldots T^{a_n})_i}^{\bar{\jmath}}\!\!&=&\!\!
{(T^{a_1}\ldots T^{a_k}T^{a_{k+1}}\ldots T^{a_n})_i}^{\bar{\jmath}}\nonumber \\
\!\!&-&\!\!\!\!\!\frac{1}{N_c}\trace(T^{a_1}\ldots T^{a_k}){(T^{a_{k+1}}\ldots T^{a_n})_i}^{\bar{\jmath}}.
\end{eqnarray}

In Eq.~(\ref{trid}) the $1/N_c$ term corresponds to the subtraction of the
trace of the $U(N_c)$ group in which $\SU(N_c)$ is embedded and thus ensures
tracelessness of the $T^a$. This trace couples directly only to quarks and commutes with
$\SU(N_c)$. As such the terms involving it disappear after one sums over all the
permutations present - a fact which is easy to check directly.
We can thus see that we are ultimately left with either sums of single traces of generators
if we only have external gluons as in Eq.~(\ref{trsimple1}) or sums of strings of generators
terminated by fundamental indices as in Eq.~(\ref{trsimple3}) if we also have
external quarks \cite{manganorev,dixon}.\footnote{Note that Eq.~(\ref{trsimple3}) is appropriate for the
case where we have just one $q\bar{q}$ pair. With more pairs there will be products of strings
 with each string terminated by fundamental and anti-fundamental indices giving
terms like ${(T^a\ldots T^b)_i}^{\bar{\jmath}}\ldots{(T^c\ldots T^{d})_k}^{\bar{l}}$. In
the final expression, each
generator will appear only once in any given term of course.}
In most of what we do we will only be
concerned with gluon scattering and can therefore write the colour decomposition of
amplitudes as
\begin{eqnarray}
\label{colourorder} A_n^{\textrm{tree}}(a_i)=
g^{n-2}\sum_{\sigma\,\in\,
S_n/Z_n}\trace(T^{a_{\sigma(1)}}T^{a_{\sigma(2)}}\ldots
T^{a_{\sigma(n)}})\mathcal{A}_n^{\textrm{tree}}(\sigma(1),\sigma(2),\ldots,\sigma(n))
\ ,
\end{eqnarray}
where $S_n$ is the set of permutations of $n$ objects and $Z_n$ is the
subset of cyclic permutations. $g$ is the coupling constant of the theory.
The $\mathcal{A}_n^{\textrm{tree}}$ sub-amplitudes are
colour-stripped and depend only on one ordering of the external
particles. It is therefore sufficient to consider
$\mathcal{A}_n^{\textrm{tree}}(1,2,\ldots,n)$ - the `reduced
colour-ordered amplitude' - and sum over all $(n-1)!$ non-cyclic permutations at
the end.

It is interesting to note that the same conclusion can be arrived
at from string theory in a somewhat more natural way \cite{mpx87,bk91}. This
arises because of the observation that in an open string theory the full on-shell amplitude for the
scattering of $n$ vector mesons can be written as a sum over non-cyclic permutations
of external legs carrying Chan-Paton factors \cite{chanpaton} multiplied by
Koba-Nielsen partial amplitudes \cite{kobanielsen69}. For the scattering of external
gluons that we are interested in we need not worry about fundamental matter because
at tree-level the Feynman rules forbid it from appearing as internal lines. In the
infinite-tension limit $(T\rightarrow\infty\, ;\, \alpha'\rightarrow 0)$ the $\U(N_c)$ string theory reduces to
a $\U(N_c)$ gauge theory and the trace part of this decouples as we have seen. We can
thus immediately conclude that the gauge theory scattering amplitudes decompose as
Eq.~(\ref{colourorder}).

For one-loop amplitudes a similar colour decomposition exists \cite{bk91}. In this case, however,
there are up to two traces over $\SU(N_c)$ generators and one must sum over the spins
of the different particles that can circulate in the loop. In an expansion in $N_{c}$, the
leading (as $N_c\rightarrow\infty$) contributions to the amplitudes are planar and the
colour structure is simply a single trace - in fact it is $N_c$ times the tree-level colour
factor when there are no particles in the fundamental representation propagating in the loop. In this
case an almost identical formula to (\ref{colourorder}) can be written down for a decomposition
of one-loop amplitudes of external gluons \cite{bk91}:
\begin{eqnarray}
\label{loopcolourorder}
\!\!\!\!\!\!\!A_n^{\textrm{1-loop}}(a_i)=g^n\!\!\!\!&\Big[&\!\!\!\!\!\sum_{\sigma\,\in\, S_n/Z_n}
\!N_c\trace(T^{a_{\sigma(1)}}\ldots
T^{a_{\sigma(n)}})\mathcal{A}_{n;1}^{\textrm{1-loop}}(\sigma(1),\ldots,\sigma(n))
\nonumber \\
&+&\sum_{c=2}^{\lfloor n/2\rfloor+1}\!\!\sum_{\sigma\,\in\,S_n/S_{n;c}}
\!\!\trace(T^{a_{\sigma(1)}}\ldots
T^{a_{\sigma(c-1)}})\,\trace(T^{a_{\sigma(c)}}\ldots
T^{a_{\sigma(n)}})\nonumber \\
& &{}\qquad\qquad\qquad\qquad\qquad\qquad\times\mathcal{A}_{n;c}^{\textrm{1-loop}}
(\sigma(1),\ldots,\sigma(n))\Big]\ ,
\end{eqnarray}
and we have left the sum over spins as being implicit in the
definitons of the colour-ordered partial amplitudes $\mathcal{A}_{n;1}^{\textrm{1-loop}}$ and
$\mathcal{A}_{n;c}^{\textrm{1-loop}}$. $\lfloor r\rfloor$ is the largest integer less
than or equal to $r$ and $S_{n;c}$ is the subset of permutations of $n$ objects
leaving the double trace structure invariant.

It is a remarkable result of Bern, Dixon, Dunbar and Kosower that
at one-loop, non-planar (multi-trace) amplitudes are simply obtained as a sum
over permutations of the planar (single-trace) ones. This is discussed in
Section 7 of \cite{Bern:zx}
where it was also noted that this applies to a generic $\SU(N_c)$ theory
(both supersymmetric and non-supersymmetric) with external particles and those running in the loop
both in the adjoint representation. As far as loop amplitudes go we will only be concerned
with particles that are in the adjoint, so it will be enough for us to consider only one
cyclic ordering (\emph{i.e.} only $\mathcal{A}_{n;1}^{\textrm{1-loop}}$, which we will generally
abbreviate to $\mathcal{A}_{n}^{\textrm{1-loop}}$) and then sum over all
the relevant permutations at the very end. We
will not actually perform this summation in what follows but leave it as something which
can easily be implemented to obtain the full amplitude.

The colour-ordered sub-amplitudes obey a number of identities such as gauge invariance, cyclicity,
order-reversal up to a sign, factorization properties and more. This means that there
isn't a huge proliferation in the number of partial amplitudes that have to be computed. For
5-point gluon scattering for example, there are only 4 independent tree-level sub-amplitudes and it
turns out that 2 of these vanish identically because of a `hidden' supersymmetry (see \S \ref{SWI}).
For a more complete list of identities see \cite{manganorev}.

\section{Spinor helicity formalism}
\label{shsection}

So far we have seen that we can reduce some of the complexity of our task by removing the
colour structure and considering only colour-ordered amplitudes. We'll also only consider
massless particles and this restricts us further, though there are still a large
number of things that $\mathcal{A}_n$ can depend on. For spinless particles (scalars), the situation is clear
and $\mathcal{A}_n=\cA_n(p_i)\,\delta^{(4)}\!\left(\sum_{i=1}^n p_i\right)$, where the $p_i$ are the momenta of the
external particles obeying $p^2=p_{\mu}p^{\mu}=0$ and we have written the delta function of
momentum conservation explicitly \cite{witten,wittenonassis04}. In fact the momentum dependence only appears in terms of
Lorentz-invariant quantities such as $p_i\cdot p_j$.

For massless particles with spin the situation is more complicated and we have to
consider their wavefunctions $\psi_i$, giving $\cA_n=\cA_n(p_i,\psi_i)\,\delta^{(4)}\!\left(\sum_{i=1}^n p_i\right)$. Textbook definitions
have the $\psi_i$ being different depending on the spin being considered. For example in the case of
spin $1/2$ electrons and positrons in QED the wavefunctions are usually taken to be the familiar
$u(p)$ and $v(p)$ and their conjugates (see \emph{e.g.} Section (3.3) of \cite{pc}), while
in the case of spin 1 gauge bosons the polarisation vectors $\epsilon^{\mu}$ in a
suitably chosen basis are common. A more unifying description would be highly desirable
and in fact one can be found using the so-called \emph{spinor helicity formalism} \cite{Jacob:1959at}.

\subsection{Spinors}
We start with the fact that,
when complexified, the Lorentz group is locally isomorphic to
\be
\label{complexlorentz}
\SO(1,3,\mathbb{C})\cong \SL(2,\mathbb{C})\times \SL(2,\mathbb{C})\ ,
\ee
and thus the finite-dimensional representations are classified as
$(p,q)$, where $p$ and $q$ are integers or half-integers.\footnote{Note that this section is based largely on the
spinor helicity reviews of \cite{witten,wittenonassis04,csreview}. See also
Appendix \ref{traceappendix} for more
details and identities and \cite{vvkreview} for another good review covering many
aspects of this chapter.} Negative-
and positive-chirality spinors transform in the $(1/2,0)$ and $(0,1/2)$
representations respectively. For a generic negative-chirality spinor we
write $\l_\a$ with $\a=1,2$ and for a generic positive-chirality spinor we
write $\lt_{\da}$ with $\da=1,2$.

The spinor indices introduced here are raised and lowered with the antisymmetric
tensors $\epsilon_{\a\b}$ and $\epsilon^{\a\b}$ as $\l^\a=\epsilon^{\a\b}\l_\b$ and
$\l_\a=\l^\b\epsilon_{\b\a}$ with $\epsilon^{12}=1$ and
$\epsilon_{\a\b}\epsilon^{\b\gamma}=-{\delta_\a}^\gamma$ (and likewise for dotted indices). Given
two spinors $\l$ and $\mu$ of negative chirality we can then form a Lorentz-invariant scalar product as
\be
\label{angbrack}
\langle\l,\,\mu\rangle=\l_\a\mu_\b\epsilon^{\b\a}\ ,
\ee
from which it follows that $\langle\l,\,\mu\rangle=-\langle\mu,\,\l\rangle$.
Similar formul{\ae} apply for positive-chirality spinors except that we use
square brackets to distinguish the two:
$[\lt,\,\tilde{\mu}]=\lt_{\da}\tilde{\mu}_{\db}\epsilon^{\db\da}$.
It is worth noting in-particular that $\langle\l,\,\mu\rangle=0$ implies $\l^\a=c\mu^\a$
where $c$ is a complex number and similarly for $\lt$ and $\tilde{\mu}$. We will often
use even more compact notation for these scalar products and write
$\langle\l,\,\mu\rangle=\langle\l\,\mu\rangle=-\langle\mu\,\l\rangle$ \emph{etc}.

The vector representation of $\SO(1,3,\mathbb{C})$ is the $(1/2,1/2)$ and as such
we can represent a momentum vector $p^{\mu}$ as a bi-spinor $p_{\a\da}$. We can
go to such a representation by using the chiral representation of the Dirac
$\gamma$-matrices - a process that is well-known in supersymmetric field theories, see
\emph{e.g.} \cite{Wess:1992cp,Sohnius}. In signature $+---$ the Dirac matrices
can then be represented as
\be
\label{chiraldirac}
\gamma^{\mu}=\twobytwo{0}{\sig^{\mu}}{\bar{\sig}^{\mu}}{0}\ ,
\ee
where $(\sig^{\mu})_{\a\da}=(1,\vec{\sigma})$ and $(\bar{\sigma}^{\mu})^{\da \a}=
-(\sigma^{\mu})^{\a\da}=\epsilon^{\da\db}\epsilon^{\a\b}(\sigma^{\mu})_{\b\db}=(1,-\vec{\sigma})$ and
$\vec{\sigma}=(\sigma^1,\sigma^2,\sigma^3)$ are the Pauli matrices as given in Equation (\ref{Paulimatrices}).
For a given vector $p^{\mu}$ we then have
\begin{eqnarray}
\label{paadot}
p_{\a\da}&=&p_{\mu}\sigma^\mu_{\a\da}\nn \\
&=&p_0\one+\vec{p}\cdot\vec{\sig}\\
&=&\twobytwo{p_0+p_3}{p_1-ip_2}{p_1+ip_2}{p_0-p_3}\ ,
\end{eqnarray}
from which it follows that $p_{\mu}p^{\mu}=\det(p_{\a\da})$.
Hence $p^{\mu}$ is light-like $(p^2=0)$ if \mbox{$\det(p_{\a\da})=0$},
which in turn means that massless vectors are those for which
\be
\label{llt}
p_{\a\da}=\lambda_\a\lt_{\da}\ ,
\ee
for some spinors $\l_\a$ and $\lt_{\da}$. These spinors are unique up to
the scaling $(\l\,,\lt)\rightarrow (c\l\,,c^{-1}\lt)$ for a complex number $c$.

If we wish $p^\mu$ to be real in Lorentz signature (in which case $p_{\a\da}$ is hermitian) then we must take $\lt=\pm\bar{\l}$ where
$\bar{\l}$ is the complex conjugate of $\l$. The sign determines whether $p^{\mu}$ has
positive or negative energy. It is also possible (and sometimes useful) to consider other
signatures. In signature $+\!+--$ $\l$ and $\lt$ are real and independent while in
Euclidean signature $(+\!+\!++)$ the spinor representations are pseudoreal. Light-like
vectors cannot be real with Euclidean signature.

The formula for $p\cdot p=\det(p_{\a\da})$ generalises for any two momenta $p$ and $q$
and using the fact that $(\sigma^{\mu})_{\a\db}(\bar{\sigma}^{\nu})^{\db \a}=2\eta^{\mu\nu}$ we
can write the scalar product for two light-like vectors $p_{\a\da}=\l_\a\lt_{\da}$ and
$q_{\b\db}=\mu_{\b}\mut_{\db}$ as
\be
\label{dotprod}
2(p\cdot q)=\langle\l\,\mu\rangle[\mut\,\lt]\ .
\ee
This is the standard convention in the perturbative field theory literature and
differs from the conventions in \cite{witten,csreview} by a sign that is related to
the choice of how to contract indices using $\epsilon^{\a\b}$.

\subsection{Wavefunctions}

Once $p^{\mu}$ is given, the additional information involved in specifying $\l$
(and hence $\lt$ in complexified Minkowski space with real $p_{\alpha\dot{\alpha}}$) is equivalent to a choice of
wavefunction for a spin $1/2$ particle of momentum $p^{\mu}$. To see this, we
can write the massless Dirac equation for a negative-chirality spinor $\psi^\a$ as
\be
\label{chiraldiraceq}
i(\sig^{\mu})_{\a\da}\partial_{\mu}\psi^{\a}=0\ .
\ee
A plane wave $\psi^\a=\rho^\a e^{ip\cdot x}$ with constant $\rho^\a$ obeys this equation
iff $p_{\a\da}\rho^\a=0$ which implies that $\rho^\a=c\l^\a$.
Similar considerations apply for positive-chirality spinors and thus we can write
fermion wavefunctions of helicity\footnote{We will often use the terms chirality and helicity
interchangeably.} $\pm 1/2$ as
\begin{equation}
\label{fermiwave}
\begin{array}{cc}
\psi^{\da}=\lt^{\da} e^{ix_{\b\db}\l^\b\lt^{\db}}\ , &
\psi^{\a}=\l^{\a} e^{ix_{\b\db}\l^\b\lt^{\db}}
\end{array}
\end{equation}
respectively.

For massless particles of spin $\pm 1$ the usual method is to specify a
polarization vector $\epsilon^{\mu}$ (which we should be careful not to confuse with
 $\epsilon^{\a\b}$) in addition to their momentum and together with
the constraint $p_{\mu}\epsilon^{\mu}=0$. This constraint is equivalent to the
Lorentz gauge condition and deals with fixing the gauge invariance inherent in
gauge field theories. It is clear that if we add any multiple of $p^{\mu}$ to
$\epsilon^{\mu}$ then this condition is still satisfied and we have the
gauge invariance
\be
\label{gauginv}
\hat{\epsilon}^{\,\mu}=\epsilon^{\mu}+\omega\, p^{\mu}\ .
\ee
If one now has a decomposition of a light-like
vector particle with momentum $p_{\a\da}=\l_\a\lt_{\da}$ then one can take the
polarisation vectors to be \cite{witten} (see also \cite{Xu:1986xb,manganorev} and
references therein):
\begin{equation}
\label{polarvecs}
\epsilon^+_{\a\da}=\frac{\mu_\a\lt_{\da}}{\langle\mu\,\l\rangle}\ ,\quad
\epsilon^-_{\a\da}=\frac{\l_\a\mut_{\da}}{[\mut\,\lt]}\ ,
\end{equation}
for positive- and negative-helicity particles respectively. $\mu$ and $\mut$ are
arbitrary negative- and positive-chirality spinors (not proportional to $\l$ or $\lt$)
respectively and it is worth noting
that the positive-helicity polarization vector is proportional to the positive-helicity spinor
$(\lt_{\da})$ associated with the momentum vector $p^{\mu}$ while the negative-helicity
polarization vector is proportional to the negative-helicity one $(\l_\a)$. These polarization vectors
clearly obey the constraint $0=p^{\mu}\epsilon_{\mu}^{\pm}=p^{\a\da}\epsilon^{\pm}_{\a\da}$ since
$\langle\l\,\l\rangle=[\lt\,\lt]=0$ and are independent of $\mu$ and $\mut$ up to a gauge
transformation \cite{witten,csreview}.
The wavefunctions for positive and negative-helicity massless vector bosons can thus be written as \cite{csreview}
\begin{equation}
\label{bosewave}
A_{\a\da}^+=\epsilon^+_{\a\da}e^{ix_{\b\db}\l^\b\lt^{\db}}\ ,\quad
A_{\a\da}^-=\epsilon^-_{\a\da}e^{ix_{\b\db}\l^\b\lt^{\db}}\ .
\end{equation}
Spinless particles have wavefunction $\phi=e^{ix_{\a\da}\l^\a\lt^{\da}}$ as usual.

\subsection{Variable reduction}

One of the central motivations for all this song and dance is that we can use the results to
homogenise our description of scattering amplitudes. The plethora of variables that we had
before can simply be traded for the bi-spinors $\lambda$ and $\lt$ to yield the compact form
of a general scattering amplitude as
\be
\label{variablereduction}
\cA_{n}=\cA_n(\l_i,\lt_i,h_i)\,\delta^{(4)}\!\left(\sum_{i=1}^n\l_i^\a\lt_i^{\da}\right)\ ,
\ee
where $h_i$ is the helicity of the $i$th particle. In this scheme we can therefore
calculate amplitudes for the scattering of specific helicity configurations
of specific colour orderings of massless particles. The full amplitude is obtained by summing over all
helicity configurations and all appropriate colour orderings.

As a final remark in this section it is useful to note (and easy to show - see \cite{witten,csreview})
that under the scaling-invariance inherent in the decomposition of Eq.~(\ref{llt}), the wavefunction
of a massless particle of helicity $h$ scales as $c^{-2h}$ and thus obeys the condition
\be
\label{scalecond}
\left(\l^\a\frac{\partial}{\partial\l^\a}-\lt^{\da}\frac{\partial}{\partial\lt^{\da}}\right)
\psi(\l,\lt)=-2h\psi(\l,\lt)\ .
\ee
Similarly, the amplitude in Eq.~(\ref{variablereduction}) obeys
\be
\label{amplscale}
\left(\l^\a_i\frac{\partial}{\partial\l^\a_i}-\lt^{\da}_i\frac{\partial}{\partial\lt^{\da}_i}\right)
\cA_n(\l_i,\lt_i,h_i)=-2h_i\cA_n(\l_i,\lt_i,h_i)
\ee
for each $i$ separately.\footnote{The full expression
$\cA_n(\l_i,\lt_i,h_i)\,\delta^{(4)}\!\left(\sum_{i=1}^n\l_i^\a\lt_i^{\da}\right)$ also
obeys \eq{amplscale} \cite{witten}.}

The interested reader can find the Feynman rules for massless $\SU(N_c)$ Yang-Mills gauge theory
in the spinor helicity formalism in Appendix \ref{helicityrules}.

\section{Supersymmetric decomposition}
\label{susydecomp}

Supersymmetric field theories are in many ways very similar to the usual
Yang-Mills theories whose Feynman rules we wrote down at the start of
the chapter. The presence of this extra symmetry - supersymmetry - means
that the particles of the theories arrange themselves into supersymmetric
multiplets containing equal numbers of bosonic and fermionic degrees of
freedom and this can often give rise to great simplifications.

Maximally supersymmetric $(\cN\!=\!4)$ Yang-Mills for example, which
has the maximum amount of supersymmetry consistent with a gauge theory (\emph{i.e.} particles
with spin less than or equal to 1) in four dimensions, contains only 1 multiplet consisting
of 1 vector boson $A_{\mu}$ (2 real degrees of freedom (d.o.f.)), 6 real scalars $\phi^I$
(6 real d.o.f.) and 4 Weyl (\emph{i.e.} chiral) fermions $\chi_{\alpha}$ (8 real d.o.f.) which
lives in the adjoint of the gauge group. This multiplet is often written in a helicity-basis
 (the helicities of the particles here are $h=(-1,-1/2,0,1/2,1)$) as
$(A^-,\chi^-,\phi,\chi^+,A^+)=(1,4,6,4,1)$ and is often referred to as the
adjoint multiplet of $\cN\!=\!4$. The meaning of this notation is that one of the degrees of freedom of the
vector boson is associated with a negative-helicity $(-1)$ state and the other with a positive-helicity $(+1)$ state. Similarly,
the chiral fermions are split into two, with 4 degrees of freedom being associated with
helicity $-1/2$ and 4 with helicity $+1/2$. The scalars are of course spinless and thus associated with
helicity 0. Other common multiplets in four dimensions include the
vector multiplet of $\cN\!=\!2$ $(1,2,2,2,1)$ - which consists of 1 vector, 2 fermions and 2 scalars -
the hyper multiplet of $\cN\!=\!2$ $(0,2,4,2,0)$ and the vector $(1,1,0,1,1)$ and chiral $(0,1,2,1,0)$
multiplets of $\cN\!=\!1$ supersymmetry.

The existence of these supersymmetric multiplets generally leads to a better control of
the field theory in question, and most-importantly for us a greater control of its perturbative expansion. Heuristically, fermions
propagating in loops give terms which have the opposite sign to bosons and
the exact matching of the bosonic and fermionic degrees of freedom leads to cancellations in the
ultraviolet divergences that plague non-supersymmetric field theories. In particular, $\cN\!=\!4$ super-Yang-Mills
is believed to be completely finite in four dimensions as well as having quantum-mechanical conformal-invariance.
Massless QCD on the other hand is classically conformally-invariant, although this is broken by quantum effects as is well-known
 from the existence of its one-loop (and higher) $\beta$-function. QCD is also UV divergent at
loop-level and thus must be renormalised order-by-order in perturbation theory.

$\cN\!=\!4$ super-Yang-Mills
has the most striking features of these four-dimensional supersymmetric gauge theories and we will
 concern ourselves with this theory as well as $\cN\!=\!1$ super-Yang-Mills.
In fact, the results for $\cN\!=\!1$ amplitudes in Chapter 2 also apply to
certain $\cN\!=\!2$ amplitudes by virtue of the fact that the $\cN\!=\!2$ hyper multiplet is
twice the $\cN\!=\!1$ chiral multiplet and the $\cN\!=\!2$ vector multiplet is equal to an
$\cN\!=\!1$ vector multiplet plus an $\cN\!=\!1$ chiral multiplet.

As we have already mentioned, we will mostly be concerned with gluon scattering in
$\SU(N_c)$ Yang-Mills theories (including QCD) and thus will only consider this case here. At tree-level
it is easy to see that gluon scattering amplitudes are the same in QCD as they are in
$\cN\!=\!4$ super-Yang-Mills theory. This is because vertices connecting
gluons to fermions or scalars in these theories couple gluons to pairs of these particles.
Thus one cannot create fermions or scalars internally without also creating a loop \cite{Parke:1985pn}. These
QCD scattering amplitudes therefore have a `hidden' $\cN\!=\!4$ supersymmetry:
\be
\label{treesusy}
\cA^{\textrm{tree}}_{\textrm{QCD}}=\cA^{\textrm{tree}}_{\cN=4}\ .
\ee
The same can of course be said about any supersymmetric field
theory with adjoint fields when one is concerned with
the scattering of external gluons at tree-level. We thus have the
more general result that
\be
\label{treesusy2}
\cA^{\textrm{tree}}_{\textrm{QCD}}=\cA^{\textrm{tree}}_{\cN=4}
=\cA^{\textrm{tree}}_{\cN=2}=\cA^{\textrm{tree}}_{\cN=1}\ .
\ee

At one-loop we can of course have other particles propagating in the
loop, but where gluon-scattering only is concerned we can still find a
supersymmetric decomposition. It is:
\be
\label{loopsusy}
\cA^{\textrm{one-loop}}_{\textrm{QCD}}=\cA^{\textrm{one-loop}}_{\cN=4}
-4\cA^{\textrm{one-loop}}_{\cN=1,\textrm{chiral}}+2\cA^{\textrm{one-loop}}_{\textrm{scalar}}\ .
\ee
In words this says that an all-gluon scattering amplitude in QCD at one loop
can be decomposed into 3 terms: Firstly a term where an $\cN\!=\!4$
multiplet propagates in the loop. Secondly a term
where an $\cN\!=\!1$ chiral multiplet propagates in the loop and lastly a term in
pure Yang-Mills where we only have 2 real scalars (or one complex scalar)
in the loop. This is easily seen due to the multiplicities of the various
multiplets in question:
$(1,0,0,0,1)=(1,4,6,4,1)-4(0,1,2,1,0)+2(0,0,1,0,0)$.

As we have already discussed
many times, the LHS of \eq{loopsusy} is extremely complicated to evaluate. However, the
3 pieces on the RHS are relatively much easier to deal with. The first two pieces
 are contributions coming from supersymmetric
field theories and these extra (super)-symmetries greatly help to reduce the
complexity of the calculations there. Much of the difficulty is thus pushed into
the last term which is the most complex of the three, but is still far easier to
evaluate than the LHS.

It is therefore clear that supersymmetric field theories are not only simpler
toy models with which to try to understand the gauge theories of the standard
model, but relevant theories in themselves which contribute parts (and sometimes
the entireity in the case of certain tree-level amplitudes \eq{treesusy2}) of the
answer to calculations in theories such as QCD. These supersymmetric decompositions
will be of great assistance to us in our quest to understand the hidden simplicity of
scattering amplitudes and in order to perform actual calculations.

For more information on supersymmetric field theories see any one of
a multitude of books, papers and reviews including
\cite{Wess:1992cp,Sohnius,Alvarez-Gaume:1996mv,Figueroa-O'Farrill:2001tr}.

\section{Supersymmetric Ward identities}
\label{SWI}

As we can now see, for a large number of scattering amplitudes in gauge theories we can
reduce the complexity of our problem by considering an appropriate colour-ordered sub-amplitude
that only depends on the positive- and negative-helicity spinors associated with the
external momenta (we usually drop the $h_i$ dependence of \eq{variablereduction} and leave it
as being implicit in the definition of the amplitude being considered). Using our `hidden' (or not, depending
on the theory in question) supersymmetry we are now in a position to learn something about
the scattering amplitudes in question. The following is also nicely reviewed in a number
of places including \cite{manganorev,dixon} and was first considered in
\cite{DeWitt:1967uc,Grisaru:1976vm,Grisaru:1977px,Parke:1985pn}. See also \emph{e.g.}
\cite{Bidder:2005in} for a recent application of supersymmetric Ward identities to loop amplitudes.

\subsection{$\cN\!=\!1$ SUSY constraints}

Let us consider what is in some ways the simplest possible setup, an
adjoint (vector) multiplet in an $\cN\!=\!1$ supersymmetric field theory where the SUSY
is unbroken. This $\cN\!=\!1$ theory has only one supercharge $Q(\eta)$
that generates the supersymmetry with $\eta$ being the fermionic parameter of the
transformation \cite{Wess:1992cp}. Because supersymmetry is unbroken we know that $Q$ must annihilate
the vacuum: $Q(\eta)|0\rangle=0$. This in turn gives rise to the following
supersymmetric Ward identity (SWI)
\be
\label{unbrokensusy}
0=\langle 0|[Q(\eta),\Psi_1\ldots\Psi_n]|0\rangle =
\sum_{i=1}^n\langle 0|\Psi_1\ldots[Q(\eta),\Psi_i]\ldots\Psi_n|0\rangle\ ,
\ee
for some fields $\Psi_i$. In addition, if we use a suitable helicity basis
in which we have a massless vector $A^{\pm}$ and a massless
spin 1/2 fermion $\chi^{\pm}$, then $Q(\eta)$ acts on the doublet $(A,\chi)$
(\emph{i.e.} $(A^-,\chi^-,0,\chi^+,A^+)$ in the notation of the previous subsection) as
\cite{Grisaru:1976vm,Grisaru:1977px}:
\begin{eqnarray}
\label{susyxfms}
\left[Q(\eta),A^{\pm}(p)\right]&=&\mp\Gamma^{\pm}(p,\eta)\chi^{\pm}\ ,\nonumber \\
\left[Q(\eta),\chi^{\pm}(p)\right]&=&\mp\Gamma^{\mp}(p,\eta)A^{\pm}\ ,
\end{eqnarray}
for some momentum $p$ associated with these states. $\Gamma(\eta,p)$ is linear
in $\eta$ and can be constructed by using the Jacobi identity
\be
\label{jacobisusy}
\left[\left[Q(\eta),Q(\zeta)\right],\Psi(p)\right]+\left[\left[Q(\zeta),\Psi(p)\right],Q(\eta)\right]
+\left[\left[\Psi(p),Q(\eta)\right],Q(\zeta)\right]=0
\ee
and the SUSY algebra relation $[Q(\eta),Q(\zeta)]=-2i\bar{\eta}P\!\!\!\!\slash\zeta$,
where $P\!\!\!\!\slash=\gamma^{\mu}P_{\mu}$ as usual.
By considering \eq{jacobisusy} for any of the chiral fields ($A^+(p)$ for example),
we can readily deduce that
\be
\label{jacobisusy}
\Gamma^+(p,\eta)\Gamma^-(p,\zeta)-\Gamma^+(p,\zeta)\Gamma^-(p,\eta)=-2i\bar{\eta}
p\!\!\!\slash\zeta\ ,
\ee
which can be solved to give (in the notation of \S \ref{shsection}) \cite{Grisaru:1976vm,Grisaru:1977px,manganorev,dixon}:
\begin{eqnarray}
\label{gammasol}
\Gamma^{+}(p,q,\vartheta)=\vartheta [p\, q]\ ,\qquad \Gamma^-(p,q,\vartheta)=\vartheta \langle p\, q\rangle\ .
\end{eqnarray}
In this expression we have written $p=\lambda_p\lt_p$ and our parameter
$\eta$ in terms of a Grassmann parameter $\vartheta$ and an arbitrary
reference momentum $q=\lambda_q\lt_q$. We have also used the shorthand notation
$\langle \lambda_p\,\lambda_q\rangle=\langle p\, q\rangle$ and
$[\lt_p\,\lt_q]=[p\, q]$ which will often be employed henceforth.

Now consider \eq{unbrokensusy} with $\Psi_1=\chi_1^+$ and $\Psi_i=A_i^+$
for $i\neq 1$:
\begin{eqnarray}
\label{allplus}
0 &=&\langle 0|[Q(\eta(q,\vartheta)),\chi_1^+(p_1)A_2^+(p_2)\ldots A_n^+(p_n)]|0\rangle\nonumber \\
&=&-\,\Gamma^-(p_1,q,\vartheta)\langle 0|A_1^+(p_1)A_2^{+}(p_2)\ldots A_n^+(p_n)|0\rangle\nonumber \\
& &{}\!+\Gamma^-(p_2,q,\vartheta)\langle 0|\chi_1^+(p_1)\chi_2^{+}(p_2)\ldots A_n^+(p_n)|0\rangle\nonumber \\
& &{}\,\,\vdots\nonumber \\
& &{}\!+\Gamma^-(p_n,q,\vartheta)\langle 0|\chi_1^+(p_1)A_2^{+}(p_2)\ldots \chi_n^+(p_n)|0\rangle\nonumber \\
&=&-\,\Gamma^-(p_1,q,\vartheta)\cA_n(A_1^+,A_2^+,\ldots,A_n^+)\nonumber \\
& &{}\!+\Gamma^+(p_2,q,\vartheta)\cA_n(\chi_1^+,\chi_2^+,\ldots,A_n^+)\nonumber \\
& &{}\,\,\vdots\nonumber \\
& &{}\!+\Gamma^+(p_n,q,\vartheta)\cA_n(\chi_1^+,A_2^+,\ldots,\chi_n^+)\ .
\end{eqnarray}
As all of the couplings of fermions to vectors conserve helicity (you always get
one fermion of each helicity coupling to a vector), the $n-1$ terms involving two
fermions and $n-2$ gluons must vanish and thus the first term involving only
gluons of positive helicity must vanish too $\cA_n(A_1^+,A_2^+,\ldots,A_n^+)=0$. Since supersymmetry
commutes with colour we can write our amplitudes as colour-ordered ones
straight away and then the relations apply to each colour-ordered amplitude separately.

If we consider the case where we have one negative-helicity in our SWI so that
$\Psi_1=\chi_1^+$, $\Psi_2=A_2^-$ and $\Psi_i=A_i^+$ for $i\neq 1,2$ for example, then
we can also show that all amplitudes with one negative-helicity particle and $n-1$ positive-helicity particles vanish. This is so both for the case of all gluon scattering and the case of
$n-2$ gluons and two fermions of opposite helicities. With more than one negative-helicity (such as
$\Psi_1=\chi_1^+$, $\Psi_2=A_2^-$, $\Psi_3=A_3^-$ and $\Psi_i=A_i^+$ for $i\neq 1,2,3$) we can
start to relate non-zero amplitudes to each other. In all of these cases it is useful to
remember that the reference momentum $q$ is arbitrary and can thus be taken to be one
of the external momenta ($q=p_i$) for example at any given stage in order to simplify the
calculations and deduce useful results.

\subsection{Amplitude relations}

\label{amplituderels}

Some of the useful relations that we can obtain are:
\begin{eqnarray}
\label{vanishing}
\cA_n^{\textrm{SUSY}}(1^{\pm},2^{\pm},\ldots,n^{\pm})&=&0\ ,\\
\label{vanishing2}
\cA_n^{\textrm{SUSY}}(1^{\mp},2^{\pm},\ldots,n^{\pm})&=&0\ ,
\end{eqnarray}
for any spins of the particles involved and \cite{Bern:zx}
\begin{eqnarray}
\label{fermiscalar}
\!\!\!\!\!\!\!\!\!\!\!\cA_n^{\textrm{SUSY}}\!(A_i^{-},\ldots,\chi_r^-,\ldots,\chi_s^+,\ldots)\!\!&=&\!\!
\frac{\langle i\,s\rangle}{\langle i\,r\rangle}\cA_n^{\textrm{SUSY}}\!(A_i^{-},\ldots,A_r^-,\ldots,A_s^+,\ldots)
\, ,\\
\label{scalarfermi}
\!\!\!\!\!\!\!\!\!\!\!\cA_n^{\textrm{SUSY}}\!(A_i^{-},\ldots,\phi_r^-,\ldots,\phi_s^+,\ldots)\!\!&=&\!\!
\frac{\langle i\,s\rangle^2}{\langle i\,r\rangle^2}\cA_n^{\textrm{SUSY}}\!(A_i^{-},\ldots,A_r^-,\ldots,A_s^+,\ldots)
\, ,
\end{eqnarray}
where we have also played the same game with an $\cN\!=\!2$ Vector multiplet in order to include
scalars $\phi$.
These relations hold order-by-order in the loop
expansion of supersymmetric field theories as no perturbative
approximations were made in deriving them, and by virtue of
\eq{treesusy2} they apply directly to tree-level QCD amplitudes involving gluons. It
turns out that tree-level QCD amplitudes involving fundamental quarks can
also be obtained from \eq{fermiscalar} because of relations between
sub-amplitudes involving gluinos (\emph{i.e.} fermionic superpartners of
gluons in an adjoint multiplet such as the $\chi$ above) and those involving fundamental quarks
\cite{Mangano:1987kp,manganorev}.

Equations \eq{vanishing} and \eq{vanishing2} amount to the statement
that for any supersymmetric theory with only adjoint fields, the
`all-plus' and `all-minus' helicity amplitudes must vanish and
the amplitudes with one minus and $n\!-\!1$ plusses (or vice-versa) must also
vanish. The same statement holds for the tree-level gluon scattering amplitudes
of QCD. As a result of this, the first non-vanishing set of amplitudes in
 a supersymmetric theory are the ones with two negative helicities and
$n\!-\!2$ positive helicities. These are thus termed the Maximally Helicity Violating
(MHV) amplitudes. Their parity conjugates, the amplitudes with two positive
helicities and $n\!-\!2$ negative helicities are similarly non-vanishing and
are sometimes termed \emph{googly} MHV (or $\overline{\mathrm{MHV}}$) amplitudes \cite{witten}. Similarly,
amplitudes with three negative helicities and $n\!-3$ positive helicities
are termed next-to-MHV (NMHV) amplitudes. The next ones are thus called
next-to-next-to-MHV (NNMHV) and so on.

The tree-level MHV amplitudes for gluon scattering, proposed at $n$-point in \cite{mhv} and then
proved in \cite{bg}, are given by \eq{MHV} or by
\be
\label{mhvproper}
\cA_n(1^+,\ldots,i^-,\ldots,j^-,\ldots,n^+)=\frac{\langle i\,j\rangle^4}
{\prod_{k=1}^{n}\langle k\,k\!+\!1\rangle}\ ,
\ee
up to a factor. $i$ and $j$ are the gluons of negative helicity and the amplitude
obeys \eq{amplscale}. The amplitude is cyclic in the ordering of the gluons and so
the $n+1^{\mathrm{th}}$ spinor appearing in the denominator of \eq{mhvproper} just denotes
the spinor of the $1^{\mathrm{st}}$ gluon. Note
in particular that this function is entirely `holomorphic' in the negative-helicity
spinors $\lambda$ - \emph{i.e.} it does not depend on any of the $\lt\,$s - and this will
be important to us presently. We will not discuss NMHV and other amplitudes yet
except to mention that they \emph{do} depend on the
$\lt\,$s.

\section{Twistor space}

There is a way in which we can understand some of the properties of amplitudes
that we have discussed above such as
the vanishing of certain helicity configurations and the simple structure
of the MHV amplitudes and that is by going to twistor space \cite{witten}.
This has two primary motivations. One is that the conformal symmetry group
has a rather exotic representation in terms of the $\lambda$ and $\lt$ variables
and the other is that the scaling-invariance mentioned under equation \eq{llt}
has an opposite action on the holomorphic spinors $\lambda$ compared with the anti-holomorphic
spinors $\lt$. It would be nice to put the conformal group\footnote{The gluon
amplitudes at tree-level are invariant under the full conformal group rather than
just the Poincar\'{e} group. This is because of the classical conformal
invariance of both massless QCD and any of the other supersymmetric
field theories that we have been considering. Amplitudes in some of these
supersymmetric theories (especially $\cN\!=\!4$ Yang-Mills) also have
quantum conformal invariance.} into a more standard
representation and it may also be nice to have the same scaling for the
negative and positive-helicity spinors.

In terms of the spinors we have already introduced in
\S \ref{shsection}, the conformal generators are \cite{witten}
\begin{eqnarray}
\label{momentumop}
P_{\a\da}&=&\lambda_\a\lt_{\da}\ ,\\
\label{lorentzop}
J_{\a\b}&=&\frac{i}{2}\left(\lambda_\a\frac{\partial}{\partial\lambda^\b}
+\lambda_\b\frac{\partial}{\partial\lambda^\a}\right)\ ,\\
\label{lorentop2}
\tilde{J}_{\da\db}&=&\frac{i}{2}\left(\lt_{\da}\frac{\partial}{\partial\lt^{\db}}
+\lt_{\db}\frac{\partial}{\partial\lt^{\da}}\right)\ ,\\
\label{dilatationop}
D&=&\frac{i}{2}\left(\lambda^\a\frac{\partial}{\partial\lambda^\a}+
\lt^{\da}\frac{\partial}{\partial\lt^{\da}}+2\right)\ ,\\
\label{specialop}
K_{\a\da}&=&\frac{\partial^2}{\partial\lambda^\a\partial\lt^{\da}}\ ,
\end{eqnarray}
where $P_{\a\da}$ is the momentum operator, $J_{\a\b}$ and $\tilde{J}_{\da\db}$ the Lorentz
generators, $D$ the dilatation operator and $K_{\a\da}$ the generator of
special conformal transformations. These give rise to the algebra of the
conformal group as
\begin{eqnarray}
\label{confalg}
\big[J_{\a\b},P_{\gamma\dot{\gamma}}\big]&=&\frac{i}{2}\left(\epsilon_{\b\gamma}P_{\a\dot{\gamma}}+
\epsilon_{\a\gamma}P_{\b\dot{\gamma}}\right)\ ,\nonumber \\
\big[\tilde{J}_{\da\db},P_{\gamma\dot{\gamma}}\big]&=&\frac{i}{2}\left(\epsilon_{\db\dot{\gamma}}P_{\gamma\da}+
\epsilon_{\da\dot{\gamma}}P_{\gamma\db}\right)\ ,\nonumber \\
\big[J_{\a\b},J_{\gamma\delta}\big]&=&\frac{1}{4}\left(\epsilon_{\gamma\a}J_{\b\delta}+\epsilon_{\gamma\b}J_{\a\delta}
+\epsilon_{\delta\a}J_{\b\gamma}+\epsilon_{\delta\b}J_{\a\gamma}\right)\ ,\nonumber \\
\big[\tilde{J}_{\da\db},\tilde{J}_{\dot{\gamma}\dot{\delta}}\big]&=&\frac{1}{4}\left(\epsilon_{\dot{\gamma}\da}\tilde{J}_{\db\dot{\delta}}
+\epsilon_{\dot{\gamma}\db}\tilde{J}_{\da\dot{\delta}}+\epsilon_{\dot{\delta}\da}\tilde{J}_{\db\dot{\gamma}}+
\epsilon_{\dot{\delta}\db}\tilde{J}_{\da\dot{\gamma}}\right)\ ,\nonumber \\
\big[D,P_{\a\da}\big]&=&\frac{i}{2}P_{\a\da}\ ,\nonumber \\
\big[K_{\a\da},D\big]&=&2K_{\a\da}\ ,\nonumber \\
\big[J_{\a\b},K_{\gamma\dot{\gamma}}\big]&=&\frac{i}{2}\left(\epsilon_{\gamma\a}K_{\b\dot{\gamma}}+
\epsilon_{\gamma\b}K_{\a\dot{\gamma}}\right)\ ,\nonumber \\
\big[J_{\da\db},K_{\gamma\dot{\gamma}}\big]&=&\frac{i}{2}\left(\epsilon_{\dot{\gamma}\da}K_{\gamma\dot{\b}}+
\epsilon_{\dot{\gamma}\db}K_{\gamma\da}\right)\ ,\nonumber \\
\big[K_{\a\da},P_{\b\db}\big]&=&i\left(\epsilon_{\a\b}\tilde{J}_{\da\db}+\epsilon_{\da\db}
J_{\a\b}+\epsilon_{\a\beta}\epsilon_{\dot{\beta}\da}D\right)\ ,
\end{eqnarray}
with all other commutators being zero. However, as can be seen from \eq{momentumop}-\eq{specialop},
the momentum operator is a multiplication operator, the Lorentz generators are first order
homogeneous differential operators, the dilatation operator an inhomogeneous first order differential
operator and the special conformal generator a degree two differential operator. We have quite
a mix.

We can in fact reduce these to a more standard representation by performing a
the transformation \cite{witten,Penrose}
\begin{eqnarray}
\label{halffourier}
\lt_{\da}&\rightarrow &i\frac{\partial}{\partial\mu^{\da}}\ ,\nonumber \\
\frac{\partial}{\partial\lt^{\da}}&\rightarrow &i\mu_{\da}\ .
\end{eqnarray}
This breaks the symmetry between $\lambda$ and $\lt$ as we have chosen to transform
one rather than the other, but giving the advantage that all the generators become
first order differential operators:
\begin{eqnarray}
\label{momentumop1}
P_{\a\da}&=&i\lambda_\a\frac{\partial}{\partial\mu^{\da}}\ ,\\
\label{specialop1}
K_{\a\da}&=&i\mu_{\da}\frac{\partial}{\partial\lambda^\a}\ ,\\
\label{lorentzop1}
J_{\a\b}&=&\frac{i}{2}\left(\lambda_\a\frac{\partial}{\partial\lambda^\b}
+\lambda_\b\frac{\partial}{\partial\lambda^\a}\right)\ ,\\
\label{lorentop12}
\tilde{J}_{\da\db}&=&\frac{i}{2}\left(\mu_{\da}\frac{\partial}{\partial\mu^{\db}}+
\mu_{\db}\frac{\partial}{\partial\mu^{\da}}\right)\ ,\\
\label{dilatationop1}
D&=&\frac{i}{2}\left(\lambda^\a\frac{\partial}{\partial\lambda^\a}-\mu^{\da}\frac{\partial}{\partial\mu^{\da}}\right)\ .
\end{eqnarray}
The scaling properties of $\lambda$ and $\mu$ are also changed such that there
is an invariance under
\be
\label{twistorscaling}
(\lambda\, ,\mu)\rightarrow (c\lambda\, ,c\mu)\ ,
\ee
for a complex number $c$, and the amplitude scalings \eq{amplscale} become
\be
\label{amplscale2}
\left(\lambda^\a_i\frac{\partial}{\partial\lambda^\a_i}+
\mu^{\da}_i\frac{\partial}{\partial\mu^{\da}_i}\right)\tilde{\mathcal{A}}_n(\lambda_i,\mu_i,h_i)=
-(2h_i+2)\tilde{\cA}_n(\lambda_i,\mu_i,h_i)\ ,
\ee
where $\tilde{\cA}_n$ is the appropriately transformed amplitude.

This transformation is perhaps easiest to understand in signature \mbox{$+\!+--$}. In this
case one can consider $\lambda^{\a}$ and $\mu^{\da}$ to be real and independent and thus
they parametrise a copy of $\mathbb{R}^4$. The scaling \eq{twistorscaling} is then
 a real scaling and reduces the space to real-projective three-space $\mathbb{RP}^3$
and the transform \eq{halffourier} is implemented by a
\mbox{`1/2-Fourier'} transform analagous to that encountered in quantum mechanics \cite{witten}:
\be
\label{qmfourier}
\tilde{f}(\mu)=\int\!\!\frac{d^2\lt}{(2\pi)^2}\,e^{i\mu^{\da}\lt_{\da}}f(\lt)\ .
\ee
In other signatures (such as Minkowski space) it may be more natural to regard $\lambda$ and $\mu$ as being complex and
independent. They thus parametrise a copy of $\mathbb{C}^4$ which reduces to
$\mathbb{CP}^3$ under the scaling \eq{twistorscaling}. These spaces - $\mathbb{RP}^3$ and
$\mathbb{CP}^3$ - were called \emph{twistor} spaces by Penrose \cite{Penrose} and we will
often use coordinates $Z^I$ with $I=1\ldots 4$ on them thus combining $\lambda^\a$ and
$\mu^{\da}$ together. One should
really refer to `real/complex projective twistor space' respectively, but
we will denote them all as being twistor space ($\mathbb{T}$) and let the context
dictate what we mean by that.

In the complex cases,
the choice of a contour for the transformation as given
by \eq{qmfourier} is not necessarily clear and it seems necessary to take
the more sophisticated approach of Penrose and use Dolbeault- or sheaf-cohomology
\cite{Penrose}. Na\"{\i}vely, this interprets
the integrand and measure of \eq{qmfourier} as a $(0,2)$-form on
twistor space, while equation \eq{amplscale2} suggests that the amplitudes
are best thought of not as functions, but sections of a line bundle
$\mathcal{L}_h$ of degree $-2h-2$, $\mathcal{L}_h=\mathcal{O}(-2h-2)$ for
each $h$. The amplitudes are thus elements of $H^{(0,2)}(\mathbb{CP}^{3^\prime},\mathcal{O}(-2h-2))$ \cite{witten}.%
\footnote{Here we follow \cite{witten} and write $\mathbb{CP}^{3^\prime}$ instead of $\mathbb{CP}^3$ because
$H^{(0,2)}(\mathbb{CP}^3,\mathcal{O}(-2h-2))=0$ and we should really work with a suitable
open set of $\mathbb{CP}^3$ (which we denote with a prime) rather then all of
twistor space.}

The transformation of wavefunctions to twistor space is in some ways more complex. One
cannot perform such a na\"{i}ve `1/2-Fourier' transform in essence because the wavefunctions
are defined by being solutions to the massless free wave equations and so one must see
how one can solve these in twistor space. It turns out that these solutions can be
written as integrals of functions of degree $2h-2$ and the wavefunctions are then described by
elements of the $\bar{\partial}$-cohomology group $H^{(0,1)}(\mathbb{CP}^{3^{\prime}},\mathcal{O}(2h-2))$
- see \emph{e.g.} \cite{witten,Penrose:1972ia,Penrose:1968me,penroseoxford} for details.

In particular these descriptions mean that scattering amplitudes with specific external states
make sense in twistor space. In a usual field theory construction one would multiply
a momentum-space scattering amplitude with its momentum-space wavefunctions and
integrate over all momenta to create a scattering amplitude with specific
external states in the position-space representation. If the wavefunctions in
position-space satisfying the appropriate free wave equations are given by
\mbox{$\varphi_i(x)=\int d^4p_i\delta(p_i^2)e^{ip_i\cdot x}{\phi}_i(p_i)$}, then we have schematically
$A(\varphi_i)=\int (\prod d^4p_i\delta(p_i^2)e^{ip_i\cdot x}{\phi}_i(p_i))\tilde{A}(p_i)$.

In twistor
space, multiplying an amplitude in \mbox{$H^{(0,2)}(\mathbb{CP}^{3^\prime},\mathcal{O}(-2h-2))$}
with a wavefunction which is in \mbox{$H^{(0,1)}(\mathbb{CP}^{3^\prime},\mathcal{O}(2h-2))$} gives an
element of $H^{(0,3)}(\mathbb{CP}^{3^\prime},\mathcal{O}(-4))$. The natural measure on $\mathbb{CP}^3$
is a $(3,0)$-form of degree 4 (it is in fact the $\Omega'$ of \eq{componenthcs}),
and so the final integral will be of a $(3,3)$-form of degree 0 which makes
sense (\emph{i.e.} the integrand is a top-form on twistor space invariant under \eq{twistorscaling})
as an integral over $\mathbb{CP}^{3^{\prime}}$. Doing this for each external particle gives the
required scattering amplitude in position-space.

Following the original suggestions
of Nair \cite{Nair}, there is a similar construction which is
particularly apt for amplitudes in $\cN\!=\!4$ Yang-Mills. In this case, particles are described by $\lambda$, $\lt$ and an
additional spinless fermionic variable $\eta_A$ with $A=1,\ldots,4$ in the $\mathbf{\bar{4}}$ representation
of the $R$-symmetry group $\SU(4)_R$ of $\cN\!=\!4$ Yang-Mills. The spacetime symmetry group in this case is no-longer the
usual conformal group, but the superconformal group $\PSU(2,2|4)$ and one can write down generators
in terms of $\lambda$, $\lt$ and $\eta$ which are again in a somewhat exotic form. After a Penrose
transform to super-twistor space, which just consists of \eq{halffourier} plus
\begin{eqnarray}
\label{superhalffourier}
\eta_A &\rightarrow & i\frac{\partial}{\partial\psi^A}\nonumber\\
\frac{\partial}{\partial\eta_A} &\rightarrow & i\psi^A\ ,
\end{eqnarray}
all superconformal generators similarly become first order differential operators
and the space spanned by $\lambda^\a$, $\mu^{\da}$ and $\psi^A$ is $\mathbb{RP}^{3|4}$ or
$\mathbb{CP}^{3|4}$. The scaling invariance of super-twistor space is:
\be
\label{supertwistorscaling}
(Z^I,\psi^A)\rightarrow(cZ^I,c\,\psi^A)\ .
\ee

In this case, the helicity operator
\be
\label{helicityop}
h=1-\frac{1}{2}\eta_A\frac{\partial}{\partial\eta_A}
\ee
modifies the scaling relation \eq{amplscale2} so that it becomes
\be
\label{superscaling}
\left(Z_i^I\frac{\partial}{\partial Z_i^I}+\psi_i^A\frac{\partial}{\partial\psi_i^A}\right)
\tilde{\mathcal{A}}_n(\lambda_i,\mu_i,\eta_A,h_i)=0\ ,
\ee
and so the scattering amplitudes are elements of $H^{(0,2)}(\mathbb{CP}^{3|4^{\prime}},\mathcal{O}(0))$.

On super-twistor space, the wavefunctions are now elements of
\mbox{$H^{(0,1)}(\mathbb{CP}^{3|4^{\prime}},\mathcal{O}(0))$} and can be
given explicitly for a particle of helicity $h$ by \cite{witten,BerkWitt2,csreview,Gao:2006mw}
\be
\label{twistorwavefunction}
\phi(\lambda,\mu,h)=\bar{\delta}(\langle\lambda\, ,\pi\rangle)
\left(\frac{\lambda}{\pi}\right)^{2h-1}\!\!\!\exp\left({i[\tilde{\pi}\, ,\mu]\frac{\pi}{\lambda}}\right)
g_h(\psi)\ ,
\ee
where $g_h(\psi)$ is simply a factor of $2-2h$ $\psi\,$s.\footnote{This factor of $\psi^I$ is
precisely what converts the wavefunctions from being of
degree $2h-2$ to being of degree 0. One might also wonder why the power of $\l/\pi$ is
only $2h-1$ and not $2h-2$ given that the wavefunctions on $\mathbb{CP}^3$ (\emph{i.e.} with the factor of
$g_h$ omitted) are of degree $2h-2$. This is because the holomorphic delta function is of degree $-1$ and
thus gives the correct scaling properties overall.}  For example,
for a positive-helicity gluon $g_h$ is 1 while for a negative-helicity gluon it is $\psi^1\psi^2\psi^3\psi^4$.
(In fact it is just the factor of $\psi$ that the associated
state multiplies in the expansion of the superfield $\cA$ in \eq{Aexpansion}.) $\bar{\delta}$ is
a `holomorphic' delta function which is a $(0,1)$-form given by $\bar{\delta}(f)=\delta^{(2)}\!(f)d\bar{f}$
for any holomorphic function $f$ - see Appendix \ref{traceappendix} for a more detailed discussion.

In this case, the multiplication of scattering amplitude and wavefunction leads to
an element of $H^{(0,3)}(\mathbb{CP}^{3|4^{\prime}},\mathcal{O}(0))$ and the
volume form is a $(3,0)$-form of degree 0 (explicitly given by \eq{volform}), so the
result makes sense (again as a scaling invariant top-form) to be integrated over $\mathbb{CP}^{3|4^{\prime}}$ and gives
the scattering amplitude in position-space.

For our treatment of amplitudes, we will generally use the definition \eq{qmfourier} and signature $+\!+\!--$
and interpret our results in other signatures when necessary. It is also worth mentioning that
we have glossed over
many subtleties in the considerations above such as the real nature of momenta
already alluded to in \S \ref{shsection}, and the exclusion of the `point at
infinity' in twistor space (\emph{i.e.} the use of $\mathbb{T}^{\prime}$ rather than $\mathbb{T}$).
For more details on all these and more detailed discussions of
twistor Theory we refer the reader to
\cite{Penrose,witten,Penrose:1972ia,Penrose:1985jw,Penrose:1986ca,Huggett:1986fs,penroseoxford,Penrose:1968me}
and related references.

\subsection{Amplitude localisation}
\label{localise}

Interpreting \eq{qmfourier} as the way to transform amplitudes into
twistor space, we are now ready to see what the tree-level MHV amplitudes look like there.
If we recall that these amplitudes depend \emph{only} on the negative-helicity spinors $\lambda_i$, the transformed amplitudes are \cite{witten}:
\begin{eqnarray}
\label{mhvxfm}
\tilde{\cA}_n^{\textrm{MHV}}(\lambda_i,\mu_i)\!\!\!&=&\!\!\!\int\!\prod_{j=1}^{n}\frac{d^2\tilde{\lambda}_j}{(2\pi)^2}
e^{i\mu_{j\,\da}\lt^{\da}_j}
\,\delta^{(4)}\!\!\left(\sum_{k=1}^n\lambda_k\lt_k\right)\cA_n^{\textrm{MHV}}
(\lambda_i,\lt_i)\nonumber \\
\!\!\!&=&\!\!\!\int\!\! d^4x\int\!\prod_{j=1}^{n}\frac{d^2\tilde{\lambda}_j}{(2\pi)^2}
\,\exp{\left(i\sum_{k=1}^n\mu_{k\,\da}\lt^{\da}_k\right)}
\exp{\left(ix_{\a\da}\sum_{k=1}^n\lambda_k^\a\lt_k^{\da}\right)}\cA_n^{\textrm{MHV}}(\lambda_i)
\nonumber \\
\!\!\!&=&\!\!\!\int d^4x\prod_{j=1}^n\delta^{(2)}\!\!\left(\mu_{j\,\da}+x_{\a\da}\lambda_j^\a\right)
\cA_n^{\textrm{MHV}}(\lambda_i)\ .
\end{eqnarray}
In the second line we have used a standard position-space representation for the
delta function of momentum conservation and then in the third we have similarly
interpreted the $\lt$ integrals as delta functions. The MHV amplitudes are thus
supported only when \mbox{$\mu_{j\,\da}+x_{\a\da}\lambda_j^\a=0$} for all
$j$ and for $\da=1,2$. For each $x_{\a\da}$ these equations define a curve
of degree one and genus zero in $\mathbb{RP}^3$ or $\mathbb{CP}^3$ (depending
on whether the variables are real or complex) which is in fact an $\mathbb{RP}^1$
or a $\mathbb{CP}^1$ \cite{witten}. $x_{\a\da}$ is the parameter or modulus describing any one of these
curves and \eq{mhvxfm} is thus an integral over the moduli space of degree one
genus zero curves in $\mathbb{T}$. As there is a delta function for every
external particle, the integral is only non-zero when all $n$-points $(\lambda_i^\a,\mu_{i\,\da})$
lie on one of these curves in twistor space.\footnote{Techincally the space is really $n$ copies
of twistor space.} Thus the MHV amplitudes are localised on simple algebraic curves in
twistor space, which are (projective) straight lines in the real case and
spheres\footnote{Recall that $\mathbb{S}^2\cong\mathbb{CP}^1$.} in the complex case.

\begin{figure} [ht]
\label{figmhv}
\centerline {
\includegraphics[width=1.5in]{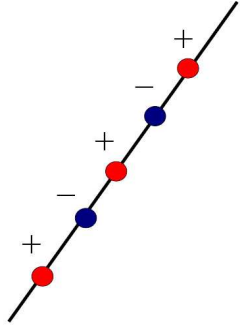}
}
\caption{\it The MHV amplitudes localise on simple straight lines in twistor space.
Here the 5-point MHV amplitude is depicted as an example.}
\end{figure}

In the maximally supersymmetric case we have an additional localisation from transforming the
fermionic variables to twistor space. As well as the delta function of
momentum conservation coming with the amplitudes, we also have a fermionic delta function
\be
\label{fermionicdelta}
\delta^{(8)}\!(\Theta)=\delta^{(8)}\!\!\left(\sum_{i=1}^n\lambda_i\eta_{i}\right)=
\int\! d^8\theta\,\exp\left(i\theta_\a^A\sum_{i=1}^n\lambda_i^\a\eta_{i\, A}\right)\ ,
\ee
and the MHV amplitudes for $\cN\!=\!4$ Yang-Mills are given by \cite{Nair,witten}
\be
\label{supermhv}
\cA_n^{\mathrm{MHV}}(\lambda_i,\lt_i,\eta_{i})=\delta^{(4)}\!(P)\delta^{(8)}\!(\Theta)
\frac{1}{\prod_{i=1}^n\langle i\,\, i\!+\!1\rangle}\ .
\ee
The transform to super-twistor space is a straightforward generalisation of \eq{mhvxfm}
and the result is \cite{witten}
\be
\tilde{\cA}_n^{\mathrm{MHV}}(\lambda_i,\mu_i,\psi_i)=
\int\! d^4xd^8\theta\prod_{j=1}^n\delta^{(2)}\!\!\left(\mu_{j\,\da}+x_{\a\da}\lambda_j^\a\right)
\delta^{(4)}\!\!\left(\psi_j^A+\theta_\a^A\lambda_j^\a\right)\frac{1}{\prod_{i=1}^n\langle i\,\, i\!+\!1\rangle}\ .
\ee
The equations $\mu_{j\,\da}+x_{\a\da}\lambda_j^\a=0$ and $\psi_j^A+\theta_\a^A\lambda_j^\a=0$ then
define (for each $j$) a $\mathbb{CP}^{1|0}$ or an $\mathbb{RP}^{1|0}$ in $\mathbb{CP}^{3|4}$ or
$\mathbb{RP}^{3|4}$ respectively on which the amplitudes lie.

The equation $\mu_{\da}+x_{\a\da}\lambda^\a=0$ is in fact of central importance
in twistor theory and is traditionally taken to be the definition of a twistor. For a given
$x$ (as in our case above), it can be regarded as an equation for $\lambda$ and $\mu$
which as we have seen defines a degree one genus zero curve that is topologically
an $\mathbb{S}^2$. A point in complexified Minkowski space is thus represented by a
sphere in twistor space and hence complexified Minkowski space is the moduli space
of such curves. Alternatively, if $\lambda$ and $\mu$ (\emph{i.e.} a point in
twistor space) are given, it can be regarded as an equation for $x$. The set of
solutions is a two complex-dimensional subspace of complexified Minkowski space that
is null and self-dual called an $\alpha$-plane. The null condition means that
any tangent vector to the plane is null, and the self-duality means that the
tangent bi-vector is self-dual in a certain sense. These $\alpha$-planes can essentially be regarded
as being light-rays and twistor space is the moduli space of $\alpha$-planes.

Other amplitudes involving more and more negative helicities can also be treated,
though in these cases performing the Penrose transform \eq{qmfourier} explicitly
becomes harder. In these cases it has been found that certain differential operators
can be constructed which help to elucidate their localisation properties in twistor space
\cite{witten,csw2}. In particular, given three points
$P_i,P_j,P_k\in \mathbb{CP}^3$ with coordinates $Z_i^I,Z_j^I$ and $Z_k^I$, the
condition that they lie on a `line' (\emph{i.e.} a linearly-embedded copy of $\mathbb{CP}^1$ as
discussed above) is that $F_{ijkL}=0$ where
\be
\label{collinear}
F_{ijkL}=\epsilon_{IJKL}Z_i^IZ_j^JZ_k^K\ .
\ee
Similarly, the condition that four points in twistor space are `coplanar' (\emph{i.e.} lie
on a linearly embedded $\mathbb{CP}^2\subset\mathbb{CP}^3$ is given by $K_{ijkl}=0$ where
\be
\label{coplanar}
K_{ijkl}=\epsilon_{IJKL}Z_i^IZ_j^JZ_k^KZ_l^L\ .
\ee
When these are explicitly used, $\mu_{\da}$ is substituted for $\partial/\partial\lt^{\da}$
and then they act on amplitudes as differential operators.

The localisation properties of many amplitudes have been checked
\cite{witten,csw2,csw3,Bena:2004ry,bdk04,BBDD,new,bbst2,freddy3,Britto,Bern:2005bb,benabern,
Bidder:2005ri,Bidder:2004vx}, and it has been found that
amplitudes with more and more negative helicities localise on
curves of higher and higher degree. For tree-level amplitudes
in particular this means that an amplitude with $q$ negative-helicity
gluons localises on a curve of degree $q-1$. In general, the
twistor version of an $n$-particle scattering amplitude is
supported on an algebraic curve in twistor space whose
degree is given by \cite{witten}
\be
\label{degree}
d=q-1+l\ ,
\ee
where $q$ is the number of negative-helicity gluons and $l$ the
number of loops. The curve is not necessarily connected and its
genus $g$ is bounded by $g\leq l$.

\begin{figure}[ht]
\begin{center}
\vspace{.1in}
\scalebox{0.5}{\includegraphics{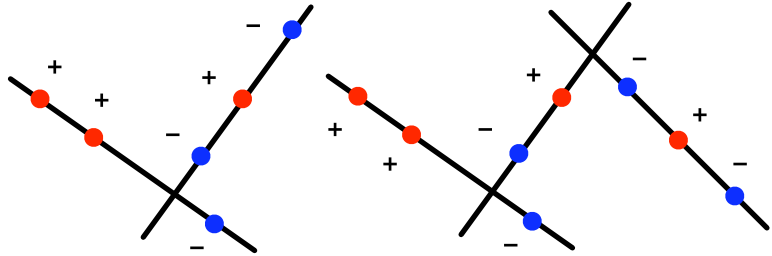}}
\caption{\it Twistor space localisation of tree amplitudes with $q=3$
and $q=4$}
\label{nonmhv}
\end{center}
\end{figure}

Tree-level next-to-MHV amplitudes for example are supported on curves of
degree 2, while NNMHV amplitudes are supported on curves of degree 3 as shown in
\mbox{Figure 1.4} above. We can
also get a geometrical understanding of the vanishing of the
all-plus amplitude and the amplitude with one minus and $n-1$ plusses at
tree-level.
By \eq{degree} these would be supported on curves of degree $d\!=\!-1$ and $d\!=\!0$ in
twistor space. In the first case, there are no algebraic curves of degree $-1$,
so these amplitudes must trivially vanish. In the second, a curve of degree $0$
is simply a point and so amplitudes of this type are supported by configurations
where all the gluons are attached to the same point $(\lambda_i,\mu_i)=(\lambda,\mu)\, \forall\ i$
in twistor space. Recalling from equation \eq{dotprod} that
$p_i\cdot p_j\propto\langle \lambda_i\,\lambda_j\rangle[\lt_j\,\lt_i]$, all
these invariants must be zero for these amplitudes. This on the other hand
is impossible for non-trivial scattering amplitudes with $n\geq 4$ particles
and thus these must vanish at tree-level.

For $n=3$ things are a bit more subtle because on-shellness, $p_i^2=0$ and
momentum conservation,
\mbox{$p_1+p_2+p_3=0$}, guarantee that
for real momenta in Lorentz signature $p_i\cdot p_j=0$. However,
for complex momenta and/or other signatures the 3-point amplitude makes
more sense. As $0=2p_i\cdot p_j=\langle \lambda_i\,\lambda_j\rangle[\lt_j\,\lt_i]$,
the independence of $\lambda_i$ and $\lt_i$ implies that either $\langle \lambda_i\,\lambda_j\rangle=0$
or $[\lt_j\,\lt_i]=0$. Thus all $\lambda_i$ are proportional or all
$\lt_i$ are proportional. As can be read-off from the Yang-Mills Lagrangian
(or seen as a special case of the googly MHV amplitudes), the $-++$ amplitude
is given by
\be
\label{minusplusplus}
\cA=\frac{[\lt_1,\lt_2]^4}{[\lt_1,\lt_2][\lt_2\,\lt_3][\lt_3\,\lt_1]}\ .
\ee
This would vanish identically if all the $\lt_i$ are proportional, so
we should pick all the $\lambda_i$ to be proportional
to ensure momentum conservation. However, $\SL(4,\mathbb{R})$ invariance
in twistor space\footnote{$\SO(3,3)\cong \SL(4,\mathbb{R})$ is the conformal group in signature
$++--$.} then implies that the $(\lambda_i,\mu_i)$ all coincide
and thus the gluons are supported at a single point in twistor space as
predicted by \eq{degree} \cite{witten}.

\section{Twistor string theory}

In this section we will give a very brief overview of a string
theory that provides a natural framework for understanding the properties
of scattering amplitudes discussed in the previous sections. We will
only describe the original approach (which has also been the one most
computationally useful to date) taken by Witten \cite{witten} though other approaches,
notably by Berkovits \cite{Berkovits,BerkMotl,Dolan:2007vv}, have been
considered. Further proposals include
\cite{Siegel:2004dj,Neitzke:2004pf,Aganagic:2004yh,Bars:2004dg}, though
these have not so far been used to calculate any amplitudes. A good
introduction to the material presented in this section can again be
found in \cite{csreview}.

It is well known that the usual type I, type II and heterotic string
theories live in the critical dimension of $d\!=\!10$, which is where they
really make sense quantum mechanically. However, there are other string
theories known as topological string theories which are typically simpler than
ordinary string theories and can make sense in other dimensions. They are
called topological because they can be obtained from certain topological
field theories which are field theories whose correlation functions
only depend on the topological information of their target space and in-particular
do not depend on the local information such as the metric of the space. Witten
introduced topological string theory in \cite{Witten:1988xj,Witten:1989ig}
as a simplified model of string theory, and it has been extensively studied
since then. We will only give a `lightning' review here and refer the
reader to the original papers and such excellent introductions as
\cite{Vonk:2005yv} for more details.

\subsection{Topological field theory}

One starts with a field theory in 2-dimensions with $\mathcal{N}\!=\!2$
supersymmetry. The supersymmetry generators usually transform as
spin $1/2$ fermions under the Lorentz group, but in 2-$d$ this is
$\SO(2)\cong \U(1)$ locally and the spin $1/2$ representation is
reducible into two representations which have opposite charge
under the $\U(1)$. Things living in these representations are often
termed left-movers and right-movers, and the supersymmetry is usually
written as being $\cN\!=\!(2,2)$ with 2 left-moving supercharges and
2 right-moving supercharges.

The symmetries of the theory consist of both the
usual Poincar\'{e} algebra as well as the $\cN\!=\!2$
supersymmetry algebra and the R-symmetry of the
theory associated with the supersymmetry. We will not write all of
these down here, but in-particular the supersymmetry generators
and their complex conjugates obey the non-zero anti-commutation relations
(in the language of \cite{Vonk:2005yv}):
\begin{eqnarray}
\label{susy2d}
\{Q_{\pm},\bar{Q}_{\pm}\}&=&\ \ \ \, P\pm H\nonumber \\
\{D_{\pm},\bar{D}_{\pm}\}&=&-(P\pm H)\ ,
\end{eqnarray}
where $H\sim d/d\xi^0$ and $P\sim d/d\xi^1$ are the Hamiltonian and
momentum operators of the 2-$d$ space with coordinates $\xi^{\alpha}$.

One thing that we can now do is to define new operators $Q_\mathrm{A}$ and
$Q_\mathrm{B}$ which are linear combinations of supercharges as
\begin{eqnarray}
\label{newcharges}
Q_\mathrm{A}&=&\bar{Q}_++Q_-\nonumber \\
Q_\mathrm{B}&=&\bar{Q}_++\bar{Q}_-\ ,
\end{eqnarray}
and then it follows from \eq{susy2d} that
\begin{eqnarray}
\label{brstlike}
Q_\mathrm{A}^2=Q_\mathrm{B}^2=0
\end{eqnarray}
and $Q_\mathrm{A}$ and $Q_\mathrm{B}$ look like BRST operators. However, $Q_{\mathrm{A/B}}$
are not scalars, so we would violate Lorentz invariance by
interpreting them as BRST operators straight away. In
fact what we can do is to make an additional modification to the Lorentz generator
of the 2-$d$ space by making linear combinations of it and the
R-symmetry generators in such a way that the $Q_{\mathrm{A/B}}$ are
scalars under the new generators. This procedure is called
\emph{twisting} and produces two different topological field theories labelled by A and B.

Now that we have a BRST operator, we can use the usual definitions
for the physical states  of our theory in terms of BRST cohomology
(see for example Chapter 16 of \cite{pc} or Chapter 15 of \cite{Weinberg:1996kr} for an introduction).
Physical states $|\psi\rangle$ are given by the condition $Q_{\mathrm{A/B}}|\psi\rangle=0$ with
states being equivalent if they differ by something which is
BRST exact such as $Q_{\mathrm{A/B}}|\phi\rangle$ for some $|\phi\rangle$.
Similarly, physical operators are taken to be those which
commute with the BRST operator modulo those which can be written as an
anti-commutator of $Q_{\mathrm{A/B}}$ with some other operator. In
particular one can show that that the stress-tensors of the twisted
theories are BRST exact as they can be written in the form
$T^{\alpha\beta}_{\mathrm{A/B}}=\{Q_{\mathrm{A/B}},\tau^{\alpha\beta}\}$ for some $\tau^{\alpha\beta}$.
This is a general property of topological field theories.

\subsection{Topological string theory}

What we have so far constructed are two 2-dimensional
topological field theories. However, we can promote these
to string theories by considering the theories to be
living on the worldsheet of a string and ensuring that
we integrate over all metrics of the \mbox{2-dimensional} space
in the path integral as well as the other fields appearing
in the action (see \emph{e.g.} Chapter 3 of \cite{gsw1} for how
this works in the usual string theory settings). The Euclidean path integral
\be
\label{PI}
Z_E=\int\! \mathcal{D}h(\xi)\,\mathcal{D}\Phi(\xi)\,e^{-S_{2d}[h,\Phi]}\ ,
\ee
where $h_{\alpha\beta}$ is the world-sheet metric, $\Phi$ are the
fields of our 2-$d$ field theory and $\xi^{\alpha}$ are the coordinates
of the 2-$d$ space then defines our topological string theory.
If we have re-defined our Lorentz generators to make $Q_\mathrm{A}$ a scalar then
the string theory is known as the A-model, while if we choose to make
$Q_\mathrm{B}$ a scalar we arrive at the B-model \cite{Witten:1988xj,Witten:1989ig}.

We can also say something about the target spaces
of these topological string theories. In `normal' string theory settings
these target spaces - the spaces in which the strings live - are known
to be 10-dimensional (or 26-dimensional for the purely bosonic string) in order
for them to be quantum-mechanically anomaly-free. The $\cN\!=\!(2,2)$
 field theories discussed above, however, naturally give rise to target
spaces which are special types of complex manifolds known as \emph{K\"{a}hler}
manifolds - even \emph{before} we perform the topological twisting.
These spaces are complex manifolds that are endowed with a Hermitian
metric (\emph{i.e.} a real metric - real in the sense that
$g_{\bar{\imath}\bar{\jmath}}=(g_{ij})^\ast$ and
$g_{\bar{\imath}j}=(g_{i\bar{\jmath}})^\ast$ - with $g_{ij}=g_{\bar{\imath}\bar{\jmath}}=0$)
and which we can write locally as the second derivative of some
function termed the K\"{a}hler potential $K(z,\bar{z})$:
\be
\label{kahler}
g_{i\bar{\jmath}}=\frac{\partial^2K(z,\bar{z})}{\partial z^i\partial \bar{z}^{\bar{\jmath}}}\ .
\ee
Here $z^i$ and $\bar{z}^{\bar{\jmath}}$ are appropriate complex coordinates on the target space. When
we do the twisting described by \eq{newcharges} it turns out that the A-model
twist can be performed for \emph{any} K\"{a}hler target space, while the B-model
twist requires the space to be of a yet more specialised form known as a
Calabi-Yau manifold.

There are many different ways to define a Calabi-Yau
manifold, but one way that is good for our purposes is that it is a K\"{a}hler
manifold that is also Ricci-flat, $R_{i\bar{\jmath}}=0$. The moduli (essentially
the parameters) describing the variety of such spaces are of two types
which are termed the K\"{a}hler moduli and the complex-structure moduli. It can
be shown that the space of K\"{a}hler moduli is locally $H^{(1,1)}(\mathcal{M}_{\mathrm{CY}})$ -
that is to say it is locally given by the Dolbeault cohomology class of $(1,1)$-forms - while
the complex-structure moduli space is locally the cohomology class of $(2,1)$-forms
$H^{(2,1)}(\mathcal{M}_{\mathrm{CY}})$. Because Calabi-Yau manifolds are automatically
K\"{a}hler manifolds to begin with and because of their high degree of symmetry,
the A-model is often also considered on a Calabi-Yau. Finally, it can be shown that
the central charge of the Virasoro algebra of the A- and B-models vanishes identically in any number of
dimensions \cite{Vonk:2005yv}, so topological strings are well-defined in target spaces of any
dimension. For more comprehensive
discussions of complex, K\"{a}hler and Calabi-Yau manifolds see \emph{e.g.}
\cite{Vonk:2005yv,Nakahara:1990th,Candelas:1987is,Horowitz:1985ft,Bouchard:2007ik}.

As for the physical operators in these models, we briefly state without
proof that in the A-model, $Q_\mathrm{A}$ can be viewed as being $Q_{\mathrm{A}}\sim d$ - the de Rham
exterior derivative - and the local physical operators are in one-to-one
correspondence with de Rham cohomology elements on the target space:
\be
\label{amodops}
\mathcal{O}_\mathrm{A}\sim A_{i_1\ldots i_p \bar{\jmath}_1\ldots\bar{\jmath}_q}(\Phi)
d\Phi^{i_1}\cdots d\Phi^{i_p}d\bar{\Phi}^{\bar{\jmath}_{1}}\cdots d\bar{\Phi}^{\bar{\jmath}_{q}}\ .
\ee
For the B-model on the other hand one can show that $Q_\mathrm{B}\sim \bar{\partial}$ -
the Dolbeault exterior derivative - and the local physical operators are now
just $(0,p)$-forms with values in the antisymmetrized product of $q$ holomorphic
tangent spaces - which we denote by $\bigwedge^qT^{(1,0)}(\mathcal{M}_{\mathrm{CY}})$:
\be
\label{bmops}
\mathcal{O}_\mathrm{B}\sim B_{\bar{\imath}_1\ldots\bar{\imath}_p}^
{\phantom{\bar{\imath}_1\ldots\bar{\imath}_p}j_1\ldots j_q}(\Phi)d\bar{\Phi}^{\bar{\imath}_1}
\cdots d\bar{\Phi}^{\bar{\imath}_p}\frac{\partial}{\partial\Phi^{j_1}}\cdots
\frac{\partial}{\partial\Phi^{j_q}}\ .
\ee

These theories also have the intruiging property of mirror symmetry \cite{Lerche:1989uy,Dixon:1987bg}
- see \emph{e.g.} \cite{Hori:2003ic} and references therein for a
comprehensive review - that the A-model on one Calabi-Yau is equivalent to the
B-model on a different Calabi-Yau which is known as its Mirror. In the
mirror map, the hodge numbers $h^{1,1}$ and $h^{2,1}$ are swapped which pertains
to the exchange of K\"{a}hler and complex-structure moduli. This is especially
useful as the B-model is generally easier to compute-with than the A-model, while the
A-model is more physically interesting in many scenarios. Hard computations
in the A-model can often be mapped to easier ones in the B-model.

\subsection{The B-model on super-twistor space}

In his original construction \cite{witten}, Witten considered the
B-model and we will do the same here. The target space on which we
will want it to live will be $\mathbb{CP}^{3|4}$, which is
 a Calabi-Yau super-manifold (with bosonic \emph{and} fermionic degrees of
 freedom) rather than a bosonic manifold as is more common.
This is fortunate because $\mathbb{CP}^3$ is \emph{not} Calabi-Yau, while
$\mathbb{CP}^{3|4}$ \emph{is}.\footnote{In fact $\mathbb{CP}^{3|\mathcal{N}}$ is Calabi-Yau iff
$\mathcal{N}\!=\!4$.} In addition, if we recall that the closed-string
sector is where gravity states arise, we would like to consider the \emph{open-string}
B-model on twistor space in order that we may end up with degrees of
freedom with spin 1 or less. In the simplest case this consists of adding $N$
bosonic-space-filling D5-branes (thus spanning all 6 bosonic directions of
$\mathbb{CP}^{3|4}$ in analogy with the purely bosonic case of \cite{Witten:1992fb}.
In addition (as Witten did), we take the D5s to wrap the fermionic directions
$\psi^I$ and $\bar{\psi}^{\bar{J}}$ in such a way that we can set
$\bar{\psi}^{\bar{J}}$ to zero. It is not entirely clear how this should be interpreted,
but one might say that the branes
wrap the $\psi$ directions while being localised in the $\bar{\psi}$ directions.
The presence of $N$ branes gives rise to a $\U(N)$ gauge symmetry as usual due to the
Chan-Paton factors of the open strings ending on them.

So far we have been considering things from a worldsheet
perspective. However, for open strings we also have the spacetime perspective
of open-string field theory \cite{Witten:1985cc}. This has a multiplication
law $\star$, an operator $Q$ obeying $Q^2=0$ and Lagrangian
\be
\label{sft}
\mathscr{L}=\frac{1}{2}\int\left({\mathscr{A}}\star Q{\mathscr{A}}
+\frac{2}{3}{\mathscr{A}}\star{\mathscr{A}}\star{\mathscr{A}}\right)\ ,
\ee
where ${\mathscr{A}}$ is the string field. In the presence of D5-branes on a 6-dimensional (bosonic)
manifold this has been shown to reduce to
holomorphic Chern-Simons theory \cite{Witten:1992fb},
where the D5-D5 modes of the string field ${\mathscr{A}}$ give a $(0,1)$-form
gauge field $A=A_{\bar{\imath}}(z,\bar{z})d\bar{z}^{\bar{\imath}}$ on the branes. On the other hand, when the target space is
the supermanifold $\mathbb{CP}^{3|4}$, ${\mathscr{A}}$ reduces
to the $(0,1)$-form gauge superfield $\cA=\cA_{\bar{I}}(Z,\bar{Z},\psi,\bar{\psi})d\bar{Z}^{\bar{I}}$, while
$Q$ becomes the $\bar{\partial}$ operator and $\star$ the usual wedge product operation $\wedge$.
The action descends to
\be
\label{hcs}
\mathcal{S}=\frac{1}{2}\int_{\mathbb{CP}^{3|4}}\Omega\wedge\trace\left(\cA\wedge\bar{\partial}\cA+
\frac{2}{3}\cA\wedge\cA\wedge\cA\right)\ ,
\ee
and with $\bar{\psi}=0$ the superfield $\cA$ can be expanded as
\be
\label{Aexpansion}
\cA(Z,\bar{Z},\psi)=A+\psi^I\lambda_I+\frac{1}{2}\psi^I\psi^J
\phi_{IJ}+\frac{1}{3!}\epsilon_{IJKL}\psi^I\psi^J\psi^K\tilde{\lambda}^L+
\frac{1}{4!}\epsilon_{IJKL}\psi^I\psi^J\psi^K\psi^LG\ ,
\ee
where $A,\lambda_I,\phi_{IJ},\tilde{\lambda}^I,G$ are all functions of
$Z$ and $\bar{Z}$ and we have suppressed the \mbox{$(0,1)$-form} structure.
$\Omega$ in \eq{hcs} is a $(3,0)$-form and is the holomorphic volume-form on $\mathbb{CP}^{3|4}$
\be
\label{volform}
\Omega=\frac{1}{4!^2}\epsilon_{IJKL}\epsilon_{MNPQ}Z^IdZ^JdZ^KdZ^L
d\psi^Md\psi^Nd\psi^Pd\psi^Q\ .
\ee

Because $dZ^I$ and $d\psi^I$ scale oppositely -- as follows from \eq{supertwistorscaling} and the
fermionic nature of $\psi^I$ ($d\psi^I\rightarrow c^{-1}d\psi^I$ under \eq{supertwistorscaling}) --
it is clear that \eq{volform} is invariant under this scaling and thus the action \eq{hcs}
is only invariant if $\cA$ is of degree zero, $\cA\in H^{(0,1)}(\mathbb{CP}^{3|4^{\prime}},\mathcal{O}(0))$.
This means that each component field in the $\psi$ expansion \eq{Aexpansion} must be of degree $2h-2$ and thus
describes a field of helicity $h$ in spacetime - \emph{c.f.} the twistor description of wavefunctions for particles of
helicity $h$ of Eq.~\eq{twistorwavefunction} and surrounding paragraphs. In addition, the fermionic nature of the $\psi^I$
restricts the number of degrees of freedom of the component fields and
it can quickly be seen that \eq{Aexpansion} describes the
$\cN\!=\!4$ multiplet,\footnote{To be more precise it is the twistor transform of the
$\cN\!=\!4$ multiplet \cite{Penrose,PopovSamann04}.} which in the notation of \mbox{\S \ref{susydecomp}} can
be written as $(A^-,\chi^-,\phi,\chi^+,A^+)\equiv(G,\lt^I,\phi_{IJ},\lambda_I,A)$,
while the action in component form can be written as
\begin{eqnarray}
\label{componenthcs}
\mathcal{S}&=&\int_{\mathbb{CP}^3}\Omega'\wedge\trace\Big(G\wedge(\bar{\partial}A+A\wedge A)
-\tilde{\lambda}^I\wedge(\bar{\partial}\lambda_I+[A,\lambda_I])\\ \nonumber &+&
\frac{1}{4}\epsilon^{IJKL}\phi_{IJ}\wedge(\bar{\partial}\phi_{KL}+
A\wedge\phi_{KL})
-\frac{1}{2}\epsilon^{IJKL}\lambda_I\wedge\lambda_J\wedge\phi_{KL}\Big)\ ,
\end{eqnarray}
where $\Omega'=\epsilon_{IJKL}Z^IdZ^JdZ^KdZ^L/4!$ is the bosonic reduction of $\Omega$ obtained after integrating out the
$\psi^I$ \footnote{Recall that for Grassman variables, $\int\! d\psi\equiv\partial/\partial\psi$ with
$\int\! d\psi^I\psi^J=\delta^{IJ}$ and $\delta(\psi)=\psi$.} and $[A,\lambda_I]=A\wedge\lambda_I+\lambda_I\wedge A$.
The equations of motion following from \eq{hcs} are $\bar{\partial}\cA+\cA\wedge\cA=0$ and the gauge invariance is
$\delta\cA=\bar{\partial}\omega+[\cA,\omega]$.

What we have arrived at is `half' of $\cN\!=\!4$ super-Yang-Mills. We have all the fields
as is apparent from \eq{Aexpansion}, but it turns out that not all the interactions
are present. One of the easiest ways to see this is to note that the symmetries
of the B-model generally leave $\Omega$ invariant \cite{witten}. However there are also interesting
transformations of the target space that leave the complex structure invariant
but transform $\Omega$ non-trivially. One such transformation is a $\U(1)_\mathrm{R}$ part
of the R-symmetry group $\U(4)_\mathrm{R}=\SU(4)_\mathrm{R}\times \U(1)_\mathrm{R}$ that acts as%
\footnote{Note that the $I$ indices on the component fields in \eq{componenthcs} are fundamental indices
of this $\SU(4)_\mathrm{R}$.}
\be
\label{u1}
S:\quad Z^I\rightarrow Z^I\, ;\quad \psi^I\rightarrow e^{i\alpha}\psi^I
\ee
with $d\psi^I\rightarrow e^{-i\alpha}d\psi^I$ because of their fermionic nature.
$\Omega\rightarrow e^{-4i\alpha}\Omega$ thus has $S=-4$ and hence so does the action \eq{hcs} as
the transformation of the $\psi^I$ inside $\cA$ are compensated by equal and
opposite transformations of the component fields: $A$ has $S=0$,
$\lambda_I$ has $S=-1$, $\phi_{IJ}$ has $S=-2$, $\lt^I$ has $S=-3$ and $G$ has
$S=-4$. In fact the component action \eq{componenthcs} is made up entirely of
terms with $S=-4$. However, the usual $\cN\!=\!4$ Yang-Mills action in component
form consists of terms which have $S=-4$ \emph{and} $S=-8$. For example the scalar
kinetic terms $(\partial_{\mu}\phi)^2$ have $S=-4$ while the scalar potential
$\phi^4$ has $S=-8$. The holomorphic Chern-Simons action \eq{hcs} thus captures all the
fields of maximally supersymmetric Yang-Mills, but not all the interactions. Although
we will not discuss it here, the theory described by \eq{hcs} is in fact
self-dual $\cN\!=\!4$ super-Yang-Mills \cite{Siegel9205} - that is, (super)-Yang-Mills theory for a gauge field $A'$ whose field strength appearing
in the action is self-dual. $A'$ is the spacetime field corresponding to the homogeneity 0 field ($A$) in
\eq{Aexpansion} and the spacetime action of this theory is $S=\int\,G'\wedge *F'\equiv\int\,G'\wedge F_{\mathrm{SD}}'$.
Here $G'$ is a self-dual 2-form whose twistor transform is the homogeneity $-4$ field ($G$) in
\eq{Aexpansion}, $F_{\mathrm{SD}}'$ is the self-dual part\footnote{Note that formally we can write the
self-dual and anti-self-dual parts of $F'$ as $F_{\mathrm{SD}}'=(F'+*F')/2$ and $F_{\mathrm{ASD}}'=(F'-*F')/2$.
Here we have taken $**F'=F'$.}
of $F'=dA'+A'\wedge A'$ and $*$ is the Hodge duality operation.

\subsection{D1-brane instantons}

Witten's solution to the aforementioned problem of the absence of the entire set of interactions was to enrich the B-model on
$\mathbb{CP}^{3|4}$ with instantons. The ones in question are Euclidean D1-branes
which wrap holomorphic curves in super-twistor space and on which the open strings can
end. These holomorphic curves are precisely the ones that we met earlier on which the
scattering amplitudes were found to localise.
We won't go into much detail here (more can be found in \cite{witten}),
but the basic idea is that these instantons have $S$-charge $-4(d+1-g)$ for the connected degree $d$ and
genus $g$ case. Thus for the `classical' tree-level MHV case\footnote{We refer to the tree-level
MHV amplitudes as being the `classical' case as it turns out that we can re-formulate perturbation
theory in terms of `MHV-vertices' - see \S \ref{cswtree} - and they are thus appropriate for consideration of
$S$-charge violation at the level of the action.} these instantons provide the terms
with $S=-8$ as we had hoped.

We can now consider other types of strings apart from D5-D5s. We also have \mbox{D1-D1}s,
D1-D5s and D5-D1s. The D1-D1 strings give rise to a $\U(1)$ gauge field on the
world-volume of an instanton which describes the motion of the instanton in $\mathbb{T}$. We
will thus ignore the D1-D1 strings from now on. Of course we \emph{do} want to involve the
D1-instantons, so we'll focus on the D1-D5 and D5-D1 strings. Witten argued that
these strings give rise to fermionic $(0,0)$-form fields living on the world-volume of the
instanton. The D1-D5 modes give rise to a fermion $\alpha^i$ and the D5-D1 modes
give a fermion $\beta_{\bar{\imath}}$, with $\bar{\imath}$ and $i$ (anti)-fundamental
$\U(N)$ indices respectively. The effective action for the low-energy modes is then
\be
\label{fermieff}
S_{\mathrm{eff}}=\int_{\mathrm{D1}}\!\!dz\,\beta(\bar{\partial}_{\bar{z}}+\cA_{\bar{z}}d\bar{z})\alpha\ ,
\ee
where $z$ and $\bar{z}$ are local complex coordinates on the D1 and
$\cA_{\bar{z}}$ (which is a background field on the D1) is the component of the superfield generated by the D5-D5 strings
lying along the D1.
The first term is the kinetic term of these modes (with $\bar{\partial}_{\bar{z}}$ the
$\bar{\partial}$ operator restricted to the D1), while the second describes their interaction
with the gauge field $\cA$ and can be written as
\be
\label{fermiinter}
S_{\mathrm{int}}=\int_{\mathrm{D1}} J\wedge (\cA_{\bar{z}}d\bar{z})\ ,
\ee
where we define $J$ to be the current $J_{\bar{\imath}}^{\phantom{\bar{\imath}}j}=\beta_{\bar{\imath}}\alpha^jdz$.

Any particular external state will contribute just one component of this superfield $\cA$ and therefore its
coupling will be
\be
\label{wfinter}
V_s=\int_{\mathrm{D1}}J_s\wedge \phi_s\ ,
\ee
where $\phi_s$ is the wavefunction of the state in twistor
space and thus a $(0,1)$-form there.\footnote{We use subscripts $s_i$ \emph{etc.} to denote the $i$th particle for the rest of this
section in order to avoid confusion with the gauge indices.} Then
if the curve which the D1 wraps were to have no moduli (\emph{i.e.} there were only one possibility for it), one would
be able to compute scattering amplitudes by evaluating the correlator $\langle V_{s_1}\ldots V_{s_n}\rangle$.
However, we know from the discussion in \S \ref{localise} that these curves \emph{do} have
moduli and thus we should integrate this correlator over their moduli space. Our prescription
for computing $n$-point scattering amplitudes whose external particles have wave functions $\phi_{s_i}$ will
then be
\be
\label{twistorcomp}
A_n=\int\! d\mathcal{M}_d\,\langle V_{s_1}\ldots V_{s_n}\rangle\ ,
\ee
where $d\mathcal{M}_d$ is an appropriate measure on the moduli space of holomorphic
curves of degree $d$ (and genus zero for our current purposes).

\subsection{The MHV amplitudes}

As an example of how \eq{twistorcomp} is implemented let us calculate the
MHV amplitudes using this prescription. From \S \ref{localise}
we saw that the MHV amplitudes lie on holomorphic curves that are embedded in
$\mathbb{CP}^{3|4}$ via the equations
\begin{eqnarray}
\label{holembedding}
\mu_{s_{k}\,\da}+x_{\a\da}\lambda_{s_k}^\a&=&0\nonumber \\
\psi^A_{s_k}+\theta_\a^A\lambda^\a_{s_k}&=&0\ .
\end{eqnarray}
$\lambda^\a_{s_k}$ are the homogeneous coordinates on the curves (with $s_k=1\ldots n$
denoting the $k^{\textrm{th}}$ particle) and their moduli are $x_{\a\da}$ and $\theta_\a^A$.
These are thus the curves that we will take the D1-instantons to be wrapping.
$x_{\a\da}$ has 4 (bosonic) degrees of freedom while $\theta_\a^A$ has 8 (fermionic) ones and
a natural measure on the moduli space is then $d\mathcal{M}_1=d^4xd^8\theta$.

For clarity let us specialise to the case of 4-particle (gluon) scattering where
particles 1 and 3 have negative-helicity and particles 2 and 4 positive-helicity. The
$n$-particle case is an easy generalisation of this. Formally we have
\begin{eqnarray}
\label{A4}
A_4&=&\int\! d\mathcal{M}_1\langle V_{s_1}V_{s_2}V_{s_3}V_{s_4}\rangle\nonumber \\
&=&\int\! d^4xd^8\theta\left\langle\int_{\mathbb{CP}^{1|0}}J_{s_1}\wedge\phi_{s_1}
\ldots\int_{\mathbb{CP}^{1|0}}J_{s_4}\wedge\phi_{s_4}\right\rangle\ ,
\end{eqnarray}
where we assume that the wavefunctions $\phi_{s_k}$ take values in the Lie-algebra of
$\U(N)$ and thus contain a generator $T^{a_k}$ in addition to \eq{twistorwavefunction}.
$(J_{s_k})_{\bar{\imath}}^{\phantom{\bar{\imath}}j}=\beta_{\bar{\imath}}(z_k)\alpha^j(z_k)dz_k$
then gives
\begin{eqnarray}
\label{A4-2}
\!\!\!\!\!\!\!\!A_4\!\!\!&=&\!\!\!\int\!d^4xd^8\theta\int\! dz_1\ldots dz_4\left\langle
\beta_{\bar{\imath}_1}(z_1)\alpha^{j_1}(z_1)\phi_{s_1}\ldots\beta_{\bar{\imath}_4}
(z_4)\alpha^{j_4}(z_4)\phi_{s_4}\right\rangle
\end{eqnarray}
up to a factor. Separating-out the Lie-algebra generators from the rest of the wavefunctions
$(\phi_{s_k}=\phi_{s_k}'T^{a_k})$ we can re-write the correlator as
\begin{eqnarray}
\label{correlator}
\int\!d^4xd^8\theta\int\! dz_1\ldots dz_4\,\phi_{s_1}'\ldots\phi_{s_4}'
\left\langle\beta_{\bar{\imath}_1}(T^{a_1})^{i_1}_{\phantom{i_1}\bar{\jmath}_1}\alpha^{j_1}\ldots
\beta_{\bar{\imath}_4}(T^{a_4})^{i_4}_{\phantom{i_4}\bar{\jmath}_4}\alpha^{j_4}\right\rangle\ .
\end{eqnarray}

This correlator has many different contributions (105 in total) coming from the possible ways of Wick contracting
the fermions $\alpha$ and $\beta$. Let us consider the cyclic one where we contract $\beta(z_1)$ with
$\alpha(z_2)$, $\beta(z_2)$ with $\alpha(z_3)$ and so on (with $\beta({z_4})$ contracted with $\alpha({z_1})$).
Because $\alpha$ and $\beta$ are fermions living on (in this case) $\mathbb{CP}^1$, their
propagator is the usual one for free fermions on the complex plane
\be
\label{fermicorrelator}
\langle\alpha^{j}(z_k)\beta_{\bar{\imath}}(z_l)\rangle=\frac{\delta^j_{\phantom{j}\bar{\imath}}}
{z_k-z_l}
\ee
and the relevant Wick contraction is
\begin{eqnarray}
\label{wick1}
W_{\textrm{cyclic}}&=&(T^{a_1})^{i_1}_{\phantom{i_1}\bar{\jmath}_1}\ldots
(T^{a_4})^{i_4}_{\phantom{i_4}\bar{\jmath}_4}\langle\alpha^{j_1}(z_1)\beta_{\bar{\imath}_2}(z_2)\rangle
\ldots\langle\alpha^{j_4}(z_4)\beta_{\bar{\imath}_1}(z_1)\rangle\nonumber \\
&=&(T^{a_1})^{i_1}_{\phantom{i_1}\bar{\jmath}_1}\ldots
(T^{a_4})^{i_4}_{\phantom{i_4}\bar{\jmath}_4}\frac{\delta^{j_1}_{\phantom{j_1}\bar{\imath}_2}}
{z_1-z_2}\ldots\frac{\delta^{j_4}_{\phantom{j_4}\bar{\imath}_1}}
{z_4-z_1}\nonumber \\
&=&\frac{\trace(T^{a_1}\ldots T^{a_4})}{(z_1-z_2)(z_2-z_3)(z_3-z_4)(z_4-z_1)}\ .
\end{eqnarray}
Dropping the single-trace colour factor for now, \eq{correlator} is
\begin{eqnarray}
\label{correlator2}
\cA_4&=&\int\!d^4xd^8\theta\int\! dz_1\ldots dz_4\,\frac{\phi_{s_1}'\ldots\phi_{s_4}'}
{(z_1-z_2)(z_2-z_3)(z_3-z_4)(z_4-z_1)}\nonumber \\
&=&\int\!d^4xd^8\theta\int\!\langle\lambda_{s_1}\, d\lambda_{s_1}\rangle\ldots\langle\lambda_{s_4}\, d\lambda_{s_4}\rangle
\frac{\phi_{s_1}'\ldots\phi_{s_4}'}
{\langle\lambda_{s_1}\, \lambda_{s_2}\rangle\ldots\langle\lambda_{s_4}\, \lambda_{s_1}\rangle}\ ,
\end{eqnarray}
where we have changed to homogeneous coordinates $\lambda_{s_k}$ on the $\mathbb{CP}^1$s
by setting $z_k=\lambda_{s_k}^2/\lambda_{s_k}^1$ with 1 and 2 indicating spinor `$\a$' indices here.

Now we must introduce the explicit form for the wavefunctions and integrate over the $\lambda_k$.
For this it is useful to note that with $z=\lambda^2/\lambda^1$ and making the more
specific choices of $\lambda^{\a}=(1,z)$ and $\zeta^{\a}=(1,b)$, \eq{holomorphicdelta2} becomes \eq{holomorphicdelta3}:
\be
\label{holomorphicdelta2}
\int\langle\lambda\,d\lambda\rangle\,\bar{\delta}(\langle\lambda\,\zeta\rangle)\,F(\lambda)=-iF(\zeta)\ .
\ee
Omitting the integral over moduli, \eq{correlator2} thus gives
\begin{eqnarray}
\label{woof}
\cA_4&=&\int_{\mathbb{CP}^{1|0}}\!\!\langle\lambda_{s_1}\, d\lambda_{s_1}\rangle\bar{\delta}(\langle\lambda_{s_1}\,\pi_{s_1}\rangle)
\left(\frac{\lambda_{s_1}}{\pi_{s_1}}\right)^{2h_1-1}\!\!\!\!e^{i[\tilde{\pi}_{s_1}\,\mu_{s_1}](\pi_{s_1}/\lambda_{s_1})}g_{h_1}(\psi_{s_1})\nonumber\\
&{}&\vdots\nonumber\\
&{}&\int_{\mathbb{CP}^{1|0}}\!\!\langle\lambda_{s_4}\, d\lambda_{s_4}\rangle\bar{\delta}(\langle\lambda_{s_4}\,\pi_{s_4}\rangle)
\left(\frac{\lambda_{s_4}}{\pi_{s_4}}\right)^{2h_4-1}\!\!\!\!e^{i[\tilde{\pi}_{s_4}\,\mu_{s_4}](\pi_{s_4}/\lambda_{s_4})}g_{h_4}(\psi_{s_4})
H(\lambda_{s_i})\nonumber\\
&=&
\psi_{s_1}^1\psi_{s_1}^2\psi_{s_1}^3\psi_{s_1}^4\psi_{s_3}^1\psi_{s_3}^2\psi_{s_3}^3\psi_{s_3}^4H(\pi_{s_i})
e^{i\sum_{s_k=1}^4[\tilde{\pi}_{s_k}\,\mu_{s_k}]}\ ,
\end{eqnarray}
where $H$ is the denominator in \eq{correlator2}.

We must now perform the integral over the moduli. For this purpose we can recall the
equations describing the embedding \eq{holembedding} and substitute
$\mu_{s_k\,\da}=-x_{\a\da}\pi_{s_k}^{\a}$ and $\psi_{s_k}^A=-\theta^{A}_\a\pi^\a_{s_k}$~\footnote{Recall
that the delta functions of \eq{woof} have set $\lambda_{s_k}=\pi_{s_k}$.} whereupon the
integral over $x$ gives
\be
\label{momcom}
\int\!\!d^4x\,\exp\left(-ix_{\a\da}\sum_{s_k=1}^4\pi_{s_k}^\a\tilde{\pi}_{s_k}^{\da}\right)
=\delta^{(4)}\!\left(\sum_{s_k=1}^4\pi_{s_k}\tilde{\pi}_{s_k}\right)\ ,
\ee
which is just the delta function of momentum conservation. For the fermionic moduli we have
(for example):
\begin{eqnarray}
\label{numerator}
\psi_{s_1}^1\psi_{s_3}^1&=&(\theta_1^1\pi_{s_1}^1+\theta_2^1\pi_{s_1}^{2})(\theta_1^1\pi_{s_3}^1+\theta_2^1\pi_{s_3}^{2})
\nonumber \\
&=&\theta_2^1\theta_1^1\pi_{s_1}^{2}\pi_{s_3}^1+\theta_1^1\theta_2^1\pi_{s_1}^1\pi_{s_3}^{2}\nonumber\\
&=&\theta_1^1\theta_2^1(\pi_{s_1}^1\pi_{s_3}^{2}-\pi_{s_1}^{2}\pi_{s_3}^1)\nonumber\\
&=&\theta_1^1\theta_2^1\langle\pi_{s_1}\,\pi_{s_3}\rangle\ .
\end{eqnarray}
After dealing with all the $\psi$s in a similar way and then integrating over the
eight $\theta$ variables gives $\langle\pi_{s_1}\,\pi_{s_3}\rangle^4$. Putting all the pieces together
we get
\begin{eqnarray}
\label{finallyMHV}
A_4(1^-,2^+,3^-,4^+)&=&\trace(T^{a_1}\ldots T^{a_4})\frac{\langle\pi_{s_1}\,\pi_{s_3}\rangle^4}{\langle\pi_{s_1}\,\pi_{s_2}\rangle
\ldots\langle\pi_{s_4}\,\pi_{s_1}\rangle}\delta^{(4)}\!\left(\sum_{s_k=1}^4\pi_{s_k}\tilde{\pi}_{s_k}\right)
\nonumber\\
&=&\trace(T^{a_1}\ldots T^{a_4})\frac{\langle 1\,3\rangle^4}{\langle 1\,2\rangle
\ldots\langle 4\,1\rangle}\delta^{(4)}\!\left(\sum_{i=1}^4\pi_{i}\tilde{\pi}_{i}\right)\ ,
\end{eqnarray}
which is precisely the formula for the MHV amplitudes that we wrote down before, though we
have kept the colour structure explicit here.

We should be careful to note that we have simply picked the particular Wick contraction that
we needed in order to get a cyclic colour ordering. All the terms with non-cyclic colour
orderings but a single trace are also present as well as multi-trace terms which in \cite{witten,BerkWitt2}
were suggested to be a sign of the presence of closed-string (and thus gravitational) states.

We have explicitly described the construction of the MHV amplitudes from the B-model in
twistor space. Other amplitudes can be calculated in this way too, though the
complexity is greater so we will not go into any detail on this. The NMHV amplitudes for
example require one to integrate over the moduli space of degree 2 curves in $\mathbb{T}$ and
some simple cases were calculated this way in \cite{rsv}. Other cases such as the $n$-point
googly MHV amplitudes (with 2 positive-helicity gluons and $n-2$ of negative helicity) were
worked out in \cite{rv} and all 6-point amplitudes in \cite{rsv2}. For these integrals over
curves of degree $d>1$, one encounters the possibility of describing these as connected curves of
degree $d$, or disconnected curves of degree $d_i$ with $\sum d_i=d$. In \cite{rsv,rv,rsv2} it
was found that the connected prescription alone reproduces the entire amplitudes in the cases considered
(at least up to a factor). However, there is also strong evidence that the same amplitudes can be
computed using the purely disconnected prescription \cite{csw}. Indeed, this disconnected prescription
led directly to the proposal of new rules for doing perturbative gauge theory which we will
describe in the next section. The authors of \cite{gmn} argued that the integrals involved in the
connected prescription localised on the subspace where a connected curve of degree $d$ degenerates
to the intersection of curves of degree $d_i$ with $\sum d_i=d$ and thus provided strong evidence that
there are ultimately $d$ different prescriptions which are all equivalent. The extreme possibilities
are that we have just one degree $d$ curve to consider, or alternatively $d$ degree one curves. This
latter case was the inspiration for \cite{csw}.

We have also not said anything about loop diagrams here except for the formal
statement that they localise on curves of degree $d=q-1+l$ with $g\leq l$. The
structure of many loop diagrams of $\cN\!=\!4$ super-Yang-Mills was elucidated in
\cite{witten,csw2,csw3,bdk04,new,benabern}, though the
situation with their calculation from the B-model is far less clear than that for trees unfortunately.
In \cite{BerkWitt2} it was shown that closed string modes give rise to states
of $\cN\!=\!4$ conformal supergravity describing deformations of the target
twistor space as well as the expected $\cN\!=\!4$ Yang-Mills states. Conformal supergravity%
\footnote{For a review see \emph{e.g.} \cite{Fradkin:1985am}.} in 4 dimensions has a Lagrangian which is the Weyl tensor of gravity squared,
\mbox{$S=\int d^4x\sqrt{\det|g_{\mu\nu}|}\,W^2$}, and is usually considered to be a somewhat unsavoury
theory as it gives rise to fourth order differential equations which are generally held to lead to a
lack of unitarity (see \emph{e.g.} \cite{Bennett:1997wj}). One might still hope to
decouple these states, but because the coupling constant is the same in both sectors
the amplitudes mix and one ends up with a theory of $\cN\!=\!4$ conformal supergravity
coupled to $\cN\!=\!4$ super-Yang-Mills, some amplitudes of which were computed in
\cite{BerkWitt2} at tree-level and more recently in \cite{Dolan:2007vv} at loop-level
(see also \cite{roibanqm}).
Despite all this, it was shown by Brandhuber, Spence and Travaglini that the proposals
of \cite{csw} can be extended to loop-level and provide a new perturbative expansion
for field theory which is valid in the quantum regime as well as the classical one. This
discovery will be a central theme in the following chapters of this thesis.

As a final remark in this section we point out that twistor string theories have also
been constructed to describe other theories with less supersymmetry and/or
product gauge groups \cite{Kulaxizi:2004pa,Giombi:2004xv,Gao:2006mw,Ahn:2004ua,Park:2004bw}
as well as more recently to describe Einstein supergravity \cite{Abou-Zeid:2006wu}. Indeed
the proposals in \cite{Abou-Zeid:2006wu} include a twistor description of $\cN\!=\!4$ SYM
coupled to Einstein supergravity which may lead to a resolution of the problem of
loops if they can be consistently decoupled.

\section{CSW rules (tree-level)}
\label{cswtree}

Motivated by the findings we have so far discussed, Cachazo, Svr\v{c}ek and Witten
proposed a set of alternative graphs for tree-level amplitudes in Yang-Mills
theory based on the MHV vertices \cite{csw}. The essential idea is the observation
that one can seemingly compute tree-level amplitudes from the totally
disconnected prescription alluded to above by gluing $d$ disconnected
lines together (on each of which there is an MHV amplitude localised)
for an amplitude involving $d+1$ negative-helicity gluons. The
gluing procedure is made concrete by connecting the lines with twistor space
propagators. In field theory terms this corresponds to the use of
MHV amplitudes as the fundamental building blocks - because their localisation properties
in twistor space translates to a point-like interaction in Minkowski space - and gluing these together with simple scalar
propagators $1/P^2$. The two ends of any propagator must have opposite helicity
labels because an incoming gluon of one helicity is equivalent to an outgoing gluon of
the opposite helicity.

\subsection{Off-shell continuation}

\label{offshellsection}

In order to glue MHV vertices together we must continue them off-shell
since one or more of the legs
must be connected to the off-shell propagator $1/P^2$. For this
purpose, consider a generic off-shell momentum vector, $L$.
On general grounds, it can be decomposed as
\cite{Bena:2004ry,dk}
\beq
\label{off}
L \ = \ l \, + \, z \eta \ ,
\eeq
where $l^2=0$, and $\eta$ is a fixed and arbitrary
null vector, $\eta^2=0$; $z$ is a real number.
Equation \eqref{off} determines $z$
as a function of $L$:
\beq
z \ = \
{L^2 \over  2 (L\! \cdot\eta)}
\ .
\eeq
Using spinor notation, we can write $l$ and $\eta$
as $l_{\a \da} = l_{\a} \llt_{\da}$,
$\eta_{\a \da} = \eta_{\a} \tilde{\eta}_{\da}$.
It then follows that%
\beqa
\label{1}
l_\a & = & {L_{\a \da} \tilde{\eta}^{\da}
\over [ \llt \, \tilde{\eta}]
}
\ ,
\\
\label{2}
\llt_{\da} & = & {\eta^\a L_{\a \da} \over \lan l \, \eta
\ran} \ .
 \eeqa
We notice that  \eqref{1} and \eqref{2} coincide
with the CSW prescription proposed in \cite{csw}
to determine the spinor variables
$l$ and $\llt$ associated with the non-null,
off-shell four-vector $L$ defined in \eqref{off}.
The denominators on the right
hand sides of \eqref{1} and \eqref{2} turn out to be
irrelevant for our applications since the expressions
we will be dealing with are homogeneous in the spinor
variables $l_{\alpha}$; hence we will usually discard them.
This defines our off-shell continuation.

\subsection{The procedure: An example}

The CSW rules for joining these MHV amplitudes together
are probably best illustrated with an example. It is clear that
a tree diagram with $v$ MHV vertices has $2v$ negative-helicity legs,
$v-1$ of which are connected together with propagators. As mentioned
above, each propagator must subsume precisely one negative-helicity
leg and thus we are left with $v+1$ external negative helicities. To
put it another way, if we wish to compute a scattering amplitude with
$q$ negative-helicity gluons we will need $v=q-1$ MHV vertices. As such
let us consider the simplest case of the 4-point NMHV amplitude
$\cA_4(1^+,2^-,3^-,4^-)$. We know from our discussions in \S \ref{amplituderels}
that this must vanish and we would thus like our calculations here to support that.
Even though we end up computing something trivial, it is a good illustration of
the procedure to follow.

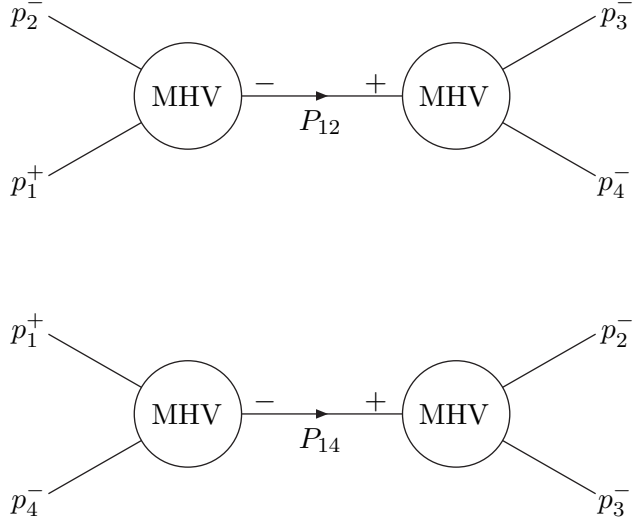
\begin{figure}[ht]
\begin{center}
\begin{picture}(450,200)(0,0)
\SetOffset(65,-40)
\put(0,120){
\Line(90,80)(38,110)
\Line(90,80)(38,50)
\Text(37,112)[r]{$p_2^-$}
\Text(37,48)[r]{$p_1^+$}
\Line(190,80)(242,110)
\Line(190,80)(242,50)
\Text(245,112)[l]{$p_3^-$}
\Text(244,48)[l]{$p_4^-$}
\ArrowLine(100,80)(180,80)
\Text(140,75)[t]{$P_{12}$}
\Text(115,85)[l]{$-$}
\Text(165,85)[r]{$+$}
\BCirc(90,80){20}
\BCirc(190,80){20}
\Text(90,84)[t]{MHV}
\Text(190,84)[t]{MHV}}
\put(0,0){
\Line(90,80)(38,110)
\Line(90,80)(38,50)
\Text(37,112)[r]{$p_1^+$}
\Text(37,48)[r]{$p_4^-$}
\Line(190,80)(242,110)
\Line(190,80)(242,50)
\Text(245,112)[l]{$p_2^-$}
\Text(244,48)[l]{$p_3^-$}
\ArrowLine(100,80)(180,80)
\Text(140,75)[t]{$P_{14}$}
\Text(115,85)[l]{$-$}
\Text(165,85)[r]{$+$}
\BCirc(90,80){20}
\BCirc(190,80){20}
\Text(90,84)[t]{MHV}
\Text(190,84)[t]{MHV}}
\end{picture}
\end{center}
\label{bddkcut}
\caption{\emph{The two MHV diagrams contributing to the $+---$ amplitude. All external momenta
are taken to be outgoing.}}
\end{figure}

As shown in Figure 1.5, there are two diagrams to consider. For
each diagram we should write down the MHV amplitudes corresponding to each vertex and join them
together with the relevant scalar propagator, remembering to use the off-shell
continuation of \eq{1} and \eq{2} to deal with the spinors associated with
the internal particles. The first (uppermost) diagram gives
\begin{eqnarray}
\label{NMHV4point}
C_1=\frac{\langle\lambda_2\,\lambda_{P_{12}}\rangle^3}{\langle\lambda_{P_{12}}\,\lambda_1\rangle
\langle\lambda_1\,\lambda_2\rangle}\frac{1}{P_{12}^2}\frac{\langle\lambda_3\,\lambda_4\rangle^3}
{\langle\lambda_4\,\lambda_{P_{12}}\rangle\langle\lambda_{P_{12}}\,\lambda_3\rangle}\ ,
\end{eqnarray}
where the momentum of the propagator is $P_{12}=-(p_1+p_2)=(p_3+p_4)$. As the external
momenta are massless, $P_{12}^2=2(p_1\cdot p_2)=\langle1\,2\rangle[2\,1]$ and the
off-shell continuation tells us that
\begin{eqnarray}
\label{offshellexplicit}
\lambda_{P_{12}}^{\a}&=&\frac{P_{12}^{\a\da}\tilde{\eta}_{\da}}{[\lt_{P_{12}}\,\tilde{\eta}]}\nonumber\\
&=&-\frac{(\lambda_1^\a\lt_1^{\da}+\lambda_2^\a\lt_2^{\da})\tilde{\eta}_{\da}}{[\lt_{P_{12}}\,\tilde{\eta}]}\nonumber\\
&=&-\lambda_1^\a\frac{a_1}{b}-\lambda_2^\a\frac{a_2}{b}\ ,
\end{eqnarray}
where we have written $\lt_i^{\da}\tilde{\eta}_{\da}=a_i$ and $[\lt_{P_{12}}\,\tilde{\eta}]=b$
for clarity and have kept the denominators of the off-shell continuation explicit
in order to demonstrate that they will drop out of the expressions. Similarly, an
appropriate form for eliminating $\lambda_{P_{12}}$ from the MHV vertex on the
right is
\be
\label{offshellexplicit2}
\lambda_{P_{12}}^\a=\lambda_3^\a\frac{a_3}{b}+\lambda_4^\a\frac{a_4}{b}\ .
\ee
On substituting
for $\lambda_{P_{12}}$ and $P_{12}^2$ \eq{NMHV4point} then becomes
\begin{eqnarray}
\label{NMHV4pointagain}
C_1&=&\frac{ba_1^3}{a_2b^3}\frac{\langle 2\,1\rangle^3}{\langle 2\,1\rangle\langle 1\,2\rangle}
\frac{1}{\langle 1\,2\rangle[2\,1]}\frac{\langle 3\,4\rangle^3}{\langle 4\,3\rangle\langle 4\,3\rangle}
\frac{b^2}{a_3a_4}\nonumber\\
&=&\frac{a_1^3}{a_2a_3a_4}\frac{\langle 3\,4\rangle}{[1\,2]}\ .
\end{eqnarray}

Going through the same procedure for the second contribution in Figure 1.5
gives
\be
\label{secondone}
C_2=\frac{a_1^3}{a_2a_3a_4}\frac{\langle 2\,3\rangle}{[4\,1]}\ ,
\ee
and the final answer is $\cA_4=C_1+C_2$. Momentum conservation is $\sum_{i=1}^4\lambda_i\lt_i=0$,
which can be applied to an expression of the form $\langle 3\,i\rangle[i\,1]$ to give
$\sum_{i=1}^4\langle 3\,i\rangle[i\,1]=\langle 3\,2\rangle[2\,1]+\langle 3\,4\rangle[4\,1]=0$ and
means that $\langle 3\,4\rangle/[1\,2]=-\langle 2\,3\rangle/[4\,1]$. Thus $C_1=-C_2$ and we get
$\cA_4=0$ as expected.

There are two essential points to note here. The first is that when we performed the
off-shell continuation all the denominators of \eq{1} cancelled out. This is in fact
generally true for the amplitudes we will be interested in and thus we will discard them
from now on. The second point is that in $C_1$ and $C_2$, the arbitrary null momentum $\eta$
of the off-shell continuation was still present, lurking as an $\tilde{\eta}_{\da}$ in the
$\alpha_i$. The contributions cancelled in the end so we didn't care too much about this, but
we might worry about the presence of this arbitrary momentum in the calculation of
amplitudes that don't vanish. In fact it tends to crop up frequently and the expressions
that one arrives at seem to depend on $\eta$ at first sight. However, it can be shown that the amplitudes are
$\eta$-independent and it can therefore sometimes be of use to set
$\eta$ to be one of the external momenta in the problem.

This procedure has been implemented both for amplitudes with more external gluons and
amplitudes with more negative helicities. In both cases the complexity grows, \emph{but}
the number of diagrams grows for large $n$ at most as $n^2$ \cite{csw} which is
a marked improvement on the factorial growth of the number of Feynman diagrams
needed to compute the same processes. Further evidence for the procedure and a
heuristic proof from twistor string theory can be found in \cite{csw}, while
a proof based on recursive techniques was given by Risager in \cite{Risager:2005vk}
which was then used to give an MHV-vertex approach to gravity amplitudes \cite{Bjerrum-Bohr:2005jr}.
Evidence for the validity of the procedure for tree \emph{and} loop amplitudes was
given in \cite{Brandhuber:2005kd}.\footnote{We will say more about loop amplitudes shortly.}

On the other
hand Mansfield found a transformation which takes the usual Yang-Mills
Lagrangian and maps it to one where the vertices are explicitly MHV vertices \cite{Mansfield:2005yd}
(see also \cite{Gorsky:2005sf}).
This involves formulating pure Yang-Mills theory in light-cone coordinates and
performing a non-local change of variables which maps the usual 3- and 4-point
vertices that arise in Feynman diagram perturbation theory into an infinite sequence of
MHV vertices starting with the 3-point $--+$ vertex. The procedure also clarifies the
origin of the null vector $\eta$ that we have used to define the off-shell continuation.
It is just the same null vector as is used to define the light-cone formulation of
the theory \cite{Mansfield:2005yd,Brandhuber:2006bf}. For further work related to
understanding the CSW rules from a Lagrangian approach see
\cite{Brandhuber:2006bf,Ettle:2006bw,Feng:2006yy,Boels:2006ir,Boels:2007qn,Boels:2007gv,Brandhuber:2007vm}.

\section{Loop diagrams from MHV vertices}
\label{loopsection}

The CSW rules at tree-level provide a new and effective way of re-organising perturbation
theory and thus lead to more efficient methods for calculating tree-level amplitudes which
often yield simpler results than more traditional approaches. Naturally we would like to be
able to extend this method beyond tree-level and consider quantum corrections which are
often a substantial contribution to the overall result. However as already mentioned the
picture from twistor string theory is not as clear at loop-level and one might expect the
CSW procedure to fail there due to the presence of conformal supergravity.

Nonetheless
Brandhuber, Spence and Travaglini showed that the CSW rules are still valid at one-loop
and provided a concrete procedure to follow from which they re-derived the one-loop
$n$-point MHV gluon scattering amplitudes in $\cN\!=\!4$ super-Yang-Mills \cite{bst}. The answers they obtained
are in complete agreement with the original results derived at 4-point by Green,
Schwarz and Brink from the low energy limit of a string theory \cite{Green:1982sw} and then at
$n$-point by BDDK \cite{Bern:zx}. We will briefly review the method proposed in \cite{bst}
and outline how it can be used to derive the $\cN\!=\!4$ amplitudes. Chapters \ref{chapter2} and
\ref{chapter3} will then be devoted to applying the same method to the $\cN\!=\!1$ amplitudes
and those in pure Yang-Mills with a scalar running in the loop respectively, thus calculating
all cut-constructible\footnote{It turns out that the CSW approach at loop-level
only calculates the cut-containing terms, thus mirroring the cut-constructibility approach
of BDDK. The rational terms are inextricably linked to these in supersymmetric theories but must be obtained in other
ways in non-supersymmetric ones. See also Appendix \ref{unitarityappendix}.}
contributions to the $n$-point MHV gluon scattering amplitudes in QCD \eq{loopsusy}.

\subsection{BST rules}

The procedure proposed in \cite{bst} can be summarised as follows \cite{bbst}:
\begin{itemize}
\item[{\bf 1.}]
Consider only the colour-stripped or partial amplitudes introduced in
\S \ref{colourordersection}. As already mentioned there, the
remarkable results discussed in Section 7 of \cite{Bern:zx} mean that
this is sufficient to re-construct the entire colour-dependent amplitude.

\item[{\bf 2.}]
Lift the MHV tree-level scattering amplitudes to vertices, by
continuing the internal lines off-shell
using the prescription described in \S \ref{offshellsection}.
Internal lines are then connected by scalar propagators
which join particles of the same spin but opposite helicity.
\item[{\bf 3.}]
Build MHV diagrams with the required external particles
at loop level using the MHV tree-level vertices
and sum over all independent diagrams
obtained in this fashion for a fixed ordering
of external helicity states.
\item[{\bf 4.}]
Re-express the loop integration measure in terms
of the off-shell parametrisation employed for
the loop momenta.
\item[{\bf 5.}]
Analytically continue to $4\!-2\epsilon$ dimensions
in order to deal with infrared divergences
and perform  all loop integrations.
\end{itemize}

\subsection{Integration measure}
\label{intmeasure}

The loop legs that we must integrate over are off-shell and in order to
proceed we must work out the integration measure used in \cite{bst}. The details of the measure were more
concretely worked-out in \cite{Brandhuber:2005kd} using the
Feynman tree theorem \cite{Feynman:1963ax,Feynman:1972mt,Feynman:1972mu}
and we use certain results from there as well as from the original construction of \cite{bst} while
following the review of Section 3 of \cite{bbst}.

We need to re-express the usual integration
measure $d^4L$ over the loop momentum $L$ in terms
of the new variables $l$ and $z$ introduced previously.
After a short calculation we find that\footnote{The $i \varepsilon$ prescription
in the left- and right-hand sides of
\eqref{mmm} was understood in \cite{bst}, and,
as stressed in \cite{Lance,Bena:2004ry,Brandhuber:2005kd}
it is essential in order to correctly perform
loop integrations.} \cite{bst,Brandhuber:2005kd}
\beq
\label{mmm}
{d^4 L \over
L^2 + i \varepsilon} \ = \
{d\cN (l) }
\, {dz \over z + i\varepsilon} \ ,
\eeq
where we define $d^4L:=\prod_{i=0}^3dL_i$ and have introduced the Nair measure  \cite{Nair}
\beq
 d\cN (l) := \, {1\over 4i}\left({\lan l
\, \, dl \ran \, d^2 \tilde{l} \ - \ [ \tilde{l} \, \, d
\tilde{l}] \, d^2 l}\right) = {d^3l\over 2l_0}\ .
\eeq
Eq.~\eqref{mmm} is key to the procedure.
It is important to notice that
the product of the measure factor with
a scalar propagator $d^4 L / (L^2 + i \varepsilon)  $
in \eqref{mmm} is independent of the reference vector $\eta$.
In \cite{Nair}, it was noticed that the Lorentz-invariant
phase space measure for a massless particle can be expressed
precisely in terms of the Nair measure:
\beq
\label{nairmeas}
 d^4l \, \delta^{(+)} (l^2)\ =  \
{d\cN (l)} \ ,
\eeq
where, as before, we write the null vector $l$ as $l_{\a \da} =
l_{\a} \tilde{l}_{\da}$, and in Minkowski space we identify
$\tilde{l} = \pm \bar{l}$ depending on whether $l_0$ is positive or
negative.

Next, we observe that in computing one-loop MHV scattering amplitudes from
MHV diagrams (shown in Figure 1.6)\footnote{We thank the authors of \cite{Brandhuber:2005kd}
for allowing the re-production of Figure 17 of that paper.},
the four-dimensional integration measure which appears is \cite{bst,Brandhuber:2005kd}
\beq
\label{dem}
d\cM \ := \ {d^4 L_1 \over {L_1^2+i\varepsilon}} {d^4 L_2 \over {L_2^2+i\varepsilon}}
\,
\delta^{(4)} (L_2 - L_1 + P_L)
\ ,
\eeq
where $L_1$ and $L_2$ are loop momenta, and $P_L$ is the
external momentum flowing outside the loop%
\footnote{In our conventions all external momenta are
outgoing.}
so that $L_2 - L_1 + P_L=0$.

\begin{figure}[ht]
\label{Figurelast}
\begin{center}
\scalebox{0.75}{\includegraphics{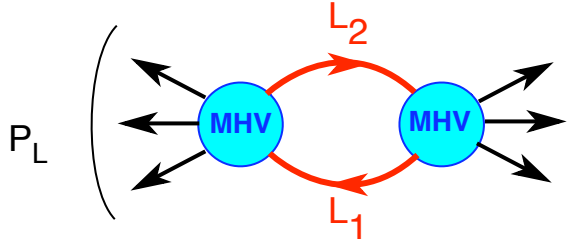}}
\end{center}
\caption{\it A generic MHV diagram contributing to a
one-loop MHV scattering amplitude.
}
\end{figure}

Now we express  $L_1$ and $L_2$ as in
\eqref{off},
\beq
\label{ells}
 L_{i; \a \da} \ =
\ l_{i \a} \llt_{i \da}  \, + \,
z_i \, \eta_{\a}\tilde{\eta}_{\da} \ ,
\qquad i=1,2 \ .
 \eeq
Using \eqref{ells},
we  re-write the argument of the delta function as
\beq
L_2 - L_1 + P_L
= l_2 - l_1 + P_{L; z} \ ,
\eeq
where we have defined
\beq
\label{plz}
P_{L;z} := P_L - z \eta
\ ,
\eeq
and
\beq
z \ := \ z_1\, - \, z_2
\ .
\eeq
Notice that we use the same $\eta$ for both the momenta
$L_1$ and $L_2$.
Using \eqref{ells}, we can then recast \eqref{dem} as
\cite{bst,Brandhuber:2005kd}
\beq
\label{gcar}
d\cM \ = \
{dz_1 \over {z_1+i\varepsilon_1}}\, { dz_2 \over {z_2+i\varepsilon_2}}
\bigg[ {d^3 l_1 \over 2 l_{10}}\,
{d^3 l_2 \over 2 l_{20}}
\,
\delta^{(4)}  (l_2 \, - \, l_1 \,  + \, P_{L; z} )
\bigg]
\ ,
\eeq
where
$\vare_{i} := {\rm sgn}(\eta_0 l_{i0}) \vare =
{\rm sgn}(l_{i0})\vare$, $i = 1, 2$ (the last equality
holds since we are assuming $\eta_0 >0$).

We now convert the integration over $z_1$ and $z_2$ into an integration
over $z$ and \mbox{$z':=z_1+z_2$} and with a careful treatment of the integrals \cite{Brandhuber:2005kd} we can
integrate out $z'$. We also make the replacement
\beq
\label{defini}
 {d^3 l_1 \over 2 l_{10}}\,
{d^3 l_2 \over 2 l_{20}}
\,
\delta^{(4)}  (l_2 \, - \, l_1 \,  + \, P_{L; z} )
\ \rightarrow  \
- \, d{\rm LIPS}  (l_2^{-} , - l_1^{+}; P_{L;z})
\ ,
\eeq
where
\beq
\label{LIPS}
d{\rm LIPS} (l_2^- , - l_1^+;P_{L;z}) \ := \
d^4 l_1 \, \delta^{(+)} (l_1^2) \
d^4 l_2 \, \delta^{(-)} (l_2^2 )\
\delta^{(4)} (l_2 - l_1 + P_{L;z})
\
\eeq
is the two-particle Lorentz-invariant phase
space  (LIPS) measure and we recall that $\delta^{\pm}(l^2):=\theta(\pm l_0)\delta(l^2)$.
Trading the final integral over $z$ for an integration over $P_{L;z}^2$, the
integration measure finally becomes \cite{bst,Brandhuber:2005kd}
\beq
\label{bst}
d\cM \ = \
 2 \pi i \
\theta (P_{L;z}^2)\
{dP_{L;z}^2 \over P_{L;z}^2 \, - \, P_L^2 \, - \, i \vare}
\
d{\rm LIPS} (l_2^{\mp} , - l_1^{\pm}; P_{L;z})
\ .
\eeq

This can now be immediately dimensionally regularised, which is accomplished by simply replacing
the four-dimensional LIPS measure by
its continuation to $D=4-2\epsilon$ dimensions:
\beq
\label{LIPSD}
d^D{\rm LIPS} (l_2^- , - l_1^+;P_{L;z}) \ := \
d^D l_1 \, \delta^{(+)} (l_1^2) \
d^D l_2 \, \delta^{(-)} (l_2^2 )\
\delta^{(D)} (l_2 - l_1 + P_{L;z})
\ .
\eeq
Eq.~\eqref{bst} was  one of the
key results of  \cite{bst}.
It gives a decomposition of the original
integration measure into a $D$-dimensional
phase space measure and a
dispersive measure.
According to Cutkosky's cutting rules
\cite {Cutkosky:1960sp}, the LIPS measure computes
the discontinuity of a Feynman diagram across
its branch cuts. Which discontinuity is evaluated
is determined by the argument of the delta function
appearing in the LIPS measure;
in \eqref{bst} this is $P_{L;z}$ (defined in
\eqref{plz}).
Notice that $P_{L;z}$ always contains a term
proportional to the reference vector $\eta$,
as prescribed by  \eqref{plz}.
Finally, discontinuities are integrated using the
dispersive measure in \eqref{bst},
thereby reconstructing the full amplitude.

As a last remark, notice that
in contradistinction with the cut-constructibility
approach of BDDK, here we sum over all the cuts --
each of which is integrated with the appropriate
dispersive measure.

\section{MHV amplitudes in $\cN\!=\!4$ super-Yang-Mills}

\label{mhvsection}

In this section we will briefly review the one-loop MHV $\cN\!=\!4$ super-Yang-Mills amplitudes
and their derivation using MHV vertices. Many more
details can be found in \cite{Bern:zx,bst}.

\subsection{General integral basis}

It is known that, at one-loop, all amplitudes in massless gauge field theories can
be written in terms of a certain basis of integral functions termed boxes, triangles and
bubbles as well as possible rational contributions (\emph{i.e.} contributions which do not
contain any branch cuts) \cite{Bern:zx,Bern:1994cg}. These functions may involve some number of loop momenta in the
numerator of their integrand, in which case they are termed tensor boxes, triangles or bubbles, though the
basic scalar integrals remain the same and at 4-, 3- and 2-point respectively are the
basic integrals arising at one-loop in scalar $\phi^3$ theory.

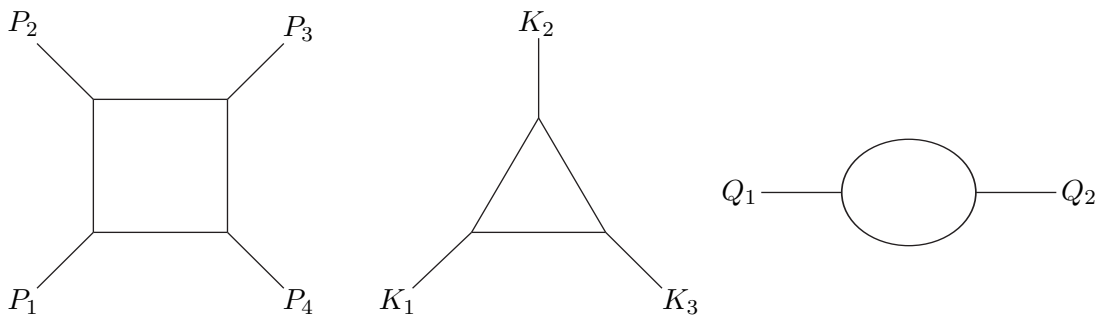
\begin{figure}[ht]
\begin{center}
\begin{picture}(450,130)(0,0)
\SetOffset(5,0)
\put(0,0){
\Text(5,5)[P_1]{$P_1$}
\Text(5,110)[P_2]{$P_2$}
\Text(108,5)[P_4]{$P_4$}
\Text(108,108)[P_3]{$P_3$}
\Line(10,10)(31,31)
\Line(31,31)(31,81)
\Line(31,31)(81,31)
\Line(31,81)(81,81)
\Line(81,31)(81,81)
\Line(31,81)(10,102)
\Line(81,31)(102,10)
\Line(81,81)(102,102)
\Text(145,5)[K_1]{$K_1$}
\Text(197,110)[K_2]{$K_2$}
\Text(251,5)[K_3]{$K_3$}
\Line(150,10)(172,31)
\Line(172,31)(222,31)
\Line(172,31)(197,74)
\Line(222,31)(197,74)
\Line(197,74)(197,104)
\Line(222,31)(243,10)
\Text(273,46)[Q_1]{$Q_1$}
\Text(400,46)[Q_2]{$Q_2$}
\Line(280,46)(310,46)
\COval(335,46)(20,25)(0){Black}{White}
\Line(360,46)(390,46)
}
\end{picture}
\end{center}
\label{boxtribub}
\caption{\emph{Boxes, Triangles and Bubbles. Here $P_i$, $K_i$ and $Q_i$ are
generic momenta representing the contribution of one or more external particles. The different functions discussed below
(1-mass, 2-mass etc.) are all special cases of these}.}
\end{figure}

A box integral is characterised by
having 4 vertices, a triangle integral by having 3 vertices while a bubble has 2. The specific
functions that occur are then characterised not-only by possible powers of loop momenta arising
in the numerator, but by the number of vertices with more than one external leg. If a vertex
has only one external leg it is called a massless vertex (as the external momentum is massless
in the theories we are considering), whilst if it has more than one external leg it is termed a
massive vertex as the external momentum emanating from it does not square to zero.

There are thus 4 generic types of box integrals: 4-mass boxes where all 4 vertices are massive; 3-mass
boxes; 2-mass `easy' boxes where the massive vertices are opposite each other; 2-mass `hard'
boxes where the massive vertices are adjacent and 1-mass boxes. At 4-point the only possible box integral is
a massless box. Similarly one can have 3-mass triangles, 2-mass triangles, 1-mass triangles, 2-mass bubbles
and 1-mass bubbles (as well as massless triangles and massless bubbles at 3- and 2-point respectively).%
\footnote{Note that 1-mass and zero-mass bubbles are usually taken to vanish in dimensional regularization
which is interpreted as a cancellation of infrared and ultraviolet divergences \cite{Bern:1994cg,Collins:1989gx}.}
Explicit forms for all these functions can be found in Appendix I of \cite{Bern:1994cg}.

\subsection{The $\cN\!=\!4$ MHV one-loop amplitudes}

\label{bddkcuts}

Concerning the above decomposition, maximally supersymmetric Yang-Mills theory is special in that its high degree of
symmetry prescribes that its one-loop amplitudes only contain scalar box integral functions (up to finite order in
the dimensional regularisation parameter $\epsilon$) \cite{Bern:zx,Bern:1994cg}. In particular, the MHV
amplitudes only depend on the 2-mass easy (2me) box functions. The full
one-loop $n$-point MHV amplitudes are proportional to the tree-level MHV amplitudes and are given by
\cite{Bern:zx}
\beq\label{fullampl}
A_{n;1}^{\cN=4 \, {\rm MHV}} =  A^{\rm
tree}_n\, V_n^g\ ,
\eeq
where \cite{Bern:zx,csw2}
\begin{eqnarray}
\label{Vfunction-bis}
 V^g_{n}   &=& \sum_{i=1}^{n} \sum_{r=1}^{[{n\over 2}]-1}
\Bigl(
1 - {1\over 2} \delta_{{n\over 2} - 1, r}
\Bigr)\,
F_{n:r;i}^{2m\,e}
\ .
\end{eqnarray}

The basic scalar box integral $I_4$ is defined by
\beq
I_4 = -i (4\pi)^{2-\epsilon}\, \int
\frac{d^{4-2\epsilon}p}{(2\pi)^{4-2\epsilon}}\,\,
\frac{1}{p^2(p-P_1)^2(p-P_1-P_2)^2(p+P_4)^2}
\ ,
 \eeq
where dimensional regularisation is used to take care of infrared divergences.
The relevant integrals arising in \eq{Vfunction-bis} are related to $I_4$
for different choices of the external momenta at each vertex $P_i$ $(i=1\ldots 4)$.
These are denoted by $I_{4:r;i}^{2m\,e}$ - see Figure 1.8 - and are given in terms of the
$F_{n:r;i}^{2m\,e}$ by
\begin{eqnarray}
I_{4:r;i}^{2m\,e}  =  \frac{2F_{n:r;i}^{2m\,e}} {
              t^{[r]}_{i}  t^{[n-r-2]}_{i+r+1}-t^{[r+1]}_{i-1}  t^{[r+1]}_{i}
               } \ ,
\end{eqnarray}
with
\beqa
\label{t}
t_i^{[r]} &=& (k_i+k_{i+1}+\dots+k_{i+r-1})^2\ ,\ \ r>0\nonumber\\
t_i^{[r]} &=& t_i^{[n-r]}\ ,\ \ r<0\ ,
\eeqa
where the $k_i$ are the external momenta.
The explicit form of $F_{n:r;i}^{2m\,e}$
is given by \cite{Bern:zx}
\begin{eqnarray}
\label{2meold}
\!\!\!\!F_{n:r;i}^{2m\,e}&=&-\frac{1}{\epsilon^2}\left[(-t_{i-1}^{[r+1]})^{-\epsilon}
+(-t_i^{[r+1]})^{-\epsilon}-(-t_i^{[r]})^{-\epsilon}-(-t_{i+r+1}^{[n-r-2]})^{-\epsilon}\right]\nonumber\\
\!\!\!\!&+& \Li2\left(1-\frac{t_i^{[r]}}{t_{i-1}^{[r+1]}}\right)+\Li2\left(1-\frac{t_i^{[r]}}{t_i^{[r+1]}}\right)
+\Li2\left(1-\frac{t_{i+r+1}^{[n-r-2]}}{t_{i-1}^{r+1}}\right)\nonumber\\
\!\!\!\!&+&\Li2\left(1-\frac{t_{i+r+1}^{n-r-2}}{t_i^{[r+1]}}\right)
-\Li2\left(1-\frac{t_i^{[r]}t_{i+r+1}^{[n-r-2]}}{t_{i-1}^{[r+1]}t_i^{[r+1]}}\right)
+\frac{1}{2}\log^2\left(\frac{t_{i-1}^{[r+1]}}{t_i^{[r+1]}}\right)\, ,
\end{eqnarray}
where $\Li2$ is Euler's dilogarithm
\be
\label{dilog}
\Li2(z):=-\int_0^{z}\! dt\,\frac{\log(1-t)}{t}\ .
\ee

\begin{figure}[ht]
\begin{center}
\begin{picture}(150,130)(0,0)
\SetOffset(5,5)
\put(0,0){
\Text(7,107)[k_i+r]{$k_{i+r}$}
\Text(112,2)[k_i-1]{$k_{i-1}$}
\Line(10,10)(31,31)
\Line(31,31)(31,81)
\Line(31,31)(81,31)
\Line(31,81)(81,81)
\Line(81,31)(81,81)
\Line(31,81)(10,102)
\Line(81,31)(102,10)
\Line(81,81)(102,102)
\Line(31,31)(01,31)
\Line(31,31)(31,01)
\Line(81,81)(111,81)
\Line(81,81)(81,111)
\Text(31,-5)[k_i]{$k_i$}
\Text(-3,31)[k_i+r-1]{$k_{i+r-1}$}
\Text(95,120)[k_i+r+1]{$k_{i+r+1}$}
\Text(124,81)[k_i-2]{$k_{i-2}$}
\Text(55,55)[I_4]{$I_{4:r;i}^{2m\,e}$}
\DashCArc(31,31)(20,180,270){2}
\DashCArc(81,81)(20,0,90){2}
}
\end{picture}
\end{center}
\label{2mebox}
\caption{\emph{The 2-mass easy box function.}}
\end{figure}
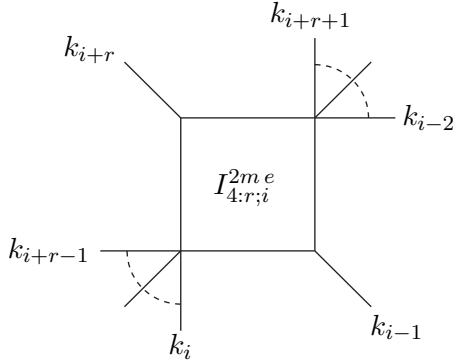

The one-loop MHV amplitudes were constructed in  \cite{Bern:zx}
from tree diagrams using
cuts. A given cut results in singularities in the relevant
momentum channels and by considering all possible cuts one can
construct the full set of possible singularities. From this and
unitarity one can deduce the amplitude as given in
(\ref{fullampl}).
More explicitly, consider a cut one-loop MHV diagram where the cut
separates the external momenta $k_{{m_1}}$ \& $k_{{m_1}-1}$, and
$k_{{m_2}}$ \& $k_{{m_2}+1}$ (i.e.~the set of external momenta
$k_{{m_1}}, k_{{m_1}+1},...,k_{{m_2}}$ lie to the left of the cut,
and the set $k_{{m_2}+1},k_{{m_2}+2},...,k_{{m_1}-1}$ lie to the
right, with momenta labelled clockwise and outgoing). This
separates the diagram into two MHV tree diagrams
connected only by two momenta $l_1$ and $l_2$
flowing across the cut, with
\beq
\label{pl}
l_1 = l_2 + P_L
\ ,
\eeq
where  $P_L = \sum_{i={m_1}}^{{m_2}}k_i$ is the sum of the
external momenta on the left of the cut. The momenta $l_1, l_2$
are taken to be null.
It is important to note that the resulting integrals
are not equal to the corresponding Feynman integrals where $l_{1}$ and
$l_{2}$ would be left off shell; however, the discontinuities in the
channel under consideration are identical
and this gives enough information to
determine the full amplitude uniquely.

However, we will now sketch how to derive the MHV amplitudes using
the method of MHV diagrams. This is quite similar, but not
identical to the approach of BDDK using cut-constructibiliy, a
brief review of which can be found in Appendix \ref{unitarityappendix}.

\subsection{MHV vertices at one-loop}

\begin{figure} [ht]
\label{fig1}
\centerline {
\includegraphics[width=4in]{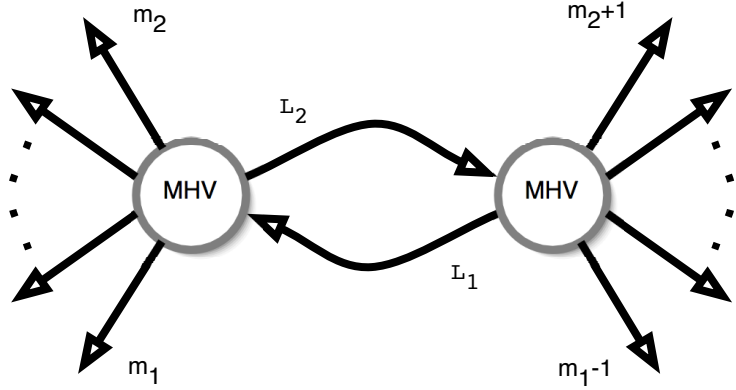}
}
\caption{\it A one-loop MHV diagram
computed using MHV amplitudes as interaction vertices.
This diagram has the
momentum structure of the cut referred to at the end of
\S \ref{bddkcuts}.}
\end{figure}

\begin{enumerate}
\item To each MHV vertex we associate the appropriate form of the MHV amplitude for that vertex,
recalling that internal lines must be taken off-shell using the prescription described in
\S \ref{offshellsection}. To each internal line we associate a scalar propagator and integrate
over the appropriate loop momentum. The generic expression for the diagram of Figure 1.9 then reads:
\begin{eqnarray}
\label{genericloop}
\cA&=&\int\frac{d^4L_1}{(2\pi)^4}\frac{d^4L_2}{(2\pi)^4}\frac{1}{L_1^2+i\varepsilon}\frac{1}{L_2^2+i\varepsilon}
\cA_L\cA_R\nonumber\\
&=&\int\frac{d^4L_1}{L_1^2+i\varepsilon}\frac{d^4L_2}{L_2^2+i\varepsilon}
\frac{iN_L\delta^{(4)}(L_2-L_1+P_L)}{D_L}\frac{iN_R\delta^{(4)}(L_1-L_2+P_R)}{D_R}\nonumber \\
&=&\delta^{(4)}(P_L+P_R)\int\frac{d^4L_1}{L_1^2+i\varepsilon}\frac{d^4L_2}{L_2^2+i\varepsilon}
\delta^{(4)}(L_2-L_1+P_L)\frac{iN_L}{D_L}\frac{iN_R}{D_R}\, .
\end{eqnarray}
Here $L$ and $R$ denote the left and right vertices respectively and we have
\newline\mbox{$P_L:=k_{m_1}+k_{m_1+1}+\ldots +k_{m_2}$} and
\mbox{$P_R:=k_{m_2+1}+k_{m_2+2}+\ldots+k_{m_1-1}$}. $N$ and $D$ denote the
functions of spinor variables describing the numerator and denominator of each
MHV vertex respectively and we have included a factor of $i(2\pi)^4$ with each
vertex in keeping with Nair's supersymmetric description \cite{Nair,bst,witten}.

\item In \cite{bst} an approach using Nair super-vertices was used.
Here we will just consider the usual MHV vertices for ease of transition to the
later chapters where we will discuss MHV amplitudes in theories with less supersymmetry.
In this case there are two possibilities to consider.
The first is where both external negative-helicity gluons lie on one MHV vertex and the second is where they lie on different
vertices (see \emph{e.g.} Figure 2.4). After some manipulation (employing the
Schouten identity stated in Appendix A), they can be shown to give the same
contribution. Extracting an overall factor of
\be
\label{n=4tree}
\cA_n^{\mathrm{tree}}:=i(2\pi)^4\delta^{(4)}(P_L+P_R)\frac{\langle i\,j\rangle^4}
{\prod_{k=1}^n\langle k\,\,k+1\rangle}\ ,
\ee
where $i$ and $j$ are the external negative-helicity gluons and regulating by
promoting the integrals to $4-2\epsilon$ dimensions, \eq{genericloop} becomes
\be
\label{mhvloop}
\cA=\frac{i}{(2\pi)^4}\cA_n^{\mathrm{tree}}\int\!d\mathcal{M}\hat{\cR}\ ,
\ee
where $d\cM$ is the measure \eq{bst} derived previously and
\beq
 \label{funct-R}
 \hat{\cR} \ := \ {
\lan m_1 - 1 \, m_1 \ran \lan l_2 \, l_1\ran \over \lan m_1 -1 \,
l_1 \ran \lan -l_1 \, m_1 \ran}\ { \lan m_2  \, m_2 + 1 \ran \lan
l_1 \, l_2 \ran \over \lan m_2  \, l_2 \ran \lan -l_2  \, m_2 + 1
\ran} \ .
 \eeq

\item
Following equations (2.11)-(2.16) of \cite{bst} we may finally
write $\hat{\cR}$ as a signed sum (\emph{i.e.} two terms come with
plus signs and two with minus signs - see Eq.~(2.13) of \cite{bst})
of terms of the form\footnote{Be careful to
note that in the following expression $i$ and $j$ refer to the
different possibilities $m_1$, $m_2$, $m_2+1$ and $m_1-1$, and
\emph{not} to the negative-helicity particles of the overall amplitude which now
only arise in the factor of $\cA_n^{\mathrm{tree}}$.}
\be
\label{cR}
\cR(i,j):=\frac{\langle i\,l_2\rangle}{\langle i\,l_1\rangle}
\frac{\langle j\,l_1\rangle}{\langle j\,l_2\rangle}\ .
\ee
Once expressed in terms of momenta by multiplying top and
bottom by appropriate anti-holomorphic spinor invariants,
 cancellations arise between different terms of the signed sum and we can schematically
write $\hat{\cR}=\sum\cR\rightarrow\sum\cR_{\mathrm{eff}}$ with \cite{bst,Brandhuber:2005kd}
\be
\label{Reff}
\cR_{\mathrm{eff}}=\frac{1}{4}\frac{P_{L;z}^2(i\,j)-2(i\,P_{L;z})(j\,P_{L;z})}{(i\,l_1)(j\,l_2)}\, .
\ee
The notation $(a\,b)$ here is shorthand for $(a\cdot b)$. \eq{mhvloop}
then becomes
\be
\label{mhveff}
\cA=\frac{i}{(2\pi)^4}\cA_n^{\mathrm{tree}}\sum\int\!d\mathcal{M}\cR_{\mathrm{eff}}\, .
\ee
It is worth mentioning that the procedure of expressing $\sum\cR\rightarrow\sum\cR_{\mathrm{eff}}$
is a clever way of cancelling the triangle and bubble contributions in $\cR$ to leave only
box functions \cite{bst,Brandhuber:2005kd} and is equivalent to the usual method of
Passarino-Veltman reduction of \cite{pv}. \eq{mhveff} is then the basic integral that we have
to work with and we will consider the specific term $\cR_{\mathrm{eff}}(m_1,m_2)$ for definiteness.

\item Recall that the measure $d\cM$ involves a dispersive part and an integral over
Lorentz-invariant phase space (\emph{d}LIPS). We wish to begin by performing the integral over
this phase space. For this we go to the centre of mass frame for $P_{L;z}$ - $P_{L;z}=P_0(1,\vec{0})$ -
and parametrize $l_1=\frac{1}{2}P_0(1,\vec{v})$ and $l_2=\frac{1}{2}P_0(-1,\vec{v})$ with
$\vec{v}:=(\sin\theta_1\cos\theta_2,\sin\theta_1\sin\theta_2,\cos\theta_1)$.
In $4-2\epsilon$ dimensions, the LIPS measure \eq{LIPSD} can be written in terms of the
angles $\theta_1$ and $\theta_2$ as\footnote{See Appendix \ref{DLIPSappendix} for
details.}
\be
\label{lips4-2e}
d^{4-2\epsilon}\mathrm{LIPS}=\frac{\pi^{{1\over2}-\epsilon}}{4\Gamma\left({1\over2}-\epsilon\right)}
\left|\frac{P_{0}^2}{4}\right|^{-\epsilon}d\theta_1d\theta_2(\sin\theta_1)^{1-2\epsilon}
(\sin\theta_2)^{-2\epsilon}\, ,
\ee
and the denominator of \eq{Reff} as
\be
\label{denominator}
(m_1\,l_1)(m_2\,l_2)=\left|\frac{P_{0}^2}{4}\right|m_{10}(1-\cos\theta_1)(A+B\sin\theta_1\cos\theta_2+
C\cos\theta_1)\, ,
\ee
where $m_1:=m_{10}(1,0,0,1)$ and $m_2:=(A,B,0,C)$ with $A^2=B^2+C^2$. The numerator of \eq{Reff} does not
involve $l_1$ or $l_2$ and we leave it as $N(P_{L;z})$ for now. We thus have
\begin{eqnarray}
\label{beenakkerint}
\Lambda_1\int\! d\mathcal{W}\Lambda_2\int\frac{d\theta_1d\theta_2(\sin\theta_1)^{1-2\epsilon}
(\sin\theta_2)^{-2\epsilon}}{(1-\cos\theta_1)(A+B\sin\theta_1\cos\theta_2+
C\cos\theta_1)}\ ,
\end{eqnarray}
where
\begin{eqnarray}
\label{stuff}
\Lambda_1&:=&\frac{i}{(2\pi)^4}\frac{\pi^{1/2-\epsilon}}{4^{1-\epsilon}\Gamma(1/2-\epsilon)}\cA_n^{\mathrm{tree}}\, ,\\
\Lambda_2&:=&\frac{N(P_{L;z})}{m_{10}P_{0}^2}(P_{0}^2)^{-\epsilon}\, ,\\
d\mathcal{W}&:=&(2\pi i)\theta(P_{L;z}^2)\frac{dP_{L;z}^2}{P_{L;z}^2-P_{L}^2-i\varepsilon}\, .
\end{eqnarray}

The integral over $\theta_1$ and $\theta_2$ has been performed in \cite{vanNeerven:1985xr} and
we borrow the result in a form from \cite{wim}. Converting $A,B,C,m_{10}$ and $P_0$ back into
Lorentz-invariants we obtain:
\begin{eqnarray}
\label{dispersint}
4\pi\Lambda_1\frac{1}{\epsilon}\int\!d\mathcal{W}(P_{L;z}^2)^{-\epsilon}
\mbox{}_{2}F_1 \left( 1, -\e, 1- \e; a_z P_{L;z}^2 \right)\ .
\end{eqnarray}
In Equations (\ref{lips4-2e}), (\ref{stuff}) and (\ref{dispersint}) above, $\Gamma$ is the gamma function and $\mbox{}_2F_1$ the Gauss hypergeometric
function. They can be defined by
\begin{eqnarray}
\label{gammahyper}
\!\!\!\!\!\Gamma(z)&:=&\int_0^{\infty}\!dt\,\,t^{z-1}e^{-t}\ ;\ \Re[z]>0\, ,\\
\!\!\!\!\!\mbox{}_{2}F_1 \left( a, b, c; z \right)&:=&\frac{\Gamma(c)}{\Gamma(b)\Gamma(c-b)}
\int_0^1\!dt\,\,t^{b-1}(1-t)^{-b+c-1}(1-tz)^{-a}
\end{eqnarray}
where the second definition holds when $\Re[c]>\Re[b]>0$ and $|\arg(1-z)|<\pi$. $a_z$
is defined to be $a_z:=(i\,j)/N(P_{L;z})$ and so is equal to $(m_1\,m_2)/N(P_{L;z})$ in this
case.

\item Finally, we would like to evaluate this dispersive integral \eq{dispersint}.
In \cite{bst}, this was done by combining different terms coming from different
$\cR_{\mathrm{eff}}$ to give a convergent integral. In fact the different
$\cR_{\mathrm{eff}}$ that one must combine come \emph{not} from different $i$ and $j$
in $\cR_{\mathrm{eff}}(i\,j)$ as obtained from $\hat{\cR}=\sum\cR\rightarrow\sum\cR_{\mathrm{eff}}$,
but from $\cR_{\mathrm{eff}}$ with the \emph{same} $i$ and $j$ coming from different terms in
the overall summation over all the cuts of the one-loop integral mentioned at the end of
\S \ref{intmeasure} and not so far alluded to in \emph{this} section. This summation
is just a summation over all cyclic partitions of the external particles between the two
MHV vertices, but at the level of the integrals we have arrived at in \eq{dispersint}
the summation over $\cR_{\mathrm{eff}}(i\,j)$ with the same values of $i$ and $j$ from
different orderings of the external particles serves to re-construct the 2me box functions
from their different cuts.

The integrals are explicitly done by expanding the hypergeometric functions above in an
expansion in $\epsilon$ in terms of
polylogarithms (generalisations of $\Li2$) and then combining different cuts of the
same box function to give a convergent answer. A key ingredient in all this is the
knowledge that the final result will be independent of $\eta$. $\eta$ has already been
eliminated from the dispersive integration measure by converting the integral over
$z$ and $z'$ into an integral over $P_{L;z}$, so one may expect that even before we
evaluate this dispersive integral we should be able to pick a particular value for
$\eta$ to simplify the calculation. However, in \cite{bst} a stronger gauge invariance
was proposed; namely that one may choose $\eta$ separately for each box function. This
was checked numerically in \cite{bst} and independently (also numerically) in \cite{Lance}
and further evidence was provided in \cite{Brandhuber:2005kd}.\footnote{See also Appendix \ref{triangleappendix}
for an analytic
proof of the same statement for triangle functions.} It means that
one can write \mbox{$N(P_{L;z})=N(P_{L})$} if one chooses $\eta=m_1$ or $\eta=m_2$ in all four
$\cR_{\mathrm{eff}}(m_1\,m_2)$ which contribute to that particular box function.

The final result (up to finite order in $\epsilon$) given in Equation (5.16) of \cite{bst} is that the contribution of
a particular box function (say a generic box function such as that in Figure 1.8, which would
come from combining the four terms with $m_1=k_{i+r}$ and $m_2=k_{i-1}$) is
\begin{eqnarray}
\label{2menew}
F^{2m\,e}_{n:r;i}&=&-\frac{1}{\epsilon^2}\left[(-t_{i-1}^{[r+1]})^{-\epsilon}
+(-t_i^{[r+1]})^{-\epsilon}-(-t_i^{[r]})^{-\epsilon}-(-t_{i+r+1}^{[n-r-2]})^{-\epsilon}\right]\nonumber\\
&+&\Li2\left(1-a\,t_i^{[r]}\right)+\Li2\left(1-a\,t_{i+r+1}^{[n-r-2]}\right)\nonumber\\
&-&\Li2\left(1-a\,t_{i-1}^{[r+1]}\right)-\Li2\left(1-a\,t_{i}^{[r+1]}\right)\ ,
\end{eqnarray}
where
\be
\label{ais}
a=\frac{t_i^{[r]}+t_{i+r+1}^{[n-r-2]}-t_{i-1}^{[r+1]}-t_i^{[r+1]}}
{t_i^{[r]}t_{i+r+1}^{[n-r-2]}-t_{i-1}^{[r+1]}t_i^{[r+1]}}\ .
\ee
Equation \eq{2menew} is in fact equal to \eq{2meold} but is an alternative form which was
discovered in \cite{Duplancic:2000sk} and independently derived in \cite{bst} and involves one less dilogarithm and one less logarithm than
\eq{2meold}. After summing over all partitions of the external particles between the
two MHV vertices we recover \eq{fullampl}.
\end{enumerate}

The calculation outlined above is essentially what we will follow in Chapters
2 and 3 for the $\cN\!=\!1$ and $\cN\!=\!0$ MHV amplitudes. For full details of the
amplitudes in $\cN\!=\!4$ see \cite{bst} and for a short discussion on the overall
$\epsilon$-normalisation of the result obtained there compared with the one obtained originally
in \cite{Bern:zx} see Appendix \ref{DLIPSappendix}.

\chapter{MHV amplitudes in $\cN\!=\!1$ super-Yang-Mills}

\label{chapter2}

In Chapter 1 we described some of the hidden simplicity of perturbative gauge theory -
in particular in the context of maximally supersymmetric Yang-Mills - and saw how it
may be applied to simplifying the calculation of perturbative quantities such as scattering
amplitudes. The many techniques available to illuminate the perturbative structure included
colour stripping, the use of a helicity scheme and supersymmetric decompositions. A perturbative
duality with a twistor string theory highlighted the unexpected compactness of the MHV amplitudes
at tree-level and provided motivation for a new perturbative expansion of gauge theory - the CSW
rules.

The CSW rules have been shown to be valid even at loop level - despite the failure of the
duality with twistor string theory - and the MHV amplitudes in $\cN\!=\!4$ super-Yang-Mills
were derived using these rules in \cite{bst} and shown to be identical to the original
derivation of \cite{Bern:zx} using 2-particle cuts. As a bonus, the CSW rules also
gave rise to a representation of the 2-mass easy box functions that is simpler to that originally
used in \cite{Bern:zx}. However, at the time it was far from certain that these remarkable
techniques would be applicable to other gauge theories. One might not have been surprised
if such results only held for a theory with an extremely high amount of symmetry such as
$\cN\!=\!4$ SYM.

In \cite{bbst,quig} a first step towards establishing the general validity of the MHV-vertex formalism was taken and it was shown independently
by Bedford, Brandhuber, Spence \& Travaglini and Quigley \& Rosali that the CSW rules correctly
calculate the MHV amplitudes in theories with less supersymmetry such as $\cN\!=\!1$ and
$\cN\!=\!2$ super-Yang-Mills. In particular the MHV amplitudes for scattering of external
gluons with an $\cN\!=\!1$ chiral multiplet running in the loop was calculated and it was
found that the results exactly agree with those originally obtained by BDDK in \cite{Bern:1994cg}.
This chapter follows \cite{bbst} and shows how the $\cN\!=\!1$ MHV amplitudes may be
obtained from MHV vertices.


\section{The $\cN=1$ MHV amplitudes at one-loop}
\label{n=1reviewsection}

The expression for the MHV amplitudes
at one-loop in $\cN=1$ SYM was obtained for
the first time by BDDK in \cite{Bern:1994cg}
using the cut-constructibility method.
We will shortly give their explicit result
and then re-write it
by introducing appropriate functions. This
turns out to be useful when we compare
the BDDK result to that which we will derive by using
MHV diagrams.

In order to obtain the one-loop MHV amplitudes in
$\cN\!=\!1$ and $\cN\!=\!2$ SYM it is sufficient to compute
the contribution  $\cA^{{\cN}\!=\!1, {\rm chiral}}_{n}$
to the one-loop MHV amplitudes coming from
a single $\cN\!=\!1$ chiral multiplet.
This was calculated in \cite{Bern:1994cg},
and the result turns out to be proportional
to the Parke-Taylor MHV tree amplitude \cite{mhv}
\beq
\cA_n^{\rm tree} \ := \
{\lan i \, j \ran^4
\over
\prod_{k=1}^{n}
\lan k\, k+1 \ran
}
\ ,
\eeq
as is also the case with the one-loop
MHV  amplitudes in $\cN\!=\!4$ SYM.
However, in contradistinction with that case,
the remaining part of the $\cN\!=\!1$ amplitudes
depends non-trivially on the
position of the negative-helicity gluons $i$ and $j$.
The result obtained in \cite{Bern:1994cg} is:
\beqa
\label{BDDKNeq1}
\cA^{{\cN}=1, {\rm chiral}}_{n}
&=&
\cA^{\rm tree}_{n}\, \cdot \,
\biggl\{ \sum_{m=i+1}^{j-1}
\sum_{s=j+1}^{i-1} b_{m,s}^{i,j}
\,
B(t_{m+1}^{[s-m]},t_m^{[s-m]},
t_{m+1}^{[s-m-1]},t_{s+1}^{[m-s-1]})
\nonumber
\\
&+&
\sum_{m=i+1}^{j-1} \sum_{a\in \cD_m}
\, c^{i,j}_{m,a}
\,
{\log (t_{m+1}^{[a-m]}/t_m^{[a-m+1]})
\over
t_{m+1}^{[a-m]}-
t_m^{[a-m+1]}}\nonumber\\
\ &+& \
\sum_{m=j+1}^{i-1} \sum_{a\in \cC_m}
\,
c_{m,a}^{i,j}
\,
{\log (t_{a+1}^{[m-a]}/t_{a+1}^{[m-a-1]})
\over
t_{a+1}^{[m-a]}-t_{a+1}^{[m-a-1]}}
\nonumber \\
&+&{c_{i+1,i-1}^{i,j}\over
t_{i}^{[2]}} K_0(t_i^{[2]})
\, + \,
{c_{i-1,i}^{i,j}
\over
t_{i-1}^{[2]}}
K_0(t_{i-1}^{[2]})\nonumber\\
\, &+& \,
{c_{j+1,j-1}^{i,j}
\over t_j^{[2]}}
K_0(t_j^{[2]})\, + \,
{c_{j-1,j}^{i,j}
\over t_{j-1}^{[2]}}
K_0(t_{j-1}^{[2]})
\biggr\}
\ ,
\eeqa
where $t_i^{[k]} := (p_i + p_{i+1} + \cdots + p_{i+k-1})^2$
for $k\geq 0$, and $t_i^{[k]} =t_i^{[n-k]}$ for $k <0$.
The sums in the second and third line of
\eqref{BDDKNeq1} cover the ranges
$\cC_m$ and $\cD_m$ defined by
\beqa
\cC_m &=&
\left\{
\begin{tabular}{ll}
$\{i,i+1, \ldots ,j-2 \}, \hskip 1.4cm   m=j+1,$
\cr \cr
$ \{i,i+1, \ldots ,j-1 \} , \hskip 1.4cm
j+2 \leq m \leq i-2 , $
\cr \cr
$\{i+1,i+2, \ldots ,j-1 \} , \hskip .7cm m =i-1,$
\end{tabular}\right.
\eeqa
and
\beqa
\cD_m &=&
\left\{
\begin{tabular}{ll}
$\{j,j+1, \ldots ,i-2 \}, \hskip 1.4cm  m=i+1,$
\cr \cr
$ \{j,j+1, \ldots ,i-1 \} , \hskip 1.4cm  i+2
\leq m \leq j-2 , $
\cr \cr
$\{j+1,j+2, \ldots ,i-1 \} , \hskip .7cm m =j-1.$
\end{tabular}\right.
\eeqa
The coefficients
$b_{m,s}^{i,j}$ and $c_{m,a}^{i,j}$ are
\beq
\label{bdefn1}
b_{m, s}^{i, j} \ :=  \
-2 \, \frac{
\tr_{+} \left( \kslash_{i} \kslash_{j} \kslash_{m}
\kslash_{s} \right)
\tr_{+} \left( \kslash_{i} \kslash_{j} \kslash_{s}
\kslash_{m} \right) }
{ [ (k_i + k_j)^{2}]^2
\,
[( k_{m} + k_s)^2]^2}
\ ,
\eeq
\beq
\label{cdef}
c_{m, a}^{i, j} \ := \
\left[
{\tr_{+} ( \kslash_{m}\kslash_{a+1}\kslash_{j} \kslash_{i}   )
\over (k_{a+1} + k_m )^2}
 \, - \,
{\tr_{+} ( \kslash_{m}\kslash_{a}\kslash_{j} \kslash_{i}   )
\over (k_{a} + k_m )^2}
\right]
\,
{ {   \tr_+ ( \kslash_{i}\kslash_{j} \kslash_{m}
\qslash_{m,a} ) - \tr_+ ( \kslash_{i} \kslash_{j}
\qslash_{m,a} \kslash_{m} )
\over [(k_i + k_j)^2]^2}}
\, ,
\eeq
where $q_{r,s}:=\sum_{l=r}^{s} k_l$.
Notice that both coefficients $b_{m, s}^{i, j}$
and $c_{m, a}^{i, j}$ are symmetric under
the exchange of $i$ and $j$.
In the case of $b$ this is evident; for $c$ it is also manifest as $c$ is expressed as the product of
two antisymmetric quantities.
\begin{figure} [ht]
\label{fig1}
\vspace{.2in}
\centerline {
\includegraphics[width=4in]{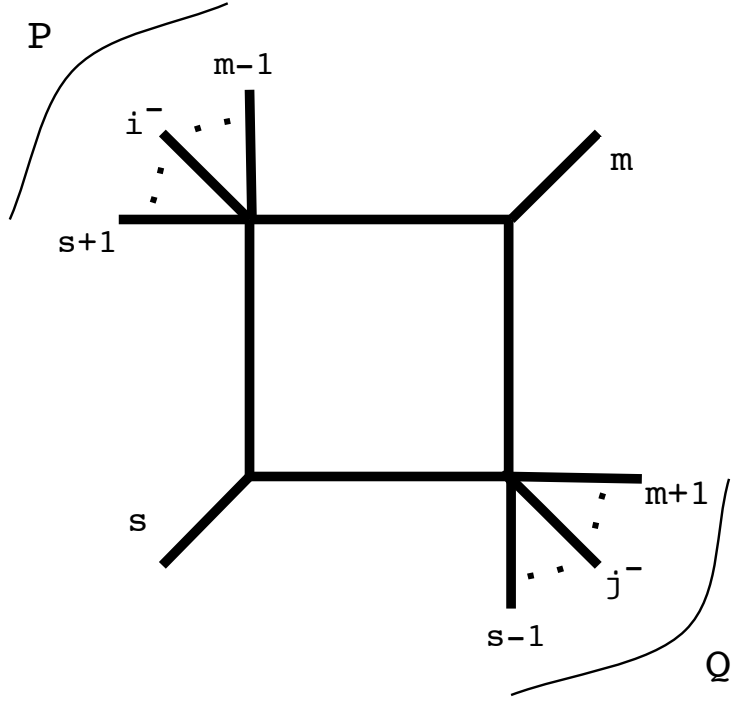}
}
\vspace{.2in}
\caption{\it
The box function $F$ of \eqref{F}, whose finite part
$B$, Eq.~\eqref{Bniceonecyril},
appears in the $\cN=1$ amplitude \eqref{BDDKNeq1}.
The two external gluons with negative helicity
are labelled by $i$ and $j$.
The legs labelled by $s$ and $m$ correspond to the null momenta
$p$ and $q$ respectively
in the notation of \eqref{Bniceonecyril}.
Moreover, the quantities
$t_{m+1}^{[s-m]}$, $t_{m}^{[s-m]}$,
$t_{m+1}^{[s-m-1]}$, $t_{s+1}^{[m-s-1]}$
appearing in the box function $B$ in
\eqref{Neq1}
correspond to the  kinematical invariants
$t:=(Q+p)^2$, $s:=(P+p)^2$, $Q^2$, $P^2$ in the notation of
\eqref{Bniceonecyril}, with $p+q+P+Q=0$.
}
\end{figure}
\begin{figure} [ht]
\label{fig2}
\vspace{.2in}
\centerline {
\includegraphics[width=4in]{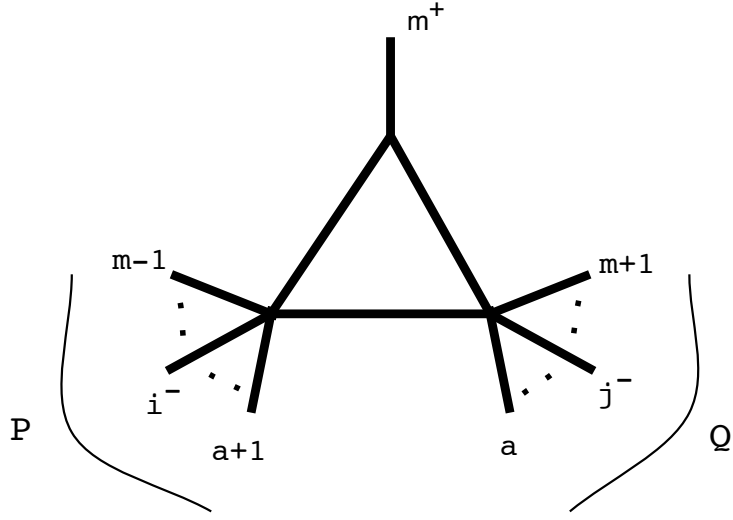}
}
\vspace{.2in}
\caption{\it
A triangle function, corresponding to the first term
$T_{\epsilon} ( p_m, q_{a+1,m-1}, q_{m+1, a})$
in the second line of \eqref{Neq1}.
$p$, $Q$ and $P$ correspond to $p_m$, $q_{m+1, a}$
and $q_{a+1, m-1}$ in the notation of Eq.~\eqref{Neq1},
where $j\in Q$, $i\in P$.
In particular, $Q^2 \to   t_{m+1}^{[a-m]}$ and
$P^2 \to t_{m}^{[a-m+1]}$.
}
\end{figure}
\begin{figure} [ht]
\label{fig3}
\vspace{.2in}
\centerline {
\includegraphics[width=4in]{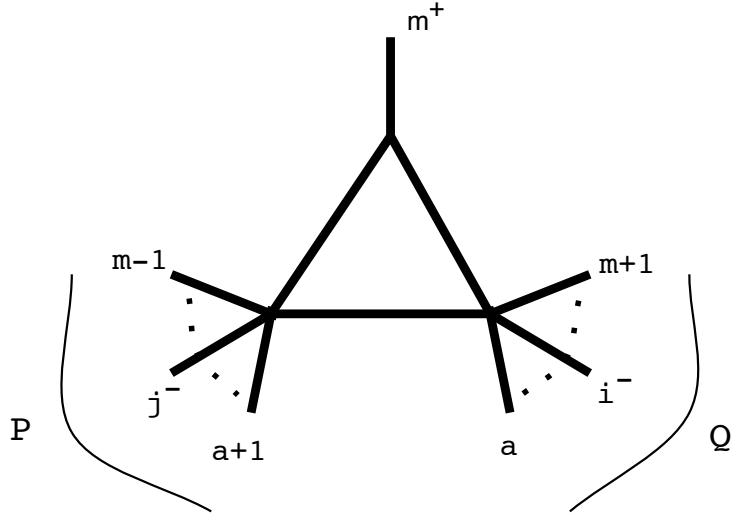}
}
\vspace{.2in}
\caption{\it
This  triangle function  corresponds to the second term
in the second line of \eqref{Neq1} -- where $i$ and $j$
are swapped.
As in Figure 2.2, $p$, $Q$ and $P$ correspond to
$p_m$, $q_{m+1, a}$ and $q_{a+1, m-1}$ in
the notation of Eq.~\eqref{Neq1},
where now $i\in Q$, $j\in P$.
In particular, $Q^2 \to   t_{a+1}^{[m-a]}$ and
$P^2 \to t_{a+1}^{[m-a-1]}$.}
\end{figure}
The function $B$ in the first line of  \eqref{BDDKNeq1}
is the ``finite'' part of the
easy two-mass  (2me) scalar box function
$F (s,t,P^2,Q^2)$, with
\beq
\label{F}
F (s,t,P^2,Q^2) \ := \
-{1\over \epsilon^2} \Bigl[ (-s)^{-\epsilon} \, +\,
(-t)^{-\epsilon} \, -\, (-P^2)^{-\epsilon} \, -\,
(-Q^2)^{-\epsilon} \Bigr] \ + \
B (s,t,P^2,Q^2) \ .
\eeq
As in \cite{bst} we have introduced the following
convenient kinematical invariants:
\beq
s\ := \ (P+p)^2 \, , \qquad t \ := \ (P+q)^2
\ ,
\eeq
where $p$ and $q$ are null momenta and $P$ and $Q$ are
in general massive. We also have momentum conservation in the form \mbox{$p+q+P+Q=0$}.%
\footnote{
The kinematical invariant $s=(P+p)^2$ should not be confused with
the label $s$ which is also used to label an external leg (as in Figure 2.1 for example).
The correct meaning will be clear from the context.}
In \cite{bst} the following new expression for
$B$ was found:
 \beq
\label{Bniceonecyril}
 B (s,t,P^2, Q^2) \ = \
\Li2(1-aP^2)\, + \, \Li2(1-aQ^2)  \, -\,  \Li2(1-as)
\,  -\,  \Li2(1-at)
\ ,
\eeq
where
\beq
\label{aagainagain} a \ = \
\frac{P^2+Q^2-s-t}{P^2Q^2-st}
\ .
\eeq
The expression \eqref{Bniceonecyril}
contains one less dilogarithm and one less logarithm
than  the traditional form used by BDDK,
\beqa
 B (s,t,P^2, Q^2) & = &
\nonumber
{\rm Li}_2 \Bigl( 1 - {P^2\over s } \Bigr) \, +\, {\rm Li}_2
\Bigl( 1 - {P^2\over t } \Bigr) +
{\rm Li}_2 \Bigl( 1 - {Q^2\over
s} \Bigr) \, + \, {\rm Li}_2 \Bigl( 1 - {Q^2\over t } \Bigr)
\\ 
&-& {\rm Li}_2 \Bigl( 1 - {P^2 Q^2\over s\, t } \Bigr)
\,
+\,
{1\over 2} \log^2 \Bigl( {s\over t} \Bigr)
\ .
\label{Bboxcsw22}
\eeqa
The agreement of \eqref{Bniceonecyril} with
\eqref{Bboxcsw22}  was discussed and proved in
Section 5 of \cite{bst}.%
\footnote{More precisely, this agreement holds only in certain kinematical
regimes \emph{e.g.}~in the Euclidean region where all kinematical invariants are negative.
More care is needed when
analytically continuing the amplitude to the physical region.  The usual prescription of
replacing a kinematical invariant $s$ by $s + i\varepsilon$ and continuing $s$ from negative
to positive values gives the correct result only for our form of the
box function \eqref{Bniceonecyril},
whereas \eqref{Bboxcsw22} has to be amended by correction terms
\cite{binoth}.}
In Figure 2.1 we give a pictorial representation of
the box function  $F$ defined in \eqref{F} (with the leg labels identified
by $s \to p$, $m \to q$).

Finally, infrared divergences
are contained in the bubble functions
$K_0 (t)$, defined by
\beq
K_0 (t) \ := \
{\, {(-t)}^{-\epsilon}\over \epsilon (1-2\epsilon)}
\ .
\eeq
We  notice that in order to re-express
\eqref{BDDKNeq1} in a simpler form, it is useful
to introduce the triangle function \cite{csw2}
\beq
\label{trian}
T(p, P, Q) \ := \
{ \log (Q^2 / P^2) \over Q^2 - P^2}
\ ,
\eeq
with $p+P+Q=0$.
A diagrammatic representation of this function
is given in \mbox{Figure 2.2}  (with $m^+ \to p$).
We also find it useful to introduce
an $\epsilon$-dependent triangle function,%
\footnote{The function $T_{\epsilon} (p, P, Q)$
defined in \eqref{epstrianglen1} arises naturally
in the twistor-inspired approach
which will be developed in  \S \ref{n=1mhvsection}.}
\beq
\label{epstrianglen1}
T_{\epsilon} (p, P, Q) \ := \
{1\over \epsilon}
{ (-P^2)^{-\epsilon} - (-Q^2)^{-\epsilon} \over Q^2 - P^2}
\ .
\eeq
As long as $P^2$ and $Q^2$ are non-vanishing, one has
\beq
\label{T1n1}
\lim_{\epsilon \to 0} T_{\epsilon} (p, P, Q) \ = \
T (p, P, Q) \ , \qquad
P^2 \neq 0 \, , \, Q^2 \neq 0
\ .
\eeq
If either of the invariants vanishes, one has a different
limit. For example, if
$Q^2=0$ one has
\beq
\label{T2n1}
\left. T_{\epsilon} (p, P, Q)
\right|_{Q^2=0}  \ \longrightarrow \
- {1\over \epsilon} \, {(-P^2)^{-\epsilon }\over P^2}\, ,
\qquad {\epsilon \to 0}
\ .
\eeq
We will call these cases ``degenerate triangles''.

The usefulness of the  previous remark stems from
the fact that precisely the quantity
$(1 / \epsilon)\cdot \bigl[(-P^2)^{-\epsilon}/ P^2 \bigr]$
appears in the last line  of \eqref{BDDKNeq1} -- the
bubble contributions.
Therefore, these can be equivalently obtained
as  degenerate triangles \emph{i.e.}~triangles where
one of the massive legs becomes massless.

Specifically, we notice that
the four degenerate triangles (bubbles) in the last line
of \eqref{BDDKNeq1} can be precisely obtained by
including the ``missing'' index assignments
in $\cD_m$ and   $\cC_m$:
\beq
(m=i+1,\,  a=i-1) \, ,
\ \ \ (m=j-1,\,  a=j) \qquad
{\rm for} \ \cD_m \, ,
\
\eeq
which correspond to two degenerate triangles,
and
\beq
(m=j+1, \, a=j-1)\, ,
\ \ \ (m=i-1,\,  a=i) \qquad
{\rm for} \ \cC_m \, ,
\eeq
corresponding to two more degenerate triangles.

In conclusion, the previous remarks
allow us to rewrite \eqref{BDDKNeq1} in a
more  compact form as follows:
\beqa
\label{Neq1}
\!\!\!\!\!\!\!\!\cA^{{\cN}=1, {\rm chiral}}_{n}
\!&=&\!
\cA^{\rm tree}_{n}\, \cdot \,
\biggl\{ \sum_{m=i+1}^{j-1}
\sum_{s=j+1}^{i-1} b_{m,s}^{i,j}
\,
B(t_{m+1}^{[s-m]},t_m^{[s-m]},
t_{m+1}^{[s-m-1]},t_{s+1}^{[m-s-1]})
\\
&+&
{1\over 1-2\epsilon}
\bigg[ \sum_{m=i+1}^{j-1} \sum_{a=j}^{i-1}
\, c^{i,j}_{m,a}
\, T_{\epsilon} ( p_m, q_{a+1,m-1}, q_{m+1, a} )
\ \, + \ \,  ( i \leftrightarrow j )
\bigg]
\biggr\}
\nonumber
\ .
\eeqa
In this expression it is understood that we only keep terms that survive in
the limit $\epsilon \to 0$. This means that the factor
$1/(1-2\epsilon) $ can be replaced by $1$ whenever the term in the sum
is finite, \emph{i.e.}~whenever the triangle is non-degenerate.
However,  in the case of degenerate triangles, which contain infrared-divergent
terms, we have to expand this factor to linear order in $\epsilon$.
In the notation of \eqref{Neq1},
$q_{m+1, a}^2 =  t_{m+1}^{[a-m]}$ and
$q_{a+1, m-1}^2 =t_{m}^{[a-m+1]} $;
in Figure 2.2, these invariants
correspond to $Q^2$ and $P^2$ respectively,
where $j\in Q$, $i\in P$.
In the sum with $i\leftrightarrow j$, one would have
$q_{m+1, a}^2= t_{a+1}^{[m-a]}$,
$q_{a+1, m-1 }^2=t_{a+1}^{[m-a-1]}$, corresponding
respectively to ${Q}^2$ and ${P}^2$
in Figure 2.3,
with $i\in {Q}$, $j\in {P}$.
It is the expression \eqref{Neq1} for the $\cN=1$
chiral multiplet
amplitude which we will derive using MHV diagrams.


\section{MHV one-loop amplitudes in $\cN=1$ SYM
from MHV vertices}
\label{n=1mhvsection}

In \S \ref{loopsection} we reviewed how MHV vertices can be sewn
together into one-loop diagrams,
and how a particular decomposition
of the loop momentum measure leads to a representation of the
amplitudes strikingly similar to traditional dispersion formul{\ae}.
This method was tested successfully in
\cite{bst} for the case of
MHV one-loop amplitudes in $\cN\!=\!4$ SYM as reviewed in \S \ref{mhvsection}.
In the following we will apply the same philosophy to amplitudes
in $\cN\!=\!1$ SYM, in particular to the infinite sequence of
MHV one-loop amplitudes, which were obtained using
the cut-constructibility approach \cite{Bern:1994cg},
and whose twistor space picture has been analysed in \cite{csw2}.

Similarly to the $\cN\!=\!4$ case, the one-loop
amplitude has an overall factor
proportional to the MHV tree-level amplitude,
but, as opposed to the $\cN\!=\!4$ case,
the remaining one-loop factor depends non-trivially
on the positions $i$ and $j$ of the two
external negative-helicity gluons.
This is due to the fact that
a different set of fields is allowed
to propagate in the loop.

The MHV diagrams contributing to
MHV one-loop amplitudes consist of two
MHV vertices connected by two
off-shell scalar propagators. If both
negative-helicity gluons are on one
MHV vertex, only gluons of a particular
helicity can propagate in the loop. This is
independent of the number of supersymmetries.
On the other hand, for diagrams with one negative-helicity gluon on one MHV vertex and the other
negative-helicity gluon on the other
MHV vertex, all components of the
supersymmetric multiplet propagate in the
loop. In the case of $\cN\!=\!4$ SYM
this corresponds to helicities $h=-1,-1/2,0,1/2,1$
with multiplicities $1,4,6,4,1$, respectively;
for the $\cN\!=\!1$ vector multiplet
the multiplicities are $1,1,0,1,1$.
Hence, we can obtain the $\cN\!=\!1$ (vector) amplitude
by simply taking the $\cN\!=\!4$ amplitude
and subtracting three times the contribution
of an $\cN\!=\!1$ chiral multiplet,
which has multiplicities $0,1,2,1,0$.%
\footnote{We can also obtain the $\cN\!=\!2$ amplitude
in a completely similar way.}

This supersymmetric decomposition of
general one-loop amplitudes is useful
as it splits the calculation into pieces of
increasing difficulty, and allows one to reduce
a one-loop diagram with gluons
circulating in the loop to a combination
of an $\cN\!=\!4$ vector amplitude,
an  $\cN\!=\!1$ chiral amplitude and finally
a non-supersymmetric amplitude with
a scalar field running in the loop as in Equation (\ref{loopsusy}).

In our case, the supersymmetric
decomposition takes the form
\begin{equation}
\label{susydec}
\cA_n^{\cN=1, {\rm vector}} \ = \
\cA_n^{\cN=4}\, - \,
3 \, \cA_n^{\cN=1,{\rm chiral}}
\ ,
\end{equation}
where $n$ denotes the number of external lines.
Since the $\cN=4$ contribution is known,  one only needs to
determine $\cA_n^{\cN=1,{\rm chiral}}$ using MHV diagrams.
To be more precise, we are solely addressing the computation of
the planar part of the amplitudes.
However, this is sufficient since at one-loop level
the non-planar partial amplitudes are obtained
as appropriate sums of permutations of
the planar partial amplitudes \cite{Bern:zx},
as discussed in \S \ref{colourordersection}.
\begin{figure} [ht]
\label{fig4}
\vspace{.2in}
\centerline {
\includegraphics[width=4in]{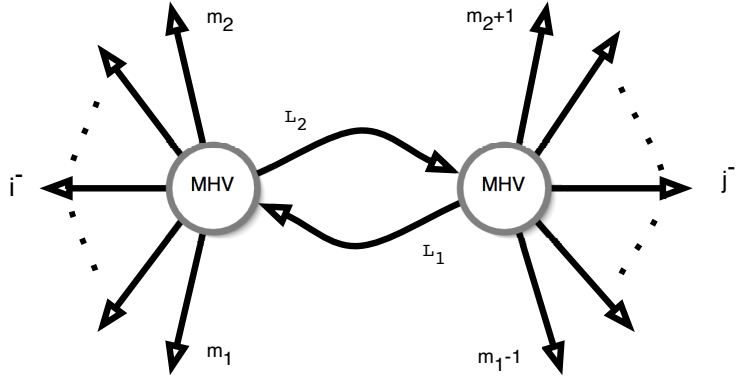}
}
\vspace{.2in}
\caption{\it A one-loop MHV  diagram,
computed in (\ref{loopintn1}) using MHV amplitudes
as interaction vertices,
with the CSW off-shell prescription.
The two external gluons with negative helicity
are labelled by $i$ and $j$.}
\end{figure}

\subsection{The procedure}

Our task therefore consists of:
\begin{itemize}
\item[{\bf 1.}]
Evaluating the class of diagrams
where we allow all the helicity states
of a chiral multiplet,
\beq
\label{rangeh}
h \in \{-1/2,0,0,1/2\}
\ ,
\eeq
to run in the loop.
We depict the prototype of such diagrams in
Figure 2.4.
\item[{\bf 2.}]
Summing over all diagrams such that each of
the two MHV vertices always has one external
gluon of negative helicity. Assigning
$i^-$ to the left and $j^-$ to the right,
the summation range of $m_1$ and $m_2$ is determined to be:
\beq
\label{rangen1}
j +1\leq m_1 \leq i \ , \qquad
i \leq m_2 \leq j-1 \ .
\eeq
\end{itemize}
Hence we get
\beqa
\label{loopintn1}
\cA_n^{\cN=1,{\rm chiral}} & = & \sum_{m_1,m_2,h}
\int\!d\cM \,
\cA(-l_1,m_1,\ldots,i^{-},\ldots,m_2,l_2) \nonumber \\
& & \,\,\,\,\,\,\,\,\,\,\,
\cdot \,
\cA(-l_2,m_2+1,\ldots,j^{-},
\ldots,m_1-1,l_1)
\,\, ,
\eeqa
where the summation ranges of $h$, $m_1$ and $m_2$ are
given in \eqref{rangeh}, \eqref{rangen1}.
Notice that, in order to compute the loop amplitude
\eqref{loopintn1}, we make use of the integration measure
$d\cM$ given in \eqref{bst}.

After some spinor algebra and after performing the sum
over the helicities $h$, the integrand of
(\ref{loopintn1}) becomes
\beq
\label{integrandn1}
-i\cA_n^{\rm tree} \, \cdot \,
\frac{\langle m_2\,  (m_2 \! + \! 1)
\rangle \, \langle (m_1 \! - \! 1) \, m_1 \rangle
\,
\langle i\,  l_1 \rangle
\,
\langle j \, l_1 \rangle \langle i \, l_2 \rangle
\langle j \, l_2 \rangle}{\langle i\,  j\rangle^2
\,
\langle m_1 \, l_1 \rangle
\,
\langle (m_1 \! - \! 1) \, l_1 \rangle
\,
\langle m_2 \, l_2 \rangle
\langle (m_2 \! + \! 1) \, l_2\rangle}
\,\, .
\eeq
The focus of the remainder of this section
will be to evaluate  the
integral in \eqref{loopintn1} explicitly.
Since $-i \cA_n^{\rm tree}$ factors out
completely, we will now drop it
and only reinstate it at the very end
of the calculation.

The integrand (without this factor)
can be rewritten in terms of dot products of
momentum vectors,
\beq
\label{redintegrand}
{\mathcal I} \ = \
\frac{\mathscr N}{(i \cdot j)^{2}
\left( m_1 \cdot l_1 \right)
\left( (m_1 \!- \! 1 ) \cdot l_1 \right)
\left( m_2 \cdot l_2 \right)
\left( (m_2 \! + \! 1) \cdot l_2 \right) } \,\, ,
\eeq
with
\beq
\label{numerator}
{\mathscr N} = \tr_+ \left( \lslash_1 \kslash_{m_1-1} \kslash_{m_1}
\lslash_1 \kslash_j \kslash_i \right)
\tr_+ \left( \lslash_2 \kslash_{m_2} \kslash_{m_2+1}
\lslash_2 \kslash_j \kslash_i \right)
\,\, .
\eeq
$\mathscr{N}$ is a product of Dirac traces,
where the $\tr_+$ symbol indicates that the
projector $(1+\gamma^{5})/2$ has been inserted.

Next, notice that each of these Dirac
traces involving six momenta
can be expressed in terms of simpler
Dirac traces involving only four momenta.
For the first factor
of (\ref{numerator}) we find
\beq
\label{6to4trace}
\tr_+ \left( \lslash_1 \kslash_{m_1-1} \kslash_{m_1}
\lslash_1 \kslash_j \kslash_i \right) \, = \,
2 (m_1 \cdot l_1) \tr_+  \left( \kslash_{i}
\kslash_{j} \kslash_{m_1-1}
\lslash_1 \right) \, - \,
2 ((m_1 \!- \!1) \cdot l_1) \tr_+
\left( \kslash_{i} \kslash_{j} \kslash_{m_1}
\lslash_1 \right) \, ,
\eeq
where
\beq
\label{4trace}
\tr_+ \left( \kslash_{a} \kslash_{b} \kslash_{c}
\kslash_d \right) = 2 \bigl[ (a \cdot b) (c \cdot d)-
(a \cdot c) (b \cdot d) +
(a \cdot d) (b \cdot c)\bigr] - 2i  \varepsilon(a,b,c,d) \, .
\eeq
The second factor in \eqref{numerator} takes a similar form.
Consequently, the integrand
becomes a sum of four terms, one of which is
\beq
\label{partintegrand}
\frac{\tr_{+} \left( \kslash_{i} \kslash_{j} \kslash_{m_{1}}
\lslash_1 \right)   \tr_{+} \left( \kslash_{i}
\kslash_{j} \kslash_{m_{2}}
\lslash_2 \right) }{(i \cdot j)^{2}
\left( m_{1} \cdot l_{1}\right) \left( m_{2} \cdot l_{2}\right)}
\, .
\eeq
The other three terms are obtained by replacing
$m_{1}$ with $m_{1}-1$ and/or
$m_{2}$ with $m_{2}+1$ in \eqref{partintegrand} and come with alternating signs.
Note that the original expression \eqref{integrandn1}
is symmetric in $i$, and $j$,
although when we make use of the decomposition
\eqref{partintegrand} this symmetry
is no longer manifest. We will
symmetrize over $i$ and $j$ at the end of the calculation in order to make this
exchange symmetry manifest in the final expression.

In the next step we have to perform the
phase space integration,  which is equivalent
to the calculation of a unitarity cut with momentum
$P_{L;z}=\sum_{l=m_{1}}^{m_{2}}
k_{l} - z \eta$ flowing through the cut.
Note that, as explained in \S \ref{offshellsection}, the momentum is
shifted by a term proportional to the reference
momentum $\eta$.
The term $(l_1 \cdot m_1) (l_2 \cdot m_2)$ in the
denominator of  \eqref{partintegrand} corresponds
to two propagators, hence the denominator by itself
corresponds to a cut box diagram.
However, the numerator of \eqref{partintegrand}
depends non-trivially on the loop momentum,
so that in fact  \eqref{partintegrand} corresponds
to a tensor box diagram,  not simply a scalar box diagram.
Using the Passarino-Veltman method  \cite{pv},
we can reduce the expression \eqref{partintegrand},
integrated with the LIPS measure,  to a sum of cuts
of scalar box diagrams, scalar and vector triangle diagrams,
and scalar bubble diagrams.
This procedure is somewhat technical and details
are collected in Appendix \ref{neq1appendix}.
Luckily, the final result takes a less intimidating
form  than the intermediate expressions.
We will now present the result of
these calculations after the LIPS integration.

\subsection{Discontinuities}

We first observe that loop integrations
are performed in $4-2\epsilon$ dimensions.
It turns out that singular $1 / \epsilon$ terms
appearing at intermediate steps of the phase space integration
cancel out completely.
Notice that this does not mean that the final result will be
free of infrared divergences.
In fact the dispersion integral can and does give rise to
$1/ \epsilon$ divergent terms but there cannot be any
$1 / \epsilon^2$ terms, as expected
for the contribution of
a chiral multiplet \cite{Bern:1994cg}.
The $1 / \epsilon$ divergences
in the scattering amplitude correspond to
the bubble contributions in \eqref{BDDKNeq1},
or degenerate triangles contributions in \eqref{Neq1},
as explained in \S \ref{n=1reviewsection}.
In Appendix \ref{neq1appendix} we show that the finite terms of the phase
space integral combine into the following simple expression:
\beq
\label{Neq1cut}
\hat{{\mathcal C}}  \ = \
{\mathcal C}(m_{1}-1,m_{2})-{\mathcal C}(m_{1},m_{2})+
{\mathcal C}(m_{1},m_{2}+1)-{\mathcal C}(m_{1}-1,m_{2}+1)
\ ,
\eeq
with%
\footnote{In \eqref{partNeq1cut} we omit an overall,
finite numerical factor
that depends on $\epsilon$. This factor,
which can be read off from \eqref{ubin1},
is irrelevant for our discussion.}
\beqa
\label{partNeq1cut}
{\mathcal C}(m_{1},m_{2}) & = & \frac{2 \pi}{1-2 \epsilon}
\frac{ (P_{L;z}^2)^{-\epsilon}}{(i \cdot j)^{2} (m_{1} \cdot m_{2})}
\left[
\frac{{\mathcal T}(m_{1},m_{2},P_{L;z})}
{\left( m_{1} \cdot P_{L;z}\right)}
\,  +  \,
\frac{{\mathcal T}(m_{2},m_{1},P_{L;z})}
{\left( m_{2} \cdot P_{L;z}\right)}
\right] \nonumber \\[12pt]
& -&
\frac{2 \pi {\mathcal T}(m_{1},m_{2},m_{2})}
{(i \cdot j)^{2}(m_{1} \cdot m_{2})^{2}}
\, (P_{L;z}^2)^{-\epsilon}\, \log \left( 1 -
a_z \, P^{2}_{L;z}\right)
\ ,
\eeqa
where
\beqa
\label{bla}
{\mathcal T}(m_{1},m_{2},P) & := &
\tr_{+} \left( \kslash_{i} \kslash_{j} \kslash_{m_{1}}
\Pslash \right)
\,
\tr_{+} \left( \kslash_{i} \kslash_{j} \kslash_{m_{2}}
\kslash_{m_{1}} \right) \,\, ,
\nonumber \\ \cr
a_z &:=& \frac{m_{1} \cdot m_{2}}{ N(P_{L;z})}
\ ,
\eeqa
and
\beq
\label{N}
N(P) \ := \ (m_{1} \cdot m_{2}) P^{2} - 2 (m_{1} \cdot P)
(m_{2} \cdot P) \,\, .
\eeq
A closer inspection of \eqref{partNeq1cut}
reveals that the first line of that expression
corresponds to two cuts of scalar triangle integrals,
up to an $\epsilon$-dependent
factor and the explicit $z$-dependence
of the two numerators.
The second line is a term familiar from \cite{bst},
corresponding to the $P_{L;z}^2$-cut
of the finite part $B$ of a  scalar box function,
defined in \eqref{Bniceonecyril} (see also \eqref{F}).
The full result for the one-loop MHV amplitudes is obtained
by summing over all possible MHV diagrams,
as specified in
\eqref{loopintn1} and \eqref{rangeh}, \eqref{rangen1}.

\subsection{The full amplitude}

We begin our analysis by focusing on the box function
contributions in \eqref{partNeq1cut},
and notice the following important facts:
\begin{itemize}
\item[{\bf 1.}]
By taking into account the four terms in
\eqref{Neq1cut} and summing over Feynman diagrams,
we see each fixed finite box function $B$
appears in exactly four phase space integrals,
one for each of its  possible cuts,
in complete similarity with \cite{bst}.
It was shown in Section 5 of that paper
that the corresponding dispersion integration
over $z$ will then yield the finite $B$ part of the
scalar box functions $F$.
It was also noted in \cite{bst} that one
can make a particular gauge choice for $\eta$
such that the $z$-dependence in $N$ disappears.
This happens when $\eta$ is chosen to be equal to one
of the massless external legs of the
box function. The question of gauge invariance is further discussed in
Appendix \ref{triangleappendix}.
\item[{\bf 2.}]
The coefficient multiplying the finite box function
is precisely equal to $b_{m_1, m_2}^{i, j}$
defined in \eqref{bdefn1}.
\item[{\bf 3.}]
Finally, the  functions $B$ generated by
summing over all MHV Feynman diagrams with the range
dictated by \eqref{rangen1} are precisely those included
in the double sum for the finite box functions
in the first line of \eqref{BDDKNeq1} (or \eqref{Neq1})
upon identifying  $m_1$ and $m_2$ with $s$ and $m $.
To be precise, \eqref{rangen1} includes the case where
the indices $s$ and/or $m$ (in the notation of
\eqref{BDDKNeq1} and \eqref{Neq1})
are equal to either $i$ or $j$; but for any of
these choices, it is easy to check that
the corresponding coefficient  $b_{m, s}^{i, j}$
vanishes.
\end{itemize}
This settles the agreement between
the result of our computation with MHV vertices
and \eqref{Neq1} for the part corresponding to
the box functions.
Next we have to collect the cuts contributing to
particular triangles, and show that the
$z$-integration  reproduces the expected triangle
functions from \eqref{Neq1},
each with the correct coefficient.

To this end, we notice that for each fixed triangle
function $T(p, P, Q)$, exactly four phase space integrals appear,
two for each of the two possible cuts of the function.
Moreover, a gauge invariance similar to that
of the box functions also exists for triangle
cuts (see Appendix \ref{triangleappendix}), so that we can choose $\eta$ in a way
that the ${\mathcal T}$ numerators
in \eqref{partNeq1cut}
become independent of $z$. A particularly convenient
choice is $\eta = k_{i}$, since it can
be kept fixed for all possible cuts.
Choosing this gauge, we see that a sum, ${\sf T}$, of
terms proportional to cut-triangles is generated
from  \eqref{Neq1cut}
(up to a common normalisation):
\beq
\label{popp}
{\sf T} \ := \ T_A \, + \,  T_B \, + \,  T_C \, + \,  T_D \,
\  ,
\eeq
where
\beqa
\label{Michelangeli}
T_A & := &
\left[ {\cS (i, j, m_1, m_2) \over (m_1 \cdot m_2 )}
\,  -\,
{\cS ( i , j , m_1 -1 , m_2) \over (( m_1 - 1)\cdot m_2)}
\right] \,
\cS (i, j, m_2 , P_L ) \,
\Delta_A
\ ,
\\ \nonumber \cr
T_B &:= &
 \left[ {\cS (i, j, m_2, m_1) \over (m_1 \cdot m_2 )}
 \, -\,
{\cS ( i , j , m_2 +1 , m_1) \over (( m_2 + 1)\cdot m_1)}
\right] \,
\cS (i, j, m_1 , P_L ) \,
\Delta_B
\ ,
\\ \nonumber \cr
T_C &:= &
\left[ {\cS (i, j, m_2+1 , m_1-1)
\over ((m_2+1) \cdot (m_1-1) )}
 \, -\,
{\cS ( i , j , m_2  , m_1-1 ) \over (m_2 \cdot (m_1-1))}
\right] \,
\cS (i, j, m_1-1 , P_L ) \,
\Delta_C
\ ,
\\ \nonumber \cr
T_D  &:= &
\left[ {\cS (i, j, m_1-1, m_2+1) \over ((m_1-1) \cdot (m_2+1) )}
 \, - \,
{\cS ( i , j , m_1 , m_2+1) \over ( m_1 \cdot (m_2 +1))}
\right]\,
\cS (i, j, m_2+1 , P_L ) \,
\Delta_D
\ .
\eeqa
Here we have defined
\beq
\cS (a,b,c,d) \:= \
\tr_{+} \left( \kslash_{a} \kslash_{b} \kslash_{c}
\kslash_{d}
 \right)
\ ,
\eeq
and $\Delta_I$, $I = A, \ldots , D$,
are the following cut-triangles, all in the
$P_{L;z}$-cut :
\beqa
\label{listofct}
\Delta_A &:=& {1\over (m_2 \cdot P_{L;z} ) }
\ = \ Q^{2}\mbox{{\rm -cut}} \ {\rm of}
\ \ -T\big(m_2, P_{L;z} - m_2, -P_{L;z}\big)
\ ,
\\ \nonumber 
\Delta_B &:=& {1\over (m_1 \cdot P_{L;z} )}
\ = \ P^{2}\mbox{{\rm -cut}} \ {\rm of}
\ \ -T\big(m_1,- P_{L;z}, P_{L;z}-m_1\big)\ ,
\\ \nonumber 
\Delta_C &:=& {1\over ((m_1 - 1)  \cdot P_{L;z} )}
\ = \ Q^{2}\mbox{{\rm -cut}} \ {\rm of}
\ \ T\big( m_1 -1,- P_{L;z} - (m_1 -1), P_{L;z}\big)
\ ,
\\ \nonumber 
\Delta_D &:=& {1\over ((m_2+ 1)  \cdot P_{L;z} )}
\ = \ P^{2}\mbox{{\rm -cut}} \ {\rm of}
\ \ T\big( m_2 + 1 , P_{L;z} , -P_{L;z}-(m_2 + 1)\big)
\ .
\eeqa

\begin{figure} [ht]
\label{fig5}
\vspace{.2in}
\centerline {
\includegraphics[width=4in]{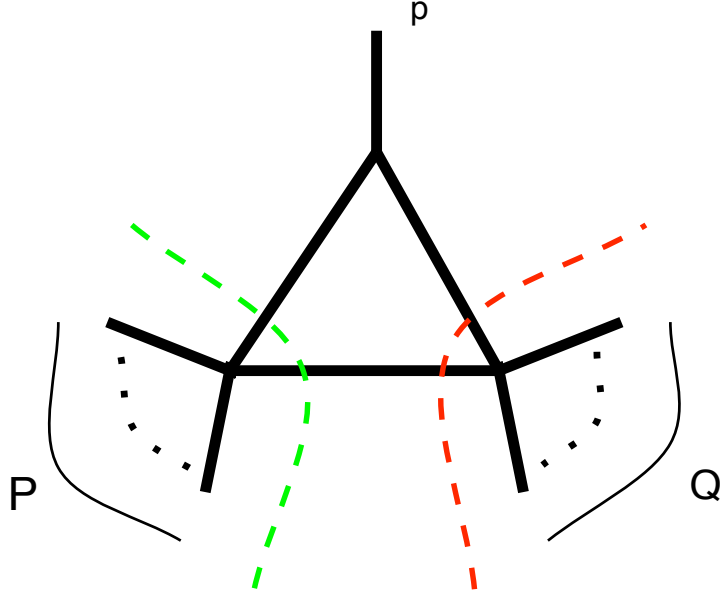}
}
\vspace{.2in}
\caption{\it A triangle function with massive
legs labelled by $P$ and $Q$, and massless leg $p$.
This function is reconstructed by summing two
dispersion integrals, corresponding to the
$P^2_z$- and $Q^2_z$-cut.
}
\end{figure}
\begin{figure} [ht]
\label{fig6}
\vspace{.2in}
\centerline {
\includegraphics[width=4in]{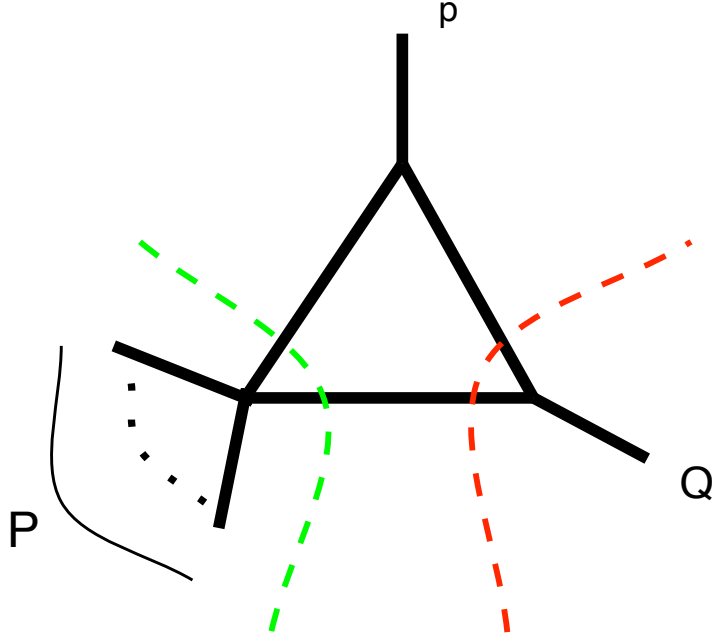}
}
\vspace{.2in}
\caption{\it
A degenerate triangle function. Here the leg labelled by $P$ is
still massive, but that labelled by $Q$ becomes massless.
This function is also reconstructed by summing over two
dispersion integrals, corresponding to the
$P^2_z$- and $Q^2_z$-cut.
}
\end{figure}

Next, we notice that the prefactors multiplying
$\D_B$, $\D_C$ become the same, up to a minus sign,
upon shifting $m_1-1 \to m_1$ in the second prefactor;
and  so do the prefactors of $\D_A$, $\D_D$
upon shifting $m_2 \to m_2 + 1$.
Doing this, $-\D_B$ and the shifted  $\D_C$ become
the two cuts of the same triangle function
$T (m_1,  -P_{L;z} , P_{L;z}-m_1 )$, and similarly,
 $-\D_A$ and  $\D_D$ give the two cuts of the function
$T (m_2,  P_{L;z}-m_2, -P_{L;z}  )$.
Furthermore,
in Appendix \ref{triangleappendix} we will show that summing the
two dispersion integrals of the two different cuts
of a triangle indeed generates the triangle function
-- in fact this procedure gives a novel way
of obtaining the triangle functions.%
\footnote{A remark is in order here.
In our procedure the momentum appearing
in each of the possible cuts is always shifted
by an amount proportional to $z \eta$;
the triangle is then reproduced by performing
the appropriate dispersion integrals.
Because of the above mentioned shift,
we produce a non-vanishing cut (with shifted momentum)
even when the cut includes only one external (massless) leg,
say $\tilde{k}$, as the momentum flowing in the cut is
effectively $\tilde{k}_z= \tilde{k}- z \eta$,
so that  $\tilde{k}_z^2 \neq 0$.
}
Specifically, the result derived in Appendix \ref{triangleappendix}
is
\beq
\label{dfd}
\int\! {dz \over z} \left[
{(P_{z}^{2})^{-\epsilon}\over ( P_{z}p) } \, + \,
{(Q_{z}^{2})^{-\epsilon}\over (Q_{z}p)}
\right]
 \ = \
2 \bigl[\pi \epsilon \csc (\pi \epsilon) \bigr]
\, T_{\epsilon}(p, P, Q)
\ ,
\eeq
where the $\epsilon$-dependent triangle function
$ T_{\epsilon}(p, P, Q) $ (with $p+P+Q=0$)
was introduced in  \eqref{epstrianglen1} and gives,
as $\epsilon \to 0$, the triangle function
\eqref{trian} (as well as the bubbles when either
$P^2$ or $Q^2$ vanish).
The result \eqref{dfd} holds for a generic
choice of the reference vector $\eta$,
see \eqref{bpbq}-\eqref{finaltriang}.
We give a pictorial representation of
the non-degenerate and degenerate triangle functions
in Figures 2.5 and 2.6, respectively.

At this point, it should be noticed  that
for a  gauge choice different from
$\eta = k_i$ adopted so far, the
numerators $\cT$ in \eqref{partNeq1cut} do acquire
an $\eta$-dependence.
This gauge dependence should not be present
in the final result for the scattering amplitude.
Indeed, it is easy to check that, thanks to
\eqref{bpbq}, the coefficient of the
$\eta$-dependent terms actually vanishes.

Using \eqref{popp}-\eqref{dfd} and collecting
terms as specified above,
we see that the generic term produced by this procedure
takes the form
\beq
\label{thisterm}
\left[
{\cS ( i , j , a  , p_m) \over (k_a \cdot p_m )}
 \, - \,
{\cS ( i , j , a+1  , p_m) \over (k_{a+1} \cdot p_m )}
\right]
\,
\cS ( i , j , p_m , Q) \
T(p_m, P, Q)
\ ,
\eeq
with
$P=q_{a+1, m-1}$ and $Q=q_{m+1 , a}$.

Finally, we implement the symmetrization
of the indices $i$, $j$, as explained earlier,
and convert \eqref{thisterm} into
\beq
c_{m, a}^{i, j} \, T(p_m, P, Q)
\ ,
\eeq
where the coefficient $c_{m, a}^{i, j}$ is%
\footnote{In writing \eqref{artn1},
we make also use of the fact that
$\cS( i, j, q_{m-1,a}, p_m)
= \cS( i, j, q_{m,a}, p_m)$.}
\beq
\label{artn1}
c_{m, a}^{i, j} \ := \
{1\over 2}  \left[
{\cS ( i , j , a+1  , p_m) \over (k_{a+1} \cdot p_m )}
 \, - \,
{\cS ( i , j , a  , p_m) \over (k_{a} \cdot p_m )}
\right]
\,
{ {\cS (i, j, p_m, q_{m,a}) - \cS( i, j, q_{m,a}, p_m)
\over [(k_i + k_j)^2]^2}}
\ ,
\eeq
which  coincides with the definition of
$c_{m, a}^{i, j}$  given in \eqref{cdef}.
Lastly, it is easy to see that in summing over the range
given by \eqref{rangen1}, we produce exactly
all the triangle functions  appearing in
the second line of \eqref{Neq1}.
It is also important to notice that the bubbles, which
appear in the last line of \eqref{BDDKNeq1},
are actually obtained as particular cases of triangle functions
where one of the massive legs becomes massless,
as observed at the end of \S \ref{n=1reviewsection}.

In conclusion, we have seen that all the terms
in \eqref{Neq1}, \emph{i.e.}~finite box contributions
and triangle contributions -
which include the bubbles as special
(degenerate) cases - are precisely reproduced
in our diagrammatic approach.


\chapter{Non-supersymmetric MHV amplitudes}

\label{chapter3}

Having seen that the CSW rules can be applied at loop level
in supersymmetric gauge theories, the obvious question is whether
the same also holds in non-supersymmetric gauge theories. To this
end the one-loop MHV amplitudes in pure Yang-Mills with a
scalar running in the loop were computed in \cite{bbst2}.
This is the last contribution to the MHV
amplitudes for gluon scattering in QCD in the supersymmetric decomposition
of Eq.~\eq{loopsusy} and has only been computed previously in certain
special cases in \cite{bdk9302280,Bern:1994cg}.

In this chapter we follow \cite{bbst2} and apply the CSW rules to this scalar amplitude in the
general case of $n$-gluon MHV scattering where the two negative-helicity
gluons sit at arbitrary positions. We find that the results agree perfectly with
those already obtained in \cite{bdk9302280,Bern:1994cg} and we go on to
present the general result for the cut-constructible part of the one-loop
MHV amplitudes in pure Yang-Mills. It turns out that the CSW rules only compute
this cut-constructible part and the rational terms (which do not contain cuts)
are not found. This is discussed in \cite{bbst2} and \S \ref{rationalsection} below.
They \emph{can} and \emph{have}, however, been recently computed using an on-shell
unitarity bootstrap \cite{Berger:2006vq} which thus completely determines the one-loop
MHV $n$-gluon amplitudes in QCD.

\section{The scalar amplitude}
In complete similarity with the ${\cal N}\! = \! 4$ and
${\cal N}\! = \! 1$ cases - see Chapters \ref{chapter1} \& \ref{chapter2}
and \emph{e.g.}~\cite{bst,bbst} -
we can immediately write down the expression for the scalar
amplitude in terms of MHV vertices as
\beqa
\label{loopint}
\cA_n^{{\rm scalar}}  & = &  \sum_{m_1, m_2, \pm }
\int\! d\cM \
\cA(-l_1^{\mp},m_1,\ldots,i^{-},\ldots,m_2,l_2^{\pm})\
\nonumber \\
&&
\qquad \qquad \qquad
\!\cdot \, \cA(-l_2^{\mp},m_2+1,\ldots,j^{-},
\ldots,m_1-1,l_1^{\pm})
\  ,
\eeqa
where the ranges of summation of $m_1$ and $m_2$ are
\beq
\label{range}
j +1\leq m_1\leq i \ , \qquad
i \leq m_2 \leq j-1 \ .
\eeq
A typical MHV diagram contributing to $\cA_n^{{\rm scalar}}$,
for fixed $m_1$ and $m_2$, is depicted in Figure 3.1.
\begin{figure} [ht]
\label{scalardiagram}
\vspace{.2in}
\centerline {
\includegraphics[width=4in]{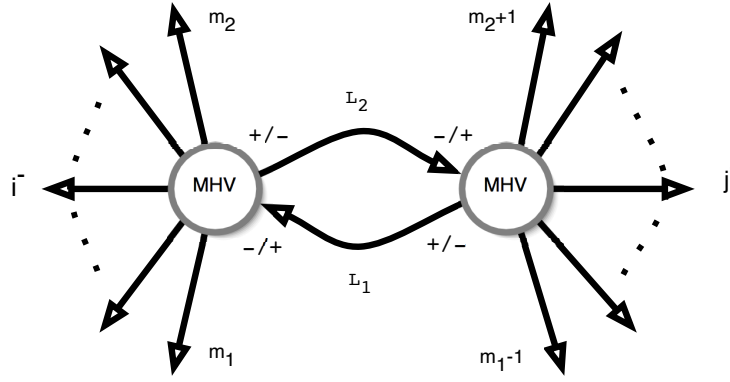}
}
\vspace{.2in}
\caption{\it A one-loop MHV diagram with a complex scalar running in the loop, computed in
Eq.~\eqref{loopint}.
We have indicated the possible helicity assignments
for the scalar particle.}
\end{figure}
The off-shell vertices  $\cA$ in \eqref{loopint}
correspond to having complex scalars running in the loop.
It follows that there are two possible
helicity assignments\footnote{For scalar fields, the  ``helicity''
simply distinguishes
particles from  antiparticles (see, for example, \cite{dixon}).}
for the scalar particles in the loop which have to be summed over.
These two possibilities are denoted
by $\pm$ in \eqref{loopint} and in
the internal lines in Figure 3.1.
It turns out that each of them
gives rise to the same integrand for \eqref{loopint}:
 \beq \label{integrand}
  -i\cA_n^{\rm tree} \,
\cdot \, \frac{\langle m_2\,  m_2 \! + \! 1 \rangle \,
\langle m_1 \! - \! 1
\, m_1 \rangle
\, \langle i\,  l_1 \rangle^2 \, \langle j \,
l_1 \rangle^2 \langle i \, l_2
\rangle^2 \langle
j \, l_2 \rangle^2}{\langle i\,  j\rangle^4 \,
\langle m_1 \, l_1 \rangle \,
 \langle m_1 \! -
\! 1 \, l_1 \rangle \, \langle m_2 \, l_2 \rangle \,
\langle m_2 \! + \! 1
 \, l_2\rangle \,
\langle l_1 \, l_2 \rangle^2} \,\, .
\eeq

A crucial ingredient in \eqref{loopint} is (as before in
Chapters \ref{chapter1} \& \ref{chapter2}) the
integration measure $d \cM$.
This measure was constructed in
\cite{bst,Brandhuber:2005kd}  using the decomposition
$L := l  + z  \eta$ for a non-null four-vector
$L$ in terms of a null vector $l$ and a real parameter $z$ as reviewed in
\S \ref{intmeasure}.
We refer the reader to \S \ref{intmeasure} and \cite{bst,Brandhuber:2005kd}
for the construction of this measure, and here we merely quote the result:
\beq
\label{lint}
d\cM \ = \
 2 \pi i \
\theta (P_{L;z}^2)\
{dP_{L;z}^2 \over P_{L;z}^2 \, - \, P_L^2 \, - \, i \vare}
\
d^{4-2\epsilon}{\rm LIPS} (l_2^{\mp} , - l_1^{\pm}; P_{L;z})
\ .
\eeq
In order to calculate \eqref{loopint},
we will first integrate the expression \eqref{integrand}
over the Lorentz-invariant phase space (appropriately
regularised to $4-2\e$ dimensions),
and then perform the dispersion integral.

For the sake of clarity,
we will separate the analysis into two parts.
Firstly, we will present the (simpler) calculation
of the amplitude in the case where
the two negative-helicity gluons
are adjacent. This particular amplitude  has already been computed
by Bern, Dixon, Dunbar and Kosower in
\cite{Bern:1994cg} using the cut-constructibility approach;
the result we will derive here
will be in precise agreement with the result in that approach.
Then, in \S \ref{generalcasesection} we will move on to address the general case,
deriving new results.

\section{The scattering amplitude in the adjacent case}
\label{adjacentsection}

The adjacent case corresponds to choosing
$i = m_1$, $j = m_1-1$ in Figure 3.1.
Therefore we  now have a single sum over MHV diagrams,
corresponding to the possible choices of $m_2$.
We will also set $i=2$, $j=1$ for the sake of definiteness,
and $m_2 = m$.

After conversion into traces,
the integrand of (\ref{loopint})
takes on the form:
\beqa
\label{reallyfinally}
\frac{
 {\tr}_{+}(\kslash_{1}\, \kslash_{2}\,
\Pslash_{L;z} \, \lslash_{2})
\ {\tr}_{+}(\kslash_{1}\, \kslash_{2}
\, \lslash_{2} \, \Pslash_{L;z})}{2^5 \, (k_1 \cdot k_2)^3 \,
(l_1 \cdot l_2)^2}
\Bigg\{\frac{{\tr}_{+}(\kslash_1 \, \kslash_2\,
\kslash_{m+1} \, \lslash_2)}{(l_2 \cdot m \! + \! 1)}-
\frac{{\tr}_{+}(\kslash_1 \, \kslash_2\,
\kslash_{m} \, \lslash_2)}{(l_2 \cdot m)}\Bigg\} \,  ,
\eeqa
where we note that
$(l_1 \cdot l_2) = -P_{L;z}^2/2$ by momentum conservation.

The next step consists of performing
the Passarino-Veltman reduction \cite{pv}
of the Lorentz-invariant phase space integral of
\eqref{reallyfinally}.
This requires the calculation of the
three-index tensor integral
\beq
\cI^{\m \n \r} (m, P_{L;z})
\ = \
\int\! d{\rm LIPS}(l_2,-l_1;P_{L;z})
\,
\frac{l_2^{\mu}\, l_2^{\nu} \, l_2^{\r} }{
(l_2\cdot m)}
\ .
\eeq
This calculation is performed in Appendix \ref{nonsusyappendix}.
The result of this procedure gives the following
term at $\cO(\epsilon^0)$, which we will
later integrate with the dispersive measure:
\beqa
\nonumber
\tilde{\cA}_n^{{\rm scalar}} \ &=& \
 \frac{\pi}{3}\,
(-P_{L;z}^2)^{-\e} \,
\frac{\left[{\tr}_{+}(\kslash_{1}\, \kslash_{2}\,
\kslash_{m} \, \Pslash_{L;z})\right]^2}{2^5 \,
(k_1\cdot k_2)^3}\,
\Bigg\{\frac{{\tr}_{+}(\kslash_{1}\, \kslash_{2}\,
\Pslash_{L;z} \, \kslash_{m})}{(m\cdot P_{L;z})^3} \ + \
\frac{2(k_1 \cdot k_2)}{(m\cdot P_{L;z})^2}\Bigg\}
\\
\ &-& \ (m \leftrightarrow m + 1) \ ,
\label{atlast}
\eeqa
and we have dropped a factor of $4\pi \hat{\l} \cA^{\rm tree}$
on the right hand side of
\eqref{atlast}, where $\hat{\lambda}$ is defined in \eqref{ubiquitous}.
We can reinstate this factor at the end of the calculation.
We also notice that  \eqref{atlast} is a finite expression,
\emph{i.e.}~it is free of infrared poles.

\subsection{Rational terms}
\label{rationalsection}

An important remark is in order here.
On general grounds, the result of a phase space
integral in, say, the $P^2$-channel, is of the form
\beq
\label{phasespaceint}
\cI( \e) \ = \ (-P^2)^{-\e} \cdot f (\e)
\ ,
\eeq
where
\beq
\label{effe}
f ( \e) \ = \ {f_{-1} \over  \e} \, + \,
f_0 \, + \, f_1 \e \, + \, \cdots
\ ,
\eeq
and $f_i$ are rational coefficients.
In the case at hand, infrared poles generated
by the phase space integrals cancel completely,
so that we can in practice replace
\eqref{effe} by
$f(\e)  \to  f_0 \, + \, f_1 \e
\, + \, \cdots $.
The amplitude $\cA$ is then obtained
by performing a dispersion integral, which converts
\eqref{phasespaceint} into an expression of the form
\beq
\label{disppint}
\cA( \e) \ = \ {(-P^2)^{-\e} \over \e}  \cdot g (\e)
\ = \ {g_0 \over \e} \, - \, g_0 \log ( -P^2)
\, + \, g_1 \, + \, \cO( \e)
\ ,
\eeq
where $g (\e) = g_0 + g_1 \e + \cdots$, and the coefficients
$g_i$ are rational functions, \emph{i.e.}~they are free of cuts.
Importantly, errors can be generated in the evaluation
of phase space integrals if one contracts
$(4-2 \e)$-dimensional vectors with ordinary four-vectors.
This does not affect
the evaluation of the coefficient $g_0:=g(\e=0)$,
and hence the part of the amplitude containing cuts
is reliably computed;
but the coefficients $g_i$ for
$i \geq 1$, in particular $g_1$, are in general affected. This implies that
{\it rational} contributions to the scattering amplitude
 cannot be detected \cite{Bern:1994cg} in this construction.
A notable exception to this is provided by the phase space integrals
which appear in supersymmetric theories. These are
``four-dimensional cut-constructible'' \cite{Bern:1994cg},
in the sense that the rational parts are unambiguously
linked to the discontinuities across cuts,
and can therefore be uniquely determined.\footnote{For more details about cut-constructibility,
see the detailed analysis  in Sections 3-5 of \cite{Bern:1994cg} and Appendix
\ref{unitarityappendix} of this thesis for a brief review.}
This occurs, for example,
in the calculation of the $\cN \! = \! 4$
MHV amplitudes at one-loop performed in \cite{bst} and reviewed in
\S \ref{mhvsection} and the $\cN\!=\!1$ MHV amplitudes at one-loop in
Chapter \ref{chapter2}.
In the present case, however,
the relevant phase space integrals violate
the cut-constructibility criteria given in
\cite{Bern:1994cg}%
\footnote{An example of an
integral violating the power-counting criterion of \cite{Bern:1994cg} is provided
by \eqref{imunu}.}, since we encounter tensor triangles with up to
three loop momenta in the numerator.
Hence, we will be able to compute the part
of the amplitude containing cuts, but not the rational terms.
In practice this means that we will compute
all phase space integrals up to $\cO (\e^0 )$
and discard $\cO (\e)$ contributions, which would generate rational terms
that cannot be determined correctly.

\subsection{Dispersion integrals for the adjacent case}

After this digression,
we now move on to the dispersive integration.
In the center of mass frame, where
$P_{L;z}\! :=\! P_{L;z}(1,\vec{0})$,
all the dependence on $P_{L;z}$ in \eqref{atlast}
cancels out, as there are equal powers of $P_{L;z}$
in the numerator as in the denominator
of any term. As a consequence, the dependence on the
arbitrary reference vector $\eta$ disappears (see \cite{quig} for
the application of this argument to the $\cN=1$ case).
We are thus left with dispersion integrals
of the form
\beqa
\label{disperse}
I(P_L^2) \ := \ \int\! \frac{ds'}{s'-P_L^2}\,
(s')^{-\epsilon} \ = \
\frac{1}{\epsilon}[\pi\epsilon
\csc(\pi\epsilon)]\, (-P_L^2)^{-\epsilon}
\ .
\eeqa
Taking this into account,
the dispersion integral of  \eqref{atlast}
then gives
\beqa
\nonumber
\!\!\!\tilde{\cA}_n^{{\rm scalar}} \ &=& \
\big[ {\pi \e \csc ( \pi \e) }\big]
\,
\frac{\pi}{3}\,
{(-P_{L}^2)^{-\e}\over \e} \,
\frac{\left[{\tr}_{+}(\kslash_{1}\, \kslash_{2}\,
\kslash_{m} \, \Pslash_{L})\right]^2}{2^5 \, (k_1\cdot k_2)^3}\,\nonumber\\
&\cdot&\Bigg[\frac{{\tr}_{+}(\kslash_{1}\, \kslash_{2}\,
\Pslash_{L} \, \kslash_{m})}{(m\cdot P_{L})^3}
+ \frac{2(k_1 \cdot k_2)}{(m\cdot P_{L})^2}\Bigg]
\ - \ (m \leftrightarrow m + 1) \ .
\label{atlast2}
\eeqa
The momentum flow can be conveniently represented as in
Figure 3.2, where we define
\beq
P \ := \ q_{2,m-1} \ , \qquad Q \ := \ q_{m+1, 1}
\ = \ -\, q_{2,m}
\ ,
\eeq
and $q_{p_1, p_2} := \sum_{l=p_1}^{p_2} k_l$.
We also have $P_L := q_{2,m} = -Q$.

Now we wish to combine the terms written explicitly in
\eqref{atlast2} with those that arise under $m\leftrightarrow m+1$.
Since \eqref{atlast2} is summed over $m$,
we simply shift  $m+1 \to m$ in these latter terms.
Let us now focus our attention on
the second  term in \eqref{atlast}
(similar manipulations
will be applied to the first term).
Writing the $m \lrar m+1$ term explicitly,
we obtain a contribution proportional to
\beq
\label{typical}
(-P_L^2 )^{-\epsilon}
\bigg[
\frac{\left[{\tr}_{+}(\kslash_{1}\, \kslash_{2}\,
\kslash_{m} \, \Pslash_L)\right]^2}{(m\cdot P_L )^2} \ - \
\frac{\left[
{\tr}_{+}(\kslash_{1}\, \kslash_{2}\,
\kslash_{m+1} \, \Pslash_{L})\right]^2}{((m+1)\cdot P_{L})^2}
\biggr]
\ .
\eeq
By shifting  $m+1 \to m$ in the second term
of \eqref{typical},
we change its $P_L$  so that \mbox{$P_L \to q_{2,m-1} = P$}
(whereas, in the non-shifted term, $P_L = -Q$).
The expression \eqref{typical} then reads
\beq
\frac{\left[{\tr}_{+}(\kslash_{1}\, \kslash_{2}\,
\kslash_{m} \, \Qslash)\right]^2}{(m\cdot Q )^2}
\,
\Big[ (-Q^2)^{-\e} - (-P^2)^{-\e}\Big]
\ ,
\eeq
where we used ${\tr}_{+}(\kslash_{1}\, \kslash_{2}\,
\kslash_{m} \Qslash)  = - {\tr}_{+}(\kslash_{1}\, \kslash_{2}\,
\kslash_{m}\Pslash)$ and
$Q\cdot m= - P \cdot m$.
Notice also that $m\cdot Q  = - (1/2) (Q^2 - P^2)$.

\begin{figure} [ht]
\label{adjacent}
\vspace{.2in}
\centerline {
\includegraphics[width=4in]{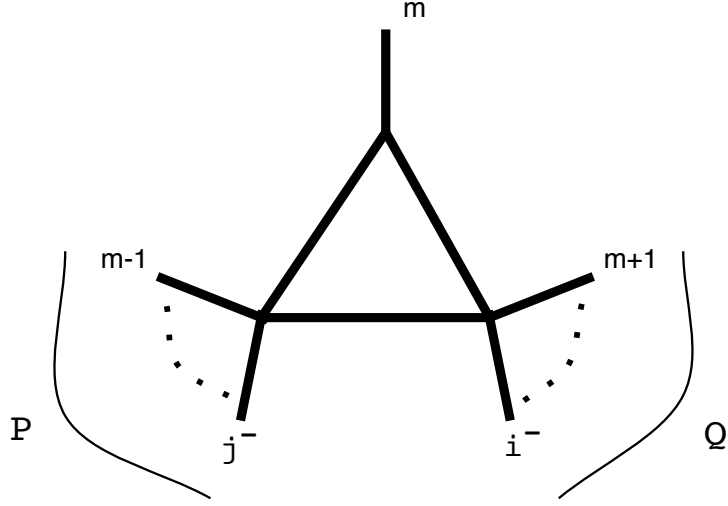}
}
\vspace{.2in}
\caption{\it A triangle function contributing to the amplitude
in the case of adjacent negative-helicity gluons.
Here we have defined $P :=  q_{j,m-1}$,
$Q  :=  q_{m+1, i}=  - q_{j,m}$
(in the text we set $i=1$, $j=2$ for definiteness).
}
\end{figure}
Next we re-instate the antisymmetry of the amplitudes under
the exchange of the indices $1 \lrar 2$ (which is manifest from
equation \eqref{integrand}).
Doing this we get
\beqa
\big[
{\tr}_{+}(\kslash_{1}\, \kslash_{2}\, \kslash_{m} \Qslash)
\big]^2 &\longrightarrow&
{1\over 2}
\Big[
\big({\tr}_{+}(\kslash_{1}\, \kslash_{2}\, \kslash_{m} \, \Qslash)
\big)^2 \, - \,
\big({\tr}_{+}(\kslash_{1}\, \kslash_{2}\,  \Qslash \, \kslash_{m})
\big)^2
\Big]
\\ [6pt]\nonumber
&=& 2 (k_1 \cdot k_2) (m\cdot Q)
\Big[
{\tr}_{+}(\kslash_{1}\, \kslash_{2}\, \kslash_{m} \, \Qslash)
\, - \,
{\tr}_{+}(\kslash_{1}\, \kslash_{2}\,  \Qslash \, \kslash_{m})
\Big]
\ .
\eeqa
Following similar steps for the first term in
\eqref{atlast2}, we arrive at the following expression
for the amplitude before taking the $\e \to 0$ limit:
\beq
\cA_\e \ = \ \cA_{1, \e} \, + \, \cA_{2, \e}
\ ,
\eeq
where
\beqa
\nonumber
\cA_{1, \e} & =&  - {\cA^{\rm tree} \over  t_1^{[2]}} \cdot
{\pi \over 6}
\cdot
\Bigl[
{\tr}_{+}(\kslash_{1}\, \kslash_{2}\, \kslash_{m} \,
\qslash_{m,1})
\, - \,
{\tr}_{+}(\kslash_{1}\, \kslash_{2}\,
\qslash_{m,1}\, \kslash_{m} )
\Bigr] \, T_{\e} (m, q_{2, m-1} , q_{2 , m})
\ ,
\\ \nonumber \cr
\label{adue}
\cA_{2, \e}  &=&
- {\cA^{\rm tree} \over (t_1^{[2]})^3} \cdot {\pi \over 3}
\cdot
\Bigl[
\big[
{\tr}_{+}(\kslash_{1}\, \kslash_{2}\, \kslash_{m} \,
\qslash_{m,1})\bigl]^2
{\tr}_{+}(\kslash_{1}\, \kslash_{2}\, \qslash_{m,1}
\,\kslash_{m}  )
\, - \,
\\ [6pt]  
&-&{\tr}_{+}(\kslash_{1}\, \kslash_{2}\, \kslash_{m} \,
\qslash_{m,1})
\big[ {\tr}_{+}(\kslash_{1}\, \kslash_{2}\,
\qslash_{m,1}\, \kslash_{m} )\big]^2
\Bigr] \,
T_{\e}^{(3)} (m, q_{2, m-1} , q_{2 , m})
\ ,
\eeqa
and $t_{1}^{[2]}$ follows from the definition of equation \eq{t}.
In order to write \eqref{adue}
in a compact from, we have introduced
$\epsilon$-dependent triangle functions
\cite{bbst} as in the previous chapter (\emph{c.f.} Eq.~\eq{epstrianglen1})
\beq
\label{epstriangle}
T_{\epsilon}^{(r)} (p, P, Q) \ := \
{1\over \epsilon}
{ (-P^2)^{-\epsilon} - (-Q^2)^{-\epsilon} \over (Q^2 - P^2)^r}
\ ,
\eeq
where  $p+P+Q=0$, and $r$ is a positive integer.%
\footnote{For $r=1$ we will omit the superscript $(1)$
in $T^{(1)}$. }

We can now take the $\e \to 0$ limit.
As long as $P^2$ and $Q^2$ are non-vanishing, one has
\beq
\label{TT1}
\lim_{\epsilon \to 0} T_{\epsilon}^{(r)} (p, P, Q) \ = \
T^{(r)} (p, P, Q) \ , \qquad
P^2 \neq 0 \, , \, Q^2 \neq 0
\ ,
\eeq
where the $\epsilon$-independent triangle functions are defined by
\beq
\label{trianr}
T^{(r)} (p, P, Q) \ := \
{ \log (Q^2 / P^2) \over ( Q^2 - P^2)^r}
\ .
\eeq
If either of the invariants vanishes,
the limit of the $\e$-dependent triangle
gives rise to an infrared-divergent term
(which we call a ``degenerate'' triangle -
this is one with two massless legs).
For example, if
$Q^2=0$, one has
\beq
\label{TT2}
\left. T_{\epsilon} (p, P, Q)
\right|_{Q^2=0}  \ \longrightarrow \
- {1\over \epsilon} \, {(-P^2)^{-\epsilon }\over P^2}\, ,
\qquad {\epsilon \to 0}
\ .
\eeq
The two possible configurations which give rise to
infrared-divergent contributions
correspond to the following two possibilities:
\begin{itemize}
\item[{\bf a.}]
$q_{2, m-1} = k_2 $ (hence $q_{2, m-1}^2 = 0$).
In this case we also have
$q_{2,m}^2 = t_{2}^{[2]}$.
\item[{\bf b.}]
$- q_{2, m} = k_1$ (hence $q_{2, m}^2 = 0$).
Therefore  $q_{2, m-1}^2 = t_{n}^{[2]}$.
\end{itemize}
We notice that infrared poles will appear only in terms
corresponding to the triangle function  $T$.
Indeed, whenever one of the
kinematical invariants contained in $T^{(3)}$ vanishes,
the combination of traces multiplying this function in
\eqref{adue} vanishes as well.

In conclusion we arrive at the following result,
where we have explicitly separated-out the infrared-divergent
terms:%
\footnote{A factor of
$-4 \pi \hat{\l}$ will be understood on the right hand sides
of Eqs.~\eqref{sz}, \eqref{endresult} and \eqref{chiral},
where  $\hat{\l}$ is defined in \eqref{ubiquitous}.}
\beq
\label{sz}
\cA_{n}^{\rm scalar} \ = \
\cA_{\rm poles} \, + \,
\cA_1 \, + \, \cA_2
 \ ,
\eeq
where
\beqa
\cA_{\rm poles} & = &
{1\over 6} \cA^{\rm tree} {1 \over \e }
\Bigl[ (-t_2^{[2]})^{-\e} \, + \, (-t_n^{[2]})^{-\e}
\Bigr]
\ ,
\\ [6pt] \nonumber
\cA_{1} & = &
 {1\over 6} \cA^{\rm tree}
{1\over t_1^{[2]}}
\sum_{m=4}^{n-1} \Bigl[
{\tr}_{+}(\kslash_{1}\, \kslash_{2}\, \kslash_{m} \,
\qslash_{m,1})
 -
{\tr}_{+}(\kslash_{1}\, \kslash_{2}\,
\qslash_{m,1}\, \kslash_{m} )
\Bigr]  T  (m, q_{2, m-1} , q_{2 , m})
\ ,
\\ [6pt] \nonumber
\cA_2  &=&
{1\over 3}
\cA^{\rm tree}
{1 \over (t_1^{[2]})^3}
\sum_{m=4}^{n-1}
\Bigl[
\big[ {\tr}_{+}(\kslash_{1}\, \kslash_{2}\, \kslash_{m} \,
\qslash_{m,1})\bigl]^2
{\tr}_{+}(\kslash_{1}\, \kslash_{2}\, \qslash_{m,1}
\,\kslash_{m}  )
\\ [6pt]\nonumber \cr
&-&{\tr}_{+}(\kslash_{1}\, \kslash_{2}\, \kslash_{m} \,
\qslash_{m,1})
\big[ {\tr}_{+}(\kslash_{1}\, \kslash_{2}\,
\qslash_{m,1}\, \kslash_{m} )\big]^2
\Bigr] \,
T^{(3)} (m, q_{2, m-1} , q_{2 , m})
\ .
\eeqa
More compactly, we can recognise that
$\cA_{\rm poles}$ and $\cA_{1}$  reconstruct
the contribution of an $\cN\!=\!1$
chiral multiplet,  and rewrite
\eqref{sz} as
\beqa
\label{endresult}
\cA_n^{\textrm{scalar}} \ = \
\frac{1}{3}\cA_{12}^{\cN = 1,\, \textrm{chiral}}
\ + \ \frac{1}{3}\cA_{12}^{\textrm{tree}}
\frac{1}{(t_1^{[2]})^3}\sum_{m=4}^{n-1}
\cB_{12}^{m}\,
T^{(3)}  (m, q_{2, m-1} , q_{2 , m})
\ ,
\eeqa
where
\beqa
\label{scalar2}
\cB_{12}^{m} &=&
 \bigl[{\tr}_{+}(\kslash_{1}\, \kslash_{2}\,
\kslash_{m} \, \qslash_{m,1})\bigr]^2{\tr}_{+}(\kslash_{1}\,
\kslash_{2}\, \qslash_{m,1} \, \kslash_{m})
\\ \nonumber \cr
&-&
\bigl[{\tr}_{+}(\kslash_{1}\, \kslash_{2}\,
\qslash_{m,1} \, \kslash_{m})\bigr]^2{\tr}_{+}(\kslash_{1}\,
\kslash_{2}\,
\kslash_{m} \, \qslash_{m,1})
\ .
\eeqa
and
\begin{eqnarray}
\label{chiral}
\cA_{12}^{\cN = 1,\, \textrm{chiral}} \ &=& \
\frac{1}{2}\cA_{12}^{\textrm{tree}}\frac{1}{t_1^{[2]}}\
\sum_{m=3}^{n}\Big\{\big[{\tr}_{+}(\kslash_{1}\, \kslash_{2}\,
\kslash_{m} \, \qslash_{m,1}) \, - \, {\tr}_{+}(\kslash_{1}\,
\kslash_{2}\,
\qslash_{m,1} \, \kslash_{m})\big]\nonumber\\
&{}&\qquad\qquad\qquad\quad\cdot\ T (m, q_{2, m-1} , q_{2 , m})\Big\}
\ .
\end{eqnarray}
This is our result for the cut-constructible part of
the $n$-gluon MHV scattering amplitude with adjacent
negative-helicity gluons in positions 1 and 2.
This expression was first derived
by Bern, Dixon, Dunbar and Kosower in \cite{Bern:1994cg},
and our result agrees precisely with this.
A remark is in order here.
In \cite{Bern:1994cg}, the final result is
expressed in terms of a function
\beq
L_2 (x) \ := \ {\log x - (x - 1/x)/2 \over (1-x)^3}
\ ,
\eeq
which contains a rational part $ -  (x - 1/x) / 2(1-x)^3$
which removes a spurious
third-order pole from the amplitude.
With our approach however we did not expect to detect rational
terms in the scattering amplitude, and indeed
we do not find  such terms.\footnote{In our notation $L_2$
corresponds to $T^{(3)}$, which, however, lacks a rational term.}
Furthermore, we do not find the other rational terms
which are known to be present in the one-loop
scattering amplitude \cite{bdk9302280,Berger:2006vq}.


\section{The scattering amplitude in the general case}
\label{generalcasesection}

The situation where the negative-helicity gluons are not
adjacent is technically more challenging.
Our starting point will be
\eqref{integrand}, to which we will apply the Schouten identity
(see Appendix \ref{traceappendix} for a collection
of spinor identities used).
Eq.~\eqref{integrand}
can then be written as a sum of four terms:%
\footnote{We drop the factor of
$-i\cA_n^{\rm tree}$ from now on and
reinstate it at the end of the calculation.}
\beqa
\label{4integrand} \hat{\mathscr{C}}={\mathscr C}(m_1, m_2  +  1) \ - \
{\mathscr C}(m_1,m_2)
\ - \  {\mathscr C}(m_1-1, m_2 +  1)\  +\ {\mathscr C}(m_1  - 1, m_2)
\eeqa
where
\beqa
\label{basicC}
{\mathscr C}(a,b) \ := \
\frac{\langle i \, l_1 \rangle \,
\langle j \, l_1 \rangle^2 \,
\langle i \, l_2 \rangle^2 \,
\langle j \, l_2 \rangle}{\langle i \, j \rangle^4 \,
\langle l_1 \, l_2 \rangle^2}
\, \cdot \, \frac{\langle i \, a \rangle \, \langle j \, b \rangle}
{\langle l_1 \, a \rangle \, \langle l_2 \, b \rangle}\ .
\eeqa
The calculation of the phase space integral of this expression is
discussed in \mbox{Appendix \ref{nonsusyappendix}}.
The result is
\begin{eqnarray}
 \label{pspaceresult}
\nonumber
&&  \!\!\!\!\!\!\!\!\!\!\!\!\!\!
\int \!d^{4-2\epsilon}{\rm LIPS}(l_2 , -l_1 ; P_{L;z})\ \
{\mathscr C}(a,b)   \\ [6pt]
\nonumber \cr
&& \quad = \ \frac{1}{3} \
\frac{\ijba}{(a\cdot b)}\ \Bigg[ \ijPLza^2\bigg[
\frac{\ijaPLz}{(P_{L;z}\cdot a)^3}
+ \frac{2(i\cdot j)}{(P_{L;z}\cdot a)^2} \bigg]  -
( a \leftrightarrow b)   \Bigg]\\ [6pt]\nonumber \cr
   &&  \quad\qquad + \
\frac{1}{2}\
\frac{\ijba\ijab}{(a\cdot b)^2}
      \  \Bigg[ \frac{\ijPLza^2}{(P_{L;z}\cdot a)^2} +
( a \leftrightarrow b) \Bigg]
\\ [6pt]\nonumber  \cr
&&   \qquad \qquad - \ \frac{\ijab\ijba}{(a \cdot b)^3}\
\Bigg[  \frac{\ijab\ijPLza}{(P_{L;z}\cdot a)}
               + ( a \leftrightarrow b)  \Bigg]
\\ [6pt] \cr
    && \qquad\qquad\qquad +\
%
%
\frac{\ijab^2\ijba^2}{(a\cdot b)^4} \
    \log\bigg( 1 - \frac{(a\cdot b)}{N}P_{L;z}^2 \bigg)
\ ,
\end{eqnarray}
where $N \equiv N(P) := (a\cdot b)P^2 - 2(P\cdot a)(P \cdot b)$,
and we have suppressed a factor of
$-4 \pi \hat{\l} (-P_L^2)^{-\e}
\cdot [2^8(i\cdot j)^4]^{-1}$
on the right hand side of \eqref{pspaceresult}, where
$\hat{\l}$ is defined in \eqref{ubiquitous}.
We notice that  \eqref{pspaceresult} is symmetric under the
simultaneous exchange of $ i$ with $j$ and $a$ with $b$.
This symmetry is manifest in the coefficient multiplying the
logarithm -- the last term in \eqref{pspaceresult};
for the remaining terms, nontrivial gamma matrix identities
are required.
For instance, consider the terms in the second line of
 \eqref{pspaceresult}. These terms are present in
the adjacent gluon case \eqref{atlast}, and it is therefore
natural to expect that the trace structure of this term
is separately invariant when $i \lrar j$ and $a \lrar b$.
Indeed this is the case thanks to the identity
\beqa
32 (i \cdot j )^3 & = & \ijPLza^2\bigg[
\frac{\ijaPLz}{(P_{L;z}\cdot a)^3}
\, + \, \frac{2(i\cdot j)}{(P_{L;z}\cdot a)^2} \bigg]
\\ \nonumber
& + &
\ijaPLz^2\bigg[
\frac{\ijPLza}{(P_{L;z}\cdot a)^3}
\, + \, \frac{2(i\cdot j)}{(P_{L;z}\cdot a)^2} \bigg]
 \ .
\eeqa
Similar identities show that the third and fourth line of
\eqref{pspaceresult} are invariant under
the simultaneous exchange $i \lrar j$ and $a \lrar b$.

The next step is to perform the dispersion integral of
\eqref{pspaceresult}, \emph{i.e.}~the integral
over the variable $z$ which has been converted to an integral over $P_{L;z}$.
The relevant terms are thus those involving
$P_{L;z}$ in \eqref{pspaceresult},
and in an overall factor $(P_{L;z}^2)^{-\epsilon}$
arising from the dimensionally regulated measure.

The integral over the term involving the
logarithm has been evaluated in \cite{bst}, with the
result
\beqa
\label{dilogint}
\int \frac{dP_{L;z}^2 }{P_{L;z}^2 - P_{L}^2 }
(P_{L;z}^2)^{-\epsilon}
\log\bigg( 1 - \frac{(a\cdot b)}{N}P_{L;z}^2 \bigg)
\ =\
\Li2\left( 1- \frac{(a\cdot b)}{N(P)} P_L^2 \right)
\ + \ \cO ( \e )
\ .
\eeqa
Notice that these terms were not present in the
adjacent negative-helicity gluon case considered in \S \ref{adjacentsection}.

Next we move  on to the remaining terms
in \eqref{pspaceresult}.
Inspecting their $z$-dependence,
we see that, in complete similarity
with the adjacent case of \S \ref{adjacentsection},
in each term there are the same powers of $P_{L;z}$
in the numerator as in the denominator.
Hence, in the centre of mass frame in which
$P_{L;z}\! :=\! P_{L;z} (1,\vec{0})$, one finds that
$P_{L;z}$ cancels completely.
Note that this also immediately resolves
the question of gauge invariance for these
terms -- this occurs only through the $\eta$ dependence in
$P_{L;z} = P_L -z\eta$.
Furthermore, the  box functions coming from \eqref{dilogint}
are separately gauge-invariant \cite{bst}.
The conclusion is that our expression for the amplitude below,
built from sums over MHV diagrams of the
dispersion integral
of \eqref{pspaceresult}, will be gauge-invariant.
Moreover, apart from \eqref{dilogint},
the only other dispersion integral we will need
is that computed in \eqref{disperse}.

It follows from this discussion that the result
of the dispersion integral of
\eqref{pspaceresult} is (suppressing a factor of
$-4 \pi \hat{\l} (-P_L^2)^{-\e}
\cdot [2^8(i\cdot j)^4]^{-1} \cdot
[\pi\epsilon \csc(\pi\epsilon)]$):
\begin{eqnarray}
 \label{dispspaceresult} \nonumber
&&  \!\!\!\!
\int \frac{dz}{z}\, \int \!
d^{4-2\epsilon}{\rm LIPS}(l_2 , -l_1 ; P_{L;z})\ \
{\mathscr C}(a,b)
\\ [6pt] \nonumber \cr
&&   =
  \frac{1}{\epsilon} {(-P_L^2)^{-\epsilon}}
 \Bigg\{
    \ \frac{1}{3} \ \frac{\ijba}{(a\cdot b)}\
\Bigg[ \ijPLa^2\bigg[  \frac{\ijaPL}{(P_L\cdot a)^3}
+ \frac{2(i\cdot j)}{(P_L\cdot a)^2} \bigg]  -
( a \leftrightarrow b)   \Bigg]
\\ [6pt]\nonumber \cr
   &&  \quad\qquad + \ \frac{1}{2}\ \frac{\ijba\ijab}{(a\cdot b)^2}
      \  \Bigg[ \frac{\ijPLa^2}{(P_L\cdot a)^2} +
( a \leftrightarrow b) \Bigg]
\\ [6pt]\nonumber  \cr
&&   \qquad \qquad - \ \frac{\ijab\ijba}{(a\cdot b)^3}\
\Bigg[  \frac{\ijab\ijPLa}{(P_L \cdot a)}
               + ( a \leftrightarrow b)  \Bigg]  \  \Bigg\}
\\ [6pt]\cr
    && \qquad\qquad\qquad +\
\frac{\ijab^2\ijba^2}{(a\cdot b)^4} \
    \Li2\left( 1 - \frac{(a\cdot b)}{N(P_L)}P_L^2 \right)\ .
\end{eqnarray}

Now, due to the four terms in \eqref{4integrand},
the sum over MHV diagrams will include
a signed sum over four expressions like
\eqref{dispspaceresult}.
Let us begin by considering
the last line of \eqref{dispspaceresult}.
This is a term familiar from \cite{bst} and
\cite{bbst} and corresponds to
one of the four dilogarithms in
the novel expression found in
\cite{bst} for the finite part $B$
of a  scalar box function,
\beqa
\label{boxf}
\nonumber
B(s,t,P^2,Q^2) & = &
\Li2\left(1-\frac{(a\cdot b)}{N(P)}P^2\right) \, + \,
\Li2\left(1-\frac{(a\cdot b)}{N(P)}Q^2\right)
\\  \cr
&-&
\Li2\left(1-\frac{(a\cdot b)}{N(P)}s\right)
\, - \,
\Li2\left(1-\frac{(a\cdot b)}{N(P)}t\right)
\ ,
\eeqa
with $s:= (P + a)^2$,  $t:= (P + b)^2$, and
$P + Q + a + b = 0$.
By taking into account the four terms in
\eqref{4integrand} and summing over MHV  diagrams
as specified in \eqref{loopint} and \eqref{range},
one sees that each of the four terms in any finite
box function $B$ appears exactly once,
in complete similarity with \cite{bst}
and \cite{bbst}. The final contribution of this term
will then be%
\footnote{We multiply our final results by a factor of $2$,
which takes into account the two possible helicity
assignments for the scalars in the loop.}
\beq
\label{finbox}
\sum_{m_1=j+1}^{i-1}\sum_{m_2=i+1}^{j-1}
{1\over 2} \, \big[ b_{m_1  m_2}^{ij} \big]^2
\,
B( q_{m_1, m_2 -1}^2 , q_{m_1+1,  m_2}^2,
q_{m_1+1 , m_2 -1}^2, q_{m_2+1, m_1 -1}^2)
\ ,
\eeq
where $t_i^{[k]} := (p_i + p_{i+1} + \cdots + p_{i+k-1})^2$
for $k\geq 0$, and $t_i^{[k]} =t_i^{[n-k]}$ for $k <0$.
In writing \eqref{finbox},
we have taken into account
that the dilogarithm in \eqref{dispspaceresult}
is multiplied by a coefficient proportional to
the square of $b_{m_1 m_2}^{i j}$, where
\beq
\label{bdef}
b_{m_1 m_2}^{i j} \ :=  \
-2 \, \frac{
\tr_{+} \left( \kslash_{i} \kslash_{j} \kslash_{m_1}
\kslash_{m_2} \right)
\tr_{+} \left( \kslash_{i} \kslash_{j} \kslash_{m_2}
\kslash_{m_1} \right) }
{ [ (k_i + k_j)^{2}]^2
\,
[( k_{m_1} + k_{m_2})^2]^2}
\ .
\eeq
We notice that $b_{m_1 m_2}^{i j}$ is the coefficient
of the box functions in the one-loop
$\cN \! = \! 1$ MHV amplitude, originally calculated by
Bern, Dixon, Dunbar and Kosower in \cite{Bern:1994cg}, and
derived in \cite{bbst, quig}
using the MHV diagram approach
for loops proposed in \cite{bst}.
\begin{figure} [ht]
\label{scalarbox}
\vspace{.2in}
\centerline {
\includegraphics[width=4in]{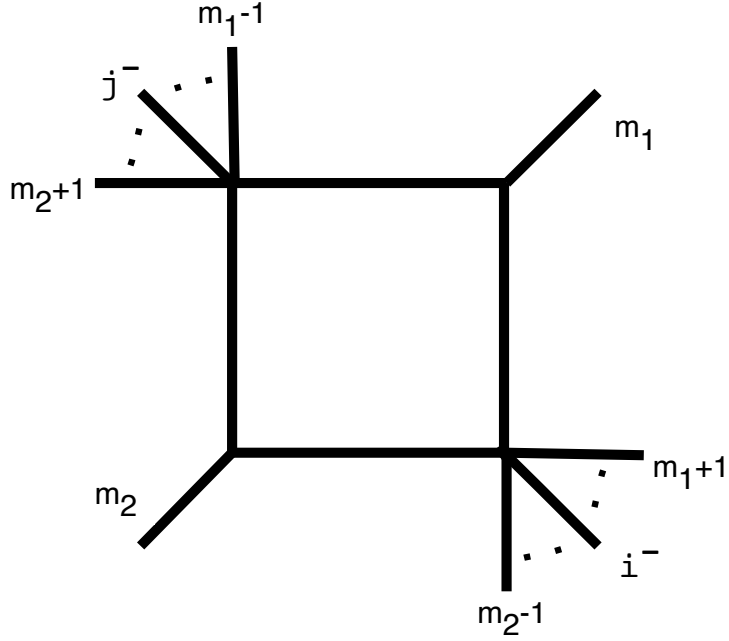}
}
\vspace{.2in}
\caption{\it A box function contributing to the amplitude
in the general case.
The negative-helicity gluons, $i$ and $j$,
cannot be in adjacent positions, as the figure
shows.}
\end{figure}
Furthermore, we observe that $b_{m_1 m_2}^{i j}$ is holomorphic
in the spinor variables, and as such has simple localisation
properties in twistor space.
Indeed, from \eqref{bdef} it follows that
\beq
\label{bdefholo}
b_{m_1 m_2}^{i j} \ =  \
2\,
{ \lan i  \, m_1 \ran \, \lan i  \, m_2 \ran \,
\lan j  \, m_1 \ran \, \lan j  \, m_2 \ran
\over
\lan i  \, j \ran^2  \, \lan m_1  \, m_2 \ran^2
}
\ .
\eeq
Summing over the four terms for the remainder of
\eqref{dispspaceresult} can be done in complete
similarity with \S \ref{n=1mhvsection} (and Section 4 of \cite{bbst}).%
\footnote{In \S \ref{adjacentsection} we have illustrated in detail
how this sum is performed for the
simpler case of adjacent negative-helicity gluons.}
We will skip the details of this derivation
and now present our result.

In order to do this, we find it convenient to define
the following expressions:
\beqa
A^{ij}_{m_1 m_2} & := &
{(i\, j\, m_2+1\,  m_1) \over ((m_2 +1) \cdot m_1)} \, - \,
{(i\, j\, m_2\,  m_1) \over (m_2  \cdot m_1)}
\\ [4pt]\nonumber
& = & -2\, [i \, j] \, \lan m_1 \, i \ran
\lan m_1 \, j \ran \,
{\lan m_2 \, m_2 + 1 \ran \over
\lan m_2+1  \, m_1  \ran
\,
\lan m_1 \, m_2  \ran}
\ ,
\\ [4pt]\cr
S^{ij}_{m_1 m_2} & := &
{(i\, j\, m_1\,  m_2 +1)(i\, j\, m_2+1\,  m_1)
\over ((m_2 +1) \cdot m_1)^2} \, - \,
{(i\, j\, m_1\,  m_2) (i\, j\, m_2\,  m_1)
\over (m_2  \cdot m_1)^2}
\ ,
\\[4pt]
I^{ij}_{m_1 m_2} & := &
{(i\, j\, m_1\,  m_2 +1)^2 (i\, j\, m_2+1\,  m_1)
\over ((m_2 +1) \cdot m_1)^3} \, - \,
{(i\, j\, m_1\,  m_2)^2 (i\, j\, m_2\,  m_1)
\over (m_2  \cdot m_1)^3}
\, ,
\eeqa
where for notational simplicity we set
$(a_1\, a_2\, a_3\, a_4) :=
{\tr}_{+} (\aslash_1 \aslash_2 \aslash_3  \aslash_4)$ in the above.
We also note the symmetry properties
\beq
A^{ji}_{m_1 m_2} \, = \, - A^{ij}_{m_1 m_2}
\ ,
\qquad
S^{ji}_{m_1 m_2} \, = \,  S^{ij}_{m_1 m_2}
\ .
\eeq
The momentum flow is best described using the triangle
diagram in Figure 3.4, where we use the following definitions:
\beqa
P & := & q_{m_2 + 1 , m_1 -1} \ = \ - q_{m_1 , m_2}
\ ,
\\ \nonumber
Q &: = & q_{m_1 + 1 ,  m_2}
\ .
\eeqa
The triangle in Figure 3.5 also appears in the calculation,
and can be  converted into a triangle as in  Figure 3.4 -
but  with $i$ and $j$ swapped -
if one shifts $m_1\!-1 \to m_1$,
and then swaps  $m_1 \lrar m_2$.

Next we introduce the coefficients
\beqa
\label{alfredina}
\cA^{ij}_{m_1 m_2} &:= &
2^{-8}
(i\cdot j)^{-4}
\, A^{ij}_{m_1 m_2}
\, \Big[
(i\, j\, m_1\,  Q)^2 (i\, j\,  Q \, m_1) \,\nonumber\\
 &{}&\qquad\qquad\qquad\quad\ -\ \,
(i\, j\, m_1\,  Q) (i\, j\,  Q \, m_1)^2\Big]
,
\\ \cr
\tilde{\cA}^{ij}_{m_1 m_2} &:= &
2^{-8} (i\cdot j)^{-4}  \, A^{ij}_{m_1 m_2}
\, \Big[
(i\, j\, m_1\,  Q)^2  \, - \,  (i\, j\,  Q \, m_1)^2\Big]
\, ,
\\ \cr
\cS^{ij}_{m_1 m_2} & = &
2^{-8} {(i\cdot j)^{-4}} \, S^{ij}_{m_1 m_2}
\, \Big[
(i\, j\, m_1\,  Q)^2 \, + \,
(i\, j\,  Q \, m_1)^2\Big]
\ ,
\\ \cr
\cI^{ij}_{m_1 m_2} &:= &
2^{-8} (i\cdot j)^{-4}  \, \Big[
I^{ij}_{m_1 m_2}
\, (i\, j\, Q \,  m_1)  \, + \,
I^{ji}_{m_1 m_2}(i\, j\,  m_1 \, Q )\Big]
\label{giuseppina}
\ .
\eeqa
We will also make use of the
$\epsilon$-dependent triangle functions introduced
in \eqref{epstriangle}, whose $\e \to 0$ limits have been
considered in \eqref{TT1}--\eqref{TT2}.
This is in order to write a compact expression
which also incorporates the infrared-divergent terms.%
\footnote{The infrared-divergent terms will be described below
and used to check that
our result has the correct infrared pole structure.}

We can now present our result for the one-loop
MHV amplitude:
\beqa
\nonumber
\frac{\cA_{\rm scalar}}{\cA_{\rm tree}}\!\! & = &\!\!\!
\sum_{m_1=j+1}^{i-1}\sum_{m_2=i+1}^{j-1}
{1\over 2} \big[ b_{m_1 m_2}^{ij} \big]^2
\,
B( q_{m_1, m_2 -1}^2 , q_{m_1+1,  m_2}^2,
q_{m_1+1,  m_2 -1}^2, q_{m_2+1, m_1 -1}^2)
\\ [6pt]\nonumber
& -  & \bigg(
 \,{8 \over 3}
\sum_{m_1=j+1}^{i-1}\sum_{m_2=i}^{j-1}
\Big[  \cA_{m_1 m_2}^{ij}\, T^{(3)} (m_1 , P , Q)
\, - \, (i \cdot j)  \tilde{\cA}^{ij}_{m_1 m_2}\,
T^{(2)} (m_1 , P , Q) \Big]
\\ [4pt]\nonumber
&+& 2
\sum_{m_1=j+1}^{i-1}\sum_{m_2=i}^{j-1}
\Big[  \cS^{ij}_{m_1 m_2}\, T^{(2)} (m_1 , P , Q)
\, + \,  \cI^{ij}_{m_1 m_2}\, T (m_1 , P , Q)
\Big] +  ( i \leftrightarrow j) \biggr) \, ,\\
\label{final}
\eeqa
where on the right hand side of \eqref{final}
a factor of $-4 \pi \hat{\l} $ is understood
and $\hat{\l}$ is defined in \eqref{ubiquitous}.
We can also introduce the coefficient
\beq
\label{art}
c_{m_1 m_2}^{i j} \ := \
{1\over 2}  \bigg[
{( i\, j \, m_2+1  \, m_1) \over \big( (m_2+1) \cdot m_1 \big)}
 \, - \,
{( i \, j \, m_2  \, m_1) \over (m_2 \cdot m_1 )}
\bigg]
\,
{ (i\, j\, m_1 \, Q) \, - \, ( i\, j\, Q\, m_1)
\over [(i \, + \, j)^2]^2}
\ ,
\eeq
which already appears
as the coefficient multiplying the triangle function $T$
in the $\cN \! = \! 1$ amplitude,
(see \emph{e.g.}~Eq.~\eq{Neq1}),
and rewrite \eqref{final} as
\beqa
\nonumber
\cF&=&
\sum_{m_1=j+1}^{i-1}\sum_{m_2=i+1}^{j-1}
{1\over 2}
\big[ b_{m_1 m_2}^{ij} \big]^2
\,
B( q_{m_1, m_2 -1}^2 , q_{m_1+1 , m_2}^2,
q_{m_1+1 , m_2 -1}^2, q_{m_2+1, m_1 -1}^2)
\\ [6pt]\nonumber
& - &
\bigg(
\,
 {1\over 2}
\sum_{m_1=j+1}^{i-1}\sum_{m_2=i}^{j-1}
 {1\over 3} \, c^{ij}_{m_1 m_2}\,
\Big[ {
(i\, j\, m_1\,  Q) \,
(i\, j\,  Q \, m_1) \over 2 (i \cdot j )^2 }
\,  T^{(3)} (m_1 , P , Q) +   \
T  (m_1 , P , Q) \Big]\nonumber\\ [6pt]
&+& 2
\sum_{m_1=j+1}^{i-1}\sum_{m_2=i}^{j-1}
\Big[  \cS^{ij}_{m_1 m_2}\, T^{(2)} (m_1 , P , Q)
\, + \,  \cI^{ij}_{m_1 m_2}\, T (m_1 , P , Q)
\Big] +\ ( i \leftrightarrow j) \ \ \biggr)\ ,\nonumber\\
\label{final2}
\eeqa
where $\cF=\cA_{\rm scalar}/\cA_{\rm tree}$.
%
\begin{figure} [ht]
\label{scalartri1}
\vspace{.2in}
\centerline {
\includegraphics[width=4in]{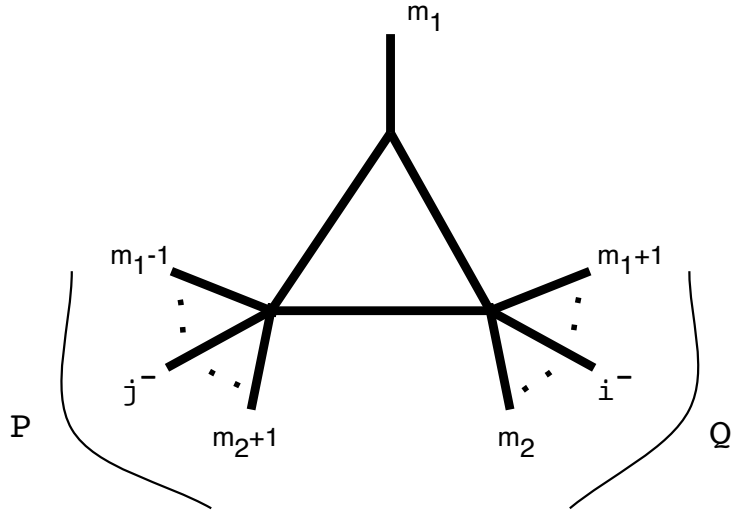}
}
\vspace{.2in}
\caption{\it One type of triangle function contributing to the amplitude
in the general case, where $i \in Q$, and $j \in P$.}
\end{figure}
\begin{figure} [ht]
\label{scalartri2}
\vspace{.2in}
\centerline {
\includegraphics[width=4in]{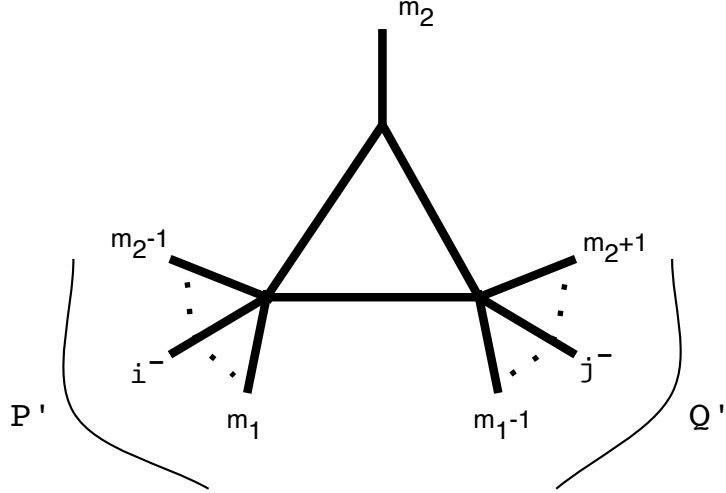}
}
\vspace{.2in}
\caption{\it Another type of triangle function contributing to the amplitude
in the general case.
By first shifting $m_1-1 \to m_1$,
and then swapping $m_1 \lrar m_2$,
we convert this into
a triangle function as in Figure 3.4 -- but with
$i$ and $j$ swapped. These are the triangle functions
responsible for the $i\lrar j$ swapped terms in
\eqref{final} -- or \eqref{final2}.
}
\end{figure}

Several remarks are in order.
\begin{itemize}
\item[{\bf 1.}]
As usual, the variables $q_{m_1, m_2 -1}^2$,
$q_{m_1+1 , m_2}^2$ correspond to the $s$- and $t$-channel
of the finite part of the ``easy two-mass'' box function with
massless legs $m_1$ and $m_2$, and massive legs
$q_{m_1+1 , m_2 -1}^2$, $q_{m_2+1, m_1 -1}^2$
(Figure 3.3).
\item[{\bf 2.}]
Compared to the ranges of $m_1$ and $m_2$
indicated in \eqref{range},
we have omitted $m_1 = i$ in the summation
of the triangles as for this value the coefficients
$A$, $S$, $I$ defined in \eqref{alfredina}--\eqref{giuseppina}
vanish. Notice also that we have  $i \in Q$ and $j \in P$.
\item[{\bf 3.}]
In the case of adjacent negative-helicity gluons,
the only surviving terms are those  containing the
coefficient $c^{ij}_{m_1 m_2}$, on the
second line of \eqref{final} or \eqref{final2}.
We will return to this point in \S \ref{checksection}.
\item[{\bf 4.}]
We comment that, in contrast to the
adjacent case  (see \eqref{endresult}),
in the general case the $\cN=1$ chiral amplitude
does not separate out naturally in
the final result. One can quickly see this from
the coefficient of the box function
$B$ in \eqref{final} for example.
\end{itemize}

Next we wish to explicitly separate out the infrared divergences
from \eqref{final}.
We can immediately anticipate that there will be four
infrared-divergent terms, corresponding to the four possible
degenerate triangles.
Two of these degenerate triangles occur when either
$P^2$ or $Q^2$ happen to vanish.
The other two
originate from the $i \lrar j$ swapped terms.

Let us first consider the terms arising from the summation with
 $i \lrar j$ unswapped.
When  $Q^2 =0$, it follows that $m_1 = i -1$ and $m_2 = i$
(see Figure 3.4).
When  $P^2 =0$, it follows that $m_1 = j+1 $ and $m_2 = j-1$
(see Figure 3.4).
Hence
\beqa
T^{(r)} (p, P, Q) & \to &
(-)^r \, {1\over \e} \, {(- t_{i-1} ^{[2]})^{-\e} \over
(t_{i-1}^{[2]})^r}
\ , \qquad Q^2 \to 0 \ ,
\\ \nonumber
T^{(r)} (p, P, Q) & \to &
- \, {1\over \e} \, {(- t_j^{[2]})^{-\e} \over (t_j^{[2]})^r}
\ , \qquad \quad \ \ P^2 \to 0 \ .
\eeqa
The infrared-divergent terms coming from $Q^2 =0$
are then easily extracted, and are
\beqa
\label{ir1}
- {1 \over 2 \e}  &\cdot & (- t_{i-1} ^{[2]})^{-\e} \,
4 (i \cdot j) \,
{(i\, j \, i-1 \, i+1) \over \big( (i+1) \cdot (i-1) \big)}
\\ \nonumber
&\cdot &
\bigg[
{8 \over 3} (i\cdot j)^2 \, - \,
2 \, {(i\, j \, i+1 \, i-1) \over \big( (i+1) \cdot (i-1) \big)}
(i \cdot j) \, + \,
 {(i\, j \, i+1 \, i-1) (i\, j \, i-1 \, i+1)
\over \big( (i+1) \cdot (i-1) \big)^2}
\bigg]
\ ,
\eeqa
and from $P^2 =0$
\beqa
\label{ir2}
- {1 \over 2 \e}  &\cdot & (- t_{j} ^{[2]})^{-\e} \,
4 (i \cdot j) \,
{(i\, j \, j-1 \, j+1) \over \big( (j+1) \cdot (j-1) \big)}
\\ \nonumber
&\cdot &
\bigg[
{8 \over 3} (i\cdot j)^2 \, - \,
2 \, {(i\, j \, j+1 \, j-1) \over \big( (j+1) \cdot (j-1) \big)}
(i \cdot j) \, + \,
 {(i\, j \, j+1 \, j-1) (i\, j \, j-1 \, j+1)
\over \big( (j+1) \cdot (j-1) \big)^2}
\bigg]
\ .
\eeqa
Likewise, from the ``swapped'' degenerate triangles we obtain
the following infrared-divergent terms:
\beqa
\label{ir3}
- {1 \over 2 \e}  &\cdot & (- t_{j-1} ^{[2]})^{-\e} \,
4 (i \cdot j) \,
{(i\, j \, j+1 \, j-1) \over \big( (j+1) \cdot (j-1) \big)}
\\ \nonumber
&\cdot &
\bigg[
{8 \over 3} (i\cdot j)^2 \, - \,
2 \, {(i\, j \, j-1 \, j+1) \over \big( (j+1) \cdot (j-1) \big)}
(i \cdot j) \, + \,
 {(i\, j \, j-1 \, j+1) (i\, j \, j+1 \, j-1)
\over \big( (j+1) \cdot (j-1) \big)^2}
\bigg]
\ ,
\eeqa
and
\beqa
\label{ir4}
- {1 \over 2 \e}  &\cdot & (- t_{i} ^{[2]})^{-\e} \,
4 (i \cdot j) \,
{(i\, j \, i+1 \, i-1) \over \big( (i+1) \cdot (i-1) \big)}
\\ \nonumber
&\cdot &
\bigg[
{8 \over 3} (i\cdot j)^2 \, - \,
2 \, {(i\, j \, i-1 \, i+ 1) \over \big( (i+1) \cdot (i- 1)
\big)}
(i \cdot j) \, + \,
 {(i\, j \, i-1 \, i+ 1) (i\, j \, i+1 \, i-1)
\over \big( (i+1) \cdot (i-1) \big)^2}
\bigg]
\ .
\eeqa


\subsection{Comments on twistor space interpretation}


We would like to make some brief comments on the interpretation in
twistor space of our result \eqref{final2}.
\begin{itemize}
\item[{\bf 1.}]
As noticed earlier, the coefficient $b^{ij}_{m_1 m_2}$ appears
already in the $\cN \! = \! 1$ chiral multiplet
contribution to a one-loop MHV amplitude, where it multiplies
the box function. It was noticed in
Section 4 of \cite{csw2} that $b^{ij}_{m_1 m_2}$
is a holomorphic function and hence it does not affect
the twistor space localisation of the finite box function.
\item[{\bf 2.}]
The coefficient $c^{ij}_{m_1 m_2}$ also
appears in the $\cN \! = \! 1$ amplitude as the coefficient
of the triangles
(see \emph{e.g.}~Eq.~(2.19) of \cite{bbst}).
Its twistor space interpretation was considered in
Section 4 of \cite{csw2}, where it was found that
$c^{i j}_{m_1 m_2}$ has support on two lines in twistor space.
Furthermore, it was also found that the corresponding term
in the amplitude has a derivative of a delta function support
on coplanar configurations.
\item[{\bf 3.}]
The combination
$c^{ij}_{m_1 m_2}\,
(i\, j\, m_1\,  Q) \,
(i\, j\,  Q \, m_1) /  (i \cdot j )^2$
already appears in the case of adjacent negative-helicity gluons.
The localisation properties of the corresponding term in the
amplitude were considered in Section 5.3 of \cite{csw2}
and found to have, similarly to the previous case,
derivative of a delta function support
on coplanar configurations.
\item[{\bf 4.}]
On general grounds, we can argue that the remaining terms
in the amplitude have a twistor space interpretation
which is similar to that of the terms already considered.
The gluons whose momenta sum to  $P$
are contained on a line; likewise,   the gluons
whose momenta sum to $Q$ localise on another line.
\end{itemize}

We observe that the rational parts of the amplitude are not
generated from the MHV diagram construction
presented here.
Such rational terms were not present
for the $\cN=1$ and $\cN=4$ amplitudes
derived in \cite{bst,bbst,quig}.
However, for the amplitude studied here,
rational terms are required to ensure the correct factorisation
properties \cite{Bern:1994cg}. These terms have recently been computed using an
on-shell unitarity bootstrap \cite{Berger:2006vq} which makes use of the
cut-constructible part \eq{final} (or \eq{final2}) as input.


\section{Checks of the general result}
\label{checksection}

In this section we present three consistency checks that we have
performed for the result \eqref{final}
(or \eqref{final2}) for the one-loop scalar contribution
to the MHV scattering amplitude. These checks are:
\begin{itemize}
\item[{ \bf 1.}]
For adjacent negative-helicity gluons,
the general expression \eqref{final} should reproduce the
previously calculated form \eqref{endresult}.
\item[{ \bf 2.}]
In the case of five gluons in the configuration $(1^-2^+3^-4^+5^+)$,
the  result \eqref{final}
should reproduce the known amplitude given in \cite{bdk9302280}.
\item[{\bf 3.}]
The result \eqref{final} should have
the correct infrared-pole structure.
\end{itemize}
We next discuss these requirements in turn.

\subsection{Adjacent case}
The amplitude where the
two negative-helicity external gluons are adjacent
is given in Section 7 of \cite{Bern:1994cg}
and was explicitly rederived in \S \ref{adjacentsection} of this thesis
by combining MHV vertices, see
Eq.~\eqref{endresult}. It is easy to show
that our general result \eqref{final2} reproduces
\eqref{endresult} correctly as a special case.

To start with, recall that our result \eqref{final2}
is expressed in
terms of box functions and triangle functions, see
Figure 3.3  and Figures 3.4, 3.5 respectively.
In the adjacent case,
the box functions are not present.
Indeed, in the sum \eqref{finbox} the negative-helicity gluons
can never be in adjacent positions (see Figure 3.3).

Next, we focus on the triangles of Figure 3.4.
In terms of these triangles, requiring
$i$ and $j$ to be adjacent eliminates the sum over $m_2$,
as we must have $m_2=i$ and $m_2 + 1 = j$.
Moreover, in this case $Q= q_{m_1 + 1, i}$, $P = q_{j, m_1 -1}$
and one has
\beqa
A_{ij}^{m_1 m_2} & = & -4 \, (i \cdot j)\ ,
\nonumber \\ \cr
S_{ij}^{m_1 m_2} & = & 0 \ , \qquad
I_{ij}^{m_1 m_2} \ = \ 0 \ ,
\eeqa
(for $m_2=i$, and $m_2 + 1 = j$).
Similar simplifications occur for the swapped triangle.
Hence the only surviving terms
are those in the second line of \eqref{final}
(or \eqref{final2}), and it is then easy to see that
they generate the same amplitude
\eqref{atlast2} already calculated in
\S \ref{adjacentsection}.

\subsection{Five-gluon amplitude}
The other special case is the non-adjacent
five-gluon amplitude $(1^-2^+3^-4^+5^+)$, given in
Equation~(9) of \cite{bdk9302280}.
This amplitude may be written as\footnote{$c_{\Gamma}=r_{\Gamma}/(4\pi)^{2-\epsilon}$ is given in terms
of Eq.~\eq{rgamma}.}
$c_{\Gamma} \cA_{\rm tree}$ times%
\footnote{The derivation in \cite{bdk9302280} used string-based
methods which affect the coefficient of the pole term.
In Eq.~\eqref{BDK}
we have written the pole coefficient
which matches the adjacent case.}
\begin{eqnarray}
 \label{BDK}\nonumber \\  \nonumber
   \frac{1}{6\epsilon} &-& \frac{1}{6} \log(-s_{34})
   \ + \ \frac{\tofi^2\ \fito^2}{2^7(2\cdot 5)^4(1\cdot 3)^4}\
    B(s_{51}, s_{12}, 0, s_{34})
      \\ [8pt] \nonumber \cr
&-&\ \frac{1}{3}\ \frac{\tofi\ \fito}{2^4(2\cdot 5)(1\cdot 3)^4}
\ \left[
  \ \fito^2 \ \frac{\log(s_{12}/s_{34})}{(s_{12}-s_{34})^3}\right.\nonumber\\
&&\qquad\qquad\qquad\qquad\qquad\quad\ \ \,
+\left.\tofi^2\ \frac{\log(s_{34}/s_{51})}{(s_{34}-s_{51})^3}\ \right]\nonumber
     \\ [8pt] \nonumber \cr
&+&    \ \frac{1}{3}\ \frac{1}{2^3(1\cdot 3)^3} \left[\ \foto\tofi^2
   \ \frac{\log(s_{34}/s_{51})}{(s_{34}-s_{51})^3}  \  \right]
    \\ [8pt] \nonumber \cr
&-&  \ \frac{\tofi^2\ \fito^2}{2^6(2\cdot 5)^2(1\cdot 3)^4} \left[\
\  \frac{\log(s_{12}/s_{34})}{(s_{12}-s_{34})^2}
    - \frac{\log(s_{34}/s_{51})}{(s_{34}-s_{51})^2}\ \right]
\\ [8pt]\nonumber \cr
&+&   \ \frac{\tofi^2\ \fito^2}{2^6(2\cdot 5)^3(1\cdot 3)^4} \left[\
   \frac{\log(s_{12}/s_{34})}{(s_{12}-s_{34})}
    + \frac{\log(s_{34}/s_{51})}{(s_{34}-s_{51})}\ \right]
 \\ [8pt]\nonumber \cr
&-&     \ \frac{1}{3} \ \frac{1}{2^2(1\cdot 3)} \left[\ \tofi \
    \frac{\log(s_{34}/s_{51})}{(s_{34}-s_{51})}  \  \right]
  \\ [8pt] \cr
&+&     (1,4) \leftrightarrow (3,5)
\ \ ,
\end{eqnarray}
where the interchange on the last line applies
to all terms above it in this equation,
including the first two terms.
The box function $B$ is defined in \eqref{boxf}.
In deriving \eq{BDK} from  \cite{bdk9302280},
 we have used the dilogarithm identity
\begin{displaymath}
\begin{array}{ccc}
\Li2(1\!-r) + \Li2(1\!-s)+ \log(r)\log(s) & = &
\Li2\left(\frac{1\!-r}{s}\right) +
\Li2\left(\frac{1\!-s}{r}\right)-
\Li2\left(\frac{1\!-s}{r}\frac{1\!-r}{s}\right)\ .
\end{array}
\end{displaymath}
%
%
We have checked explicitly that our expression for the
$n$-gluon non-adjacent amplitude \eqref{final},
when specialised to the case with five gluons
in the configuration $(1^-2^+3^-4^+5^+)$,
yields precisely the result
\eqref{BDK} above.
For the terms involving dilogarithms, this is
easily done. For the remaining terms,
which contain logarithms, a more involved calculation
is necessary using various spinor identities
from Appendix \ref{traceappendix}.  A straightforward method of
doing this calculation begins
with the explicit sum over MHV diagrams in this case,
isolating the coefficients of each logarithmic
function such as \emph{e.g.}~$\log(s_{12})$,
and then checking that these coefficients
match those in \eqref{BDK}.
The remaining $1/\epsilon$ term arises
from the following discussion.


\subsection{Infrared-pole structure}

The infrared-divergent terms (poles in $ 1 / \e$)
can easily be extracted  from \eqref{ir1}--\eqref{ir4}
by simply replacing $ ( - t_r^{[2]})^{- \e} \to 1$
($r = i-1,  i,  j-1,  j$).
Consider first the terms in \eqref{ir2} and \eqref{ir3}.
After a little algebra, and using
\beq
(i \, j \, j+1 \, j-1) \, + \, (i \, j \, j-1 \, i-1)
\ = \
4\,  (i \cdot j) \, \big( (j-1) \cdot (j +1) \big)
\ ,
\eeq
one finds that these two contributions
add up to
\beq
- {64 \over 3\, \e} \ (i \cdot j)^4
\ .
\eeq
Similarly, the pole contribution arising from
\eqref{ir1} and \eqref{ir4} gives an additional contribution
of  $ - ({64 /  3\, \e})\  (i \cdot j)^4$.
Reinstating a factor  of
$-2 \cdot 2^{-8} (i\cdot j)^{-4} \cdot  \cA_{\rm tree}$, we see that
the pole part of \eqref{final} is simply given by
\beq
\left. \cA_{\rm scalar}\right|_{\rm \e-pole}
 \ = \ {\cA_{\rm tree} \over 3}
\ .
\eeq
Hence our result \eqref{final} has the expected
infrared-singular behaviour.

\section{The MHV amplitudes in QCD}

We conclude by mentioning that the full one-loop $n$-gluon MHV amplitudes (with arbitrary positions for the
negative-helicity gluons)
in QCD can now be constructed. These are given by:
\be
\label{loopsusyagain}
\cA^{\textrm{MHV}}_{\textrm{QCD}}=\cA^{\textrm{MHV}}_{\cN=4}
-4\cA^{\textrm{MHV}}_{\cN=1,\textrm{chiral}}+\cA^{\textrm{MHV}}_{\textrm{scalar}}\ ,
\ee
where in contradistinction with \eq{loopsusy} we have written the scalar contribution in terms
of a complex scalar rather than a real scalar. The individual pieces (to finite order in
$\epsilon$) can be found as follows:
\begin{itemize}
\item $\cA^{\textrm{MHV}}_{\cN=4}$ was first computed in \cite{Bern:zx} and can be found there
as Equation~(4.1). Alternatively it is given as Equation \eq{fullampl} in Chapter 1 of this thesis.
Note that an alternative form to Eq.~\eq{2meold} for the 2me box functions is given by Eq.~\eq{2menew}.

\item $\cA^{\textrm{MHV}}_{\cN=1,\textrm{chiral}}$ was first computed in \cite{Bern:1994cg} and can
be found there as Equation~(5.12) or more compactly as Equation~\eq{Neq1} in Chapter 2 of this thesis.

\item In contrast to the $\cN\!=\!4$ and $\cN\!=\!1$ cases, $\cA^{\textrm{MHV}}_{\textrm{scalar}}$
is an amplitude in a non-supersymmetric theory and as such its cuts are not uniquely determined
by its cut-constructible part (${\mathcal{A}}^{\textrm{MHV}}_{\textrm{s-cut}}$). $\cA^{\textrm{MHV}}_{\textrm{s-cut}}$
was first computed in
\cite{bbst2} and can be found there as Equation~(4.20) or Equation~(4.22). Alternatively it can be found
earlier in this chapter as Equation \eq{final} or Equation \eq{final2}.

\item Building on the results of \cite{bbst2}, the rational part of $\cA^{\textrm{MHV}}_{\textrm{scalar}}$
($\cA^{\textrm{MHV}}_{\textrm{s-rational}}$) was computed in \cite{Berger:2006vq}. In doing
this it was found that it is useful to `complete' the cut-constructible parts obtained in \cite{bbst2}
by introducing certain preliminary rational terms in order
to remove spurious singularities. The cut-completion of $\cA^{\textrm{MHV}}_{\textrm{s-cut}}$
is given by Equation (A1) of Appendix A of \cite{Berger:2006vq} and the full amplitude is then obtained
by adding the remaining rational terms as given in Equation (5.30) of that paper. Explicitly, the full scalar amplitude is
given by Equation (5.1) (for negative-helicity gluons 1 and $m$), where $\hat{C}$ is given by (A1) and
$\hat{R}$ by Equation (5.30) of \cite{Berger:2006vq}.

\item $\cA^{\textrm{MHV}}_{\textrm{QCD}}$ can then be found using the decomposition \eq{loopsusyagain}
and
\begin{eqnarray*}
\cA^{\textrm{MHV}}_{\cN=4} & = & \textrm{Eq.~(4.1) of \cite{Bern:zx}}\\
\cA^{\textrm{MHV}}_{\cN=1,\textrm{chiral}} & = & \textrm{Eq.~(5.12) of \cite{Bern:1994cg}}\\
\cA^{\textrm{MHV}}_{\textrm{scalar}} & = & \textrm{Eq.~(5.1) of \cite{Berger:2006vq}}\ .
\end{eqnarray*}
\end{itemize}

\chapter{Recursion relations in gravity}
\label{chapter4}

The proposal of a twistor string dual to perturbative Yang-Mills in \cite{witten}
led not-only to the advances described in Chapters 1-3 of the so-called
MHV rules for perturbation theory, but to many others as well. The
support of many quantities such as scattering amplitudes, their integral functions
and the coefficients of these functions in twistor space has led to many insights
\cite{witten,csw2,csw3,Bena:2004ry,bdk04,BBDD,new,bbst2,freddy2,freddy3,Britto,Bern:2005bb,benabern,
Bidder:2005ri,Bidder:2004vx} as
has the use of signature $++--$ (or equivalently the restriction of momenta to be complex
rather than real). In particular, this second technique of using complex momenta has proved very
powerful, leading to the idea of generalised unitarity \cite{Britto,Brandhuber:2005jw} and then
to the tree-level on-shell recursion relations \cite{bcf,bcfw} which will be central to this chapter.

Recursion relations have been known for some time in field theory since Berends and Giele
proposed them in terms of off-shell currents \cite{bg}. However, the gluon recursion relations
introduced by Britto, Cachazo and Feng in \cite{bcf} (stemming from observations in
\cite{rsv04}) and then proved in \cite{bcfw} are in some ways much more powerful. They apply
directly to on-shell scattering amplitudes and are particularly apt when the amplitudes are
written in the spinor helicity formalism, which as we have seen in the preceding chapters
is a formalism which tends to favour simple and compact expressions.

The proof of the on-shell recursion relation for gluons presented in \cite{bcfw} is very simple, only relying
(essentially) on the ability to express an amplitude as a function of a complex variable $z$ and
then the asymptotic behaviour of this function as $z\rightarrow\infty$. As such, it is natural to
ask whether such a recursive structure might persist in other field theories and even in gravity.\footnote{Here
we mean gravity as a field theory (rather than as a string theory).} This
question was answered independently in \cite{bbst3} and \cite{Cachazo:2005ca} in the affirmative, where
the authors of \cite{bbst3} (including the present author) used it to present a new compact formula
for $n$-graviton MHV amplitudes at tree-level in general relativity (GR). Such compact formul{\ae} are particularly
interesting as gravity is very-much more complicated than Yang-Mills - the 3-point vertex of GR for example
contains 171 terms in total, while the 4-point vertex has 2850 altogether \cite{DeWitt:1967uc}.

In this chapter we will follow \cite{bbst3} and describe the recursion relation in Einstein gravity at
tree-level. We will not summarise the proof of the relation in Yang-Mills as it is almost identical
to that in gravity. Any differences between the two are pointed out in what follows.


\section{The recursion relation}
In this section we closely follow the proof of the recursion
relation in Yang-Mills \cite{bcfw}, which we will extend to the
case of gravity amplitudes. As we shall see, the main new
ingredient is that gravity amplitudes depend on more kinematical
invariants than the corresponding Yang-Mills amplitudes, namely
those which are sums of non-cyclically adjacent momenta; hence,
more multi-particle channels should be considered.

To derive a recursion relation for scattering amplitudes, we start
by introducing a one-parameter family of scattering amplitudes,
$\cM(z)$ \cite{bcfw}, where we choose $z$ in such a way that
$\cM(0)$ is the amplitude we wish to compute. We  work in
complexified Minkowski space and regard $\cM(z)$ as a complex
function of $z$ and the momenta. One can then consider the contour
integral \cite{Bern:2005hs} \beq \label{int} \cC_{\infty} \ := \
{1\over 2\pi i} \oint\! dz \  { \cM(z) \over z } \ , \eeq where
the integration is taken around the circle at infinity in the
complex $z$ plane. Assuming that $\cM (z)$ has only simple poles
at $z=z_i$, the integration gives \beq \cC_{\infty} \ = \ \cM(0)
\, + \, \sum_i { [{\rm Res} \, \cM(z)]_{z=z_i} \over z_i} \ . \eeq
In the important case of Yang-Mills amplitudes, $\cM(z) \to 0$ as
$z \to \infty$, and hence $\cC_{\infty}=0$ \cite{bcfw}.

Notice that up to this point the definition of the family of
amplitudes  $\cM(z)$ has not been given -- we have not even
specified the theory whose scattering amplitudes we are computing.

There are some obvious requirements for $\cM(z)$. The main point
is to define $\cM(z)$ in such a way that poles in $z$ correspond
to multi-particle poles in the scattering amplitude  $\cM(0)$. If
this occurs then the corresponding residues
can be computed from factorisation properties of scattering
amplitudes (see, for example, \cite{dixon,wein}). In order to
accomplish this, $\cM(z)$ was defined in \cite{bcf,bcfw} by
shifting the momenta of two of the external particles in the
original scattering amplitude. For this procedure to make sense,
we have to make sure that even with these shifts overall momentum
conservation is preserved and that all particle momenta remain
on-shell. We are thus led to define $\cM(z)$ as the scattering
amplitude $\cM (p_1 , \ldots , p_k (z), \ldots , p_l(z), \ldots ,
p_n)$, where the momenta of particles $k$ and $l$ are shifted to
\beq \label{shifts} p_k (z) \ := \  p_k + z \eta \ ,  \qquad
p_l(z)\  := \ p_l - z \eta \ . \eeq Momentum conservation is then
maintained. As in \cite{bcf}, we can solve $p_k^2 (z) = p_l^2 (z)
= 0$ by choosing $\eta = \l_l \lt_k$ (or  $\eta = \l_k \lt_l$),
which makes sense in complexified Minkowski space. Equivalently,
\beq \label{shiftsspin} \l_k (z) \ := \  \l_k + z \l_l \ ,  \qquad
\lt_l (z)\  := \  \lt_l - z \lt_k \ , \eeq with $\l_l$ and $\lt_k$
unshifted.

More general families of scattering amplitudes can also be
defined, as pointed out in \cite{Bern:2005hs}. For instance, one
can single out three particles $k$, $l$, $m$, and define \beq
\label{shiftsgen} p_k (z) \ := \  p_k +  z \eta_k \ ,  \qquad
p_l(z)\  := \ p_l + z \eta_l \ ,  \qquad p_m(z)\  := \ p_m + z
\eta_m \ , \eeq where $\eta_k$, $\eta _l$ and $\eta_m$ are null
and $\eta_k + \eta_l + \eta_m =0$. Imposing $p_k^2(z)=
p_l^2(z)=p_m^2(z)=0$, one finds the solution \beq \label{genshift}
\eta_k = - \a \l_k \lt_l - \b \l_k \lt_m \ , \qquad \eta_l = \a
\l_k \lt_l \ , \qquad \eta_m = \b \l_k \lt_m \ , \eeq for
arbitrary $\a$ and $\b$. This has been used in \cite{Bern:2005hs}.
In the following we will limit ourselves to shifting only two
momenta as in \cite{bcf} and \cite{bcfw}.

At tree level, scattering amplitudes in field theory can only have
simple poles in multi-particle channels; for $\cM (z)$, these
generate poles in $z$  (unless the channel contains both particles
$k$ and $l$, or none). Indeed, if $P(z)$ is a sum of momenta
including $p_l(z)$ but not $p_k(z)$, then  $P^2 (z) = P^2 - 2 z (P
\cdot \eta)$ vanishes at $z_P = P^2 / 2  (P\cdot \eta)$
\cite{bcfw}. In Yang-Mills theory, one considers colour-ordered
partial amplitudes which have a fixed cyclic ordering of the
external legs. This implies that a generic Yang-Mills partial
amplitude can only depend on kinematical invariants made of sums
of cyclically adjacent momenta. Hence, tree-level Yang-Mills
amplitudes can only have poles  in kinematical channels made of
cyclically adjacent sums of momenta.

For gravity amplitudes this is not the case as there is no such
notion of ordering for the external legs. Therefore, the
multi-particle poles which produce poles in $z$ are those obtained
by forming all possible combinations of momenta which include
$p_k(z)$  but not $p_l(z)$. This is the only modification to the
BCFW recursion relation we need to make in order to derive a
gravity recursion relation.

For any such multi-particle channel  $P^2 (z)$, we have \beq \cM
(z) \to \sum_{h}\, {\cM_L^{h} (z_P)} \, {1 \over P^2 (z)} \,
{\cM_R^{-h} (z_P)}
 \ ,
\eeq as $P^2 (z) \to 0$ (or, equivalently, $z \to z_P$). The sum
is over the possible helicity assignments on the two sides of the
propagator which connects the two lower-point tree-level
amplitudes $\cM_L^{h}$ and $\cM_R^{-h}$. It follows that \beq
[{\rm Res} \, \cM (z) ]_{z=z_P} \ = \
 - \sum_{h}\,
\cM_L^{h} (z_P) \, {z_P\over P^2} \, \cM_R^{-h} (z_P) \ , \eeq so
that finally \beq \cM(0)  \ = \ \cC_{\infty}\, + \, \sum_{P, h} {
\cM_L^{h} (z_P)  \, \cM_R^{-h} (z_P) \over P^2} \ . \eeq The sum
is over all possible decompositions of momenta such that $p_k \in
P$ but $p_l \notin P$.

If $\cC_{\infty} = 0$ there is no boundary term in the
recursion relation and \beq \label{rec} \cM(0)  \ = \ \sum_{P, h}
{ \cM_L^{h} (z_P) \,  \cM_R^{-h} (z_P) \over P^2} \ . \eeq In
\cite{bcfw} it was shown that for Yang-Mills amplitudes the boundary
terms $\cC_{\infty}^{\rm YM}$ always vanish. Two different proofs
were presented, the first based on the use of CSW diagrams
\cite{csw} and the second on Feynman diagrams. An MHV-vertex formulation of
gravity only recently appeared \cite{Bjerrum-Bohr:2005jr}, so at the time
the authors of \cite{bbst3} could only rely
on Feynman diagrams. This is also the case for other field
theories we might be interested in (such as $\lambda \phi^4$, for
example).

As we have remarked, $\cC_{\infty} = 0$ if  $\cM (z)
\to 0$ as $z \to \infty$. $\cM (z)$ is a scattering amplitude with
shifted, $z$-dependent external null momenta. One can then try to
estimate the behaviour of $\cM (z)$  for large  $z$ by using power
counting (different theories will of course give different
results). In $\l \phi^4$ the Feynman vertices are momentum
independent and  $\cC_{\infty} = 0$ (see \S \ref{otherappssection}); in
quantum gravity, however, vertices are quadratic in momenta, and
one cannot determine \emph{a priori} whether or not a boundary term is
present.

From the previous discussion, it follows that the behaviour of
$\cM (z)$ as $z \to \infty$ is related to the high-energy
behaviour of the scattering amplitude (and hence to the
renormalisability of the theory). The ultraviolet behaviour of
quantum gravity, however, is full of surprises (for a summary, see
for example Section 2.2 of \cite{Bern:2002kj} and also more recent results of
\cite{Bjerrum-Bohr:2006sg,Bjerrum-Bohr:2006yw,Bern:2006kd,Bern:2007hh,
Green:2006gt,Green:2006yu}). We may therefore
expect a more benign behaviour of $\cM (z)$ as $z \to \infty$.
Specifically, in the next section we will focus on the $n$-graviton MHV
scattering amplitudes which have been computed
by Berends, Giele and Kuijf (BGK) in \cite{bgk}. Performing the
shifts \eqref{shifts} explicitly in the BGK formula,
one finds the surprising result%
\footnote{We have checked that $\cM(z) \sim \cO (1 / z^2 )$ as $z
\to \infty$, analytically for $n\leq 7$ legs and numerically for
$n\leq 11$ legs.} \beq \lim_{z \to \infty} \cM_{\rm MHV}(z) \ = \
0 \ . \eeq

In more general amplitudes one can (at least in
principle) use the (field theory limit of the) KLT relations
\cite{klt}, which connect tree-level gravity amplitudes to
tree-level amplitudes in Yang-Mills, to estimate
the large-$z$ behaviour of the scattering amplitude.%
\footnote{See Appendix \ref{kltappendix} for explicit examples of KLT relations
for four, five and six legs.} As an example, we have considered
the next-to-MHV gravity amplitude $\cM (1^- , 2^- , 3^- , 4^+ ,
5^+ , 6^+)$, and performed the shifts as in \eqref{shiftsspin},
with $k=1$ and $l=2$. Similarly to the MHV case, we find that \beq
\lim_{z \to \infty} \cM (1^- , 2^- , 3^- , 4^+ , 5^+ , 6^+) (z)  \
= \ 0 \ . \eeq

In \cite{Cachazo:2005ca} it was shown that $\cM(z)$ vanishes as
$z\rightarrow\infty$ for all amplitudes up to eight gravitons and also
for all $n$-point MHV and NMHV amplitudes. Further to this, recent work
\cite{Benincasa:2007qj} provides a proof of this statement for \emph{all} tree-level $n$-graviton
 amplitudes thus establishing the validity of the recursion relation
in gravity unambiguously.

In the next section we will apply the recursion relation
\eqref{rec} to the case of MHV amplitudes in  gravity and show
that it does generate correct expressions for the amplitudes. As a
bonus, we will derive a new closed-form expression for the
$n$-particle scattering amplitude.

\section{Application to MHV gravity amplitudes }
In the following we will compute the MHV scattering amplitude
$\cM(1^- , 2^- , 3^+ , \ldots , n^+)$ for  $n$ gravitons. We will
choose the two negative-helicity gravitons $1^-$ and $ 2^-$ as
reference legs. This is a particularly convenient choice as it
reduces the number of terms arising in the recursion relation to a
minimum. The shifts for the momenta of particles 1 and 2 are \beq
p_1 \to p_1 +  z \l_2 \lt_1 \ , \qquad p_2 \to p_2 -  z \l_2 \lt_1
\  . \eeq In terms of spinors, the shifts are realised as \beq
\label{shiftsonceagain} \l_1 \to \hat{\l}_1 := \l_1 +  z \l_2 \ ,
\qquad \lt_2 \to \hat{\lt}_2 := \lt_2 - z \lt_1 \  , \eeq with
$\l_2$ and $\lt_1$ unmodified.

\begin{figure}[ht]
\label{figure-vv}
\begin{center}
\scalebox{0.65}{\includegraphics{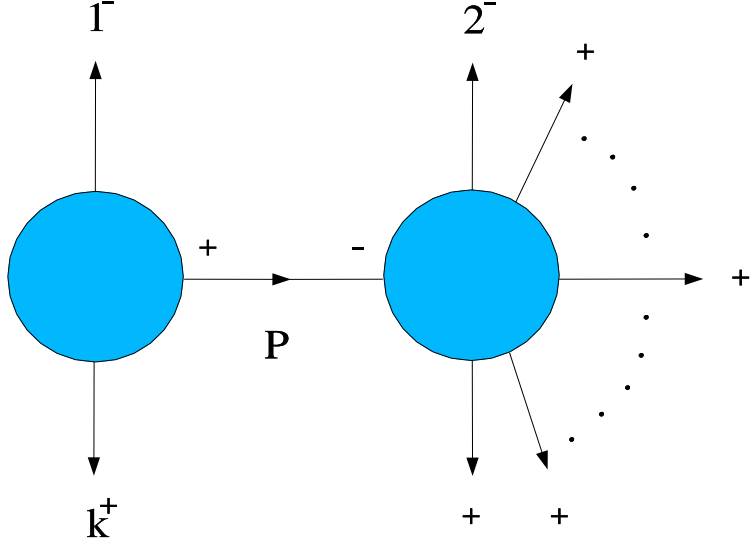}}
\end{center}
\caption{\it One of the  terms contributing to the recursion
relation for the MHV  amplitude $\cM  (1^- , 2^- , 3^+ , \ldots ,
n^+)$. The gravity scattering amplitude on the right is symmetric
under the exchange of gravitons of the same helicity. In the
recursion relation, we sum over all possible values of $k$,
i.e.~$k=3, \ldots , n$. This amounts to summing over cyclical
permutations of $(3, \ldots , n)$. }
\end{figure}

\begin{figure}[ht]
\label{figure-vv2}
\begin{center}
\scalebox{0.65}{\includegraphics{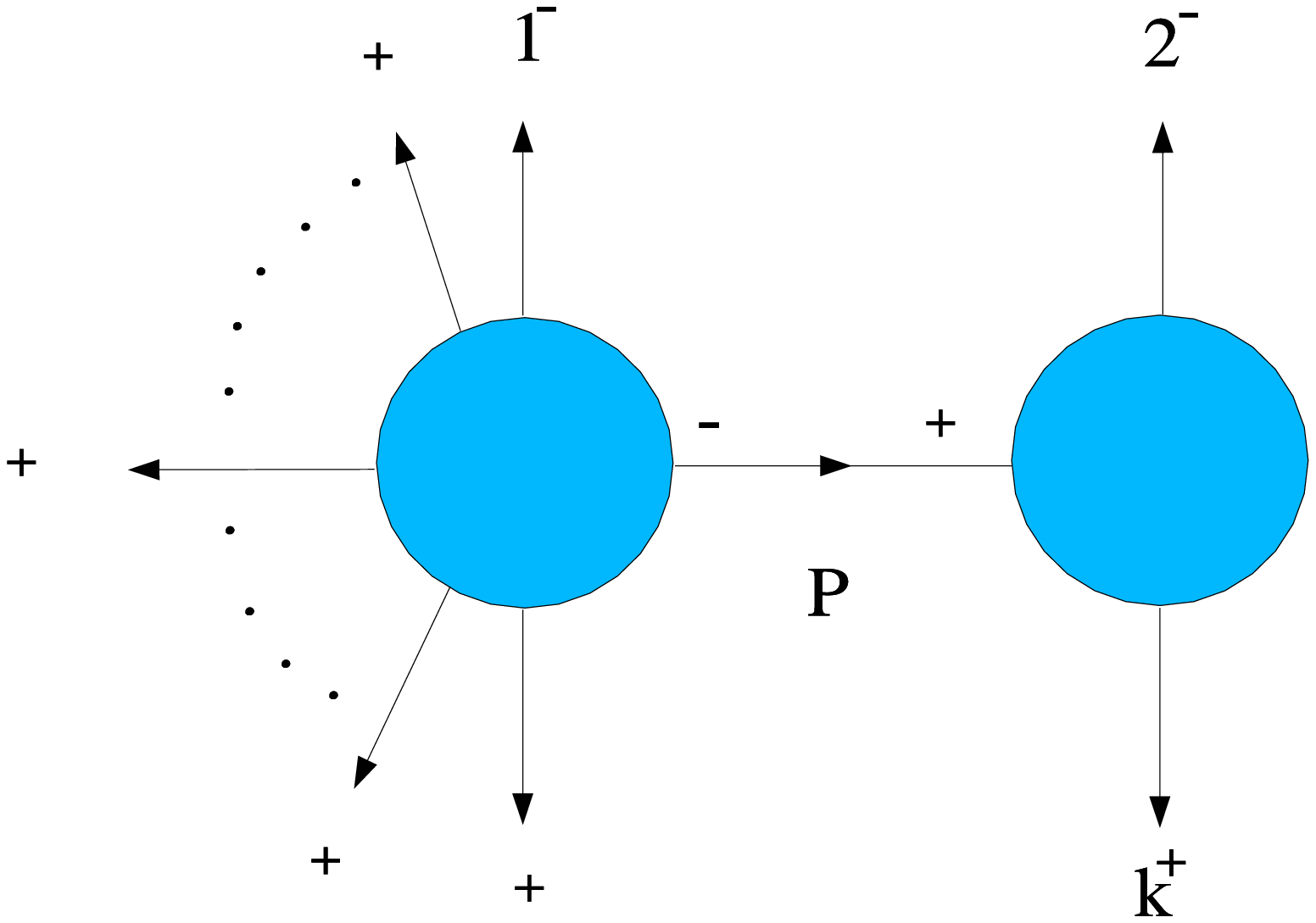}}
\end{center}
\caption{\it This class of diagrams also contributes to the
recursion relation for the MHV amplitude $\cM  (1^- , 2^- , 3^+ ,
\ldots , n^+)$; however, each of these diagrams vanishes if the
shifts \eqref{shiftsonceagain} are performed. }
\end{figure}

Let us consider the possible recursion diagrams that can arise.
There are only two possibilities, corresponding to the two
possible internal helicity assignments, $(+-)$ and $(-+)$:
\begin{itemize}
\item[{\bf 1.}] The amplitude on the left is googly $(++-)$
whereas on the right there is an MHV gravity amplitude with $n-1$
legs (see Figure 4.1). \item[{\bf 2.}] The amplitude on the right is
googly and the amplitude on the left is MHV (see Figure 4.2).
\end{itemize}
We recall that a gravity amplitude is symmetric under the
interchange of identical helicity gravitons; this implies that we
have to sum $n-2$ diagrams for each of the configurations in
Figures 4.1 and 4.2. Each diagram is then completely specified by
choosing $k$, with $k=3, \ldots, n$.

However, it is easy to see that diagrams of type {\bf 2}
actually give a vanishing contribution. Indeed, they are
proportional to \beq [k\, \hat{P} ] \, = \, { [ k | \hat{P} |
\hat{2} \ran \over \lan \hat{P} \, \hat{2} \ran} \, = \, { [ k | P
| 2 \ran \over \lan \hat{P} \, \hat{2} \ran} \, = \, 0 \ , \eeq
where the last equality follows from $P= p_k + p_2$. Hence we will
have to compute diagrams of type {\bf 1} only. We will do this in
the following.

\subsection{Four-, five- and six-graviton scattering}
To show explicitly how our recursion relation generates
amplitudes we will now derive the 4-, 5- and 6-point MHV
scattering amplitudes.

We start  with the four point case. There are two diagrams to sum,
one of which is represented in Figure 4.3; the other is obtained by
swapping the labels $4$ and $3$.
\begin{figure}[ht]
\label{figure-vvv}
\begin{center}
\scalebox{0.65}{\includegraphics{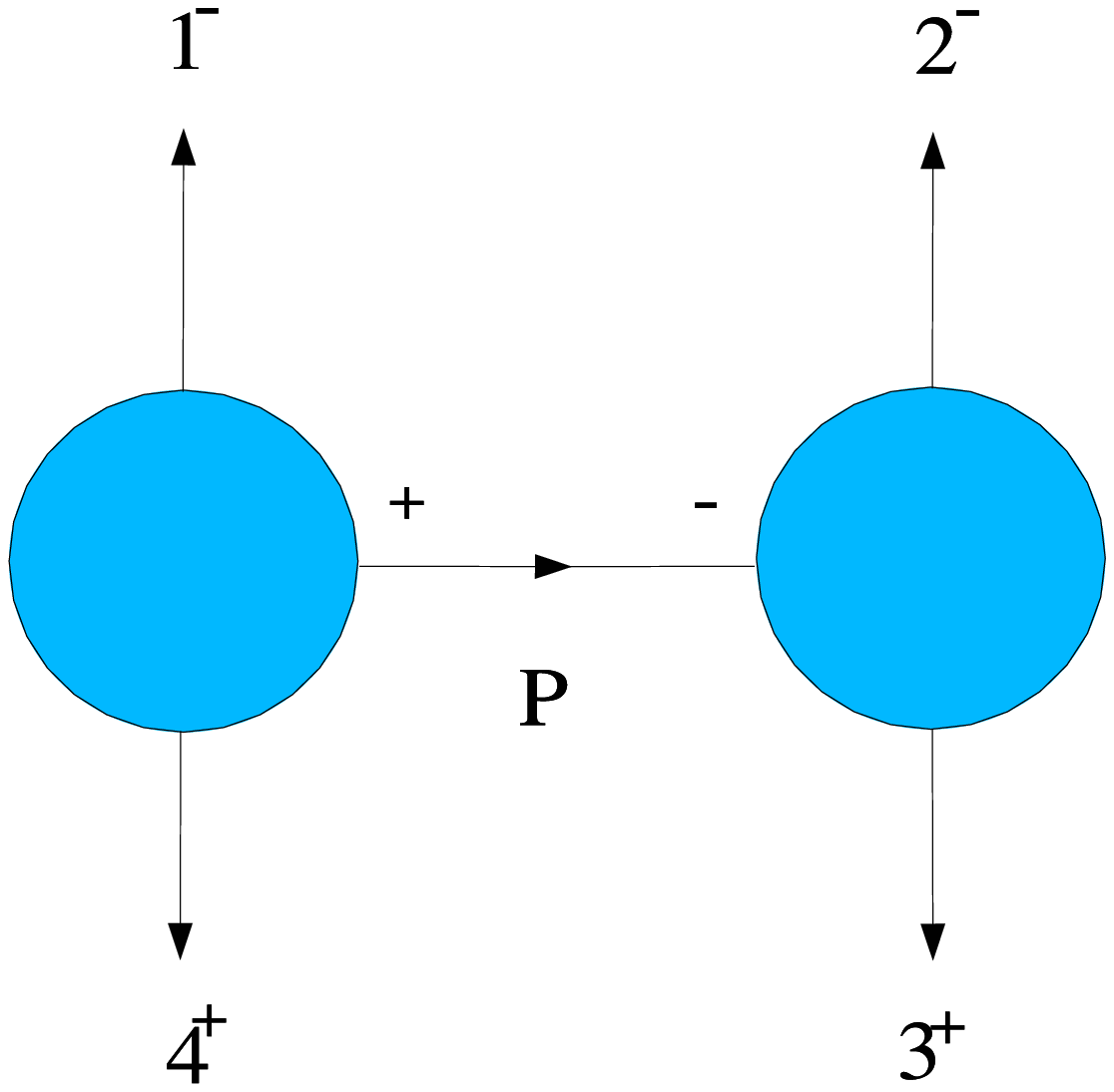}}
\end{center}
\caption{\it One of the two diagrams contributing to the recursion
relation for the MHV  amplitude $\cM  (1^- , 2^- , 3^+ , 4^+)$.
The other is obtained from this by cyclically permuting the labels
$(3,4)$ -- i.e.~swapping $3$ with $4$.}
\end{figure}
For the diagram in Figure 4.3, we have \beq \cM^{(4)} \ = \ \cM_{L}
\, {1 \over P^2} \, \cM_R \ , \eeq where the superscript denotes
the label on the positive-helicity leg in the trivalent googly MHV vertex,
\beqa \cM_L & = & \left( { [ \hat{P} \, 4]^3 \over [ 4 \, 1] [ 1
\, \hat{P} ] } \right)^2 \ ,
\\ \nonumber \cr
\cM_R & = & \left( { \lan  \hat{P} \, 2\ran^3 \over \lan 2 \,
3\ran \lan 3 \, \hat{P} \ran } \right)^2 \ , \eeqa and $P^2 = (p_1
+ p_4)^2$. Using \beq \lan i \, \hat{P} \ran \, = \, {\lan i | P |
1] \over [ \hat{P} \, 1 ] } \ , \eeq we find, after a little
algebra, \beq \cM^{(4)} \ = \ \frac {\lan 1 2 \ran^6 [14] } {\lan
14\ran \lan 23\ran^2 \lan 34\ran^2} \ . \eeq The full amplitude is
$\cM(1^-,2^-,3^+,4^+) = \cM^{(3)} +
 \cM^{(4)}$.
Thus, we conclude that the four point MHV amplitude generated by
our recursion relation is given by
\beq \cM(1^-,2^-,3^+,4^+)\  = \ \frac {\lan 1 2 \ran^6 [14] }
{\lan 14\ran \lan 23\ran^2 \lan 34\ran^2} \, + \, 3
\leftrightarrow 4 \ . \eeq
It is easy to check that this agrees with the conventional formula
for this amplitude
\beq \cM(1^-,2^-,3^+,4^+) \ = \ \frac {\lan 1 2 \ran^8 [12] }
{N(4) \lan 34 \ran} \ , \eeq
where
\beq N(n) \ := \ \prod_{1\leq i < j \leq n} \lan i\, j\ran \ ,
\eeq or, equivalently, with the expression from the
appropriate KLT relation, Eq.~\eqref{4}.

For the five-graviton scattering case our recursion relation
yields a sum of three diagrams. A calculation similar to that
illustrated previously for the four-point case leads to the result
\beq \label{5point} \cM(1^-,2^-,3^+,4^+,5^+) \ = \ \frac {\lan
12\ran^6 [15][34] } {\lan 15\ran \lan 23\ran \lan 24\ran\lan
34\ran\lan 35\ran\lan 45\ran} \, +\, \cP^{\rm c} (3,4,5) \ , \eeq
where $\cP^{\rm c} (3,4,5)$ means that we have to sum over cyclic
permutations of the labels $3,4,5$. The conventional formula for
the five graviton MHV scattering amplitude is
\beq \cM(1^-,2^-,3^+,4^+ , 5^+) = \frac {\lan 1 2 \ran^8  } {N(5)}
\Big( [12][34] \lan 13\ran\lan 24 \ran-[13][24] \lan 12 \ran\lan
34 \ran\Big) \ . \eeq
Using standard spinor identities and momentum conservation, it is
straightforward to check that our expression \eqref{5point} agrees
with this (alternatively, one can use the KLT relation \eqref{5}).

For the six graviton scattering amplitude, our recursion relation
yields a sum of four terms,
\beqa \label{6point} \quad\cM(1^-,2^-,3^+,4^+,5^+,6^+) & = & \frac
{\lan 12\ran^6 [16] }{\lan 16\ran} \cdot {1 \over \lan 2 \, 6\ran
\lan 3 \, 4\ran \lan 3 \, 5\ran \lan 4 \, 5\ran }
\\ \nonumber
&& \hspace{-3.5cm} \cdot\ \bigg( {[3\, 4] \over \lan 2 \, 3\ran  \lan 2
\, 4\ran} \, \frac{\lan 2 | 3+4 |5]}{\lan 56\ran }
 +
{[4\, 5] \over \lan 2 \, 4\ran  \lan 2 \, 5\ran}
 \,
\frac{\lan 2 | 4+ 5 |3]}{\lan 36\ran }
 +
{[5\, 3] \over \lan 2 \, 3\ran  \lan 2 \, 5\ran}
 \,
\frac{\lan 2 | 5 + 3 |4]}{\lan 46\ran } \bigg)
\nonumber \\
&+&  \ \cP^{\rm c} (3,4,5,6) \, . \nonumber
\eeqa
The known formula for this amplitude is
\beq \cM_{\mathrm{MHV}}^{6\textrm{-point}}=\lan 12\ran^8 \bigg(
      \frac { [12][45][3|4+5|6\ran  }{ \lan 15\ran
\lan 16\ran\lan 12\ran\lan 23\ran\lan 26\ran\lan 34\ran\lan 36\ran
\lan 45\ran\lan 46\ran\lan 56\ran}
                     + \cP(2,3,4) \bigg)
\ , \eeq where $\cP (2,3,4)$ indicates permutations of the labels
$2,3,4$.
We have checked  numerically that the  formula \eqref{6point}
agrees with this expression.


\subsection{General formula for MHV scattering}
Recursion relations of the form given in \cite{bcf}, or the
graviton recursion relation given here, naturally produce general
formul{\ae} for scattering amplitudes. For a suitable choice of
reference spinors, these new formul{\ae} can often be simpler than
previously known examples. For the choice of reference spinors
$1,2,$ which we have made above, the graviton recursion relation
is particularly simple as it produces only one term at each step.
This immediately suggests that one can use it to generate an
explicit expression for the $n$-point amplitude. This turns out to
be the case, and experience with the use of our recursion relation
leads us to propose the following new general formula for the
$n$-graviton MHV scattering amplitude. This is (labels $1,2$ carry
negative helicity, the remainder carry positive helicity)
\beqa \label{npoint} \cM(1, 2, {i_1}, \cdots , {i_{n-2}}) &=&
\frac{\lan 1\, 2 \ran^6 [1\,  i_{n-2}]}{\lan 1 \, i_{n-2}\ran}\,
\, G(i_1,i_2,i_3)\, \prod_{s=3}^{n-3} \frac{\lan 2 |
i_1+...+i_{s-1} | i_s]} {\lan i_s i_{s+1}\ran \lan 2 i_{s+1}\ran }
\nonumber\\
&+& \cP(i_1,...,i_{n-2}), \eeqa
where
\beq G(i_1,i_2,i_3) = {1\over 2} \frac{[i_1 i_2]}{\lan
2i_1\ran\lan 2i_2\ran \lan i_1i_2\ran\lan i_2i_3\ran\lan
i_1i_3\ran} \ . \eeq
(For $n=5$ the product term is dropped from \eqref{npoint}). It is
straightforward to check that this amplitude satisfies the
recursion relation with the choice of reference legs $1^-$ and
$2^-$.

The known general MHV amplitude for two negative-helicity
gravitons, 1 and 2, and the remaining $n-2$ with positive
helicity is given by \cite{bgk} \beq \cM (1, 2, 3, \cdots , n) =
\lan 12\ran^8 \Biggl[ {[12] [n-2 ~n-1] \over \lan 1~n-1\ran}
{1\over N(n)} \prod_{i=1}^{n-3} \prod_{j=i+2}^{n-1} \lan ij\ran ~F
\ +\ {\cal P} (2,\ldots ,n-2)\Biggr], \label{15} \eeq where \beq
F= \left\{ \begin{array}{ll} \prod_{l=3}^{n-3}\,  [l| (p_{l+1} +
p_{l+2}+\cdots +p_{n-1}) | n\ran &
n\geq 6\\
1& n=5\\
\end{array}
\right.\label{16} \eeq
We have checked numerically, up to $n=11$, that our formula
\eqref{npoint} gives the same results as \eqref{15}.

It is interesting to note that the very existence of this recursion
relation in gravity - described here and in \cite{bbst3,Cachazo:2005ca} -
has something to say about the divergences of quantum gravity. A central
feature of the recursion relation is that it requires $\cM(\infty)=0$, and
the behaviour of $\cM(z)$ as $z\rightarrow\infty$ is related to the
high-energy behaviour (and hence the renormalisability) of the theory. It is
not \emph{a priori} clear that gravity has this behaviour, though the analyses of
\cite{bbst3,Cachazo:2005ca} and more recently the complete analysis of \cite{Benincasa:2007qj}
show that indeed $\cM(\infty)=0$ for any tree-level amplitude in gravity. This
means that at tree-level, gravity has divergences in the UV that are perhaps better than
one might expect. This supports recent arguments that
gravity may not be as divergent as previously thought and more specifically that
4-dimensional $\cN\!=\!8$ supergravity may be finite \cite{Bjerrum-Bohr:2006sg,Bjerrum-Bohr:2006yw,Bern:2006kd,Bern:2007hh,
Green:2006gt,Green:2006yu}.

\section{Applications to other field theories}
\label{otherappssection}
One of the striking features of the BCFW proof of the BCF
recursion relations is that the specification of the theory with which
one is dealing is almost unnecessary. Indeed in \cite{bcfw} the only
step where specifying the theory \emph{did} matter was in the estimate of
the behaviour of the scattering amplitudes $\cM (z)$ as $z \to
\infty$, which was important to assess the possible existence of
boundary terms in the recursion relation. This leads us to
conjecture that recursion relations could be a more generic
feature of massless (or spontaneously broken)
field theories in four dimensions.%
\footnote{This was also suggested in \cite{Bern:2005hs}.} After
all, the BCF recursion relations - as well as the recursion
relation for gravity amplitudes discussed in this chapter and in \cite{Cachazo:2005ca} - just
reconstruct a tree-level amplitude (which is a rational function)
from its poles.

Let us focus on massless $\l (\phi \phi^{\dagger})^2$ theory in four
dimensions. We use the spinor helicity formalism, meaning that
each momentum will be written as $p_{a \dot{a}} = \l_a
\lt_{\dot{a}}$. A scalar propagator $1/P^2$ connects states of
opposite ``helicity'', which here just means that the propagator
is $\lan \phi (x) \phi^\dagger (0) \ran$, with
 $\lan \phi (x) \phi (0) \ran =
\lan \phi^\dagger (x) \phi^\dagger  (0) \ran = 0$. Now consider a
Feynman diagram contributing to an $n$-particle scattering
amplitude, and let us shift the momenta of particles $k$ and $l$
as in \eqref{shifts}. As for the Yang-Mills case discussed in
\cite{bcfw}, there is a unique path of propagators going from
particle $k$ to particle $l$. Each of these propagators
contributes $1/z$ at large $z$, whereas vertices are independent
of $z$. We thus expect Feynman diagrams contributing to the
amplitude to vanish in the large-$z$ limit.

An exception to the above reasoning is represented by those
Feynman diagrams where the shifted legs belong to the same vertex;
these diagrams are $z$-independent, and hence not suppressed as $z
\to \infty$. In order to deal with this problematic situation, and
ensure that the full amplitude $\cM(z)$ computed from Feynman diagrams
vanishes  as $z \to \infty$ we propose two alternatives.

Firstly, if one considers  $(\phi \phi^\dagger)^2$ theory without
any group structure, one can remove the problem by performing
multiple shifts. This possibility has already been used in the
context of the rational part of one-loop amplitudes in pure
Yang-Mills \cite{Bern:2005hs}. In our case, it is sufficient to
shift at least four external momenta.

Alternatively, we can consider $(\phi \phi^\dagger)^2$ theory with
global symmetry group $\U(N)$ and $\phi$ in the adjoint. In this
case we can group the amplitude into colour-ordered partial
amplitudes, as in the Yang-Mills case. Then, for any
colour-ordered amplitude one can always find a choice of shifts
such that the shifted legs do not belong to the same Feynman
vertex.
The procedure can be repeated for any colour ordering, and the
complete amplitude is obtained by summing over non-cyclic
permutations of the external legs.

In this way, the appearance of a boundary term $\cC_{\infty}$ can
be avoided, and one can thus derive a recursion relation for
scattering amplitudes akin to \eqref{rec}. A similar analysis
can be carried out in other theories, possibly in the presence of
spontaneous symmetry breaking \emph{etc}. We expect this to play an
important r\^{o}le in future studies.

\section{CSW as BCFW}

Finally, we would like to point out the connection between the CSW rules
at tree-level \cite{csw} and the BCFW recursion relation introduced in
\cite{bcf,bcfw} and discussed for gravity in this chapter. This was hinted at in
\cite{bcfw} where it was noted that the existence of BCFW recursion (which can
construct any gluon amplitude solely from a knowledge of its singularities) provides an
indirect proof of the CSW rules since the CSW rules provide results which are Lorentz-invariant,
gauge-invariant and have the correct singularities. It was also briefly
touched on in \cite{bbst3} where some formal observations were made regarding the
relation between the way that the shifts of Eq.~(\ref{shifts}) are performed - so as
to keep the corresponding momenta on-shell in the BCFW recursion relation - and the way that the internal legs
in the CSW rules are shifted (Eq.~(\ref{off})).

However, Risager showed that the CSW rules are in fact a special case
of the BCFW recursion relation when specific shifts of momenta are made \cite{Risager:2005vk}. The most natural
shifts to make when using the recursion relations are those which minimise the number of terms
appearing and thus the work that one has to do. In \cite{Risager:2005vk}, however, a different set of
shifts was employed which affects every propagator that may appear in a CSW diagram. The propagators
are defined by the momenta that flow through them and thus by a set of consecutive external particles. In the
case of CSW diagrams, the vertices are MHV vertices and thus this set of consecutive particles (and its
compliment on the other vertex to-which the propagator is attached) must contain at least one gluon of
negative helicity each. Exactly this set of propagators is affected if every external negative-helicity
gluon is shifted, provided that the sum of any subset of the shifts does not vanish. In addition, the shifts
must all involve the anti-holomorphic spinors so that all 3-point googly amplitudes drop out.

Using these shifts (see Eq.~(5.1) of \cite{Risager:2005vk} for an explicit example of the shifts for an NMHV amplitude),
Risager used induction to prove the CSW rules directly thus highlighting their connection with the BCFW recursion
relation. In \cite{Bjerrum-Bohr:2005jr} these ideas were then used to construct an MHV-vertex formalism for
gravity, thus accentuating the remarkable similarities between gauge theory and gravity despite the
latter's more complicated structure.

\chapter{Conclusions and outlook}
\label{chapter5}

In the previous chapters we have studied gluon scattering amplitudes in perturbative
gauge theory and have seen how they can be stripped of colour and written in terms of
spinor variables to illuminate their basic structure in a unified context. Their twistor-space localisation
then allows for an understanding
of the unexpected simplicity of many $n$-point processes. The tree-level MHV
amplitudes were seen to lie on simple straight lines in twistor space and it was
shown how they could be calculated from a topological string theory
as an integral over the moduli space of holomorphically embedded, degree 1, genus 0 curves.
This in turn motivated a new perturbative expansion of Yang-Mills gauge theory where tree-level MHV amplitudes
are taken off-shell and joined with scalar propagators to create tree-level amplitudes with
successively greater numbers of negative-helicity particles. The MHV vertices effectively combine many
Feynman diagrams into one and thus provide a great simplification which aids calculation and
highlights the underlying geometrical structure.

We saw how these techniques could be applied at loop-level to calculate the MHV amplitudes in
$\cN\!=\!4$ super-Yang-Mills, which is a slightly surprising result as the duality with the twistor
string theory constructed in \cite{witten} (and also that in \cite{Berkovits}) fails at loop-level.
These string theories contain conformal supergravity states which do not decouple at
one-loop and this suggests that the application of the CSW rules to loops might fail or
simply calculate amplitudes in some theory of Yang-Mills coupled to conformal supergravity. Indeed,
a recent calculation of various loop amplitudes in Berkovits' twistor string theory
appears to give amplitudes in such a theory \cite{Dolan:2007vv}.

One might also expect that such a surprising result would only apply to maximally supersymmetric
Yang-Mills. However in Chapter 2 we saw that MHV vertices can be used to calculate amplitudes at
loop-level in theories with less supersymmetry such as $\cN\!=\!1$ super-Yang-Mills. There we
calculated the one-loop MHV amplitudes and found complete agreement with the known results in
\cite{Bern:1994cg}. The calculation itself is more involved than the corresponding
one in $\cN\!=\!4$ presented in \cite{bst} and reviewed in Chapter 1 because the reduction in
supersymmetry leads to fewer cancellations. Happily though, this does not
spoil the technique of using MHV amplitudes as effective vertices.

In Chapter 3 we applied the loop-level CSW rules to pure Yang-Mills with a scalar running in the
loop. Pure
Yang-Mills is a non-supersymmetric theory and as such the calculation is even more involved than
before. This still does not invalidate the process, although it was found that the use of
MHV vertices only calculates the cut-constructible part of the amplitude. The rational
parts, which are intrinsically linked to the cuts for supersymmetric theories, were thus missed.
Nonetheless, the results obtained match perfectly with the known (special) cases \cite{Bern:1994cg,bdk9302280}
and provide the cut-constructible part of the MHV amplitude in pure Yang-Mills with arbitrary positions
for the negative-helicity gluons for the first time. The rational part of the amplitude has since been
calculated in \cite{Berger:2006vq} building on the results described in Chapter 3.

In Chapter 4 we turned our attention to gravity and another interesting development stemming from
twistor string theory, namely that of on-shell recursion relations. Recursion relations have been
used before in the construction of amplitudes \cite{bg}, but it wasn't until recently that they
were used to recursively turn on-shell amplitudes into amplitudes with a larger number of external legs.
They were introduced in \cite{bcf} at tree-level and have since been used in a bootstrap approach
to loop amplitudes which was in fact one of the techniques applied in \cite{Berger:2006vq}.

We saw that on-shell recursion relations can also be applied at tree-level in gravity and
it is a beauty of the proof of these relations in gauge theory \cite{bcfw} which means
that they can be proved in gravity without too much (!) extra work. The main additional
ingredient is a proof of the behaviour of tree-level $n$-graviton amplitudes as a function of a
complex variable $z$ as $z\rightarrow\infty$. In Chapter 4 we argued the case for many
amplitudes of interest, but a recent proof that $\lim_{z\rightarrow\infty}\cM_n(z)=0$
establishes that the recursion relation in gravity can construct any tree-level $n$-graviton
amplitude \cite{Benincasa:2007qj}.

We showed how this recursion relation could be used to construct MHV amplitudes with
successively more external gravitons and as a by-product constructed a new
compact form for the $n$-graviton MHV amplitudes which provides an interesting alternative to
the previously-known form in \cite{bgk}. We finished by commenting on the relation between the
tree-level CSW rules and on-shell recursion relations both in field theory and in gravity and
also made some observations on the existence of recursion relations in other theories such as
scalar $\phi^4$ theory.

Unsurprisingly, this is not the end of the story. In the introduction we already mentioned some
of the directions that have been explored following from and related to the material presented
here. This includes the
construction of twistor string theories describing $\cN\!=\!4$ Yang-Mills as well as
ones describing other field theories such as a recent description of Einstein supergravity \cite{Abou-Zeid:2006wu},
the use of on-shell recursion relations at loop level in both gauge theory and gravity
\cite{Berger:2006cz,Brandhuber:2007up} and
improvements to the unitarity method \cite{Britto}. It may be particularly interesting to note that
in \cite{Abou-Zeid:2006wu}, one of the theories for which a twistor description is found is
$\cN\!=\!4$ Yang-Mills coupled to $\cN\!=\!4$ Einstein supergravity. It appears that there
exists a decoupling limit for this theory which gives pure Yang-Mills and thus opens the
door to the possibility of understanding loops in Yang-Mills from twistor strings.

From the point of view of the MHV diagram formulation of gauge theory there has also been
some considerable progress. Their use at tree-level is already well-established
and a Lagrangian formulation now exists \cite{Gorsky:2005sf,Mansfield:2005yd,Ettle:2006bw}. In
this scenario, a non-local change of variables is made to the light-cone Yang-Mills Lagrangian
which yields a kinetic term describing a scalar propagator connecting positive and negative helicities
and interaction terms consisting of the infinite sequence of MHV amplitudes.

Quantisation of this Lagrangian, however, is still an open problem. One of the main points here is
the fact that - as demonstrated in Chapter \ref{chapter3} - the use of MHV diagrams alone
is not enough to generate a complete amplitude at the quantum level in non-supersymmetric theories and rational terms
are missed. As such, one might ask how one could compute the one-loop all-plus (and $-+\ldots +$) amplitude in
pure Yang-Mills from MHV diagrams. At tree-level this vanishes, but at one-loop it is a purely
rational function - see \emph{e.g.} Equation~(3.4) of \cite{Brandhuber:2005jw}. Construction of a
one-loop amplitude from MHV diagrams will always give $q$ negative-helicity gluons
that satisfies $q\geq 2$, and thus the all-plus amplitude (and also the $-+\ldots +$ amplitude)
cannot be constructed from MHV vertices alone. In \cite{csw2} it was conjectured that
perhaps the all-plus amplitude could be elevated to the status of a vertex to generate these
missing amplitudes, but at the time an appropriate off-shell continuation for this
amplitude could not be found.

Recently, however, more progress has been made in this direction
\cite{Brandhuber:2006bf,Brandhuber:2007vm,Ettle:2007qc}. It appears that the
all-plus amplitude is intimately connected with the regularisation procedure needed to evaluate loop diagrams as was initially
hinted-at by the fact that the parity conjugate of this amplitude, the all-minus amplitude, arises from
an $\epsilon\times 1/\epsilon$ cancellation in dimensional regularisation \cite{Brandhuber:2006bf}.
Inspired by this, Brandhuber, Spence, Travaglini and Zoubos showed in \cite{Brandhuber:2007vm}
that a certain one-loop two-point Lorentz-violating counterterm is the generating function for
the infinite sequence of one-loop all-plus amplitudes in pure Yang-Mills although there must be another
contribution in this story to correctly explain the origin of the
$-+\ldots +$ amplitude. In their
approach it was found that a certain four-dimensional regularisation scheme (rather than
dimensional regularisation) \cite{Thorn:2005ak,Chakrabarti:2005ny,Chakrabarti:2006mb} was
most useful. It may be interesting and insightful to see if a light-cone approach and such a
regularisation scheme is also helpful for computing the cut-constructible terms of amplitudes using MHV diagrams.

Despite these advances, the MHV diagram technique is still practically-speaking limited to
tree-level amplitudes and the cut-constructible part of one-loop MHV amplitudes. This is largely because of the intrinsic
complexity of loop calculations, though there are other complications. For example, the
topologies involved in calculating the cut-constructible part of amplitudes with more than
2 negative-helicity gluons can include (in the case of the NMHV amplitude say) triangle diagrams where each
vertex is an MHV vertex. In such diagrams one has 3 different internal particles to take off-shell
and it is not clear whether the measure can be found in terms of LIPS integrals and
dispersion integrals such as that described in \cite{bst,Brandhuber:2005kd} which has been so
instrumental in the application of the CSW rules at loop-level so far. Such issues are common to
one-loop amplitudes which have $q>2$ negative-helicity gluons and higher loops as well.
It would be desirable from both a theoretical and a phenomenological perspective to understand
how the MHV rules can be used to calculate such quantities and would also help to give the
MHV rules a more solid footing.

Another interesting avenue of exploration is the suggestion that (planar)
higher-loop amplitudes in $\cN\!=\!4$ Yang-Mills may be expressed (essentially)
as an exponential of certain one-loop amplitudes
\cite{Anastasiou:2003kj,Bern:2005iz,Bern:2006ew,Cachazo:2006mq,Cachazo:2006tj,Cachazo:2006az}.
Such expressions are termed cross-order relations and may be remarkably powerful if more generally
applicable than has been found to date. They could allow the summation of amplitudes in $\cN\!=\!4$ Yang-Mills to all
orders in perturbation theory and so to non-perturbative information which
may be connected to perturbative string theory via the AdS/CFT correspondence.\footnote{A very recent paper \cite{Alday:2007hr} by
Alday and Maldacena appears to have taken a step in this direction. They show how to calculate
gluon scattering amplitudes at strong coupling from a classical string configuration via the AdS/CFT correspondence.
As a result the full finite form of the four-gluon scattering amplitude in $\cN\!=\!4$ super-Yang-Mills is presented.
See also \cite{Abel:2007mw} which addresses the $n$-point case.}
It would be interesting to see how the known cross-order relations
arise from MHV diagrams. It is possible that the different terms in the cross-order
relations may arise naturally from MHV diagrams which might then provide a framework for
proving their validity more generally.

The situation for gravity is in some ways even more exciting, with the possibility that there
may exist UV-finite \emph{field} theories of gravity. Such proposals have recently been made for
$\cN\!=\!8$ supergravity \cite{Bern:2005bb,Bjerrum-Bohr:2005xx,Bjerrum-Bohr:2006sg,Bjerrum-Bohr:2006yw,Bern:2006kd,Bern:2007hh,
Green:2006gt,Green:2006yu} and it would be interesting to make contact between this and the
twistor approach. One such point of contact may be the recent proposal of a twistor string theory
describing $\cN\!=\!8$ supergravity \cite{Abou-Zeid:2006wu}. Another possibility is that of
loop amplitudes from MHV vertices in ($\cN\!=\!8$ super-) gravity. These have not yet been understood and
their explication would provide a new prescriptive method for the
calculation of loop amplitudes in gravity which could shed light on their UV properties.%
\footnote{Shortly after the completion of this work \cite{adele} appeared which deals with precisely
this point.}

A further possibility that has not been explored so far (in either gauge theory or gravity) is a more
direct connection between recursion relations and loop amplitudes than those
already mentioned. Risager \cite{Risager:2005vk} showed that the CSW rules at tree-level are really just a specific
case of the on-shell recursion relations proposed by Britto, Cachazo and Feng \cite{bcf}. As we have
seen throughout this thesis, the CSW rules can naturally be extended to loop amplitudes which begs the question
of whether the same can be done for other cases of the on-shell recursion relation in either Yang-Mills
or in gravity.

\pagebreak

In this thesis we have seen some of the improvements that can be made to
perturbative techniques in field theory and gravity and the power that they can have. These also hint at new underlying
structures whose elucidation could prove extremely interesting
if not revolutionary in our understanding. Could such structures presage the existence of new
symmetries and will they end up replacing Feynman diagrams entirely in the future? Whatever the
outcome, these are exciting timely developments that are sure to aid the discovery of new physics
at colliders such as the LHC and deepen our understanding of nature.


\begin{appendix}

\chapter{Spinor and Dirac-trace identities}
\label{traceappendix}

In this appendix we
present some useful identities pertaining to the spinor helicity formalism
and also to help in dealing with Dirac traces.

\section{Spinor identities}

We take the metric to be the usual field-theory one $\eta_{\mu\nu}=(1,-1,-1,-1)$
and the epsilon tensors with which we raise and lower indices to be
\be
\epsilon^{\a\b}=\epsilon^{\da\db}=i\sigma^2=\twobytwo{0}{1}{-1}{0}\ ,
\ee
with $\epsilon_{\a\b}=\epsilon^{\a\b}\Rightarrow \epsilon_{\a\b}\epsilon^{\b\gamma}=-\delta_\a^{\phantom{\b}\gamma}$ and
\begin{eqnarray}
\label{Paulimatrices}
\sigma^1&=&\twobytwo{0}{1}{1}{0}\ ,\nonumber \\
\sigma^2&=&\twobytwo{0}{-i}{i}{0}\ ,\nonumber \\
\sigma^3&=&\twobytwo{1}{0}{0}{-1}\ .
\end{eqnarray}
We also have $\sigma^{\mu}_{\a\da}=(1,\vec{\sigma})$
(with $\vec{\sigma}=(\sigma^1,\sigma^2,\sigma^3)$), giving
\begin{eqnarray}
P_{\a\da}&=&P_{\mu}\sigma^\mu_{\a\da}\nonumber \\
&=&\twobytwo{P_0+P_3}{P_1-iP_2}{P_1+iP_2}{P_0-P_3}\nonumber \\
&=&\twobytwo{P^0-P^3}{-P^1+iP^2}{-P^1-iP^2}{P^0+P^3}\ ,
\end{eqnarray}
and $\bar{\sigma}^{\mu\,\da \a}=-\sigma^{\mu\,\a\da}=\epsilon^{\da\db}
\epsilon^{\a\b}\sigma^{\mu}_{\b\db}=(1,-\vec{\sigma})$, giving
\begin{eqnarray}
P^{\da \a}&=&P_{\mu}{\bar{\sigma}^{\mu\,\da \a}}\nonumber\\
&=&\twobytwo{P_0-P_3}{-P_1+iP_2}{-P_1-iP_2}{P_0+P_3}\nonumber\\
&=&\twobytwo{P^0+P^3}{P^1-iP^2}{P^1+iP^2}{P^0-P^3}\ .
\end{eqnarray}

Some useful identities involving $\sigma$ and $\bar{\sigma}$ are:
\begin{eqnarray}
\label{usefulsigma}
\sigma^{\mu}_{\a\db}\bar{\sigma}^{\nu\,\db \a}&=&2\eta^{\mu\nu}\\
\sigma^{\mu}_{\a\da}\bar{\sigma}_{\mu}^{\b\db}&=&2\delta_\a^{\phantom{\a}\b}
\delta_{\da}^{\phantom{\da}\db}\ ,
\end{eqnarray}
which means that we can interpret $\eta_{\mu\nu}$ as acting as $2\epsilon_{\da\db}\epsilon_{\a\b}$
and $\eta^{\mu\nu}$ as acting as $\epsilon^{\da\db}\epsilon^{\a\b}/2$ in spinor space,
giving $\eta_{\mu\nu}\eta^{\mu\nu}=\epsilon_{\da\db}\epsilon^{\db\da}\epsilon_{\a\b}\epsilon^{\b\a}=4$
as it should.

We are concerned with massless particles for which we can write
\begin{eqnarray}
p_{\a\da}&=&\lambda_{\a}\lt_{\da}\nonumber\\
&=&\doublet{\lambda_1}{\lambda_2}\row{\lt_{\dot{1}}}{\lt_{\dot{2}}}\nonumber\\
&=&\twobytwo{\lambda_1\lt_{\dot{1}}}{\lambda_1\lt_{\two}}{\lambda_2\lt_{\dot{1}}}
{\lambda_2\lt_{\two}}\ ,
\end{eqnarray}
which implies (by raising indices) that
\begin{eqnarray}
p^{\da \a}&=&-\lt^{\da}\lambda^{\a}\nonumber\\
&=&-\doublet{\lt^{\dot{1}}}{\lt^{\two}}\row{\lambda^1}{\lambda^2}\nonumber\\
&=&-\twobytwo{\lt^{\dot{1}}\lambda^1}{\lt^{\dot{1}}\lambda^2}{\lt^{\two}\lambda^1}
{\lt^{\two}\lambda^2}\nonumber\\
&=&\twobytwo{\lt_{\two}\lambda_2}{-\lt_{\two}\lambda_1}{-\lt_{\dot{1}}\lambda_2}
{\lt_{\dot{1}}\lambda_1}\ ,
\end{eqnarray}
which follows from having $\lambda^{\a}=(\epsilon^{\a\b}\lambda_{\b})^T
=-\lambda^T_{\b}\epsilon^{\b\a}$
and $\lt^{\da}=(\lt_{\db}\epsilon^{\db\da})^T=-\epsilon^{\da\db}\lt_{\db}^T$

For scalar products we take
\begin{eqnarray}
\label{angle}
\langle\lambda\,\mu\rangle&=&\lambda^{\a}\mu_{\a}\nonumber\\
&=&\row{\lambda^1}{\lambda^2}\doublet{\mu_1}{\mu_2}\nonumber\\
&=&\epsilon^{\a\b}\lambda_{\b}^T\mu_{\a}\ ,
\end{eqnarray}
and
\begin{eqnarray}
\label{square}
[\tilde{\lambda}\,\tilde{\mu}]&=&\lt_{\da}\mut^{\da}\nonumber\\
&=&\row{\lt_{\dot{1}}}{\lt_{\two}}\doublet{\mut^{\dot{1}}}{\mut^{\two}}\nonumber\\
&=&\lt_{\da}\mut_{\db}^T\epsilon^{\db\da}
\end{eqnarray}
Note that $\lambda_{\a}$ and $\lt^{\da}$ are most naturally
associated with column vectors, while $\lt_{\da}$ and $\lambda^{\a}$ are
most naturally associated with row vectors.

For spinor manipulations, the Schouten identity is very useful:\footnote{Recall
that we have the shorthand notation $\langle\lambda_i\,\lambda_j\rangle=\langle i\,j\rangle$ \emph{etc}.}
\begin{eqnarray}
\label{schouten}
\langle i\, j\rangle\langle k\, l\rangle =
\langle i\, k\rangle\langle j\, l\rangle + \langle i\,
l\rangle\langle k\, j\rangle \, ,\nonumber\\
\left[i\, j\right]\left[k\, l\right]=\left[i\, k\right]\left[j\, l\right]+\left[i\, l\right]\left[k\, j\right]\ .
\end{eqnarray}

For other introductions to the spinor helicity formalism see \emph{e.g.}
\cite{dixon,manganorev}.

\section{The holomorphic delta function}

Consider the $x\!-\!y$ plane in real coordinates and let $(x,y)=(x^1,x^2)$. Now change to
complex coordinates by letting
\begin{eqnarray}
\label{zs}
z&=&x^1+ix^2\\
\bar{z}&=&x^1-ix^2\ .
\end{eqnarray}
Also define derivatives
\begin{eqnarray}
\label{derivs}
\partial_z&=&{\partial\over\partial z}=\frac{1}{2}\left(\frac{\partial}{\partial x^1}-i\frac{\partial}{\partial x^2}\right)
=\frac{1}{2}\left(\partial_1-i\partial_2\right)\\
\partial_{\bar{z}}&=&{\partial\over\partial \bar{z}}
=\frac{1}{2}\left(\frac{\partial}{\partial x^1}+i\frac{\partial}{\partial x^2}\right)
=\frac{1}{2}\left(\partial_1+i\partial_2\right)\ ,
\end{eqnarray}
which have the properties that
\begin{displaymath}
\begin{array}{cccc}
\partial_zz=1\ ;&\partial_z\bar{z}=0\ ;&\partial_{\bar{z}}\bar{z}=1\ ; &\partial_{\bar{z}}z=0\ .
\end{array}
\end{displaymath}

We take the area element in the $x\!-\!y$ plane to be $d^2x=dx^1dx^2=|dx^1\wedge dx^2|$,
where $|\,|$ just indicates that one picks a plus sign to define the orientation. For the
area element in the $z\!-\!\bar{z}$ plane we take $d^2z=i|dz\wedge d\bar{z}|$ so that we have
$d^2z=2d^2x$.

We normalise delta functions in the $x\!-\!y$ plane as
\be
\label{deltaxy}
\int\!d^2x\,\delta^{(2)}\!(x-a)=1\ ,
\ee
where $\delta^{(2)}(x-a):=\delta(x^1-a^1)\delta(x^2-a^2)$, and after transforming
this to the $z\!-\!\bar{z}$ plane (with $b=a^1+ia^2$ and $\bar{b}=a^1-ia^2$) we have
\be
\label{deltazz}
\int\!d^2z\,\delta^{(2)}\!(z-b)=1\ ,
\ee
where
\begin{eqnarray}
\label{deltazzxy}
\delta^{(2)}\!(z-b)&:=&\delta(z-b)\delta(\bar{z}-\bar{b})\nonumber\\
&=&\frac{1}{2}\delta^{(2)}\!(x-a)\ .
\end{eqnarray}

We now define a \emph{holomorphic} delta function as
\be
\label{holomorphicdelta}
\bar{\delta}(z-b)=\delta^{(2)}\!(z-b)d\bar{z}\ ,
\ee
which gives us
\begin{eqnarray}
\label{holomorphicdelta2}
\int\!dz\,\bar{\delta}(z-b)&=&\int\!dz\wedge d\bar{z}\,\delta^{(2)}\!(z-b)\nonumber\\
&=&-i\int\!d^2z\,\delta^{(2)}\!(z-b)\nonumber\\
&=&-i\ .
\end{eqnarray}
As can be seen the holomorphic delta function is a $\bar{\partial}$-closed
$(0,1)$-form and is defined for a general holomorphic function $f$ by
$\bar{\delta}(f)=\delta^{(2)}\!(f)d\bar{f}$.

A representation of this holomorphic delta function which will be particularly
useful for us is the following \cite{csw}. Consider a momentum-vector described by $\lambda$ and
$\lt$ with $\lt=\bar{\lambda}$ in order to ensure that $p_{\a\da}=\l_\a\lt_\da$
is real. Go to coordinates where $\lambda^{\alpha}=(1,z)$ and choose an
arbitrary spinor $\zeta^\a=(1,b)$ with $b$ a complex number. The tilded
spinors are then
\begin{displaymath}
\begin{array}{cc}
\lt^{\da}=\doublet{1}{\bar{z}} & \tilde{\zeta}^{\da}=\doublet{1}{\bar{b}}\ ,
\end{array}
\end{displaymath}
and we have $\langle\zeta\,\lambda\rangle=z-b$ and $\langle\l\, d\l\rangle=dz$. So,
a more covariant statement of (\ref{holomorphicdelta2}) is
\begin{eqnarray}
\label{holomorphicdelta3}
\int\langle\l\, d\l\rangle\,\bar{\delta}(\langle\zeta\,\l\rangle)\,F(\lambda)
&=&\int\!dz\,\bar{\delta}(z-b)\,F(z)\nonumber\\
&=&-iF(b)\nonumber\\
&=&-iF(\zeta)\ .
\end{eqnarray}

\section{Dirac traces}
Some basic formul{\ae} for converting between spinor invariants and
Dirac traces are
\begin{eqnarray}
\label{spinor1}
\langle i\, j\rangle\left[j\,
i\right]
&=&
{\tr}_{+}(\kslash_i \kslash_j)
\ ,
\\ \cr
\label{spinor2}
\langle i\, j\rangle\left[j\, l\right]\langle l\,
m\rangle\left[m\, i\right]
&=&
{\tr}_{+}(\kslash_i \kslash_j
\kslash_{l} \kslash_m)
\ ,
\\ \cr
\label{spinor3}
\langle i\, j\rangle\left[j\, l\right]\langle l\,
m\rangle\left[m\, n\right]\langle n\, p\rangle\left[p\, i\right]
&=&
{\tr}_{+}(\kslash_i \kslash_j \kslash_{l} \kslash_m \kslash_n
\kslash_p)\ ,
\end{eqnarray}
for momenta
$k_i,k_j,k_l,k_m,k_n,k_p$ and where the $+$ sign indicates the insertion of
\newline\mbox{$(1+\gamma_5)/2$}:
\be
\label{gamma5}
{\tr}_{+}(\kslash_i\kslash_j):=\frac{1}{2}{\tr}_{+}((1+\gamma_5)\kslash_i\kslash_j)\ .
\ee
We also note that
\begin{eqnarray}
\label{traceof2}
{\tr}_{+}(\kslash_i \kslash_j)&=&2(k_i\cdot k_j)\\
\label{4trace}
\tr_+ \left( \kslash_{a} \kslash_{b} \kslash_{c}
\kslash_d \right) &=& 2  (k_a \cdot k_b) (k_c \cdot k_d)-
2(k_a \cdot k_c) (k_b \cdot k_d)\nonumber\\ &+&
 2(k_a \cdot k_d) (k_b \cdot k_c)
 - 2i  \varepsilon(k_a,k_b,k_c,k_d)\ .
\end{eqnarray}

The following identities are additionally useful:
\begin{eqnarray}
\label{trace1} {\tr}_{+}(\kslash_i \kslash_j \kslash_{l}
\kslash_m)
&=&
{\tr}_{+}(\kslash_m \kslash_l \kslash_{j}
\kslash_i)\ = \
{\tr}_{+}(\kslash_l \kslash_m \kslash_{i}\kslash_{j})
\ ,
\\ \cr
\label{trace2}
{\tr}_{+}(\kslash_i \kslash_j \kslash_{l}
\kslash_m)
&=&
4(k_i\cdot k_j)(k_l\cdot k_m)-{\tr}_{+}(\kslash_j
\kslash_i \kslash_{l} \kslash_m)
\ ,\\ \cr
\label{trace3}
\ijmuP \, \ijmum &=& 0
\ ,
\\ \cr
\label{trace4}
\ijmuP \, \ijmmu &=& 4(i\cdot j)\, \ijmP
\end{eqnarray}
for similarly generic momenta and where we use the shorthand
$\trace_+(\kslash_i\kslash_j)=\trace_+(i\!\!\!\slash j\!\!\!\slash)$ \emph{etc}.
If $k_i,k_j,k_{m_1}$ and $k_{m_2}$
are massless, while $P_L$ is not necessarily so,
then we have the remarkable identity:
\begin{eqnarray}
\label{remarkable}
2(k_{m_1}\cdot k_{m_2}){\tr}_{+}(\kslash_i
\kslash_j \kslash_{m_1} \Pslash_L){\tr}_{+}(\kslash_i \kslash_j
\kslash_{m_2} \Pslash_L)
\nonumber
\\  [3pt] +
P_L^2\,{\tr}_{+}(\kslash_i \kslash_j \kslash_{m_1}
\kslash_{m_2}){\tr}_{+}(\kslash_i
\kslash_j \kslash_{m_2} \kslash_{m_1})
\nonumber\\ [3pt]
-2(k_{m_1}\cdot P_L){\tr}_{+}(\kslash_i \kslash_j \kslash_{m_1}
\kslash_{m_2}){\tr}_{+}(\kslash_i \kslash_j \kslash_{m_2}
\Pslash_L)
\nonumber \\ [3pt]
-2(k_{m_2}\cdot P_L){\tr}_{+}(\kslash_i
\kslash_j \kslash_{m_1} \Pslash_L){\tr}_{+}(\kslash_i \kslash_j
\kslash_{m_2} \kslash_{m_1})
&=&0
\ .
\end{eqnarray}
We also have, for null momenta $i,j,k,a,b,$
\beq
\label{simplifier}
\frac{\ijab\jakb}{(j\cdot a)} = - \frac{\ijba\iakb}{(i\cdot a)}
\ .
\eeq

\chapter{Feynman rules in the spinor helicity formalism}
\label{helicityrules}

In this appendix we present the Feynman rules for massless $\SU(N_c)$
Yang-Mills theory in Feynman gauge written in the spinor helicity formalism for
comparison with those laid out at the start of Chapter 1. As mentioned in a
footnote in \S \ref{colourordersection} we will use the normalisation
$\trace(T^aT^b)=\delta^{ab}$ for the Lie-algebra in order to reduce the
proliferation of factors of 2.

\section{Wavefunctions}

\begin{itemize}
\item \textbf{External Scalar:}
\be
\phi = 1
\ee
\item \textbf{External outgoing fermion $i$, helicity plus:}
\be
\psi^+_i =  \lt_{i\,\da} = [i|
\ee
\item \textbf{External outgoing fermion $i$, helicity minus:}
\be
\psi^-_i  =\lambda_i^{\a} = \langle i|
\ee
\item \textbf{External outgoing anti-fermion $j$, helicity plus:}
\be
\bar{\psi}^+_j = \lt_{j}^{\da} = |j]
\ee
\item \textbf{External outgoing anti-fermion $j$, helicity minus:}
\be
\bar{\psi}^-_j = \lambda_{j\,\a} =|j\rangle
\ee
\item \textbf{External outgoing vector $p=\lambda\lt$, helicity plus:}
\be
\epsilon^+_{\a\da}=\sqrt{2}\,\frac{\mu_{\a}\lt_{\da}}{\langle \mu\, \lambda\rangle}
=\sqrt{2}\,\frac{|\mu\rangle[\lt|}{\langle \mu\, \lambda\rangle}
\ee
\item \textbf{External outgoing vector $p=\lambda\lt$, helicity minus:}
\be
\epsilon^-_{\a\da}=\sqrt{2}\,\frac{\lambda_{\a}\mut_{\da}}{[\mut\,\lt]}
=\sqrt{2}\,\frac{|\lambda\rangle[\mut|}{[\mut\, \lt]}\ ,
\ee
\end{itemize}
where $q=\mu\mut$ is an arbitrary reference spinor that can be chosen independently
for each external particle. All the above wavefunctions are
understood to be multiplied by a factor of $\exp(ix_{\beta\dot{\beta}}\lambda^\beta\lt^{\dot{\beta}})$,
where $p_{\beta\dot{\beta}}=\lambda_\beta\lt_{\dot{\beta}}$ is the momentum of the particle.

\section{Propagators}

\begin{itemize}
\item \textbf{Scalars with kinetic term $(\partial\phi)^2/2$:}
\be
\label{sprop}
\frac{i}{p^2}
\ee

\item \textbf{Fermions with $p=\lambda\lt$ and kinetic term $\bar{\psi}i\bar{\sigma}^{\mu}\partial_{\mu}\psi$:}
\be
\label{fprop}
\frac{i}{\bar{\sigma}^{\mu}p_{\mu}}=\frac{ip_{\a\da}}{2p^2}=
\frac{i|\lambda\rangle[\lt|}{2p^2}
\ee

\item \textbf{Vectors with kinetic term $-(\partial A)^2/4$:}
\be
\label{vprop}
\frac{-2i\epsilon_{\da\db}\epsilon_{\a\b}}{p^2}
\ee
\end{itemize}

\section{Vertices}

\vspace{-10pt}\hfill\\
\begin{picture}(300,100)(0,47)
\thicklines
\Text(80,50)[tbv]{Fermion Vertex:}
\Photon(215,100)(215,50){5}{3}
\ArrowLine(215,50)(172,25)
\ArrowLine(258,25)(215,50)
\Vertex(215,50){2}
\Text(215,110)[amu]{$\db\b$}
\Text(170,19)[crho]{$\da$}
\Text(270,19)[bnu]{$\a$}
\end{picture}
$\displaystyle =\,\, ig\sqrt{2}\,\delta_\a^{\phantom{\a}\b}\delta_\da^{\phantom{\da}\db}$
\\
\begin{picture}(300,100)(0,47)
\thicklines
\Text(80,50)[tbv]{3-Boson Vertex:}
\Photon(215,100)(215,50){5}{3}
\Photon(172,25)(215,50){5}{3}
\Photon(258,25)(215,50){5}{3}
\Vertex(215,50){2}
\Text(215,108)[amu]{$\da \a$}
\Text(170,19)[crho]{$\dot{\gamma}\gamma$}
\Text(270,19)[bnu]{$\db \b$}
\LongArrow(225,65)(225,85)
\Text(233,75)[p1]{$p_1$}
\LongArrow(198,51)(180,39)
\Text(186,50)[p3]{$p_3$}
\LongArrow(221,36)(240,25)
\Text(227,25)[p2]{$p_2$}
\end{picture}
\begin{tabular}{l}
$\displaystyle \!\!\!\!\!\!\!\!\!\!=\,\, \frac{-g}{2\sqrt{2}}
[\epsilon^{\da\db}\epsilon^{\a\b}(p_1\!-\!p_2)^{\dot{\gamma}\gamma}$\\
$\displaystyle\!\!\!\!\!\!\!\!\!\!\,\quad\quad\quad\!+\,
\epsilon^{\db\dot{\gamma}}\epsilon^{\b\gamma}(p_2\!-\!p_3)^{\da \a}$\\
$\displaystyle\!\!\!\!\!\!\!\!\!\!\,\quad\quad\quad\!+\,
\epsilon^{\dot{\gamma}\da}\epsilon^{\gamma\a}(p_3\!-\!p_1)^{\db \b}]$
\end{tabular}
\\
\begin{picture}(300,100)(0,47)
\thicklines
\Text(80,50)[fbv]{4-Boson Vertex:}
\Photon(250,85)(215,50){5}{3}
\Photon(180,85)(215,50){5}{3}
\Photon(180,15)(215,50){5}{3}
\Photon(250,15)(215,50){5}{3}
\Vertex(215,50){2}
\Text(255,87)[bnu]{$\db \b$}
\Text(176,94)[amu]{$\da \a$}
\Text(168,10)[dsig]{$\dot{\delta} \delta$}
\Text(267,13)[crho]{$\dot{\gamma}\gamma$}
\end{picture}
\begin{tabular}{l}
$\displaystyle =\frac{ig^2}{8}[2\epsilon^{\da\dot{\gamma}}\epsilon^{\a\gamma}
\epsilon^{\db\dot{\delta}}\epsilon^{\b\delta}$\\
$\displaystyle \qquad-\,\,\,\epsilon^{\da\dot{\delta}}\epsilon^{\a\delta}
\epsilon^{\db\dot{\gamma}}\epsilon^{\b\gamma}$\\
$\displaystyle \qquad-\,\,\,\epsilon^{\da\db}\epsilon^{\a\b}
\epsilon^{\dot{\gamma}\dot{\delta}}\epsilon^{\gamma\delta}]$\ .
\end{tabular}
\vspace{10pt}\hfill \\
\vspace{1cm}
\begin{figure}[ht]
\label{spinorvertices}
\caption{\emph{The Vertices of the colour-stripped scheme in terms of spinors.}}
\end{figure}

For more details on how these arise see for example \cite{manganorev}.

\section{Examples}

\paragraph{4-Point MHV gluon scattering}\mbox{}

Let us consider how we get the $\cA(1^-_g,2^-_g,3^+_g,4^+_g)$ gluon amplitude. The diagrams contributing
to this amplitude are shown in Figure B.2

\begin{figure}[ht]
\begin{center}
\begin{picture}(440,120)(0,0)
\put(0,10){
\Photon(25,20)(95,80){3}{8}
\Photon(25,80)(95,20){3}{8}
\Text(23,85)[r]{$p_2^-$}
\Text(23,15)[r]{$p_1^-$}
\Text(100,17)[l]{$p_4^+$}
\Text(100,85)[l]{$p_3^+$}
\Vertex(60,50){2}}
\put(130,60){\Text(0,0)[l]{+}}
\put(180,10){
\SetOffset(-30,0)
\Photon(40,50)(80,50){3}{4}
\Photon(25,80)(40,50){3}{4}\Text(23,85)[r]{$p_2^-$}
\Photon(40,50)(25,20){3}{4}\Text(23,15)[r]{$p_1^{-}$}
\Photon(95,20)(80,50){3}{4}\Text(100,17)[l]{$p_4^+$}
\Photon(80,50)(95,80){3}{4}\Text(100,85)[l]{$p_3^+$}
\Vertex(40,50){2} \Vertex(80,50){2}
}
\put(280,60){\Text(0,0)[l]{+}}
\put(360,0){
\SetOffset(-60,0)
\Photon(50,40)(50,80){3}{4}
\Photon(25,100)(50,80){3}{4}\Text(23,100)[r]{$p_2^-$}
\Photon(50,40)(25,20){3}{4}\Text(23,20)[r]{$p_1^-$}
\Photon(75,20)(50,40){3}{4}\Text(80,20)[l]{$p_4^+$}
\Photon(50,80)(75,100){3}{4}\Text(80,100)[l]{$p_3^+$}
\Vertex(50,40){2} \Vertex(50,80){2}
}
\end{picture}
\label{4ptmhv}
\caption{\emph{The diagrams contributing to the 4-gluon MHV tree-amplitude. All external momenta
are taken to be outgoing.}}
\end{center}
\end{figure}
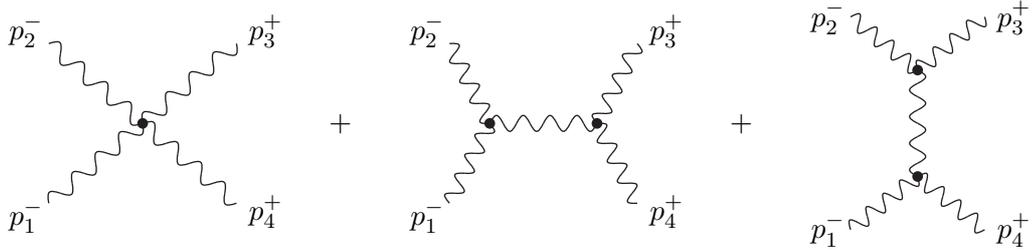

In order to calculate the amplitude we need to specify external wavefunctions as
prescribed by the Feynman rules and for gluons this includes a choice of reference
momentum. In order to minimise the number of terms we need to consider we will
make the choices $q_1=q_2=p_4$ and $q_3=q_4=p_1$. This means that the wavefunctions are
\begin{displaymath}
\begin{array}{cc}
\epsilon_{1\,\a\da}^-=\sqrt{2}{\lambda_\a^1\lt_\da^4\over [4\,1]} &
\epsilon_{2\,\a\da}^-=\sqrt{2}{\lambda_\a^2\lt_\da^4\over [4\,2]} \\
\\
\epsilon_{3\,\a\da}^+=\sqrt{2}{\lambda_\a^1\lt_\da^3\over \langle 1\,3\rangle} &
\epsilon_{4\,\a\da}^+=\sqrt{2}{\lambda_\a^1\lt_\da^4\over \langle 1\,4\rangle}\ ,
\end{array}
\end{displaymath}
while momentum conservation for the second (going from left to right) two diagrams reads $P_{12}=-(p_1+p_2)=p_3+p_4$
and $P_{14}=-(p_1+p_4)=p_2+p_3$ where $P_{12}$ and $P_{14}$ are the momenta of the
propagators of the respective diagrams. By writing down the Feynman rules for the
different diagrams it can quickly be seen that the contributions of the 1st and the
3rd diagrams both vanish. The 2nd diagram gives:
\begin{eqnarray}
\label{2nddiagram}
\cA_4&=&\left(-g\over 2\sqrt{2}\right)^2\left(\sqrt{2}\right)^4
\frac{\l_\a^1\lt_\da^4\l_\b^2\lt_\db^4\l_{\sigma}^1\lt_{\dot{\sigma}}^3\l_{\tau}^1\lt_{\dot{\tau}}^4}
{[4\,1][4\,2]\langle 1\,4\rangle\langle 1\,3\rangle}\Big[\epsilon^{\da\db}\epsilon^{\a\b}
(p_1\!-\!p_2)^{\dot{\gamma}\gamma}\nonumber\\
&+&\epsilon^{\db\dot{\gamma}}\epsilon^{\b\gamma}(p_2\!-\!P_{12})^{\da\a}
+\epsilon^{\dot{\gamma}\da}\epsilon^{\gamma\a}(P_{12}-p_1)^{\db\b}\Big]\times
\frac{-2i\epsilon_{\dot{\gamma}\dot{\delta}}\epsilon_{\gamma\delta}}{P_{12}^2}\nonumber\\
&\times &\Big[\epsilon^{\dot{\sigma}\dot{\tau}}\epsilon^{\sigma\tau}(p_3-p_4)^{\dot{\delta}\delta}
+\epsilon^{\dot{\tau}\dot{\delta}}\epsilon^{\tau\delta}(p_4+P_{12})^{\dot{\sigma}\sigma}
+\epsilon^{\dot{\delta}\dot{\sigma}}\epsilon^{\delta\sigma}(-P_{12}-p_3)^{\dot{\tau}\tau}\Big]\nonumber\\
&=&-\frac{4ig^2}{\langle 1\,2\rangle[2\,1]}\frac{(\l_2^\a\l_{\a}^1)(\lt_{\da}^4\lt_2^\da)(\l_3^{\tau}\l_{\tau}^1)
(\lt_{\dot{\tau}}^4\lt_3^{\dot{\tau}})(\lt_{\dot{\sigma}}^3\lt_{\dot{\beta}}^4\epsilon^{\db\dot{\sigma}})
(\epsilon^{\b\sigma}\l_{\sigma}^1\l_{\b}^2)}
{[4\,1][4\,2]\langle 1\,4\rangle\langle 1\,3\rangle}\nonumber\\
&=&-4ig^2\frac{\langle 1\,2\rangle[3\,4]^2}{\langle 1\,4\rangle[1\,2][4\,1]}\ .
\end{eqnarray}
This is our answer, though it is in a rather unfamiliar form! We can convert it into something
more familiar by multiplying both top and bottom by $\langle 2\,3\rangle\langle 3\,4\rangle$.
We then use momentum conservation in the numerator in the form
$\langle 2\,3\rangle[3\,4]=-\langle 2\,1\rangle[1\,4]$ and recognise that
$s_{34}:=2(p_3\cdot p_4)=\langle 3\,4\rangle[4\,3]=s_{12}=\langle 1\,2\rangle[2\,1]$ to give
\begin{eqnarray}
\label{mhv4point}
\cA_4&=&-4ig^2\frac{\langle 1\,2\rangle(\langle 2\,3\rangle[3\,4])(\langle 3\,4\rangle[4\,3])}
{[1\,2]\langle 2\,3\rangle\langle 3\,4\rangle\langle 4\,1\rangle[41]}\nonumber\\
&=&-4ig^2\frac{\langle 1\,2\rangle^3}{\langle 2\,3\rangle\langle 3\,4\rangle\langle 4\,1\rangle}\ ,
\end{eqnarray}
which is the usual form for the Parke-Taylor amplitude at 4-point.

\paragraph{MHV $\mathbf{q\bar{q}\rightarrow gg}$}\mbox{}

\begin{figure}[ht]
\begin{center}
\begin{picture}(440,120)(0,0)
\put(0,10){
\SetOffset(60,0)
\Photon(40,50)(80,50){3}{4}
\ArrowLine(25,80)(40,50)\Text(23,85)[r]{$p_2^+$}
\ArrowLine(40,50)(25,20)\Text(23,15)[r]{$p_1^{-}$}
\Photon(95,20)(80,50){3}{4}\Text(100,17)[l]{$p_4^+$}
\Photon(80,50)(95,80){3}{4}\Text(100,85)[l]{$p_3^-$}
\Vertex(40,50){2} \Vertex(80,50){2}
}
\put(205,60){\Text(0,0)[l]{+}}
\put(180,0){
\SetOffset(60,0)
\ArrowLine(50,80)(50,40)
\ArrowLine(25,100)(50,80)\Text(23,100)[r]{$p_2^+$}
\ArrowLine(50,40)(25,20)\Text(23,20)[r]{$p_1^-$}
\Photon(75,20)(50,40){3}{4}\Text(80,20)[l]{$p_4^+$}
\Photon(50,80)(75,100){3}{4}\Text(80,100)[l]{$p_3^-$}
\Vertex(50,40){2}\Vertex(50,80){2}}
\end{picture}
\label{4ptmhv}
\caption{\emph{The diagrams for $\tilde{\cA}(1_q^-,2_{\bar{q}}^+,3_g^-,4_g^+)$.}}
\end{center}
\end{figure}

Again we take all momenta to be outgoing. Momentum conservation is the same as for diagrams 2 and 3 of the
previous example and for the gluon wavefunctions we take $q_3=p_4$ and $q_4=p_1$. This
gives polarisation vectors
\begin{displaymath}
\begin{array}{cc}
\epsilon_{3\,\a\da}^-=\sqrt{2}\frac{\l^3_\a\lt^4_\da}{[4\,3]} &
\epsilon_{4\,\a\da}^+=\sqrt{2}\frac{\l^1_\a\lt^4_\da}{\langle 1\,4\rangle}\ .
\end{array}
\end{displaymath}
The second diagram can be seen to vanish while the first gives:
\begin{eqnarray}
\label{qqgg}
\tilde{\cA}_4&=&(\lt_2^\da\epsilon^{\a\gamma}\l^1_{\gamma})(ig\sqrt{2}\delta_\a^{\phantom{\a}\b}
\delta_{\da}^{\phantom{\da}\db})\left(\frac{-2i\epsilon_{\db\dot{\delta}}\epsilon_{\b\delta}}{P_{12}^2}\right)
\left(\frac{-g}{2\sqrt{2}}\right)\Big(\epsilon^{\dot{\sigma}\dot{\tau}}\epsilon^{\sigma\tau}(p_3\!-\!p_4)^{\dot{\delta}\delta}
\nonumber\\
&+&\epsilon^{\dot{\tau}\dot{\delta}}\epsilon^{\tau\delta}(p_4\!+\!P_{12})^{\dot{\sigma}\sigma}
+\epsilon^{\dot{\delta}\dot{\sigma}}\epsilon^{\delta\sigma}(-P_{12}\!-\!p_3)^{\dot{\tau}\tau}\Big)
(\sqrt{2})^2\frac{\l^3_{\sigma}\lt^4_{\dot{\sigma}}\l^1_{\tau}\lt^4_{\dot{\tau}}}{[4\,3]\langle 1\,4\rangle}
\nonumber\\
&=&-4g^2\frac{(\lt_2^{\db}\epsilon_{\db\dot{\delta}}\epsilon^{\dot{\delta}\dot{\sigma}}\lt^4_{\dot{\sigma}})
(\l_1^\b\epsilon_{\b\delta}\epsilon^{\delta\sigma}\l^3_{\sigma})(\l^1_{\tau}\l_3^{\tau})(\lt^4_{\dot{\tau}}
\lt_3^{\dot{\tau}})
}{\langle 1\,2\rangle[2\,1][4\,3]\langle 1\,4\rangle}\nonumber\\
&=&-4g^2\frac{\langle 1\,3\rangle^2[4\,2]}{\langle 1\,2\rangle [2\,1]\langle 4\,1\rangle}\nonumber\\
&=&4g^2\frac{\langle 1\,3\rangle^3}{\langle 1\,2\rangle\langle 3\,4\rangle\langle 4\,1\rangle}\nonumber\\
&=&i\frac{\langle 2\,3\rangle}{\langle 1\,3\rangle}\cA(1_g^-,2_g^+,3_g^-,4_g^+)\ ,
\end{eqnarray}
thus verifying the relations between amplitudes that we derived from supersymmetric Ward identities
in \S \ref{SWI}.

\paragraph{MHV $\mathbf{q\bar{q}\rightarrow q\bar{q}}$}\mbox{}

As a final example let us consider the amplitude $\hat{\cA}(1^-_q,2^+_{\bar{q}},3^-_q,4^+_{\bar{q}})$. This time
both of the diagrams
\begin{figure}[ht]
\begin{center}
\begin{picture}(440,120)(0,0)
\put(0,10){
\SetOffset(60,0)
\Photon(40,50)(80,50){3}{4}
\ArrowLine(25,80)(40,50)\Text(23,85)[r]{$p_2^+$}
\ArrowLine(40,50)(25,20)\Text(23,15)[r]{$p_1^{-}$}
\ArrowLine(95,20)(80,50)\Text(100,17)[l]{$p_4^+$}
\ArrowLine(80,50)(95,80)\Text(100,85)[l]{$p_3^-$}
\Vertex(40,50){2} \Vertex(80,50){2}
}
\put(205,60){\Text(0,0)[l]{+}}
\put(180,0){
\SetOffset(60,0)
\Photon(50,80)(50,40){3}{4}
\ArrowLine(25,100)(50,80)\Text(23,100)[r]{$p_2^+$}
\ArrowLine(50,40)(25,20)\Text(23,20)[r]{$p_1^-$}
\ArrowLine(75,20)(50,40)\Text(80,20)[l]{$p_4^+$}
\ArrowLine(50,80)(75,100)\Text(80,100)[l]{$p_3^-$}
\Vertex(50,40){2}\Vertex(50,80){2}}
\end{picture}
\label{4ptmhv}
\caption{\emph{The diagrams for $\hat{\cA}(1_q^-,2_{\bar{q}}^+,3_q^-,4_{\bar{q}}^+)$.}}
\end{center}
\end{figure}
are non-zero. The first one gives
\begin{eqnarray}
\label{qqqq}
\hat{\cA}_4^1&=&(\lt^\da_2\epsilon^{\a\gamma}\l_{\gamma}^1)(ig\sqrt{2}\delta_\a^{\phantom{\a}\delta}
\delta_\da^{\phantom{\da}\dot{\delta}})\left(\frac{-2i\epsilon_{\dot{\delta}\dot{\sigma}}\epsilon_{\delta\sigma}}
{P_{12}^2}\right)(ig\sqrt{2}\delta_{\beta}^{\phantom{\beta}\sigma}\delta_{\dot{\beta}}^{\phantom{\dot{\beta}}\dot{\sigma}})
(\lt_4^\db\epsilon^{\b\tau}\lambda_{\tau}^3)\nonumber\\
&=&-4ig^2\frac{(\lt^2_{\db}\lt_4^{\db})(\l_3^\b\l_\b^1)}{\langle 1\,2\rangle[2\,1]}\nonumber\\
&=&-4ig^2\frac{[2\,4]\langle 3\,1\rangle}{\langle 1\,2\rangle[2\,1]}\nonumber\\
&=&4ig^2\frac{\langle 1\,3\rangle^2}{\langle 1\,2\rangle\langle 3\,4\rangle}\ .
\end{eqnarray}
A similar calculation - or equivalently the realisation that diagrams two is simply the same as
diagram one with $2\leftrightarrow 4$ - gives
\be
\label{qqqq2}
\hat{\cA}_4^2=4ig^2\frac{\langle 1\,3\rangle^2}{\langle 2\,3\rangle\langle 4\,1\rangle}
\ee
for the second diagram and thus the total is
\begin{eqnarray}
\label{qqqqtotal}
\hat{\cA}_4&=&\hat{\cA}_4^1+\hat{\cA}_4^2\nonumber\\
&=&4ig^2\frac{\langle 1\,3\rangle^2}{\langle 1\,2\rangle\langle 2\,3\rangle\langle 3\,4\rangle\langle 4\,1\rangle}
(\langle 1\,2\rangle\langle 3\,4\rangle+\langle 2\,3\rangle\langle 4\,1\rangle)\nonumber\\
&=&-\cA(1_g^-,2_g^+,3_g^-,4_g^+)\left(\langle 1\,2\rangle\langle 3\,4\rangle+\langle 2\,3\rangle\langle 4\,1\rangle\over
\langle 1\,3\rangle^2\right)\ .
\end{eqnarray}

\chapter{D-Dimensional Lorentz-Invariant Phase Space}
\label{DLIPSappendix}

In this appendix we expound on the $D$-dimensional measure for
Lorentz-invariant two-body phase space, ultimately focussing on the case $D=4-2\epsilon$.

\section{D-spheres}
One thing that we will need to consider is the volume of a $D$-dimensional unit sphere $V(\mathbb{S}^D)$. We mean
this in the sense of a $D$-sphere regarded as a manifold. Thus the volume we are talking about is
the volume of that manifold rather than the volume enclosed by it when it is regarded as being embedded in one-dimension
higher. Thus $V(\mathbb{S}^1)=2\pi$ - the circumference of a circle - and $V(\mathbb{S}^2)=4\pi$, the surface area of a
sphere such as the Earth.

In fact we can parametrize a round $D$-sphere in terms of $D$ angles $\theta_i$.
In this case the volume element of an $\mathbb{S}^D$ is given by
\be
\label{dvsd}
dV(\mathbb{S}^D)=d\theta_1\ldots d\theta_D(\sin\theta_1)^{D-1}(\sin\theta_2)^{D-2}\ldots(\sin\theta_{D-1})^1\ ,
\ee
with the result
\begin{eqnarray}
\label{vsd}
V(\mathbb{S}^D)&=&\int_{\theta_i=0}^{\substack{\theta_j=\pi, j\neq D\\ \theta_D=2\pi}}dV(\mathbb{S}^D)\nonumber\\
&=&\frac{2\pi^{\frac{D+1}{2}}}{\Gamma\left(\frac{D+1}{2}\right)}\ .
\end{eqnarray}

\section{\emph{d}LIPS}
Recall from Chapter 1, Equation~\eq{LIPSD} that
\be
\label{LIPSD2}
d^D\mathrm{LIPS}(l_2^-,-l_1^+;P)=d^Dl_1d^Dl_2\delta^{(+)}\!(l_1^2)\delta^{(-)}\!(l_2^2)\delta^{(D)}\!(P+l_2-l_1)\ ,
\ee
where $\delta^{(\pm)}\!(l^2):=\theta(\pm l_0)\delta(l^2)$, $\theta$ is the unit step function\footnote{Not
to be confused with the angles $\theta_i$ of \eq{dvsd} and \eq{vsd}.} and $l_0$ the
0-component (energy) of $l$. If we also remember that
\be
\label{deltaff}
\int\!dx\,g(x)\delta(f(x)-a)=\left.\frac{g(x)}{\left|\frac{df}{dx}\right|}\right|_{x=x_0;\,f(x_0)=a}
\ee
then we can integrate over the 0-components of $l_1$ and $l_2$ to get
\be
d^D\mathrm{LIPS}=\frac{d^{D-1}\vec{l_1}}{2|l_{10}|}\frac{d^{D-1}\vec{l_2}}{2|l_{20}|}
\delta^{(D)}\!(P+l_2-l_1)\Bigg|_{\substack{ l_{10}=|\vec{l_1}|\\ l_{20}=-|\vec{l_2}|}}\ ,
\ee
where $\vec{l}$ represents the spatial components of the $D$-vector $l$. Furthermore, going to the
center of mass frame for the vector $P^{\mu}$, $P=(P_0,\vec{0})$ we can use $D-1$ of the remaining
delta functions to localise the integral:
\begin{eqnarray}
\label{d-1}
d^D\mathrm{LIPS}&=&\frac{d^{D-1}\vec{l_1}}{2|\vec{l_1}|}\frac{d^{D-1}\vec{l_2}}{2|\vec{l_1}|}
\delta^{(D-1)}\!(\vec{l_2}-\vec{l_1})\delta(P_0-2|\vec{l_1}|)\nonumber\\
&=&\frac{1}{2}\frac{d^{D-1}\vec{l_{1}}}{4|\vec{l_{1}}|^2}\,\delta\!\left(|\vec{l_1}|-\frac{P_0}{2}\right)\ .
\end{eqnarray}

Now, for $d^{n}\vec{l}$ we can write
\be
\label{dnl}
d^{n}\vec{l}=d|\vec{l_{\phantom{1}}}\!|\,|\vec{l_{\phantom{1}}}\!|^{n-1}dV(\mathbb{S}^{n-1})\ ,
\ee
so we have
\begin{eqnarray}
\label{gettingthere}
d^{D-1}|\vec{l_1}|&=&d|\vec{l_1}|\,|\vec{l_1}|^{D-2}d\theta_1 d\theta_2(\sin\theta_1)^{D-3}(\sin\theta_2)^{D-4}\nonumber\\
&\times &\left(d\theta_3\ldots d\theta_{D-2}(\sin\theta_3)^{D-5}\ldots(\sin\theta_{D-3})\right)\nonumber\\
&=&d|\vec{l_1}|\,|\vec{l_1}|^{D-2}d\theta_1 d\theta_2(\sin\theta_1)^{D-3}(\sin\theta_2)^{D-4}dV(\mathbb{S}^{D-4})\ .
\end{eqnarray}
For our case of a 2-particle phase space in $4-2\epsilon$ dimensions, 2 angles $\theta_1$ and $\theta_2$ are sufficient and none
of the momenta will depend on any of the other angles. We can thus integrate over them to get
\begin{eqnarray}
\label{reallynearly}
d^{D-1}|\vec{l_1}|&=&d|\vec{l_1}|\,|\vec{l_1}|^{D-2}d\theta_1 d\theta_2(\sin\theta_1)^{D-3}(\sin\theta_2)^{D-4}V(\mathbb{S}^{D-4})\nonumber\\
&=&\frac{2\pi^{\frac{D-3}{2}}}{\Gamma\left(\frac{D-3}{2}\right)}d|\vec{l_1}|\,|\vec{l_1}|^{D-2}d\theta_1 d\theta_2(\sin\theta_1)^{D-3}
(\sin\theta_2)^{D-4}\ .
\end{eqnarray}

With $D=4-2\epsilon$ this leads us to
\begin{eqnarray}
\label{d4minus2}
d^{4-2\epsilon}\mathrm{LIPS}&=&\frac{1}{2}\frac{d^{3-2\epsilon}}{4|\vec{l_1}|^2}\delta\!\left(|\vec{l_1}|-\frac{P_0}{2}\right)\nonumber\\
&=&\frac{\pi^{\frac{1}{2}-\epsilon}}{4\Gamma\left(\frac{1}{2}-\epsilon\right)}\left|\frac{P_0^2}{4}\right|^{-\epsilon}
d\theta_1 d\theta_2(\sin\theta_1)^{1-2\epsilon}(\sin\theta_2)^{-2\epsilon}\nonumber\\
&=&\frac{\pi^{\frac{1}{2}-\epsilon}}{4\Gamma\left(\frac{1}{2}-\epsilon\right)}\left|\frac{P^2}{4}\right|^{-\epsilon}
d\theta_1 d\theta_2(\sin\theta_1)^{1-2\epsilon}(\sin\theta_2)^{-2\epsilon}\ ,
\end{eqnarray}
and
\begin{eqnarray}
\label{intd4minus2}
\int\!d^{4-2\epsilon}\mathrm{LIPS}=\frac{\pi^{\frac{1}{2}-\epsilon}}{4\Gamma\left(\frac{1}{2}-\epsilon\right)}\left|\frac{P^2}{4}\right|^{-\epsilon}
\int_{\theta_1=0}^{\pi}\int_{\theta_2=0}^{\pi}d\theta_1 d\theta_2(\sin\theta_1)^{1-2\epsilon}(\sin\theta_2)^{-2\epsilon}\ .
\end{eqnarray}

\section{Overall amplitude normalisation}

In the original papers of \cite{Bern:zx,Bern:1994cg}, the one-loop amplitudes derived are normalised
with a factor of $c_{\Gamma}=r_{\Gamma}/(4\pi)^{2-\epsilon}$ where
\be
\label{rgamma}
r_{\Gamma}=\frac{\Gamma(1+\epsilon)\Gamma^2(1-\epsilon)}{\Gamma(1-2\epsilon)}\ .
\ee
In \cite{bst,bbst,bbst2} and this thesis, however, the normalisation most naturally arises as
\be
\label{bstnormalisation}
\frac{\pi\epsilon}{\sin\pi\epsilon}\frac{1}{\Gamma\left(\frac{1}{2}-\epsilon\right)}\ ,
\ee
where the gamma function comes from the LIPS measure described above and the factor of
$\pi\epsilon\csc\pi\epsilon$ comes from performing the dispersion integral (see \emph{e.g.} Section 5 of
\cite{bst}). We are mostly interested in the results of these amplitude calculations up to order $\epsilon^0$, and
as \eq{bstnormalisation} $=1/\sqrt{\pi}+\mathcal{O}(\epsilon)$ we have usually dropped it as an uninteresting overall factor.
Nonetheless, the all-orders in $\epsilon$ results can be useful and we will here show how the two are related.

To start with there is the product identity for gamma functions:
\be
\label{product}
\Gamma(z)\Gamma(1-z)=\frac{\pi}{\sin\pi z}\ ,
\ee
which can be combined with the well-known recurrence relation $z\Gamma(z)=\Gamma(z+1)$ to give
\be
\label{sinpie}
\frac{\pi\epsilon}{\sin\pi\epsilon}=\Gamma(1+\epsilon)\Gamma(1-\epsilon)\ .
\ee

There is also the Legendre duplication formula:
\be
\label{legendre}
\Gamma(z)\Gamma\left(z+1/{2}\right)=2^{1-2z}\sqrt{\pi}\ \Gamma(2z)\ ,
\ee
which implies that
\be
\label{gammahalf}
\Gamma\left({1}/{2}-\epsilon\right)=\frac{\Gamma(1-2\epsilon)\sqrt{\pi}\,2^{2\epsilon}}{\Gamma(1-\epsilon)}\ .
\ee

This therefore leads us to
\begin{eqnarray}
\label{finalnorm}
\frac{\pi\epsilon}{\sin\pi\epsilon}\frac{1}{\Gamma\left(\frac{1}{2}-\epsilon\right)}&=&
\frac{1}{4^{\epsilon}\sqrt{\pi}}\frac{\Gamma(1+\epsilon)\Gamma^2(1-\epsilon)}{\Gamma(1-2\epsilon)}\nonumber\\
&=&\frac{r_{\Gamma}}{4^{\epsilon}\sqrt{\pi}}\ ,
\end{eqnarray}
and we can see that the two are the same up to a simple factor.

\chapter{Unitarity}
\label{unitarityappendix}

Unitarity is a well-known and useful tool in quantum field theory
\cite{Gell-Mann:1954db,Landau:1959fi,Mandelstam:1958xc,Mandelstam:1959bc,Cutkosky:1960sp,bib}.\footnote{Note that some of this
appendix is based on Section 7.3 of \cite{pc}.} The unitarity of the
$S$-matrix, $S^{\dag}\!S=1$, is the basic starting point and leads to the possibility of being able to
reconstruct scattering amplitudes from the knowledge of their properties as
functions of complex momenta. In certain cases this can lead to a purely algebraic construction of
amplitudes.

It can be checked that each Feynman diagram contributing to an $S$-matrix element $\mathcal{S}$ is purely real unless
some denominator vanishes, in which case the $i\varepsilon$ prescription for treating poles becomes relevant. We thus get an
imaginary part for $\mathcal{S}$ only when virtual particles in a Feynman diagram go on-shell.

Consider now $\mathcal{S}(s)$ as an analytic function of a complex variable $s$. $s$ is the square of the
centre of mass energy, and while this is physically real we will consider it to be complex for now. If $s_0$ is the
minimum (square of the) energy for production of the lightest multiparticle state (\emph{i.e.} the minimum energy
for the creation of an intermediate multiparticle state such as when a loop is formed in a Feynman diagram), then for real $s$ lying
below $s_0$, the intermediate state cannot go on-shell. $\mathcal{S}(s)$ is thus real and we have
\be
\label{ms}
\mathcal{S}(s)=\overline{\mathcal{S}}(\bar{s})\ .
\ee
However, as we are regarding $\mathcal{S}(s)$ as an analytic function of $s$, we can analytically continue this
equation to anywhere in the complex plane. If we explicitly split $\mathcal{S}(s)$ into its real and imaginary parts,
\mbox{$\mathcal{S}(s)=\Re[\mathcal{S}(s)]+i\Im[\mathcal{S}(s)]$}, then at a point $s>s_0$ that is $\varepsilon$ away from the real line
\eq{ms} implies that
\begin{eqnarray}
\label{reims}
\Re[\mathcal{S}(s+i\varepsilon)]&=&\,\,\,\,\,\Re[\mathcal{S}(s-i\varepsilon)]\ ,\nonumber\\
\Im[\mathcal{S}(s+i\varepsilon)]&=&-\Im[\mathcal{S}(s-i\varepsilon)]\ .
\end{eqnarray}
There is thus a branch cut along the positive real axis starting at $s_0$ and the discontinuity $\mathfrak{D}$ of
$\mathcal{S}(s)$ across the cut is
\be
\label{discontinuity}
\mathfrak{D}[\mathcal{S}(s)]=2i\Im[\mathcal{S}(s+i\varepsilon)]\ .
\ee
It turns out that this discontinuity - which only arises because we have intermediate multiparticle states and
thus loop contributions to Feynman diagrams - can be related to simpler amplitudes which may be known already
or more easily computed. This is the content of the optical theorem which we review below.

\section{The optical theorem}

The $S$-matrix is a unitary operator which evolves the initial states $k_a$ so that one may compute their overlap with
the final states $p_i$ in a scattering process:
\be
\label{inout}
\mbox{}_{\mathrm{out\,}}\!\langle p_i|k_a\rangle_{\mathrm{in}}=\langle p_i|S|k_a\rangle\ .
\ee
It is conventional to split $S$ into the part that describes unimpeded propagation of the initial particles
and a part $T$ due to interactions, $S=1+iT$. The matrix element \eq{inout} taken with the interacting part of
$S$ is what then gives a scattering amplitude. More concretely, we can write
\be
\label{M}
\langle p_i|T|k_a\rangle=(2\pi)^4\delta^{(4)}\!\left(\sum(p_i+k_a)\right)\mathcal{S}(k_a\rightarrow p_i)\ ,
\ee
where we have taken all particles to be outgoing.

Unitarity of $S$, $S^{\dag}\!S=1$ implies
\be
\label{unitarity}
-i(T-T^{\dag})=T^{\dag}T\ ,
\ee
and we may extract some useful information by taking the matrix element of this between some
particle states $p_i$ and $k_a$. The LHS of \eq{unitarity} gives
\begin{eqnarray}
\label{LHSun}
-i(\langle p_i|T|k_a\rangle-\langle p_i|T^{\dag}|k_a\rangle)\!\!&=&\!\!
-i\left(\langle p_i|T|k_a\rangle-\overline{\langle k_a|T|p_i\rangle}\right)\nonumber\\
\!\!&=&\!\!-i(2\pi)^4\delta^{(4)}\!\left(\sum(p_i+k_a)\right)
\left(\mathcal{S}(k_a\!\rightarrow\! p_i)-\overline{\mathcal{S}}(p_i\!\rightarrow\! k_a)\right)\nonumber\\
\!\!&=&\!\!-i(2\pi)^4\delta^{(4)}\!\left(\sum(p_i+k_a)\right)\mathfrak{D}[\mathcal{S}(p_i;k_a)]\ .
\end{eqnarray}
On the RHS of \eq{unitarity} we can insert the identity operator as a sum over a
complete set of intermediate states to obtain
\begin{eqnarray}
\label{RHSun}
\langle p_i|T^{\dag}T|k_a\rangle &=&\sum_n\left(\prod_{j=1}^n\int\!\frac{d^4l_j}{(2\pi)^4}\delta(l_j^2-m_j^2)\right)
\langle p_i|T^{\dag}|l_j\rangle\langle l_j|T|k_a\rangle\nonumber\\
&=&(2\pi)^4\sum_n\left(\prod_{j=1}^n\int\!\frac{d^4l_j}{(2\pi)^4}\delta(l_j^2-m_j^2)\right)
\overline{\mathcal{S}}(p_i\rightarrow l_j)\mathcal{S}(k_a\rightarrow l_j)\nonumber\\
&{}&\,\,\times\,\, \delta^{(4)}\!\left(\sum(p_i+l_j)\right)\delta^{(4)}\!\left(\sum(k_a-l_j)\right)\nonumber\\
&=&(2\pi)^4\delta^{(4)}\!\left(\sum(p_i+k_a)\right)
\sum_n\int\!d\mathrm{LIPS}(n)\overline{\mathcal{S}}(p_i\rightarrow l_j)\mathcal{S}(k_a\rightarrow l_j)\ ,\nonumber\\
\end{eqnarray}
where $d\mathrm{LIPS}(n)$ is the $n$-body Lorentz-invariant phase space measure. Putting the LHS and RHS of
\eq{unitarity} back together again we find\footnote{In fact the optical theorem is usually
stated in terms of the forward scattering amplitude, in which case we have $k_a=p_i$. The theorem is more general than this though and
can be applied to generic asymptotic states.}
\begin{eqnarray}
\label{optic}
-i\mathfrak{D}[\mathcal{S}(p_i;k_a)]=\sum_n\int\!d\mathrm{LIPS}(n)\overline{\mathcal{S}}(p_i\rightarrow l_j)\mathcal{S}(k_a\rightarrow l_j)\ .
\end{eqnarray}

Equation \eq{optic} says that the discontinuity of a scattering amplitude may be obtained as
a sum of integrals over the phase spaces of intermediate multiparticle states of the amplitudes for
scattering of the initial and final states into these intermediate states. In particular, for
a one-loop process, the amplitudes arising on the RHS of \eq{optic} are tree-level amplitudes and the
phase space is a 2-particle one.

\section{Cutting rules}

Cutkosky showed that using some \emph{cutting} rules, one may compute the physical discontinuity of
any Feynman diagram and prove the optical theorem to all orders in perturbation theory
\cite{Cutkosky:1960sp}. The rules are as follows \cite{pc}:
\begin{enumerate}
\item Cut through a diagram in all possible ways such that the cut propagators may be put on-shell.

\item For each cut (massive) propagator replace $1/(p^2-m^2+i\varepsilon)$ with a delta function
\mbox{$-2\pi i\delta(p^2-m^2)$}.
This explicitly provides the delta functions which generate the $d$LIPS measure in \eq{RHSun}. The off-shell
vertices that are separated by the cut are thus put on-shell. For massless
momenta the replacement is simply $1/(p^2+i\varepsilon)\rightarrow-2\pi i\delta(p^2)$.

\item Sum the contributions of all possible cuts.
\end{enumerate}

For example, for a Feynman diagram in massless $\lambda\phi^3$ theory such as Figure D.1,
\begin{figure}[ht]
\begin{center}
\begin{picture}(450,100)(0,0)
\SetOffset(-125,0)
\put(0,0){
\Text(273,46)[r]{$p_1$}
\Text(400,46)[l]{$p_2$}
\Line(280,46)(310,46)
\COval(335,46)(20,25)(0){Black}{White}
\DashLine(335,71)(335,21){3}
\Text(320,72)[r]{$l_2$}
\Text(350,22)[l]{$l_1$}
\Line(360,46)(390,46)}
\end{picture}
\end{center}
\label{fibub}
\caption{\emph{The cut of a bubble diagram in massless $\lambda\phi^3$ theory.}}
\end{figure}
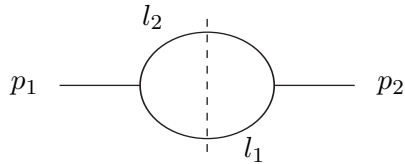
the Feynman rules would give
\begin{eqnarray}
\label{phi3}
\cA\ \propto\ \delta^{(4)}(p_1+p_2)\int\!\frac{d^4l_1}{(2\pi)^4}\frac{d^4l_2}{(2\pi)^4}\lambda\frac{1}{l_1^2}\frac{1}{l_2^2}\lambda
\delta^{(4)}(p_1+l_1-l_2)\ .
\end{eqnarray}
Cutkosky's rules on the other hand would give
\begin{eqnarray}
\label{phicut}
\mathfrak{D}[\cA]&\propto &
\delta^{(4)}(p_1+p_2)\int\!\frac{d^4l_1}{(2\pi)^4}\frac{d^4l_2}{(2\pi)^4}\lambda\delta(l_1^2)\delta(l_2^2)\lambda\delta^{(4)}(p_1+l_1-l_2)\nonumber\\
&\propto&\lambda^2\delta^{(4)}(p_1+p_2)\int\!d\mathrm{LIPS}(l_2,-l_1;p_1)\ ,
\end{eqnarray}
which allows one to calculate the discontinuity of the diagram concerned.

\subsection{BDDK's unitarity cuts}

In \cite{Bern:zx,Bern:1994cg} Cutkosky's rules were applied at the level of amplitudes to derive
one-loop MHV amplitudes in supersymmetric and non-supersymmetric gauge theories. In this case the
factors on either side of the cut are not vertices (\emph{e.g.} the $\lambda$ factors of \eq{phicut}),
but full amplitudes. In fact for the one-loop MHV amplitudes these
factors are tree-level MHV amplitudes.

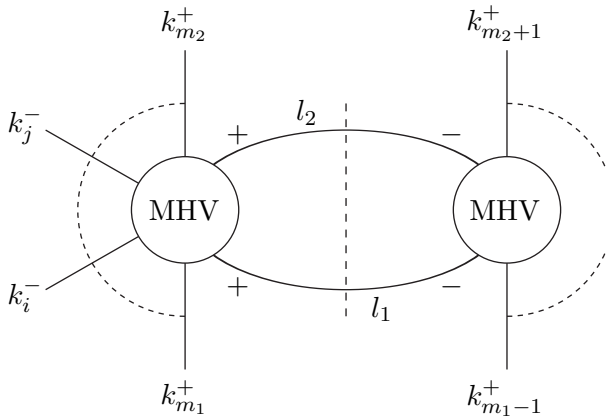
\begin{figure}[ht]
\begin{center}
\begin{picture}(450,160)(0,0)
\SetOffset(65,0)
\put(0,0){
\COval(140,80)(30,60)(0){Black}{White}
\DashLine(140,120)(140,40){3}
\Text(130,117)[r]{$l_2$}
\Text(150,43)[l]{$l_1$}
\Line(80,80)(28,110)
\Text(27,112)[r]{$k_j^-$}
\Text(27,48)[r]{$k_i^-$}
\Line(80,80)(28,50)
\Line(200,100)(200,140)
\Text(200,144)[b]{$k_{m_2+1}^+$}
\Text(200,16)[t]{$k_{m_1-1}^+$}
\Line(200,60)(200,20)
\BCirc(80,80){20}
\BCirc(200,80){20}
\Text(80,84)[t]{MHV}
\Text(200,84)[t]{MHV}
\Text(100,105)[b]{$+$}
\Text(100,55)[t]{$+$}
\Text(180,105)[b]{$-$}
\Text(180,55)[t]{$-$}
\DashCArc(80,80)(40,90,270){2}
\DashCArc(200,80)(40,270,90){2}
\Line(80,100)(80,140)
\Text(80,144)[b]{$k_{m_2}^{+}$}
\Line(80,60)(80,20)
\Text(80,16)[t]{$k_{m_1}^+$}}
\end{picture}
\end{center}
\label{bddkcut}
\caption{\emph{The cut of a one-loop MHV amplitude in the $t_{m_1}^{[m_2-m_1+1]}$ channel.}}
\end{figure}

Consider for concreteness the $n$-point one-loop MHV amplitudes for gluon scattering in $\cN\!=\!4$ super-Yang-Mills as
reviewed in \S \ref{mhvsection}. We would like to see how these can be obtained from 2-particle
cuts as in \cite{Bern:zx}.

By analogy with the Cutkosky rules, the procedure is to consider `cuts' in every possible kinematical
channel and then add the contributions without overcounting. We are then left with LIPS integrals as above (but this time
with non-trivial kinematic factors in the integrand) which can in-principle be
evaluated to reveal the discontinuities of the amplitude. However, BDDK recover the amplitude by using an
algebraic procedure which means that these dispersion integrals do not actually need to be done. This involves
replacing the delta functions associated with the cuts with propagators (a procedure that is known as `reconstruction
of the Feynman integral') which then produces Feynman integrals rather than LIPS
integrals. These integrals contain cuts in the channel being considered (as well as cuts in other
channels too) and by considering all channels and avoiding over-counting the amplitude can be
re-constructed.

When we cut the amplitudes, we must assign
helicities to the particles that were in the loop. Since we use conventions in which
all particles are outgoing, the helicities of these internal particles are reversed.
For the one-loop MHV amplitudes there are two distinct cases. Case $(a)$ is where the negative-helicity external particles $i$ and $j$ are on the same side of the cut, and case $(b)$ is where
they are on opposite sides of the cut. Case $(a)$, is \emph{a priori} the simpler of the two as the
two internal particles must have the same helicities and thus amplitude relations of equations
(\ref{fermiscalar}) and (\ref{scalarfermi}) mean that only gluons can circulate in the loop. This is the situation
regardless of the amount of supersymmetry present. Case
$(b)$ involves the entire multiplet circulating in the loop and for maximally supersymmetric Yang-Mills
it turns out that this case is the same as case $(a)$ after applying identities such as the
Schouten identity \eq{schouten}. For the case being considered of $\cN\!=\!4$ Yang-Mills it is thus
enough for us to treat case $(a)$ only.

Consider now a cut in the channel where $P_L$, the momentum on the left of the cut, is given
by $P_L^2=(k_{m_1}+k_{m_1+1}+\ldots+k_{m_2-1}+k_{m_2})^2=t_{m_1}^{[m_2-m_1+1]}$ and where $k_i,k_j\in P_L$.
This situation is shown in Figure D.2 and the rules that we have outlined above give
\begin{eqnarray}
\label{bddkdisp}
\mathfrak{D}[\cA(t_{m_1}^{[m_2-m_1+1]})]&=&\int\!\frac{d^{4}l_1}{(2\pi)^4}\frac{d^{4}l_2}{(2\pi)^4}\,\cA_{\mathrm{tree}}^{\mathrm{MHV}}
(-l_1^+,m_1^+,\ldots,i^-,\ldots,j^-,\ldots,m_2^+,l_2^+)\nonumber\\
&{}&\!\!\times\,\,\,\,\,\delta(l_1^2)\,\,\delta(l_2^2)\,\,\,\cA_{\mathrm{tree}}^{\mathrm{MHV}}
(-l_2^-,{m_{2}+1}^+,\ldots,{m_{1}-1}^+,l_1^-)\nonumber\\
&\equiv &\frac{i}{(2\pi)^4}\,\cA_{\mathrm{tree}}^{\mathrm{MHV}}(i^-,j^-)\int\!d\mathrm{LIPS}(l_2,-l_1;P_{L})\,\,\hat{\mathcal{R}}\\
\label{bdkcomplete}
&\rightarrow &\frac{i}{(2\pi)^4}\,\cA_{\mathrm{tree}}^{\mathrm{MHV}}(i^-,j^-)\int\!d^4l_1d^4l_2\frac{1}{l_1^2}\frac{1}{l_2^2}
\hat{\mathcal{R}}\ ,
\end{eqnarray}
where
\be
\label{Ragain}
\hat{\cR} \ := \ {
\lan m_1 - 1 \, m_1 \ran \lan l_2 \, l_1\ran \over \lan m_1 -1 \,
l_1 \ran \lan -l_1 \, m_1 \ran}\ { \lan m_2  \, m_2 + 1 \ran \lan
l_1 \, l_2 \ran \over \lan m_2  \, l_2 \ran \lan -l_2  \, m_2 + 1
\ran}
\ee
as in \eq{funct-R} and the MHV amplitudes for negative-helicity gluons $l,s$ are defined as in \eq{n=4tree}:
\be
\label{n=4treeagain}
\cA_{\mathrm{tree}}^{\mathrm{MHV}}(l^-,s^-):=i(2\pi)^4\delta^{(4)}\left(\sum_ik_i\right)\frac{\langle l\,s\rangle^4}
{\prod_{r=1}^n\langle r\,\,r+1\rangle}\ .
\ee
Note that Equation~\eq{bdkcomplete} is a Feynman integral rather than a LIPS integral.

Now recall from \S \ref{mhvsection}
and \cite{Bern:zx,Bern:1994cg}
that the basis of integral functions at one-loop is known and the
Feynman integrals can be done to give explicit expressions (see \emph{e.g.} Appendix~I of \cite{Bern:1994cg}). The
Feynman integrals generated in \eq{bdkcomplete} (and for other channels) can then be compared with the Feynman integrals for the known
integral functions and the amplitude recreated. Since the integral functions are already known one
can reconstruct the amplitude in a purely algebraic manner. As a strong check of the final expression, the results can be compared with
the known behaviour (on general grounds) for the collinear ($p_a\, ,p_b\rightarrow p_a \parallel p_b$) and soft ($p_a\rightarrow 0$)
limits of such an amplitude.

For supersymmetric theories any terms which do not contain cuts are uniquely linked to the cut-containing terms
and thus the entire amplitude is reconstructed. In particular, the $\cN\!\!=4$ amplitudes discussed above can be completely
constructed in this way leading to \eq{fullampl}. In non-supersymmetric theories more information is needed to
get the rational (cut-free) terms and thus only the cut-constructible part may be obtained this way.

\section{Dispersion relations}

Imagine now that we stop at \eq{bddkdisp} and proceed to do the LIPS integral rather than uplift to
Feynman integrals. If we can actually do this integral we can calculate the discontinuity of the
amplitude directly. However, we would
really like to know the whole amplitude rather than just the imaginary part of it and the natural
question is whether it is possible to arrive at this from what we have so far. For a function with a branch cut,
it is in fact possible to reconstruct the real part from the
imaginary part and the relations which allow one to do this are known as dispersion relations (or sometimes
Kramers-Kronig relations).

By considering a function $\mathscr{A}(z)$ which is analytic in the complex plane with a branch cut
along the positive real axis starting at $x_0$, it is possible to show using complex analysis that
\begin{eqnarray}
\label{kk}
\Re[\mathscr{A}(x)]=\frac{1}{\pi}\,P\!\int_{x_0}^{\infty}\!\frac{dx'}{x'-x}\,\Im[\mathscr{A}(x')]+\frac{1}{2\pi i}I_{\infty}\ ,
\end{eqnarray}
where $x\in\mathbb{R}$ in the range\footnote{For purely massless theories, $x_0=0$.} $(x_0,\infty)$ and $P$ denotes the
Cauchy principal value prescription (\emph{i.e.} the value of the integral without consideration of the pole at $x'=x$).\footnote{
See \emph{e.g.} \cite{aw} for a fuller explanation of these ideas.}
$I_{\infty}$ is the contribution from the contour at
infinity which represents the ambiguity due to possible rational terms (\emph{i.e.}
terms which are cut-free functions of the kinematic invariants).

$I_{\infty}$ vanishes in any supersymmetric gauge
theory, and while these \emph{do} contain rational terms they are fixed uniquely by the
supersymmetry once one knows the cut-containing terms \cite{Bern:zx,Bern:1994cg}. Such theories are said to be cut-constructible (in
4 dimensions). Non-supersymmetric theories are not cut-constructible in 4 dimensions, but \emph{are}
 in $4-2\epsilon$ dimensions with $\epsilon\neq 0$ \cite{vanNeerven:1985xr,Bern:1995db,Bern:1996ja}.
While this is a powerful statement, it does mean that one has to consider the prospect of
using amplitudes with particles continued to $4-2\epsilon$ dimensions which are not simple.

In a sense, the one-loop CSW rules make BDDK's approach prescriptive for the MHV amplitudes. The imaginary part of the
amplitude is constructed as a phase space integral and then the dispersion integral over $P_{L;z}^2$ in \eq{bst} performs
\eq{kk} with $I_{\infty}$ absent. For supersymmetric theories this is sufficient to construct the full amplitude, while
in non-supersymmetric theories we must find other methods to calculate the rational part.

\chapter{Integrals for the $\cN\!=\!1$ Amplitude}
\label{neq1appendix}

In this appendix we give details of the integrals
needed to compute the discontinuities of the $\cN\!=\!1$ amplitude
discussed in Chapter \ref{chapter2}.

\section{Passarino-Veltman reduction}
\label{n=1appfirstsection}

In \S \ref{n=1mhvsection} we saw that a typical term in the
$\mathcal{N}=1$ amplitude is the dispersion integral
of the following phase space integral:
\begin{eqnarray}
\label{partintegraln1}
\cC (m_1 , m_2)  \ := \
\int\! d{\rm LIPS}(l_2, -l_1; P_{L;z}) \
\frac{{\tr}_{+}(\kslash_i \kslash_j
\kslash_{m_1} \lslash_1){\tr}_{+}(\kslash_{i}\kslash_{j}
\kslash_{m_2}\lslash_{2})}{(i\cdot j)^2(m_1\cdot l_1)
(m_2\cdot l_2)}
\ .
\end{eqnarray}
The full amplitude is then obtained by adding
the dispersion integrals of three more terms
similar to  \eqref{partintegraln1} but
with $m_1$ replaced by $m_1-1$ and/or
$m _2$ replaced by \mbox{$m_2+1$}.
The goal of this appendix is to perform the
Passarino-Veltman reduction \cite{pv} of
\eqref{partintegraln1}, which will lead us to
re-express $\cC (m_1 , m_2)$
in terms of cut-boxes, cut-triangles and cut-bubbles.

The explicit forms for the Dirac traces
involve Lorentz contractions over the various momenta,
so in a short-hand notation we can write these as
\beq
\label{lazy}
T(i,j,m_1)_{\mu}\, l_1^{\mu}
\ := \
{\tr}_{+}(\kslash_i \kslash_j
\kslash_{m_1} \lslash_1)
\ .
\eeq
$\cC (m_1 , m_2) $ can then be recast as
\beq
\label{partintegralii}
\cC (m_1 ,m_2)  \ = \
{T(i,j,m_1)_{\mu} \, T(i,j,m_2)_{\nu} \over (i\cdot j)^2}
\ \cI^{\m \n} (m_1, m_2, P_{L;z})
\ ,
\eeq
where%
\footnote{For the rest of this appendix we drop the subscript
$z$ in $P_{L;z}$ for the sake of brevity.}
\beq
\label{imunu}
\cI^{\mu\nu} (m_1, m_2, P_{L})
\ = \
\int\! d{\rm LIPS}(l_2,-l_1;P_{L})
\,
\frac{l_1^{\mu}\, l_2^{\nu}}{
(m_1\cdot l_1)(m_2\cdot l_2)}
\ .
\eeq
$\cI^{\mu\nu} (m_1, m_2, P_{L})$
contains  three independent momenta $m_1$, $m_2$ and  $P_L$.
On general grounds we can therefore decompose it as
\begin{eqnarray}
\label{ansatzn1}
\!\!\cI^{\mu\nu}\!\! &=&\!\! {\eta}^{\mu\nu}\, \cI_0
\ +\
m_1^{\mu}m_1^{\nu}\, \cI_1
\ + \
m_2^{\mu}m_2^{\nu}
\, \cI_2
\ + \
P_L^{\mu}P_L^{\nu}\, \cI_3\ + \
m_1^{\mu}m_2^{\nu}\cI_4
\nonumber\\ \cr
&+ &
m_2^{\mu}m_1^{\nu}\, \cI_5
\ + \ m_1^{\mu}P_L^{\nu}\, \cI_6
\ + \
P_L^{\mu}m_1^{\nu}\, \cI_7
\ + \ m_2^{\mu}P_L^{\nu}\, \cI_8
\ + \ P_L^{\mu}m_2^{\nu}\, \cI_{9}
\ ,
\end{eqnarray}
for some coefficients $\cI_i,\, i=0,\ldots ,9$. One can
then contract with different combinations of
the independent momenta in order to solve for the
$\cI_{i}$.
For instance, two of the integrals that
we will end up having to do are
${\eta}^{\mu\nu}\cI_{\mu\nu}$
and $m_1^{\mu}m_1^{\nu}\cI_{\mu\nu}$.
Using momentum conservation $l_2-l_1+P_L=0$
and the identity $a\cdot b = {(a+b)^2}/2=-{(a-b)^2}/2$
for $a$,$b$ massless momenta,
we can convert these integrals into
ones which have the general form
\beq
\label{genform}
{\tilde{\cI}}^{(a,b)}=\int \frac{d{\rm
LIPS} (l_2, -l_1; P_L)}{(l_1\cdot m_1)^{a}(l_2\cdot m_2)^{b}}
\ ,
\eeq
possibly with a kinematical-invariant coefficient,
and with $a$ and $b$
ranging over the values $1,0,-1$.
The results of these integrals are collected in
\S \ref{n=1intsection}.
As an example, we find that
\beq
\label{point}m_1^{\mu}m_1^{\nu}\cI_{\mu\nu}\ = \
\int\! d{\rm LIPS}(l_2 , -l_1 ; P_{L}) \,
\frac{(l_1\cdot m_1)}{(l_2\cdot m_2)}
\,  - \,
(m_1\cdot P_L)
\int\!
{d{\rm LIPS}(l_2 , -l_1 ; P_{L})  \over (l_2\cdot P_L)}
\ .
\eeq
Considering the values $(a,b)$, the case $(1,1)$
is a cut scalar box,
$(1,0)$ and $(0,1)$ are cut scalar triangles,
$(1,-1)$ and $(-1,1)$ are cut
vector triangles, whilst $(0,0)$ is a cut scalar bubble.

Because of the structure of
$T(i,j,m_1)_{\mu}$ and $T(i,j,m_2)_{\nu}$,
terms with coefficients such as
$T(i,j,m_1)_{\mu}T(i,j,m_2)_{\nu}{m_1}^{\mu}{m_2}^{\nu}$
are zero, and thus some of the $\cI_{i}$
do not contribute to the final answer.
The only contributing terms are found to be
$\cI_3$, $\cI_5$,  $\cI_7$ and $\cI_8$, and we find that
\begin{eqnarray}
\label{reducedintegral}
\cC (m_1 , m_2)  &=& \frac{{\tr}_{+}(\kslash_i
\kslash_j \kslash_{m_1} \Pslash_L)
{\tr}_{+}(\kslash_{i}\kslash_{j}\kslash_{m_2}
\Pslash_{L})}{(i\cdot j)^2}
\, \cI_3
\nonumber \\
&+ &
\frac{{\tr}_{+}(\kslash_i \kslash_j \kslash_{m_1}
\kslash_{m_2}){\tr}_{+}(\kslash_{i}\kslash_{j}
\kslash_{m_2}\kslash_{m_1})}{(i\cdot j)^2}
\, \cI_5
\nonumber \\
&+ & \frac{{\tr}_{+}(\kslash_i \kslash_j
\kslash_{m_1} \Pslash_L){\tr}_{+}(\kslash_{i}
\kslash_{j}\kslash_{m_2}\kslash_{m_1})}{(i\cdot j)^2}
\, \cI_7
\nonumber \\
&+ &
\frac{{\tr}_{+}(\kslash_i \kslash_j
\kslash_{m_1} \kslash_{m_2}){\tr}_{+}(\kslash_{i}
\kslash_{j}\kslash_{m_2}\Pslash_{L})}{(i\cdot j)^2}\,
\cI_8
\ .
\end{eqnarray}
The inversion of (\ref{ansatzn1}) in order to find
the coefficients
is tedious and somewhat lengthy, so we just
present the results for the relevant
$\cI_i$ in (\ref{reducedintegral}) above:
\begin{eqnarray}
\label{longeqn1}
\cI_3 &=& \frac{1}{N^2}
\left\{2(m_1\cdot m_2)P_L^2\,
{\tilde{\cI}}^{(0,0)}-N(m_1\cdot P_L)\,
{\tilde{\cI}}^{(1,0)}+N(m_2\cdot P_L)\,
{\tilde{\cI}}^{(0,1)}\right.{}
\nonumber \\
&+ & \left. 2(m_2\cdot P_L)^2\,
{\tilde{\cI}}^{(-1,1)}+2(m_1\cdot P_L)^2\,
{\tilde{\cI}}^{(1,-1)}\right\}
\ ,
\\ \cr
\label{longeqn2}
\cI_5 &=& \frac{1}{(m_1\cdot m_2)^2N^2}
\Bigg\{\bigg[
4(m_1\cdot P_L)^2(m_2\cdot P_L)^2\nonumber\\
&-&6(m_1\cdot P_L)
(m_2\cdot P_L)(m_1\cdot m_2)P_L^2+
3(m_1\cdot m_2)^2\left(P_L^2\right)^2\bigg]\,
{\tilde{\cI}}^{(0,0)}\nonumber\\
&+&\left[2(m_1\cdot P_L)^2(m_2\cdot P_L)-
\frac{3}{2}(m_1\cdot m_2)P_L^2\right] N(m_1\cdot P_L)\,
{\tilde{\cI}}^{(1,0)}
\nonumber \\
&-  &
\left[2(m_1\cdot P_L)^2(m_2\cdot P_L)-
\frac{3}{2} (m_1\cdot m_2)P_L^2\right]
N(m_2\cdot P_L)\,  {\tilde{\cI}}^{(0,1)}+\frac{N^3}{4}\,
{\tilde{\cI}}^{(1,1)}
\nonumber \\
&+ &
2\bigg[(m_1\cdot m_2)P_L^2-
(m_1\cdot P_L)(m_2\cdot P_L)\bigg]
(m_2\cdot P_L)^2\,  {\tilde{\cI}}^{(-1,1)}
\nonumber \\ &+ &
2\bigg[(m_1\cdot m_2)P_L^2- (m_1\cdot P_L)(m_2\cdot P_L)\bigg]
(m_1\cdot P_L)^2\, {\tilde{\cI}}^{(1,-1)}\Bigg\}
\ ,
\\ \cr
\label{longeqn3}
\cI_7 &=& \frac{1}{(m_1\cdot P_L)(m_1\cdot m_2)N^2}
\Bigg\{\bigg[2(m_1\cdot P_L)^2(m_2\cdot P_L)^2\nonumber\\
&-&
3(m_1\cdot P_L)(m_2\cdot P_L)(m_1\cdot m_2)P_L^2\bigg]
\nonumber \\
&\cdot &
{\tilde{\cI}}^{(0,0)}+
\frac{1}{2}(m_1\cdot m_2)P_L^2N(m_1\cdot P_L)\,
{\tilde{\cI}}^{(1,0)}-(m_1\cdot P_L)
N(m_2\cdot P_L)^2\, {\tilde{\cI}}^{(0,1)}
\nonumber \\
&- &
2(m_1\cdot P_L)(m_2\cdot P_L)^3\,
{\tilde{\cI}}^{(-1,1)}-(m_1\cdot m_2)P_L^2
(m_1\cdot P_L)^2\, {\tilde{\cI}}^{(1,-1)}\Bigg\}
\ ,
\\ \cr
\label{longeqn4}
\cI_8 &=&
\frac{1}{(m_2\cdot P_L)(m_1\cdot m_2)N^2}
\Bigg\{\bigg[2(m_1\cdot P_L)^2(m_2\cdot P_L)^2\nonumber\\
&-& 3(m_1\cdot P_L)(m_2\cdot P_L)(m_1\cdot m_2)P_L^2\bigg]
\nonumber \\ & \cdot &
{\tilde{\cI}}^{(0,0)}+
(m_2\cdot P_L)N(m_1\cdot P_L)^2\,
{\tilde{\cI}}^{(1,0)}-
\frac{1}{2}(m_1\cdot m_2)P_L^2N(m_2\cdot P_L)\,
{\tilde{\cI}}^{(0,1)}
\nonumber \\
&- & (m_1\cdot m_2)P_L^2(m_2\cdot P_L)^2\,
{\tilde{\cI}}^{(-1,1)}-2(m_1\cdot P_L)^3(m_2\cdot P_L)\,
{\tilde{\cI}}^{(1,-1)}\Bigg\}
\ ,
\end{eqnarray}
where
$N = (m_1\cdot m_2)P_L^2-2(m_1\cdot P_L)(m_2\cdot P_L)$.
The explicit expressions for the relevant
${\tilde{\cI}}^{(a,b)}$
are summarised in \S \ref{n=1intsection}.

Combining \eqref{reducedintegral} and
\eqref{longeqn1}-\eqref{longeqn4}
with the identity \eqref{remarkable}
and the explicit expressions for the integrals
$\tilde{\cI}^{(a,b)}$ in \S \ref{n=1intsection},
we arrive at the final result \eqref{partNeq1cut}.

\section{Box \& triangle discontinuities
from phase space integrals}
\label{n=1intsection}

The integrals that arise in the Passarino-Veltman reduction
in \S \ref{n=1appfirstsection}  have the general form:
\beq
\label{ab}
{\tilde{\cI}}^{(a,b)} =
\int \!\frac{d^{4-2\epsilon}{\rm LIPS} (l_2 , -l_1 ; P_{L;z})}
{(l_1\cdot m_1)^{a}(l_2\cdot m_2)^{b}}
\ ,
\eeq
where we have introduced dimensional regularisation in
dimension $D=4-2\epsilon$
\cite{'tHooft:fi} in order to deal with infrared
divergences.

There are six cases to deal with:
${\tilde{\cI}}^{(0,0)}$,
${\tilde{\cI}}^{(1,0)}$,
${\tilde{\cI}}^{(0,1)}$,
${\tilde{\cI}}^{(1,1)}$,
${\tilde{\cI}}^{(-1,1)}$,
${\tilde{\cI}}^{(1,-1)}$,
though due to symmetry we can transform
${\tilde{\cI}}^{(1,0)}$ into
${\tilde{\cI}}^{(0,1)}$, and
${\tilde{\cI}}^{(-1,1)}$ into
${\tilde{\cI}}^{(1,-1)}$,
so we only need consider four cases
overall.

Generically we will evaluate these integrals
in convenient special frames
following Appendix B of \cite{bst},
with a convenient choice for
$m_1$ and $m_2$. For instance, in the case of
${\tilde{I}}^{(1,1)}$ it is convenient
to transform to the centre
of mass frame of the vector $l_1-l_2$, so that
\beq
\label{comass}
l_1 \ = \ \frac{1}{2}P_{L;z} \bigl(  1 \, , \, {\vec{v}} \big)\ ,
\qquad l_2 \ = \   \frac{1}{2}P_{L;z} \bigl( -1 \, , \,  {\vec{v}}\big),
\eeq and write \beq \label{comass2}
{\vec{v}} \ = \
(\sin\theta_1\cos\theta_2\, , \, \sin\theta_1\sin\theta_2 \, , \,   \cos\theta_1)
\ .
\eeq
Using a further spatial rotation we write
\begin{equation}
\label{comass3}
   m_1 = (m_1,0,0,m_1) \ , \qquad
   m_2 = (A,B,0,C)
\ ,
\end{equation}
with the mass-shell condition $A^2 = B^2 + C^2$.

After integrating over all angular coordinates except $\theta_1$
and $\theta_2$, the two-body phase space measure in
$4-2\epsilon$ dimensions becomes (see Appendix \ref{DLIPSappendix})
\beq
d^{4-2\epsilon}{\rm LIPS}
(l_2, -l_1; P_{L;z}) \ = \
{ \pi^{ {1\over 2} - \epsilon} \over 4
\, \Gamma \big({1\over 2} - \epsilon\big) } \, \left| P_{L;z}
\over 2 \right|^{- 2 \epsilon} \ d\theta_1\, d\theta_2\, \,
(\sin\theta_1)^{1-2\epsilon} \, (\sin\theta_2)^{-2\epsilon}
\ .
\eeq
As a result of this and of our parametrizations of
$l_1,l_2,m_1$ and $m_2$, the integrals take the form
\beq
\label{wibble}
\tilde{\cI}^{(a,b)}\ = \
\Lambda^{(a,b)}{ \pi^{ {1\over 2} - \epsilon} \over 4 \,
\Gamma \big({1\over 2} - \epsilon\big) } \, \left| P_{L;z}
\over 2
\right|^{- 2 \epsilon}{\cal{J}}^{(a,b)}\ ,
\eeq
where
\begin{eqnarray}
\label{lambda0}
{\Lambda}^{(0,0)}&=&1
\ ,
\\ \nonumber
\label{lambda1}
{\Lambda}^{(1,0)} &=&
\frac{2}{P_{L;z}m_1}
\ ,
\\ \nonumber
\label{lambda2}
{\Lambda}^{(0,1)}
&=&-\frac{2}{P_{L;z}m_2}
\ ,
\\ \nonumber
\label{lambda3}
{\Lambda}^{(1,1)}&=&
-\frac{4}{P_{L;z}^2m_1}
\ ,
\\ \nonumber
\label{lambda4}
{\Lambda}^{(-1,1)}&=&-m_1
\ ,
\\ \nonumber
\label{lambda5}
{\Lambda}^{(1,-1)}&=&-m_2
\ ,
\end{eqnarray}
and
${\cal{J}}^{(a,b)}$ is the angular integral
\beq
\label{bigint}
{\cal{J}}^{(a,b)}\ := \
\int_{0}^{ \pi}\! d\theta_1 \int_{0}^{
\pi}d\theta_2\,\! \frac{(\sin\theta_1)^{1-2\epsilon}
(\sin\theta_2)^{-2\epsilon} }
 { (1 - \cos\theta_1)^a(A + C\cos\theta_1 +
B \sin\theta_1 \cos\theta_2)^b }
\ .
\eeq
The integrals (\ref{bigint}) have been evaluated in
\cite{vanNeerven:1985xr} for the values of $a$ and $b$ specified
above, and we borrow the results in a form from \cite{wim}:
\begin{eqnarray}
\label{jay1}
{\cal{J}}^{(0,0)}&=&
\frac{2\pi}{1-2\epsilon}
\ ,
\\ \nonumber
\label{jay2}
{\cal{J}}^{(1,0)}
&=&-\frac{\pi}{\epsilon}
\ ,
\\ \nonumber
\label{jay3}{\cal{J}}^{(1,1)}
&=&
-\frac{\pi}{\epsilon}\frac{1}{A}
\ {}_2F_1\left(1,1,1-\e,\frac{A-C}{2A}\right)
\ ,
\\ \nonumber
\label{jay4}
{\cal{J}}^{(-1,1)}
&=&
-\frac{2\pi(1-\epsilon)}
{\epsilon(1-2\epsilon)} \ {}_2F_1\left(-1,1,1-\e,
\frac{A-C}{2A}\right)
\ .
\end{eqnarray}
Here, $A$ and $C$ will differ depending on which case we are
considering and our particular parametrization for it,
but in all
cases the combinations that arise
can be re-expressed in terms of
Lorentz-invariant quantities using
suitable identities. In the
case of ${\cal{J}}^{(1,1)}$ for example,
one uses the easily verified
identities
\begin{equation}
\label{happy}
N(P_{L;z}) \ = \ -P_{L;z}^2(A+C)m_1\, ,
\quad m_1\cdot m_2 = m_1(A-C)
\ ,
\
\end{equation}
where $N(P_{L;z})$ was defined in \eqref{N}.

Eventually, after re-expressing $A$ and $C$ in this way,
and upon application of some standard hypergeometric identities
we find the following:
\begin{eqnarray}
\label{beebop1}
\l^{-1}\, {\tilde{\cI}}^{(0,0)}&=&\frac{2\pi}{1-2\epsilon}
\ ,
\\ \nonumber
\label{beebop2}
\l^{-1}\, {\tilde{\cI}}^{(1,0)}&=&
- \frac{1}{\epsilon}\, \frac{2\pi}{m_1\cdot
P_{L;z}}
\ ,
\\ \nonumber
\label{beebop3}
\l^{-1}\, {\tilde{\cI}}^{(0,1)}
&=&
\frac{1}{\epsilon}\, \frac{2\pi}{m_2\cdot
P_{L;z}}
\ ,
\\ \nonumber
\label{beebop4}
\l^{-1}\, {\tilde{\cI}}^{(1,1)}
&=&
-\frac{8\pi}{N(P_{L;z})}\left\{\frac{1}{\epsilon}+
\log{\left(1-\frac{(m_1\cdot
m_2)
P_{L;z}^2}{N(P_{L;z})}\right)}+
\mathcal{O}({\epsilon})\right\}
\ ,
\\ \nonumber
\label{beebop5}
\l^{-1}\, {\tilde{\cI}}^{(-1,1)}&=&\frac{\pi}{(m_1\cdot
P_{L;z})^2}\left\{-\frac{N(P_{L;z})}{\epsilon}\, \right.
\nonumber
\\ \nonumber
&+ &
\left. \frac{2}{1-2\epsilon}\bigl[(m_1\cdot P_{L;z})(m_2\cdot
P_{L;z})-(m_1\cdot m_2)P_{L;z}^2\bigr]\right\}
\ ,
\\ \nonumber
\label{beebop6}
\l^{-1}\, {\tilde{\cI}}^{(1,-1)}&=&\frac{\pi}{(m_2\cdot
P_{L;z})^2}\left\{-\frac{N(P_{L;z})}{\epsilon}
\right.
\nonumber \\
&+ &
\left.
\frac{2}{1-2\epsilon}
\bigl[ (m_1\cdot P_{L;z})(m_2\cdot
P_{L;z})-(m_1\cdot m_2)P_{L;z}^2\bigr]\right\},
\nonumber
\end{eqnarray}
where $\l$ is the ubiquitous factor
\beq
\label{ubin1}
\l \ := \
{ \pi^{ {1\over 2} -
\epsilon} \over 4 \, \Gamma \big({1\over 2} - \epsilon\big) } \,
\left| P_{L;z} \over 2 \right|^{- 2 \epsilon}
\ .
\eeq

\chapter{Gauge-invariant triangle reconstruction} \label{triangleappendix}

In this appendix we find a new representation of
the triangle function
\beq
\label{triangle-funct}
T(p, P, Q) \ = \
{\log (Q^{2} / P^{2}) \over Q^{2 }- P^{2}}
\ ,
\eeq
as the dispersion integral of a sum of two cut-triangles.\footnote{For
a review of dispersion relations see
\cite{bib} and Appendix \ref{unitarityappendix}.}

A comment on \mbox{gauge (in)dependence} is in order here.
Recall from \S \ref{offshellsection}, Equation \eqref{off}, that in
the approach of \cite{bst} to loop diagrams
one introduces an arbitrary null vector $\eta$
in order to perform loop integrations.
The corresponding gauge dependence should
disappear in the expression for scattering amplitudes.
In what follows we will work in  an arbitrary gauge,
and show analytically that gauge-dependent terms
disappear in the final result for the triangle function.
Perhaps unsurprisingly, this gauge invariance
will also hold for the finite-$\epsilon$ version
of $T(p, P, Q)$, which we define in
\eqref{epstrianglen1}.

\section{Gauge-invariant dispersion integrals}

To begin with, recall from \eqref{listofct}
that the basic quantity we have to compute reads
\beq
\label{crcr}
\mathscr{R}\ := \
\int\! {dz \over z} \left[
{(P_{z}^{2})^{-\epsilon}\over ( P_{z}p) } \, + \,
{(Q_{z}^{2})^{-\epsilon}\over (Q_{z}p)}
\right]
\ ,
\eeq
where $P+Q+p=0$.
We will work in an arbitrary gauge, where
\beq
P_{z}\ := \ P \, - \, z \eta \ ,
\qquad
Q_{z} \  := \ Q \, + z \eta
\ .
\eeq
A short calculation shows that
\beqa
P_{z}p & =  & Pp \Big[
1 - b_{P } (P^{2 } - P_{z}^{2})
\Big]
\ ,
\\
Q_{z}p & = & Qp \Big[
1 - b_{Q} (Q^{2} - Q_{z}^{2})
\Big]
 \ ,
 \eeqa
where
\beq
b_{P} \ := \
{\eta p \over 2 (\eta P) (pP)}
\ ,
\qquad
b_{Q} \ := \
{\eta p \over 2 (\eta Q)(pQ)}
 \ .
 \eeq
It is also useful to notice the relation
\beq
\label{bpbq}
{1\over b_{Q}}
\ = \
{1\over b_{P}} \, + \, Q^{2} -P^{2}
\ ,
\eeq
as well as $(Pp)=-(Qp)= (1/2)(Q^{2} - P^{2})$,
which trivially follows from momentum conservation.
We can then rewrite \eqref{crcr} as
\beq
\mathscr{R} \ = \ \mathscr{I}_{1  } \, - \, \mathscr{I}_{2}
\ ,
\eeq
where
\beqa
\mathscr{I}_{1} & :=  &
{1\over (Pp)}
\int\! ds' \, (s')^{-\epsilon }\
{1\over (s'-P^{2}) \big[
1-b_{P} (P^{2 } - s')
\big]}
\\ \nonumber
&=&
{\pi \csc  (\pi \epsilon) \over (Pp)}
\left[
(-P^{2})^{-\epsilon} \, - \,
\left(
{-b_{P}\over b_{P}P^{2} -1}
\right)^{\epsilon}
\right]
\ ,
\eeqa
\beqa
\mathscr{I}_{2} &:= &
{1\over (Pp)}
\int\! ds' \, (s')^{-\epsilon }\
{1\over (s'-Q^{2}) \big[
1-b_{Q} (Q^{2 } - s')
\big]}
\\ \nonumber
&=&
{\pi \csc  (\pi \epsilon) \over (Pp)}
\left[
(-Q^{2})^{-\epsilon} \, - \,
\left(
{-b_{Q}\over b_{Q}Q^{2} -1}
\right)^{\epsilon}
\right]
\ .
\eeqa
But \eqref{bpbq} implies
\beq
{-b_{P}\over b_{P}P^{2} -1} \ = \
{-b_{Q}\over b_{Q}Q^{2} -1}
\ ,
\eeq
so that we can finally recast \eqref{crcr} as:
\beq
\label{finaltriang}
\mathscr{R} \ = \
2\bigl[\pi \epsilon \csc (\pi \epsilon)\bigr]
\,
{1\over \epsilon}
{(-P^{2})^{{-\epsilon}} \, - \,
(-Q^{2})^{-\epsilon}
 \over Q^{2} - P^{2}}
\ = \
2 \bigl[\pi \epsilon \csc (\pi \epsilon) \bigr]
\, T_{\epsilon}(p, P, Q)
\ ,
\eeq
where the $\epsilon$-dependent triangle function
is%
\footnote{The $\epsilon$-dependent triangle function
already appeared in \eqref{epstrianglen1}.}
\beq
\label{epstriangleagain}
T_{\epsilon} (p, P, Q) \ := \
{1\over \epsilon}
{ (-P^2)^{-\epsilon} - (-Q^2)^{-\epsilon} \over Q^2 - P^2}
\ .
\eeq
This is the result we were after.
Notice that  all the gauge dependence,
i.e.~any dependence on the arbitrary
null vector $\eta$, has completely cancelled out
in \eqref{finaltriang}.

We now discuss the $\epsilon \to 0$ limit of the
final expression \eqref{finaltriang}.
As already discussed in \S \ref{n=1reviewsection}
(see \eqref{T1n1} and \eqref{T2n1}), in studying the
$\epsilon \to 0$ limit of $\mathscr{R}$  (and hence of
$ T_{\epsilon}(p, P, Q)$)
we need to distinguish the case where $P^2$ and $Q^2$
are both nonvanishing from the case where one of the two,
say $Q^2$, vanishes.
In the former case, we get precisely the triangle function
$T(p,P,Q)$ defined in \eqref{triangle-funct}:
\beq
\lim_{\epsilon \to  0} \mathscr{R} \ = \
 2 \, T(p, P, Q)\, , \qquad
P^2 \neq 0 \, , \, Q^2 \neq 0
\ .
\eeq
In the latter case, where $Q^2 =0$, we have instead
\beq
\lim_{\epsilon \to  0} \mathscr{R} \ = \
-{2\over \epsilon} \,
{(-P^2)^{-\epsilon}  \over P^2}
\, , \qquad
P^2 \neq 0 \, , \, Q^2 = 0
\ ,
\eeq
which corresponds to a degenerate triangle.

The final issue is that of the gauge invariance of the
contributions to the amplitude from the box functions $B$
(this is also relevant to the issue
of gauge invariance
in the $\cN=4$ calculation of \cite{bst},
and in that paper a
general argument for gauge invariance was also given - further evidence
can be found in \cite{Brandhuber:2005kd}).
We expect that an explicit analytic proof of the
gauge invariance of the box function contribution
to the amplitude
 could be constructed using identities such as those in
Appendix B of \cite{bst}.
In the meantime, numerical tests have shown that
gauge invariance is present \cite{Lance}.
Indeed, it would be surprising if this
were not the case given that
the correct, gauge-invariant, amplitudes
are derived with the choices
of gauge we have made here and in \cite{bst}.
We have also carried out the MHV diagram analysis of this paper using
the alternative gauge choice $\eta = k_{m_2}$; one obtains \eqref{Neq1}.


\chapter{Integrals for the non-supersymmetric amplitude}
\label{nonsusyappendix}

In this appendix we give details of the integrals
needed to compute the discontinuities of the non-supersymmetric amplitude
discussed in Chapter \ref{chapter3}.

\section{Passarino-Veltman reduction}

In \S \ref{generalcasesection} we saw that a typical term in the
cut-constructible part of the Yang-Mills
amplitude is the dispersion integral
of the following phase space integral:
\begin{eqnarray}
\label{partintegral}
\mathscr{C} (m)   :=
\int\!\! d{\rm LIPS}(l_2, -l_1; P_{L;z})
\frac{
 {\tr}_{+}(\kslash_{1} \kslash_{2}
\Pslash_{L;z}  \lslash_{2})
 {\tr}_{+}(\kslash_{1} \kslash_{2}
 \lslash_{2}  \Pslash_{L;z})
 {\tr}_{+}(\kslash_1  \kslash_2\,
\kslash_{m}  \lslash_2)
}
{(l_2\cdot m) \, (k_1\cdot k_2)^3 \, (P_{L;z}^2)^2
}\
\end{eqnarray}
The goal of this appendix is to perform the
Passarino-Veltman reduction \cite{pv} of
\eqref{partintegral}.
To this end, we rewrite $\mathscr{C} (m ) $ as
\beq
\label{partintegralii}
\mathscr{C} (m)  \ = \
{
{\tr}_{+}(\kslash_{1}\, \kslash_{2}\,
\Pslash_{L;z} \, \gamma_{\mu} )
\ {\tr}_{+}(\kslash_{1}\, \kslash_{2}
\, \gamma_\nu \, \Pslash_{L;z})
\
{\tr}_{+}(\kslash_1 \, \kslash_2\,
\kslash_{m} \, \gamma_{\rho})
\over
 (k_1\cdot k_2)^3 (P_{L;z}^2)^2 \,
}
\ \cI^{\m \n \r} (m , P_{L;z})
\ ,
\eeq
where%
\footnote{For the rest of this appendix we will generally drop the subscript
$z$ in $P_{L;z}$ for the sake of brevity.}
\beq
\label{imunu}
\cI^{\m \n \r} (m, P_{L})
\ = \
\int\! d{\rm LIPS}(l_2,-l_1;P_{L})
\,
\frac{l_2^{\mu}\, l_2^{\nu} \, l_2^{\r} }{
(l_2\cdot m)}
\ .
\eeq
On general grounds, $\cI^{\mu\nu\r} (m, P_{L})$
can be decomposed as
\begin{eqnarray}
\label{ansatz}
\!\!\!\!\!\!\cI^{\mu\nu\r}\! &=&\!
m^{\mu}m^{\nu}m^{\r} \, \mathscr{J}_1 \ + \
(m^{\mu}m^{\nu}P_{L}^{\r}
\, + \,
m^{\mu}P_L^{\nu}m^{\r} \, + \,
P_L^{\mu}m^{\nu}m^{\r}
) \, \mathscr{J}_2
\nonumber\\ \cr
&+ &
(m^{\mu}P_L^{\n}P_L^{\r}
\, + \,
P_L^{\mu}m^{\nu}P_L^{\r}
+
P_L^{\mu}P_L^{\nu}m^{\r} )
\, \mathscr{J}_3
\ + \
P_L^{\mu}P_L^{\nu}P_L^{\r}\, \mathscr{J}_4
\nonumber\\ \cr
&+ &
( \eta^{\m \n} m^{\r}
\, + \, \eta^{\m \r} m^{\n} \, + \, \eta^{\n \r} m^\m)\,
\mathscr{J}_5
 +
(\eta^{\m \n} P_L^{\r}
\, + \, \eta^{\m \r} P_L^{\n} \, + \, \eta^{\n \r} P_L^\m)
\mathscr{J}_6\ \
\end{eqnarray}
for some coefficients $\mathscr{J}_i,\, i=0,\ldots,6$. One can
then contract with different combinations of
the independent momenta in order to solve for the
$\mathscr{J}_{i}$. Introducing the quantities
\beqa
\label{pvstuff}
A & := &  \ m_\m m_\n m_\r \, \cI^{\m \n \r}\ ,
\cr
B & := &  \ m_\m m_\n P_{L\r} \, \cI^{\m \n \r}\ ,
\cr
C & := &  \ m_\m P_{L\n} P_{L\r} \, \cI^{\m \n \r}\ ,
\cr
D & := &  \ P_{L\m} P_{L\n} P_{L\r} \, \cI^{\m \n \r}\ ,
\cr
E & := &  \ \eta_{\m \n} m_{\r} \, \cI^{\m \n \r}\ \ = \ 0
\ ,
\cr
F & := &  \ \eta_{\m \n} P_{L\r} \, \cI^{\m \n \r}\ = \ 0 \ ,
\eeqa
the result for the Passarino-Veltman reduction of
$\{\mathscr{J}_{1},\ldots ,  \mathscr{J}_{6}\}$
in the basis $\{A, \ldots , D \}$ is:
\beqa
\label{basis}
\mathscr{J}_2 & = & \left( 5 (P_L^2)^2 /
\big( 2 (m \cdot P_L)^5\big)\, , \,
-6 P_L^2 / (m \cdot P_L)^4\,  , \,
3  / (m \cdot P_L)^3 \, ,\,  0\right)
\ ,
\cr
 \mathscr{J}_3 & = & \left( -2 P_L^2 /  (m \cdot P_L)^4\, , \,
3/ (m \cdot P_L)^3 \, ,\,  0\, ,\,  0\right)
\ ,
\cr
\mathscr{J}_4& = & \left( 1 / (m \cdot P_L)^3\, , \,
0 \, ,\,  0\, , \, 0\right)
\cr
\mathscr{J}_5 & = & \left( - (P_L^2)^2 /  \big( 2 (m \cdot P_L)^4\big)
\, ,
3 P_L^2 / \big( 2(m \cdot P_L)^3\big) \, , \,
-1  / (m \cdot P_L)^2\, ,\,  0\right)
\ ,
\cr
\mathscr{J}_6& = & \left( P_L^2 / \big(2 (m \cdot P_L)^3)\, ,
-1/(m \cdot P_L)^2 \, , 0\, , 0\right)
\ .
\eeqa
We omit the decomposition for $\mathscr{J}_1$
as the corresponding term in
\eqref{ansatz} drops out of all future expressions due to
$k_m^2 =0$.

Finally, using the methods of \cite{bbst} and the results of \S \ref{n0phaseintsection},
the integrals in (\ref{pvstuff}) are found to be,
keeping only terms to $\mathcal{O}(\epsilon^0)$,
\beqa
\label{explicit}
A \ &=& \ (m\cdot P_L)^2\, {4 \over 3} \pi \hat{\lambda} \ ,
\\ [2pt]
\cr
B \ &=& \ P_L^2(m\cdot P_L)
\,
\pi \hat{\lambda} \, ,\\ [2pt]
\cr
C \ &=& \ (P_L^2)^2\,
\pi \hat{\lambda} \, ,\\ [2pt]
\cr
D \ &=& \ -\frac{(P_L^2)^3}{8(m\cdot P_L)}\, \frac{4\pi}{\epsilon}\,
\hat{\lambda} \, ,
\eeqa
where
\beq
\label{ubiquitous}
\hat{\l} \ :=  \frac { \pi^{ \frac{1}{2} - \epsilon }
 } {  4^{1-\e} \, \Gamma \big(\frac{1}{2} - \epsilon\big) }
\ .
\eeq
%


\section{Evaluating the integral of ${\mathscr C}(a,b)$}
\label{cabappsection}


The basic expression which arises in
the MHV diagram construction in this paper is
\beqa
\label{basicintegrand}
{\mathscr C}(a,b) = \frac{\langle i \, l_1 \rangle \,
\langle j \, l_1 \rangle^2 \,
\langle i \, l_2 \rangle^2 \,
\langle j \, l_2 \rangle}{\langle i \, j \rangle^4 \,
\langle l_1 \, l_2 \rangle^2}
\frac{\langle i \, a \rangle \, \langle j \, b \rangle}
{\langle l_1 \, a \rangle \, \langle l_2 \, b \rangle}.
\eeqa
We wish to integrate this expression
over the Lorentz-invariant phase space. We begin by simplifying
it, using multiple applications of the Schouten identity.
First note that using this identity twice, one deduces that
\begin{eqnarray}
\label{basicmanip} \frac{\lrangle{i}{l_2} \, \lrangle {j}{l_1} }
                     {\lrangle{l_1}{a}\,\lrangle{l_2}{b} }
\lrangle{a}{b}^2
     &=& \lrangle{i}{a}\lrangle{b}{j} +\lrangle{i}{a}\lrangle{a}{j}
     \frac{\lrangle{l_1}{b}}{\lrangle{a}{l_1}} +\lrangle{b}{j}
\lrangle{i}{b}
     \frac{\lrangle{l_2}{a}}{\lrangle{b}{l_2}}
\\ \nonumber \cr
     &+&  \lrangle{a}{j}\lrangle{i}{b}
     -\lrangle{a}{j}\lrangle{i}{b}
\frac{\lrangle{l_1}{l_2}\lrangle{a}{b}}{\lrangle{a}{l_1}
     \lrangle{b}{l_2}}
\ .
\end{eqnarray}
Now use this identity in ${\mathscr C}(a,b)$.
This generates five terms, which we will label
(in correspondence with the ordering arising from the order of
terms in \eqref{basicmanip} above)
as $T_i,\, i=1,\dots,4$, and $U$.
The $T_i$ have dependence on the loop momenta such that
we may use the phase space integrals
of \S \ref{n0phaseintsection} to calculate them. The term $U$
is more complicated; however,
one may again use the identity \eqref{basicmanip},
generating another five terms,
which we will label $T_5,\dots,T_8$, and $V$.
Again, the expressions in $T_i,\, i=5,\ldots,8$ may be calculated using
the integrals of \S \ref{n0phaseintsection}.
Finally, the term $V$ may be simplified,
here using the identity \eqref{basicmanip}
with $i$ and $j$ interchanged.
This generates a further five terms, which we label $T_9,\dots,T_{13}$.
The explicit forms of these terms follow:
\begin{eqnarray}
\label{allterms1}
 T_1& =&
\frac{\ijba^2\ijlolt\ijltlo}{2^{8}(i\cdot j)^4
(a\cdot b)^2(l_1\cdot l_2)^2} \ ,
\\  [2pt]\cr
\label{T1}
 T_2& =&
\frac{\ijab\ijba\ijlolt\ijltlo\ibloa}{2^{10}(i\cdot j)^4
(a\cdot b)^2(l_1\cdot l_2)^2(i\cdot b)(a\cdot l_1)} \ ,
\\ [2pt]\cr
\label{T2}
 T_3& =&
\frac{\ijab\ijba\ijlolt\ijltlo\jaltb}{2^{10}(i\cdot j)^4
(a\cdot b)^2(l_1\cdot l_2)^2(j\cdot a)(b\cdot l_2)} \ ,
\\ [2pt]\cr
\label{T3}
T_4& =& -\frac{\ijab\ijba\ijlolt\ijltlo}{2^{8}(i\cdot j)^4
(a\cdot b)^2(l_1\cdot l_2)^2} \ ,
\end{eqnarray}
and
\begin{eqnarray}
\label{allterms2}
T_5& =& \frac{\ijba^2\ijab\ijltlo}{2^{8}(i\cdot j)^4(a\cdot b)^3
(l_1\cdot l_2)}\ ,
\\ [2pt]\cr
\label{T5}
T_6& =& \frac{\ijab^2\ijba\ijltlo\ibloa}
{2^{10}(i\cdot j)^4(a\cdot b)^3(l_1\cdot l_2)(i\cdot b)
(a\cdot l_1)}
\ ,
\\ [2pt]\cr
\label{T6}
T_7& =& - \frac{\ijab\ijba^2\ijltlo\ialtb}{2^{10}(i\cdot j)^4
(a\cdot b)^3(l_1\cdot l_2)(i\cdot a)(b\cdot l_2)}
\ ,
\\ [2pt]\cr
\label{T7}
T_8& =& -\frac{\ijab^2\ijba\ijltlo}{2^{8}(i\cdot j)^4
(a\cdot b)^3(l_1\cdot l_2)}
\ ,
\end{eqnarray}
and
\begin{eqnarray}
\label{allterms3}
T_{9}& =& \frac{\ijab^3\ijba}{2^{8}(i\cdot j)^4(a\cdot b)^4}\ ,
\\ [2pt]\cr
\label{T9}
T_{10}& =& \frac{\ijab^2\ijba^2
\jbloa}{2^{10}(i\cdot j)^4
(a\cdot b)^4(j\cdot b)(a\cdot l_1)}
\ ,
\\ [2pt]\cr
\label{T10}
T_{11}& =&  \frac{\ijab^2\ijba^2\ialtb}{2^{10}(i\cdot j)^4
(a\cdot b)^4(i\cdot a)(b\cdot l_2)}
\ ,
\\ [2pt]\cr
\label{T11}
T_{12}& =& -\frac{\ijab^2\ijba^2}{2^{8}(i\cdot j)^4(a\cdot b)^4}
\ ,
\\ [2pt]\cr
\label{T12}
T_{13}& =& \frac{\ijba^2\ijab^2\bltloa}{2^{10}(i\cdot j)^4
(a\cdot b)^4(a\cdot l_1)(b\cdot l_2)}
\ ,
\end{eqnarray}
The expression ${\mathscr C}(a,b)$ is
then the sum of the terms $T_i, i=1,\dots,13$.

Before performing the phase space integrals,
it proves convenient to
%
%
collect the resulting expressions in pairs as $T_1+T_2$,
$T_3+T_4$, $T_5+T_6$,
$T_7+T_8$,  $T_9+T_{11}$ and
$T_{10}+T_{12}$.
This leads us to the following decomposition:
\beqa
-\mathscr{C} (a, b) & = &
{ \ijlolt \ijltlo\ijloa\ijblt \over
2^8 (i \cdot j)^4 (l_1 \cdot l_2 )^2 (l_1 \cdot a)
(l_2 \cdot b) }
\nonumber \\ \cr
&=&
{1 \over 2^8 (i \cdot j)^4} \,
(\cH_1 \, + \, \cdots \, + \cH_4 )
\ ,
\eeqa
where
 \beqa \nonumber
\!\!\!\!\!\!\!\!\cH_1 &:=&
{ \ijba  \ijlolt \ijltlo \over (l_1 \cdot l_2)^2 \, (a \cdot b)}
\bigg[
{\ijloa \over (l_1 \cdot a)} -
{\ijltb \over (l_2 \cdot b)}
\bigg]
\ ,
\\ [5pt]\nonumber \cr
\!\!\!\!\!\!\!\! \cH_2 &:=&
{ \ijab \ijba  \ijltlo \over (l_1 \cdot l_2) \, (a \cdot b)^2}
\bigg[
{\ijloa \over (l_1 \cdot a)} -
{\ijltb \over (l_2 \cdot b)}
\bigg]
\ ,
\\ [5pt]\nonumber \cr
\!\!\!\!\!\!\!\! \cH_3 &:=&
- { ( \ijab )^2 \ijba  \over  (a \cdot b)^3 }
\bigg[
{\ijloa \over (l_1 \cdot a)} -
{\ijltb \over (l_2 \cdot b)}
\bigg]
\ ,
\\ [5pt] \cr
\!\!\!\!\!\!\!\! \cH_4 &:=&
 { ( \ijab )^2 (\ijba )^2 \loablt  \over  4 (a \cdot b)^4
\, (l_1 \cdot a) \, (l_2 \cdot b) }
\ .
\eeqa

Finally, we perform the phase space integrals of
the above expressions, using the formul{\ae}  in
\S \ref{n0phaseintsection} below. One quickly finds
that the divergent (as $\epsilon\rightarrow 0$) part
of the total expression is zero.
The finite part, after further spinor
manipulations, becomes the expression
we have given in \eqref{pspaceresult}.


\section{Phase space integrals}
\label{n0phaseintsection}


The basic method which we use for evaluating
Lorentz-invariant phase space integrals has been
outlined in \cite{bst, bbst} and also discussed in
\S \ref{mhvsection} and Appendix \ref{neq1appendix}.
Here we will just quote the results which we need.
In the following we will use a shorthand notation where
$\int \equiv
\int \!d^{4-2\epsilon}{\rm LIPS} (l_2 , -l_1 ; P_{L;z})$,
and a common factor of $ 4 \pi \hat{\l} ( -P_{L;z})^2 $
is understood to multiply all expressions, where
$\hat{\l}$ is the ubiquitous factor of \eq{ubiquitous}.
%
%
We also define
$\alpha = (a\cdot P)$, $\beta=(b\cdot P)$,
$N(P) = (a\cdot b)P^2 -2(a\cdot P)(b\cdot P)$
and drop the $L;z$ subscripts
from $P_{L;z}$ for clarity.

Firstly we quote the results from Appendix B of
\cite{bbst} up to terms of
$\mathcal{O}(\epsilon^0)$:
\begin{eqnarray}
\label{beebop1}
\int 1 &=& 1
\ , \qquad
\int \frac{1}{(a\cdot l_1)}= - \frac{1}{\epsilon\alpha}\ ,  \qquad
\int \frac{1}{(b\cdot l_2)} = \frac{1}{\epsilon\beta}
\ ,
\\ [6pt]\nonumber
\label{beebop4}
&&\int \frac{1}{(a\cdot l_1)(b\cdot l_2)}
=
-\frac{4}{N(P)}\left(\frac{1}{\epsilon}+ L\right)\ ,
\nonumber
\end{eqnarray}
where
$$
L = \log\left( 1 - \frac{(a\cdot b)}{N}P^2 \right).
$$
%
From this, we can recursively derive the following
integrals (up to $\cO (\e^0)$):
\begin{eqnarray}
\label{beebop5}
\int l_1^\mu &=&  \frac{1}{2}P^\mu, \qquad
\int l_2^\mu \ = \
- \,  \frac{1}{2}P^\mu
\ ,
\\ [4pt]\nonumber
\label{beebop6}
\int l_1^\mu l_1^\nu &=&
\int l_2^\mu l_2^\nu \ =\   \frac{1}{3}
\left( P^\mu P^\nu \, - \,
 \frac{1}{4}\eta^{\mu\nu}P^2\right)
\ ,
\\ [4pt]\nonumber
\label{beebop8}
\int \frac{l_1^\mu}{(a\cdot l_1)}
&=&
-\, \frac{P^2}{2\epsilon\alpha^2}a^\mu \, + \,
\frac{1}{\alpha}P^\mu \, - \, \frac{P^2}{\alpha^2}a^\mu
\ ,
\\ [4pt]\nonumber
\label{beebop9}
\int \frac{l_2^\mu}{(b\cdot l_2)}
&=&
-\, \frac{P^2}{2\epsilon\beta^2}b^\mu \, +
\, \frac{1}{\beta}P^\mu \, - \,
\frac{P^2}{\beta^2}b^\mu
\ ,
\nonumber
\end{eqnarray}
and
\begin{eqnarray}
\label{beebop10}
\!\!\!\!\!\!\!\!\int\!\! \frac{l_1^\mu l_1^\nu}{(a\cdot l_1)}\! &=&\!
-\frac{P^4}{4\epsilon\alpha^3}a^\mu a^\nu
+\frac{1}{2\alpha}P^\mu P^\nu +
\frac{P^2}{2\alpha^2} P^{(\mu}a^{\nu)}
-\frac{3P^4}{4\alpha^3}a^\mu a^\nu -
\frac{P^2}{4\alpha}\eta^{\mu\nu}
\, ,
\nonumber\\ [4pt]
\label{beebop11}
\int\!\! \frac{l_2^\mu l_2^\nu}{(b\cdot l_2)}\! &=&\!
\frac{P^4}{4\epsilon\beta^3}b^\mu b^\nu
-\frac{1}{2\beta}P^\mu P^\nu -
\frac{P^2}{2\beta^2} P^{(\mu}b^{\nu)}
+\frac{3P^4}{4\beta^3}b^\mu b^\nu +
\frac{P^2}{4\beta}\eta^{\mu\nu}
\, ,
\\ [4pt]\nonumber
\label{beebop12}
\int\!\! \frac{l_2^\mu }{(a\cdot l_1)(b\cdot l_2)}\! &=&\!
\frac{1}{\epsilon N}\left(
2P^\mu -\frac{P^2}{\alpha}a^\mu +
\frac{P^2}{\beta}b^\mu \right)
+ \frac{2L}{N} \left( P^\mu - \frac{\beta}{(a\cdot b)}a^\mu +
\frac{\alpha}{(a\cdot b)}b^\mu \right)
\, .
\nonumber
\end{eqnarray}
Finally, there are integrals involving
cubic powers of loop momenta in the numerator.
The first is
\begin{eqnarray}
\label{cubic2}
\!\!\!\!\!\!\!\!\!\!\!\!\!\!\int
\frac{l_1^\mu l_1^\nu l_1^\rho}{(a\cdot l_1)}\ =\
 \frac{P^4}{4\alpha^3} P^{(\mu}a^\nu a^{\rho)}  &+&
\frac{P^2}{4\alpha^2} P^{(\mu}P^\nu a^{\rho)}
 + \frac{1}{3\alpha} P^\mu P^\nu P^{\rho}\nonumber \\ &-&
\frac{P^4}{8\alpha^2} \eta^{(\mu\nu} a^{\rho)}
\ \ \  - \frac{P^2}{4\alpha} \eta^{(\mu\nu} P^{\rho)}
\, ,
\end{eqnarray}
where we have suppressed terms cubic in $a$
as they prove not to contribute
when this integral
is contracted into the products of
Dirac traces which appear in the expressions in
\S \ref{cabappsection}.
The second cubic integral required is
\begin{eqnarray}
\label{cubic1}
\!\!\!\!\!\!\!\!\!\!\!\!\!\!\int
\frac{l_2^\mu l_2^\nu l_2^\rho}{(b\cdot l_2)}\ =\
 \frac{P^4}{4\b^3} P^{(\mu}b^\nu b^{\rho)}  &+&
\frac{P^2}{4\b^2} P^{(\mu}P^\nu b^{\rho)}
 + \frac{1}{3\b} P^\mu P^\nu P^{\rho}\nonumber \\ &-&
\frac{P^4}{8\b^2} \eta^{(\mu\nu} b^{\rho)}
\ \ \ - \frac{P^2}{4\b} \eta^{(\mu\nu} P^{\rho)}
\, ,
\end{eqnarray}
again suppressing  terms cubic in $b$ which will not contribute.

\chapter{KLT relations}
\label{kltappendix}
 For completeness, in this
appendix we write  the field theory limit of the KLT relations
\cite{klt} for the cases of four, five and six points: \beqa
\label{3} \cM (1,2,3) & = & -i\cA (1,2,3)\, \cA (1,2,3) \ ,
\\  \cr
\cM (1,2,3,4) & = & -is_{12}\ \cA (1,2,3,4)\cA(1,2,4,3) \ ,
\label{4} \eeqa \beqa \label{5} \cM (1,2,3,4,5)  & = &
is_{12}s_{34}\ \cA (1,2,3,4,5)\cA(2,1,4,3,5)
\nonumber \\
&+&i s_{13}s_{24}\ \cA(1,3,2,4,5)\cA(3,1,4,2,5) \ ,
\\ \cr
\label{6} \cM (1,2,3,4,5,6)& = & -is_{12}s_{45}\ \cA(1,2,3,4,5,6)
\big[ s_{35}\cA(2,1,5,3,4,6)
\nonumber \\
 & + & (s_{34}+s_{35})\ \cA (2,1,5,4,3,6)\big]
\\ \nonumber &+ & \cP (2,3,4) \ .
\eeqa In these formul{\ae}, $\cM(i)$ ($\cA(i)$) denotes a tree-level
gravity (Yang-Mills colour-ordered) amplitude, $s_{ij} : = (p_i +
p_j)^2$, and $\cP (2,3,4)$ stands for permutations of $(2,3,4)$.
The relation for a generic number of particles can be found
in Appendix A of \cite{Bern:1998sv}.

\end{appendix}

\bibliographystyle{JHEP}
\bibliography{thesis}
\end{document}